\documentclass{article}
\usepackage{sao2,psfig}
\issue{1998, 46, 62--96}
\begin{document}
\markboth{Trushkin}{Radio spectra of complete sample of Galactic supernova remnants}
\title{Radio spectra of complete sample of galactic supernova remnants}
\author{Trushkin S.A.}
\institute{\saoname}
\date{November 6, 1998}{November 25, 1998}
\maketitle

\begin{abstract}
We present radio continuum spectra for $200$ Galactic supernova
remnants (SNRs) from 220 known and included in Green's (1998) catalog.
The spectra can be plotted only for 200 SNRs because
about 20 remaining new and weak SNRs
(Whiteoak and Green, 1996; Gray, 1994a) have only one--frequency flux density
measurements.
Spectrum plotting is an ``on--line'' procedure of the CATS database
(Verkhodanov  et al., 1997) created for some other multi-frequency catalogs.
These spectra include most of the  measurements available in literature,
as well as multi-frequency measurements of nearly $120$ SNRs with the RATAN-600
radio telescope in 1, 2 and 4 Galactic quadrants and from the Galactic plane
survey at 960 and 3900 MHz (Trushkin, 1986, 1988, 1989, 1996, 1998).
The measurements have been placed on the same absolute
flux density scale of Baars (1977)
 as in the paper by  Kassim (1989a),
using the correcting factor from the compiled catalog (Kuhr et al., 1981).
The presented compilation has given a possibility of plotting quite accurate
spectra with the thermal plasma free--free absorption in fitting the spectra
accounted for.
\end{abstract}

\section{Introduction}
The non-thermal radio spectrum  is a key property distinguishing SNRs
from extended Galactic plane radio sources.
The current catalog of SNRs (Green, 1998, see also Green, 1984, 1988, 1991, 1996)
contains 220 Galactic SNRs and some dozens of possible or probable ones,
named SNR candidates here.
While it contains information about the spectral index and flux densities at 1 GHz,
there is no complete collection of the
available data on flux density measurements,
scattered over hundreds of publications and catalogs
(for example: Kesteven, 1968; Shaver and Goss, 1970a,b; Green, 1974; Milne and Dickel, 1974, 1975; Green et al., 1975; Angerhofer et al., 1977;
Slee, 1977; Dubner et al., 1993, 1996; Grey, 1994a,b).
Reviews of
radio spectra of a few dozen SNRs are given in the early papers by Shaver and
Goss (1970a,b), Willis (1973), Dickel and  DeNoyer (1975),
the recent papers by Kassim (1989a,b), Kovalenko et al. (1994a, 1994b) and the papers
on Galactic plane sources (Reich et al., 1986, 1988; F\"{u}rst et al., 1987, 1989)
based on the 11\,cm and 21\,cm Effelsberg Galactic plane surveys
 (Reich et al., 1990a, 1990b).

The results  of new all-sky surveys  with a resolution of about $1'$ are presently
available, which comprise  much more data on extended sources in the
Galactic plane.

We used our measurements of the flux densities of nearly 120 SNRs
(Trushkin, 1986; Trushkin et al., 1988; Trushkin, 1996a,b, 1997).
Thus the flux densities of half of the SNRs  were measured at several
(up to six) frequencies, which corrected and complemented essentially the spectra of
many SNRs.

Here we attempt to collect all these data in a single file and make them
accessible in the CATS database (Verkhodanov et al., 1997).
The spectrum plotting procedure of the SNRs with option of formula of
approximation is designed to simplify the
statistical investigation of radio properties of these SNRs by providing
easy and public access to the available data.

While there are no significant correlations of global SNR parameters
with the  spectral indices (Caswell and Clark, 1975; Lerche, 1980),
Weiler (1983) proposed an extremely important classification which divided
the SNRs into groups:
shells, plerions, and composite SNRs. Probably the SNRs are intimately
related with the basic  type classification of supernova: SNIa, SNIb and SNII.
The  SNIb and SNII are the birthplace of neutron stars and black holes and thus
lead to production of  plerions or, in dense ISM, composite SNRs with
appearance of filled-center and shell structure in the X-rays and radio emission,
respectively.
On the other hand, the classical shells are created by SNIa, as is the case with
the historical SN~1604 Tycho.

Besides the morphology differences in radio emission, these three classes
of SNRs have different mean spectral indices ($S_\nu \propto \nu^{\alpha}$)
(Weiler, 1983, 1985).

Accurate spectra are very important for the classification ``shell/plerion'',
recognition of the mechanism of generation of relativistic particles,
search for possible high- or low-frequency turnovers of spectra.

Based on the historical  shell SNR data, Green (see Jones et al., 1997)
pointed out that ``there is an apparent trend
in synchrotron spectral index with remnant age''. However from the data
presented
it is evident  that there are only 11 SNRs having an angular size less than
$10'$ from a total of 36 objects with a steep spectrum: $\alpha_4 <-0.65$.
Thus it is unlikely that compact and young SNRs have steeper spectra than
extended and ``adult'' ones.

It is worth noting that the alternative point of view of  Glushak (1997)
concerning the evolution of spectral index and existence of correlation
between spectral index and surface brightness (``$\alpha-\Sigma$'' plane).
However, as long as the significance of such a collection remains unproved,
it is untimely to draw conclusions concerning the physical evolution of SNRs
on its basis.

{\small
\begin{table*}
\caption{Fragment of the spectral catalog of the Galactic SNRs}
\begin{tabular}{clrlllll}
\hline
  Name     &$\nu$(MHz) &S$_\nu(Jy)$&$\Delta${S$_\nu(Jy)$}&F$_{cor}$&Telescope&Type& Bibcode\\
\hline
\hline
G009.8+0.6 & ~960.0 &4.10&    &    &RATAN    &S &{\tt 1996BSAO...41...64Trushkin}    \\
G009.8+0.6 & 1465.0 &3.50&0.40&    &VLA      &  &{\tt 1993AJ....105.2251Dubner+}     \\
G009.8+0.6 & 2695.0 &1.00&0.40&0.99&GB 43m   &  &{\tt 1970A\&AS....1..319Altenhoff+} \\
G009.8+0.6 & 2700.0 &1.80&    &1.03&Parkes   &  &{\tt 1970AuJPA..13....3Goss \& Day} \\
G009.8+0.6 & 3900.0 &1.90&    &    &RATAN    &  &{\tt 1996BSAO...41...64Trushkin}    \\
G009.8+0.6 & 7700.0 &1.20&    &    &RATAN    &  &{\tt 1996BSAO...41...64Trushkin}    \\
G010.0$-$0.3&~~57.5*&$<$3.00&  &1.00&CLRO     &Fx&{\tt1988ApJS...68..715Kassim}      \\
G010.0$-$0.3&~~83.0 &6.0 &3.  &    &BSA,DKR  &  &{\tt 1994AR.....38...95Kovalenko+ }  \\
\hline
\end{tabular}
\end{table*}
}

\section{Flux density measurement data}

The catalog of the flux measurements is a file $SNR\_spectra.dat$
(nearly 2300 entries), which is a base for plotting
the spectra of 200 SNRs and candidates.
The ``on-line'' spectrum plotting is accessible from the homepage of
the CATS database: {\tt \verb*Chttp://cats.sao.ru/C} and its local mirror
{\tt \verb*Chttp://www.ratan.sao.ru/~catsC}.
The catalog consists of columns:
name (as in Green's catalog), frequency (MHz),
flux density (Jy), flux error (Jy), correcting factor, radio telescope,
SNR type (+indicator of compact object) and SIMBAD/NED bibcode
(YYYYJJJJJ.VVV.PPPPAuthor).
A fragment of the spectral data catalog is shown in Table 1.

In the second column the symbol `*' is used to indicate that these data are
not used for the spectrum fitting. Signs `$<$' and `$>$' are used for the upper
or lower limits and are not used for the spectrum fitting either.
If a flux point has no error, we adopt it to be equal to 10\% of the total flux
density in order to set the relevant weight factor for each point of measurements.

In the current version of the catalog the SNRs type indicator was added:
F --- filled-center or plerion, S --- shell and C --- composite type.
There are indicators of active neutron stars (NS) for 16 known pairs SNR +
NS from the paper by Frail (1998): PSR --- pulsar; SGR --- repeater
gamma-burster;  QNS --- radio-quiet NS; XRB --- X-ray
binary system; SXP --- slow X-ray pulsar. Also  ``PSR?'' shows
the possible pairs SNR+NS from Kaspi (1998). ``XPSR'' indicates
slow X-ray pulsars in G27.4+0.0 and G29.7-0.3 according to Gotthelf (1998).

For Vela XYZ SNR complex (G327.7+14.6) we give two different spectra:
for the whole SNR and for the Vela\,X.

Here we can not give a complete list of references (see CATS procedure of
spectrum plotting or original SNR catalog by D. Green (1998)).
In the  References we give some most relevant references.

For every SNR spectrum in the CATS database procedure a file GLLL.L$\pm$B.B.fit
will be created where the bibcodes of the references  are presented.

We give the spectra of the original calibrators, SNR Cas A (G111.7+1.2) for
the epoch 1965.0 and the Crab Nebula (G189.1+3.0), from the table data collected in
the paper by Baars et al. (1977).

We used our measurements of the flux densities of about 120 SNRs
(Trushkin, 1986, 1996a,b, 1997; Trushkin et al., 1988).
We added to the list of SNRs four first detected SNRs: G3.2$-$5.2, G11.2$-$1.1,
G16.0+2.7 and G356.2+4.4, and also the SNR candidates G4.7+1.3, G4.8+1.2,
G4.8+6.2, G9.7$-$0.1 and G85.2$-$1.2  from Duncan et al. (1997)
and Taylor et al. (1992).

The complete list of 350 RATAN-600 multi-frequency measurements is shown
in Table\,2. There are  new data of 1997--98 in the list.

\section{Spectrum fitting}

The thermal absorption of the foreground  describes well the spectra of
most SNRs:
$S_\nu = [S_{408}(\nu/408)^\alpha]\,{\tt\,exp}[-\tau_{408}(\nu/408)^{-2.1}] $,
(Kassim, 1989a).
Kassim (1989b) used the low-frequency data to derive spectra for 32 SNRs,
and  their turnovers at low frequencies ($<100$\,MHz) to indicate the presence
of extended ionized medium along the line of sight.

For fitting we used, as a rule, an approximation formula,
$y = A + Bx + C\,{\tt exp}(Dx)$, or a simple linear case,  $y = A + Bx$,
where $x=\log\nu$, $y=\log{S_\nu}$, $\nu$ --- frequency (MHz), $S_\nu$
--- flux density (Jy), $D=\pm1$ or $D=-2.1$. Clearly the latter case is
adequate to real thermal absorption at low frequencies and steady in fitting
the spectra with a few points.
The inverse squares of relative flux errors,
$({\Delta\,S_{\nu}/S_{\nu})^{-2}}$, are used as formal weights of
their flux points. Often we used the option ``without errors'' for the
spectrum plotting, when the scattering of points on the spectrum is larger than
the values of errors.

It is worth noting that a user can try any optional methods of approximation
``on-line'' in the procedure of the CATS database. The next version of
this procedure will include the interactive removal of the discrepant flux points
on the spectrum.

For fitting the curve we give two spectral indices at 0.4 and 4 GHz if these
frequencies are within the fitting range.
The first one is closer to $\tau=1$ since the spectra have maxima
near 80--200 MHz.
The second one is the spectral index of an optically thin spectrum and is not
influenced by propagation conditions, but there is a high-frequency turnover
of spectra because of the properties of synchrotron radiation.
Now the plot of curved spectra contains the frequency of flux
maximum, $\nu_{max}$.

\begin{figure*}
\centerline{\vbox{\psfig{figure=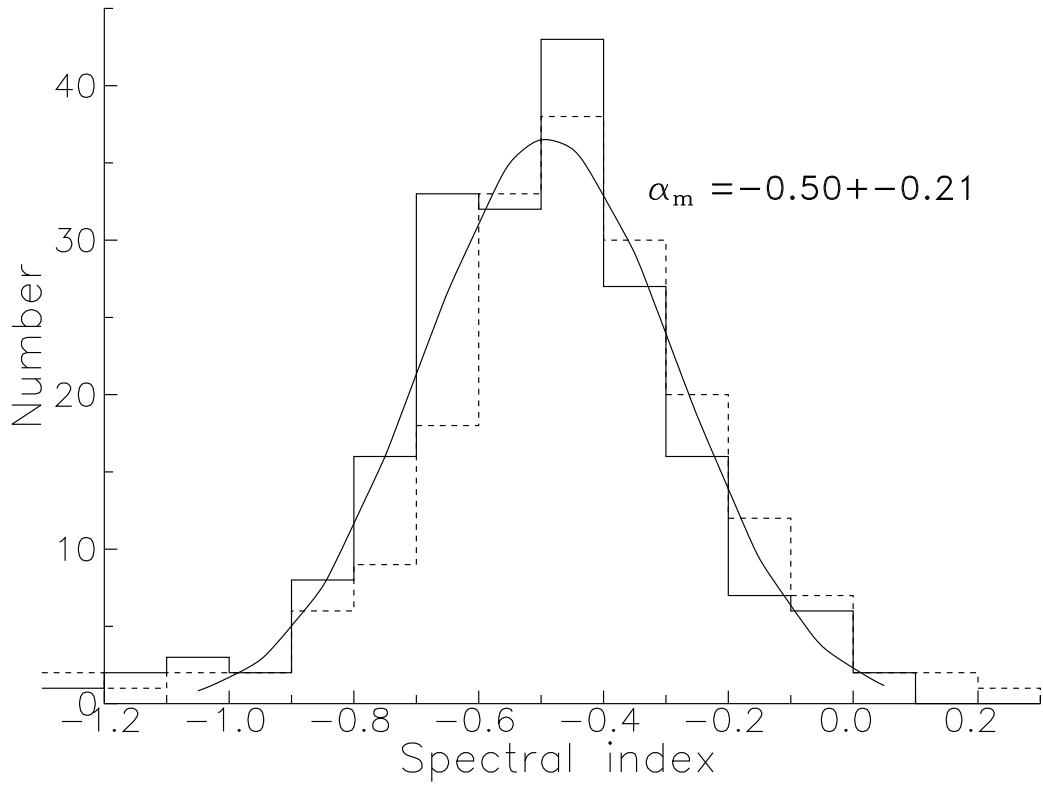,width=14cm,angle=-90}}}
\caption{
Distribution of $\alpha_{4}$ (solid line)
 and $\alpha_{0.4}$ (dashed line) for 192 SNRs
}
\label{snr_a}
\end{figure*}

\section{The atlas of spectra of 200 SNRs }
The spectra of 192 SNRs, included in Green's catalog,
and the spectra of eight new SNRs or SNR candidates,
G3.5$-$5.2, G4.7+1.3, G4.8+6.2, G9.7$-$0.1, G11.2$-$1.1, G16.0+2.7,
 G85.2$-$1.2 and
G356.2+4.4, are presented at the end of the paper.
The figures are ordered by Galactic longitude from SNR names in two
 columns on a page.

The labels under the name denote the type of approximation curve:
``lin'' --- $a+b\cdot\,x$; ``ther'' --- ${\rm lin}+{\tt exp}(-2.1x)$;
``exp$\pm$'' --- ${\rm lin}+{\tt exp}(\pm{x})$, where $x={\rm log}(\nu)$.

These spectra include  most flux density measurements from literature
and our 350 measurements of flux densities of 120 SNRs with the RATAN-600
radio telescope in 1, 2 and 4 Galactic quadrants and from
the Galactic plane survey at 0.96 and 3.9 GHz (Trushkin 1988, 1996a),
which are collected in Table\,2.

\section{ Analysis of spectra }

The sample mean spectral indices at 0.4 and 4.0 GHz are
$\alpha_{0.4}=-0.41\pm0.34$  and $\alpha_{4}=-0.50\pm0.21$.
The distributions of the low- and high-frequency spectral indices
($\alpha_{0.4}$ and $\alpha_{4}$, respectively) for 192 SNRs are shown
in Fig.\,\ref{snr_a}.
For comparison  we present the best gaussian fit of the distribution with
a dispersion $\sigma = 0.3$. Although the number of SNRs with the index
in the interval $-0.8 < \alpha_4 < -0.6$ with respect to $\alpha_{0.4}$
is slightly in excess, we could not find
significant difference in the distributions of both spectral indices.

As Lerche (1980) has shown,  Bell's (1978a,b) mechanism (as a variant
of the Fermi acceleration of the first order) of repowering
is attractive because it provides a simple explanation of the observed
spread of spectral indices: from acceleration whether on a strong adiabatic
shock wave with a compression factor $\chi=4$ or on a strong  isothermal
shock with $\chi=2.4$. Thus the spectral index of relativistic
electrons, $\gamma = \frac{2+\chi}{\chi-1}$, varies from 2 to 3.1. Therefore
the spectral indices may vary from $-0.5$ to $-1.1$ if the equation of
state in SNR is described by strong adiabatic or isothermal shocks.

We did not find considerable correlation between spectral index and
Galactic coordinates $l$ and $b$ of SNRs. (Because the distance $d$  estimates are
very uncertain, we did not search for any correlation with $d$  and
distance from Galactic plane $z$).

\begin{figure*}
\centerline{\vbox{\psfig{figure=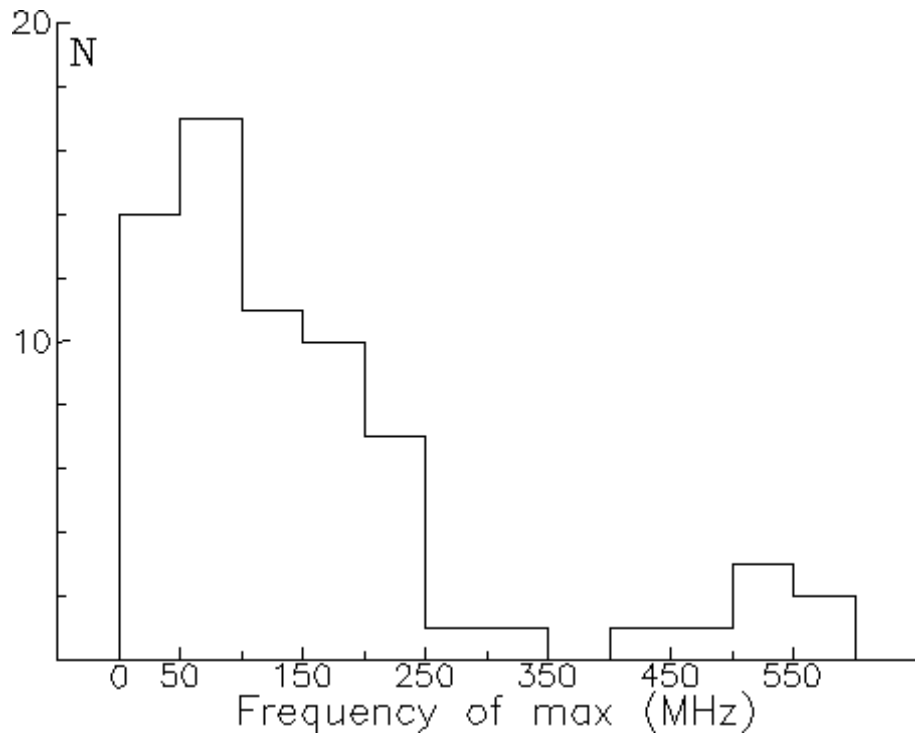,width=12.0cm,angle=0}}}
\caption{
Distribution of flux maximum frequency for 70 SNRs
}
\label{freq}
\end{figure*}

An analysis of 190 spectra showed that 70 SNRs (37\,\%)
have clear low-frequency turnover caused, apparently, by  absorption in the thermal
foreground of the Milky Way. Fig.\,\ref{freq} shows the distribution of the
maximum flux frequency $\nu_{max}$ for these SNRs.

These frequencies  $\nu_{max}$ do not correlate with the Galactic
coordinates. It should be  noted that in the region $l=67-70^\circ$
all four SNRs have a relatively high $\nu_{max}=500-700$\,MHz, while
most of the SNRs have $\nu_{max}=50-150$\,MHz. Only the SNRs in the
direction of the Galactic
center have  $\nu_{max}>250$\,MHz.

It is worth  noting that from 16 SNRs with active neutron stars (Frail, 1998),
the  spectra of 15 ones have no low-frequency turnover at $>20-50$\,MHz
which is logically associated with the contribution of neutron stars (or pulsars)
inside SNRs. Pulsars as a rule have  steeper spectra than  SNRs,
thus their contribution will be higher at low frequencies.

The catalog of SNR spectra has ten cases of clear turn-up at low frequencies.
It is interesting that five (50\,\%!) such SNRs contain  radio pulsars
(Kaspi, 1998). Of course,  it is probable that there is discrepancy of flux scales in
low- and high-frequency measurements of flux densities.
But in further  observations we should  pay special attention to the sample of
SNRs without  turnover or even with turn-up at low frequencies for
search for pulsars or active stellar supernova remnants.

The Crab-like SNR G74.1$-$1.2 is a single radio source with a clear turnover at
high frequencies. Only  G41.1$-$0.3 and G180.0$-$1.7 showed  similar turnovers.

There are few data for SNR with large angular size at frequencies
higher than 10\,GHz in our catalog. Only in these SNRs with big ages such
turnovers
are due to synchrotron losses.

\section{Other resources on SNRs in the CATS database}

In addition we note that besides the procedure of SNR spectrum plotting
``on-line'' from the WEB homepage of the CATS database:
{\tt \verb*Chttp://cats.sao.ru/C} or
{\tt \verb*Chttp://www.ratan.sao.ru/~catsC}
the atlas of the RATAN drift scans of 80 SNRs from
1, 2 and 4 Galactic quadrants and the compiled catalog of X-ray and radio maps
of more than 80\,\% Galactic SNRs are accessible.

The atlas gives the strip-distributions of radio brightness for  SNRs
in observations with a ``knife'' antenna beam of the RATAN-600 radio telescope,
realized at low elevations.
Compact sources can  easily be detected on such drift scans.
These sources are of special interest because they
could be stellar SNRs. A comparison of such distributions of radio brightness
at different frequencies can also be made.

The catalog of extended SNRs in different ranges allows us to have a detailed
comparison with the observational data obtained at the RATAN-600. The
catalog contains more than 900 maps of nearly 220 SNRs in the X-ray, optical,
and radio ranges. The catalog can be useful in new observational programs,
and also for solving  statistical problems.

\section{Conclusions}

We present radio continuum spectra for 192 Galactic supernova
remnants (SNRs) from 220 known and included in Green's (1998) catalog.
We added eight SNR candidates detected in the Galactic survey carried out
with the RATAN-600 radio telescope (Trushkin, 1996) and in the
investigations of 1997--1998.
The catalog contains about 2200 flux density measurements.
The spectra can be plotted only for 200 SNRs because
about 20 other new and weak SNRs
(Whiteoak and Green, 1996; Gray, 1994a) have only one--frequency flux density
measurements.

The procedure of spectrum plotting based on this catalog is ``on-line'' in
the CATS data base (Verkhodanov et al., 1997) developed for six multi-frequency
catalogs. The spectra are temporarily stored in GIF-files which are shown
in the figures.

These spectra include  most  flux density measurements from literature
and our measurements of  flux densities of nearly 120 SNRs with the RATAN-600
radio telescope in 1, 2, and 4 Galactic quadrants and from
the Galactic plane survey at  0.96 and  3.9 GHz (Trushkin 1988, 1996).

Where it was possible, the flux measurements were reduced to the common
flux scale of Baars (1977), as it was done in the work of Kassim (1989a).
The correcting coefficients from the compiled catalog of Kuhr et al. (1981)
were used.

The presented compiled catalog of flux density measurements enables one
to plot spectra of SNRs with allowance made for their turnover at low
frequencies due to thermal absorption in the Galaxy.

\begin{acknowledgements}
The author is grateful to his colleagues O.V. Verkhodanov, V.N. Chernenkov
and H. Andernach who created the CATS database and to the Russian Foundation for
Basic Research for financial support of the CATS project
(grant No 96-07-89075).
\end{acknowledgements}

\vspace*{10cm}

\newpage
\begin{onecolumn}
\begin{center}
\topcaption{RATAN--600 measurements of the Galactic SNRs}
\tabletail{\hline}
\tablefirsthead{\hline
 Name    &$\nu$(MHz)&S$_\nu$(Jy)&$\Delta{S}(Jy)$& Ref &  Name   &$\nu$(MHz) & S$_\nu$(Jy)&$\Delta{S}$(Jy)& Ref\\
~~~~~1~~~~& ~~~2~~~ &~~3~~   & ~~~4~~~   &  5  & ~~~~~1~~~~& ~~~2~~~ &~~3~~   & ~~~4~~~   &  5 \\
\hline }
\tablehead{%
\hline
~~~~~1~~~~& ~~~2~~~ &~~3~~   & ~~~4~~~   &  5 & ~~~~~1~~~~& ~~~2~~~ &~~3~~   & ~~~4~~~   &  5 \\ \hline }
\begin{supertabular}{lrrrc|lrrrc}
G000.0+0.0   &  960 & 330.00&  10.0~&4 &G010.0$-$0.3 & 3900 &   0.45&   0.1~&3 \\
G000.0+0.0   & 2300 & 250.00&   5.0~&4 &G010.0$-$0.3 &11200 &   0.37&   0.1~&3 \\
G000.0+0.0   & 3900 & 216.00&   5.0~&4 &G011.2$-$0.3 &  960 &  17.00&   2.0~&3 \\
G000.0+0.0   &11200 & 115.00&  10.0~&4 &G011.2$-$0.3 & 3650 &  11.0~&   1.0~&3 \\
G000.9+0.1   & 3900 &   9.30&   0.9~&3 &G011.2$-$0.3 & 3900 &   9.5~&   0.9~&3 \\
G000.9+0.1   & 7700 &   6.30&   0.5~&3 &G011.2$-$1.1 &  960 &   9.0~&   0.2~&3 \\
G001.4$-$0.1 & 3900 &   2.50&   0.5~&3 &G011.2$-$1.1 & 2300 &   7.0~&   0.5~&3 \\
G001.9+0.3   & 3900 &   0.45&   0.1~&3 &G011.2$-$1.1 & 3900 &   5.0~&   0.2~&3 \\
G001.9+0.3   &11200 &   0.15&   0.05&3 &G011.2$-$1.1 & 7700 &   3.5~&   0.4~&3 \\
G003.2$-$5.2 &  960 &   4.00&   1.0~&3 &G011.4$-$0.1 &  960 &   4.20&   0.7~&3 \\
G003.2$-$5.2 & 3900 &   2.00&   0.5~&3 &G011.4$-$0.1 & 3900 &   1.70&   0.2~&3 \\
G003.7$-$0.2 &  960 &   2.40&   0.2~&4 &G012.0$-$0.1 &  960 &   2.60&   0.3~&3 \\
G003.7$-$0.2 & 3900 &   0.90&   0.2~&4 &G012.0$-$0.1 & 3900 &   0.40&   0.05&3 \\
G004.2$-$3.5 &  960 &   4.70&   0.3~&3 &G012.0$-$0.1 &11200 &   0.15&   0.03&3 \\
G004.2$-$3.5 & 3900 &   2.45&   0.15&3 &G012.2$-$1.1 &  960 &   2.5~&   0.25&4 \\
G004.5+6.8   &  960 &  21.00&   0.5~&3 &G012.2$-$1.1 & 3900 &   1.6~&   0.15&4 \\
G004.5+6.8   & 2300 &  12.30&   0.3~&3 &G013.5+0.2   &  960 &   2.00&   0.3~&3 \\
G004.5+6.8   & 3900 &   8.30&   0.2~&3 &G013.5+0.2   & 2300 &   0.70&   0.1~&3 \\
G004.7+1.3   &  960 &   2.50&   0.3~&4 &G013.5+0.2   & 3900 &   0.46&   0.05&3 \\
G004.7+1.3   & 3900 &   1.75&   0.15&4 &G015.1$-$1.6 &  960 &   4.80&   0.5~&3 \\
G004.8+6.2   &  960 &   3.75&   0.3~&4 &G015.1$-$1.6 & 3900 &   3.90&   0.3~&3 \\
G004.8+6.2   & 3900 &   3.1~&   0.5~&4 &G015.9+0.2   &  960 &   4.00&   0.4~&3 \\
G005.2$-$2.6 &  960 &   1.80&   0.2~&4 &G015.9+0.2   & 3900 &   1.90&   0.2~&3 \\
G005.2$-$2.6 & 3900 &   2.15&   0.15&4 &G015.9+0.2   &11200 &   1.05&   0.05&3 \\
G005.4$-$1.2 &  960 &  33.0~&   2.0~&3 &G016.0+2.7   &  960 &   1.85&   0.15&4 \\
G005.4$-$1.2 & 2300 &  23.0~&   1.5~&3 &G016.0+2.7   & 3900 &   1.10&   0.1~&4 \\
G005.4$-$1.2 & 3900 &  21.0~&   1.5~&3 &G016.7+0.1   &  960 &   4.80&   0.5~&3 \\
G005.9+3.1   &  960 &   4.0~&   0.4~&3 &G016.7+0.1   & 3900 &   2.10&   0.2~&3 \\
G005.9+3.1   & 3900 &   1.3~&   0.2~&3 &G016.7+0.1   &11200 &   1.05&   0.1~&3 \\
G006.1+1.2   &  960 &   4.8~&   0.4~&3 &G016.8$-$1.1 &  960 &  15.00&   1.5~&3 \\
G006.1+1.2   & 3900 &   1.45&   0.20&3 &G016.8$-$1.1 & 3900 &   7.00&   0.5~&3 \\
G006.4+4.0   &  960 &   1.80&   0.3~&3 &G016.8$-$1.1 &11200 &   3.00&   0.25&3 \\
G006.4+4.0   & 3900 &   1.20&   0.15&3 &G017.4$-$2.3 &  960 &   5.40&   0.5~&3 \\
G006.4$-$0.1 &  960 & 300.0~&  15.0~&3 &G017.4$-$2.3 & 3900 &   2.00&   0.2~&3 \\
G006.4$-$0.1 & 3900 & 170.00&  10.0~&3 &G017.8$-$2.6 &  960 &   2.20&   0.25&3 \\
G007.7$-$3.7 &  960 &  11.80&   1.2~&3 &G017.8$-$2.6 & 3900 &   2.10&   0.2~&3 \\
G007.7$-$3.7 & 3900 &   6.70&   0.7~&3 &G018.8+0.3   &  960 &  28.00&   0.3~&3 \\
G008.7$-$0.1 &  960 &  95.00&   8.0~&3 &G018.8+0.3   & 3650 &  18.00&   0.3~&3 \\
G008.7$-$0.1 & 3900 &  63.00&   4.0~&3 &G018.8+0.3   & 3900 &  18.90&   0.3~&3 \\
G008.7$-$5.0 &  960 &   8.90&   0.8~&3 &G018.9$-$1.1 &  960 &  34.00&   3.0~&3 \\
G008.7$-$5.0 & 3900 &   4.40&   0.4~&3 &G018.9$-$1.1 & 3650 &  22.00&   2.0~&3 \\
G009.7$-$0.1 &  960 &   5.70&   0.6~&3 &G018.9$-$1.1 & 3900 &  18.90&   1.8~&3 \\
G009.7$-$0.1 & 2300 &   2.40&   0.25&3 &G018.9$-$1.1 & 7700 &  17.00&   1.5~&3 \\
G009.7$-$0.1 & 3900 &   1.70&   0.15&3 &G020.0$-$0.2 &  960 &   6.90&   0.7~&3 \\
G009.8+0.6   &  960 &   4.10&   0.4~&3 &G020.0$-$0.2 & 3900 &   7.10&   0.7~&3 \\
G009.8+0.6   & 3900 &   1.90&   0.2~&3 &G021.5$-$0.9 &  960 &   5.00&   0.4~&3 \\
G009.8+0.6   & 7700 &   1.20&   0.15&3 &G021.5$-$0.9 & 2300 &   6.00&   0.3~&3 \\
G010.0$-$0.3 &  960 &   0.94&   0.2~&3 &G021.5$-$0.9 & 3900 &   6.40&   0.3~&3 \\
G021.5$-$0.9 & 7700 &   6.50&   0.3~&3 &G039.2$-$0.3 & 7700 &   9.40&   0.9~&3 \\
G021.5$-$0.9 &11200 &   6.25&   0.3~&3 &G039.7$-$2.0 &  960 &  65.00&   4.0~&3 \\
G021.8$-$0.6 &  960 &  63.00&   6.0~&3 &G039.7$-$2.0 & 2300 &  44.00&   4.0~&3 \\
G021.8$-$0.6 & 3650 &  30.00&   3.0~&3 &G039.7$-$2.0 & 3900 &  32.00&   3.0~&3 \\
G021.8$-$0.6 & 3900 &  28.00&   3.0~&3 &G040.5$-$0.5 &  960 &  10.00&   1.0~&3 \\
G023.6+0.3   &  960 &   7.20&   0.7~&3 &G040.5$-$0.5 & 3900 &   8.60&   0.8~&3 \\
G023.6+0.3   & 2300 &   4.60&   0.4~&3 &G041.1$-$0.3 &  960 &   8.50&   0.8~&3 \\
G023.6+0.3   & 3900 &   3.30&   0.3~&3 &G041.1$-$0.3 & 3900 &   6.70&   0.7~&3 \\
G023.6+0.3   & 7700 &   2.30&   0.3~&3 &G041.1$-$0.3 &11200 &   2.66&   0.3~&3 \\
G024.7+0.6   &  960 &   6.50&   0.7~&3 &G042.8+0.6   &  960 &   4.70&   0.5~&3 \\
G024.7+0.6   & 2300 &   7.00&   0.7~&3 &G042.8+0.6   & 3900 &   1.44&   1.4~&3 \\
G024.7+0.6   & 3900 &   7.00&   0.7~&3 &G043.3$-$0.2 &  960 &  37.0~&   3.5~&3 \\
G024.7$-$0.6 &  960 &   7.50&   0.8~&3 &G043.3$-$0.2 & 3900 &  18.8~&   2.0~&3 \\
G024.7$-$0.6 & 3900 &   2.60&   0.25&3 &G043.3$-$0.2 &11200 &  10.40&   1.4~&3 \\
G027.4+0.0   &  960 &   4.00&   0.4~&3 &G043.9+1.6   &  960 &   4.50&   0.5~&3 \\
G027.4+0.0   & 3900 &   1.65&   0.15&3 &G043.9+1.6   & 3900 &   1.90&   0.2~&3 \\
G027.8+0.6   &  960 &  32.60&   3.3~&3 &G045.7$-$0.4 &  960 &  12.60&   1.3~&3 \\
G027.8+0.6   & 2300 &  25.40&   2.5~&3 &G045.7$-$0.4 & 3900 &   4.00&   0.3~&3 \\
G027.8+0.6   & 3900 &  20.00&   2.0~&3 &G045.7$-$0.4 &11200 &   1.50&   0.2~&3 \\
G029.7$-$0.3 &  960 &   9.80&   0.9~&3 &G046.8$-$0.3 &  960 &  15.80&   1.6~&3 \\
G029.7$-$0.3 & 2300 &   3.70&   0.4~&3 &G046.8$-$0.3 & 3900 &   9.00&   0.9~&3 \\
G029.7$-$0.3 & 3900 &   1.60&   0.3~&3 &G046.8$-$0.3 &11200 &   5.30&   0.5~&3 \\
G030.7+1.0   &  960 &   5.80&   0.6~&3 &G049.2$-$0.7 &  960 & 174.0~&  10.0~&3 \\
G030.7+1.0   & 3900 &   2.84&   0.3~&3 &G049.2$-$0.7 & 3900 & 124.0~&   8.0~&3 \\
G030.7$-$2.0 &  960 &   1.50&   0.2~&3 &G049.2$-$0.7 &11200 &  33.0~&   3.0~&3 \\
G030.7$-$2.0 & 3900 &   0.45&   0.15&3 &G053.6$-$2.2 &  960 &  10.80&   1.1~&3 \\
G031.9+0.0   &  960 &  26.00&   2.5~&3 &G053.6$-$2.2 & 3900 &   4.10&   0.4~&3 \\
G031.9+0.0   & 3900 &  11.30&   1.1~&3 &G054.1+0.3   &  960 &   0.40&   0.1~&3 \\
G031.9+0.0   &11200 &   5.60&   0.6~&3 &G054.1+0.3   & 3900 &   0.45&   0.1~&3 \\
G032.8$-$0.1 &  960 &   5.60&   0.6~&3 &G054.1+0.3   &11200 &   0.43&   0.1~&3 \\
G032.8$-$0.1 & 3900 &   5.50&   0.5~&3 &G055.7+3.4   & 3900 &   0.75&   0.15&3 \\
G033.2$-$0.6 &  960 &   2.00&   0.3~&3 &G057.2+0.8   &  960 &   1.80&   0.2~&3 \\
G033.6+0.1   &  960 &  15.00&   1.5~&3 &G057.2+0.8   & 2300 &   0.90&   0.1~&3 \\
G033.6+0.1   & 2300 &   8.50&   0.8~&3 &G057.2+0.8   & 3900 &   0.80&   0.05&3 \\
G033.6+0.1   & 3900 &   6.50&   0.7~&3 &G059.5+0.1   &  960 &   2.70&   0.3~&3 \\
G033.6$-$0.6 &  960 &  14.00&   1.4~&3 &G059.5+0.1   & 2300 &   1.75&   0.2~&3 \\
G033.6$-$0.6 & 2300 &   8.10&   0.8~&3 &G059.5+0.1   & 3900 &   1.40&   0.15&3 \\
G033.6$-$0.6 & 3900 &   5.30&   0.4~&3 &G059.8+1.2   &  960 &   1.50&   0.2~&3 \\
G034.7$-$0.4 &  960 & 306.00&  30.0~&3 &G059.8+1.2   & 3900 &   0.80&   0.05&3 \\
G034.7$-$0.4 & 2300 & 190.00&  20.0~&3 &G063.7+1.1   &  960 &   2.15&   0.25&3 \\
G034.7$-$0.4 & 3900 & 136.00&  13.0~&3 &G063.7+1.1   & 2300 &   1.40&   0.15&3 \\
G036.6+2.6   &  960 &   0.81&   0.1~&3 &G063.7+1.1   & 3900 &   1.30&   0.10&3 \\
G036.6+2.6   & 2300 &   0.76&   0.08&3 &G063.7+1.1   &11200 &   1.20&   0.15&3 \\
G036.6+2.6   & 3900 &   0.65&   0.07&3 &G065.7+1.2   &  960 &   5.70&   0.5~&3 \\
G036.6$-$0.7 &  960 &  11.50&   1.1~&3 &G065.7+1.2   & 3900 &   2.05&   0.2~&3 \\
G036.6$-$0.7 & 3900 &   3.00&   0.4~&3 &G067.7+1.8   &  960 &   2.20&   0.25&3 \\
G039.2$-$0.3 &  960 &  21.00&   1.5~&3 &G067.7+1.8   & 2300 &   1.05&   0.06&3 \\
G039.2$-$0.3 & 2300 &  13.40&   1.0~&3 &G067.7+1.8   & 3900 &   0.60&   0.05&3 \\
G039.2$-$0.3 & 3900 &  11.50&   1.0~&3 &G068.6$-$1.2 &  960 &   0.70&   0.15&3 \\
G068.6$-$1.2 & 3900 &   0.26&   0.05&3 &G109.1$-$1.0 & 7700 &   7.5~&   1.0~&1 \\
G069.7+1.0   &  960 &   3.00&   0.3~&3 &G112.0+1.2   &  960 &  20.0~&   2.0~&1 \\
G069.7+1.0   & 3900 &   1.30&   0.2~&3 &G112.0+1.2   & 2300 &  14.5~&   1.0~&1 \\
G073.9+0.9   &  960 &   8.90&   0.9~&3 &G112.0+1.2   & 3650 &   9.0~&   1.0~&1 \\
G073.9+0.9   & 3900 &   5.20&   0.5~&3 &G112.0+1.2   & 3900 &   7.0~&   1.0~&1 \\
G074.9+1.2   &  960 &   9.00&   0.9~&3 &G114.3+0.3   &  960 &  25.0~&   1.0~&1 \\
G074.9+1.2   & 3650 &   7.20&   0.7~&3 &G114.3+0.3   & 2300 &  20.0~&   2.0~&1 \\
G074.9+1.2   & 3900 &   6.60&   0.5~&3 &G114.3+0.3   & 3650 &  15.0~&   2.0~&1 \\
G074.9+1.2   & 7700 &   5.70&   0.4~&3 &G114.3+0.3   & 3900 &  13.0~&   2.0~&1 \\
G074.9+1.2   &11200 &   2.60&   0.3~&3 &G116.5+1.1   & 3900 &   4.8~&   0.2~&1 \\
G078.2+2.1   &  960 & 323.0~&  30.0~&3 &G116.9+0.2   &  960 &  10.1~&   0.4~&1 \\
G078.2+2.1   & 2300 & 186.0~&  18.0~&3 &G116.9+0.2   & 2300 &   5.5~&   0.4~&1 \\
G078.2+2.1   & 3900 & 150.0~&  15.0~&3 &G116.9+0.2   & 3900 &   3.8~&   0.2~&1 \\
G083.0$-$0.2 &  960 &   1.00&   0.15&3 &G117.4+5.0   &  960 &  55.0~&   5.0~&1 \\
G083.0$-$0.2 & 2300 &   1.25&   0.15&3 &G117.4+5.0   & 3900 &  35.0~&   3.0~&1 \\
G083.0$-$0.2 & 3900 &   0.95&   0.10&3 &G119.5+10.   &  960 &  36.0~&   3.0~&1 \\
G084.2$-$0.8 &  960 &  10.50&   1.1~&3 &G119.5+10.   & 3650 &  26.0~&   3.0~&1 \\
G084.2$-$0.8 & 2300 &   5.50&   0.5~&3 &G119.5+10.   & 3900 &  22.0~&   3.0~&1 \\
G084.2$-$0.8 & 3900 &   3.50&   0.35&3 &G120.1+1.4   &  960 &  54.8~&   0.8~&1 \\
G084.9+0.5   &  960 &   6.00&   0.6~&3 &G120.1+1.4   & 2300 &  32.1~&   0.5~&1 \\
G084.9+0.5   & 2300 &   2.80&   0.3~&3 &G120.1+1.4   & 3650 &  23.6~&   0.3~&1 \\
G084.9+0.5   & 3900 &   2.00&   0.2~&3 &G120.1+1.4   & 3900 &  23.1~&   0.3~&1 \\
G084.9+0.5   &11200 &   1.30&   0.2~&3 &G120.1+1.4   & 7700 &  15.1~&   0.5~&1 \\
G085.2$-$1.2 &  960 &   2.30&   0.1~&3 &G126.2+1.6   &  960 &   9.3~&   1.5~&1 \\
G085.2$-$1.2 & 2300 &   1.40&   0.15&3 &G126.2+1.6   & 3650 &   3.0~&   0.5~&1 \\
G085.2$-$1.2 & 3900 &   1.70&   0.15&3 &G126.2+1.6   & 3900 &   3.0~&   0.5~&1 \\
G089.0+4.7   &  960 & 190.0~&  20.0~&1 &G127.1+0.5   &  960 &  11.4~&   0.5~&1 \\
G089.0+4.7   & 2300 & 155.0~&  20.0~&1 &G127.1+0.5   & 2300 &  10.0~&   2.0~&1 \\
G089.0+4.7   & 3650 & 118.0~&  15.0~&1 &G127.1+0.5   & 3650 &   4.5~&   0.3~&1 \\
G089.0+4.7   & 3900 & 115.0~&  15.0~&1 &G127.1+0.5   & 3900 &   4.5~&   0.3~&1 \\
G089.0+4.7   & 7700 &  90.0~&  20.0~&1 &G130.7+3.1   &  960 &  33.3~&   1.5~&1 \\
G093.3+6.9   &  960 &   8.1~&   0.5~&1 &G130.7+3.1   & 2300 &  32.3 &   2.0~&1 \\
G093.3+6.9   & 3650 &   4.3~&   0.2~&1 &G130.7+3.1   & 3650 &  32.7~&   1.0~&1 \\
G093.3+6.9   & 3900 &   3.6~&   0.3~&1 &G130.7+3.1   & 3900 &  31.3~&   1.0~&1 \\
G093.7$-$0.2 &  960 &  45.0~&   2.0~&1 &G132.7+1.3   &  960 &  70.0~&   5.0~&1 \\
G093.7$-$0.2 & 3650 &  18.3~&   0.3~&1 &G132.7+1.3   & 3650 &  35.0~&   5.0~&1 \\
G093.7$-$0.2 & 3900 &  17.5~&   0.3~&1 &G132.7+1.3   & 3900 &  33.0~&   5.0~&1 \\
G094.0+1.0   &  960 &  14.0~&   1.4~&1 &G132.7+1.3W  & 2300 &  22.9~&   2.0~&1 \\
G094.0+1.0   &  960 &  14.00&   1.1~&4 &G132.7+1.3W  & 3650 &  14.5~&   1.5~&1 \\
G094.0+1.0   & 3650 &   9.2~&   0.3~&1 &G132.7+1.3W  & 3900 &  12.5~&   1.5~&1 \\
G094.0+1.0   & 3650 &   9.20&   0.9~&4 &G132.7+1.3W  & 7700 &   9.0~&   2.0~&1 \\
G094.0+1.0   & 3900 &   9.1~&   0.3~&1 &G160.9+2.6   &  960 & 109.0~&   5.0~&1 \\
G094.0+1.0   & 3900 &   9.10&   0.9~&3 &G160.9+2.6   & 3900 &  46.0~&   5.0~&1 \\
G094.0+1.0   & 7700 &   6.7~&   0.5~&1 &G189.1+3.0   &  960 & 150.0~&  10.0~&1 \\
G094.0+1.0   & 7700 &   6.70&   0.7~&3 &G189.1+3.0   & 3900 &  84.0~&   5.0~&1 \\
G109.1$-$1.0 &  960 &  22.8~&   0.6~&1 &G189.1+3.0   & 7700 &  51.0~&   5.0~&1 \\
G109.1$-$1.0 & 2300 &  13.8~&   0.5~&1 &G192.8$-$1.1 & 3650 &  25.0~&   2.5~&2 \\
G109.1$-$1.0 & 3650 &  12.5~&   0.5~&1 &G192.8$-$1.1 & 3900 &  30.0~&   2.5~&2 \\
G109.1$-$1.0 & 3900 &  12.3~&   0.5~&1 &G261.9+5.5   &  960 &  14.00&   1.5~&4 \\
G261.9+5.5   & 2300 &   8.10&   1.0~&4 &G355.6$-$0.0 & 3900 &   3.7~&   0.3~&4 \\
G261.9+5.5   & 3900 &   5.30&   0.5~&4 &G355.9$-$2.5 &  960 &   7.60&   0.7~&3 \\
G327.6+14.   & 2300 &  13.00&   1.3~&3 &G355.9$-$2.5 & 3900 &   4.00&   0.3~&3 \\
G327.6+14.   & 3900 &   9.20&   0.8~&3 &G356.2+4.4   &  960 &   6.86&   0.5~&4 \\
G327.6+14.   & 7700 &   6.00&   0.5~&3 &G356.2+4.4   & 3900 &   3.16&   0.3~&4 \\
G344.7$-$0.1 &  960 &   3.60&   0.5~&3 &G356.3$-$0.3 &  960 &   0.9~&   0.2~&4 \\
G344.7$-$0.1 & 2300 &   1.65&   0.2~&3 &G356.3$-$0.3 & 3900 &   0.8~&   0.2~&4 \\
G344.7$-$0.1 & 3900 &   1.40&   0.1~&3 &G357.7+0.3   &  960 &   7.20&   0.7~&3 \\
G344.7$-$0.1 &11200 &   0.70&   0.1~&3 &G357.7+0.3   & 3900 &   6.10&   0.5~&3 \\
G345.7$-$0.2 & 3900 &   0.40&   0.15&4 &G357.7$-$0.1 &  960 &  43.0~&   2.0~&3 \\
G346.6$-$0.2 &  960 &  11.40&   1.1~&3 &G357.7$-$0.1 & 3650 &  22.0~&   1.0~&3 \\
G346.6$-$0.2 & 2300 &   8.80&   0.9~&3 &G357.7$-$0.1 & 3900 &  22.0~&   1.0~&3 \\
G346.6$-$0.2 & 3900 &   6.30&   0.6~&3 &G357.7$-$0.1 & 7700 &  13.0~&   1.5~&3 \\
G346.6$-$0.2 & 7700 &   4.90&   0.5~&3 &G357.7$-$0.1 &11200 &   8.60&   1.0~&3 \\
G346.6$-$0.2 &11200 &   2.90&   0.3~&3 &G359.1$-$0.5 & 3900 &  15.50&   1.5~&3 \\
G348.5$-$0.0 &  960 &   8.0~&   1.0~&3 &             &      &       &       &  \\
G348.5$-$0.0 & 3900 &   4.5~&   0.5~&3 &             &      &       &       &  \\
G348.5+0.1   &  960 & 118.0~&  12.0~&3 &             &      &       &       &  \\
G348.5+0.1   & 3900 &  41.0~&   3.0~&3 &             &      &       &       &  \\
G348.7+0.3   & 3900 &  21.0~&   2.5~&3 &             &      &       &       &  \\
G349.7+0.2   &  960 &  25.0~&   2.5~&3 &             &      &       &       &  \\
G349.7+0.2   & 2300 &  15.50&   1.5~&3 &             &      &       &       &  \\
G349.7+0.2   & 3900 &   9.50&   0.9~&3 &             &      &       &       &  \\
G349.7+0.2   & 7700 &   8.00&   0.8~&3 &             &      &       &       &  \\
G349.7+0.2   &11200 &   4.00&   0.4~&3 &             &      &       &       &  \\
G350.0$-$2.0 &  960 &  29.0~&   0.3~&4 &             &      &       &       &  \\
G350.0$-$2.0 & 3900 &  14.70&   2.0~&4 &             &      &       &       &  \\
G350.1$-$0.3 &  960 &   6.20&   0.6~&3 &             &      &       &       &  \\
G350.1$-$0.3 & 2300 &   3.20&   0.3~&3 &             &      &       &       &  \\
G350.1$-$0.3 & 3900 &   2.10&   0.2~&3 &             &      &       &       &  \\
G350.1$-$0.3 & 7700 &   1.50&   0.2~&3 &             &      &       &       &  \\
G351.2+0.1   &  960 &   3.30&   0.4~&3 &             &      &       &       &  \\
G351.2+0.1   & 2300 &   3.90&   0.4~&3 &             &      &       &       &  \\
G351.2+0.1   & 3900 &   2.60&   0.2~&3 &             &      &       &       &  \\
G351.2+0.1   & 7700 &   2.30&   0.3~&3 &             &      &       &       &  \\
G351.2+0.1   &11200 &   2.70&   0.3~&3 &             &      &       &       &  \\
G352.7$-$0.1 & 3900 &   2.60&   0.2~&3 &             &      &       &       &  \\
G352.7$-$0.1 &11200 &   1.40&   0.2~&3 &             &      &       &       &  \\
G354.8$-$0.8 &  960 &   3.1~&   0.3~&4 &             &      &       &       &  \\
G354.8$-$0.8 & 3900 &   1.0~&   0.2~&4 &             &      &       &       &  \\
G355.4+0.7   &  960 &   6.0~&   0.6~&4 &             &      &       &       &  \\
G355.4+0.7   & 3900 &   2.4~&   0.2~&4 &             &      &       &       &  \\
G355.6$-$0.0 &  960 &   4.0~&   0.3~&4 &             &      &       &       &  \\
\end{supertabular}
\end{center}
Bibcode of the references in column 5: \\ 1  -- 1987AISAO..25...81T;
 2  -- 1989Thes........1T;  3  -- 1996BSAO...41...64T;
 4  -- this paper.
\end{onecolumn}
\newpage
\textheight=24.0cm
\begin{figure}\centerline{\vbox{\psfig{figure=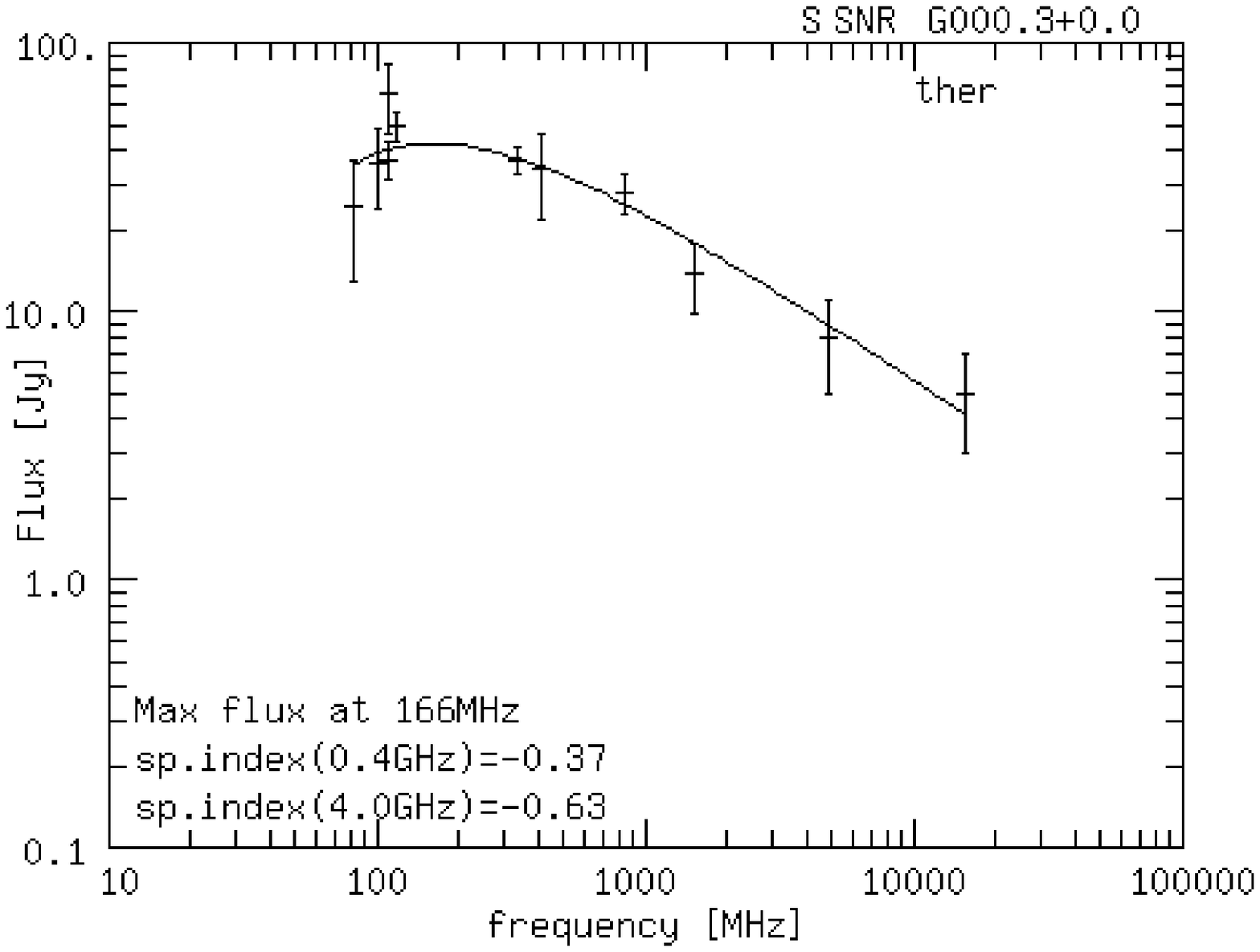,width=7.4cm,angle=0}}}\end{figure}
\begin{figure}\centerline{\vbox{\psfig{figure=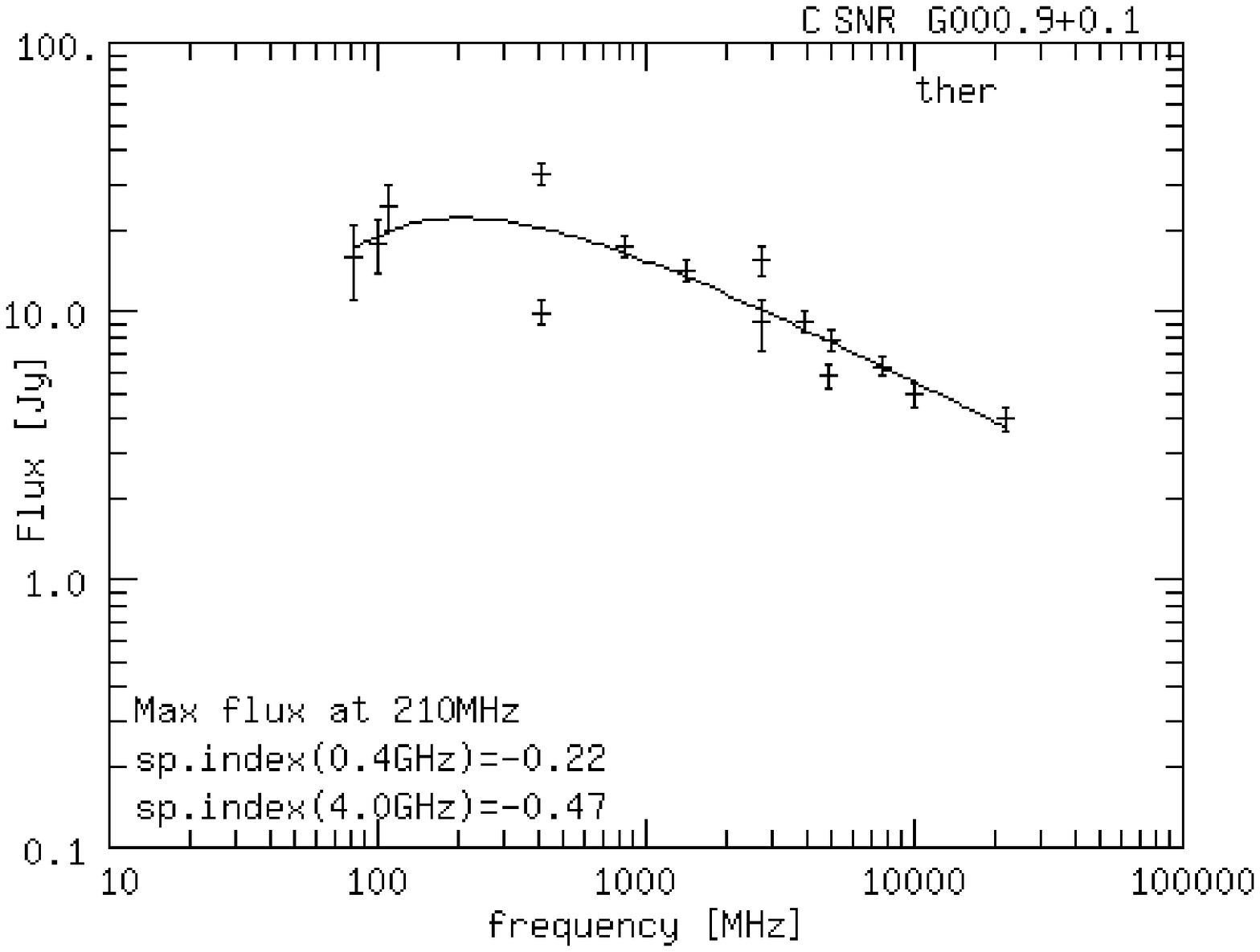,width=7.4cm,angle=0}}}\end{figure}
\begin{figure}\centerline{\vbox{\psfig{figure=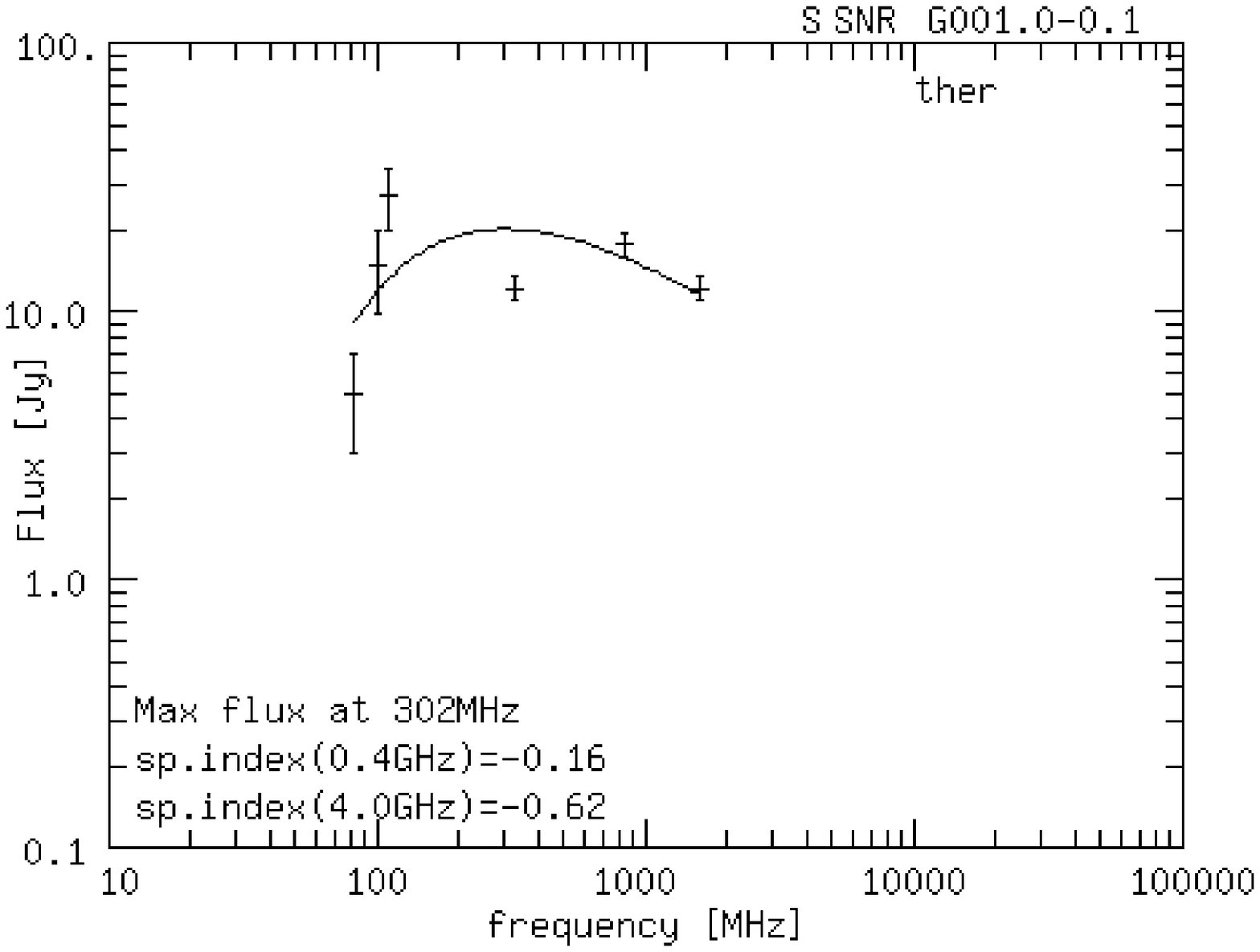,width=7.4cm,angle=0}}}\end{figure}
\begin{figure}\centerline{\vbox{\psfig{figure=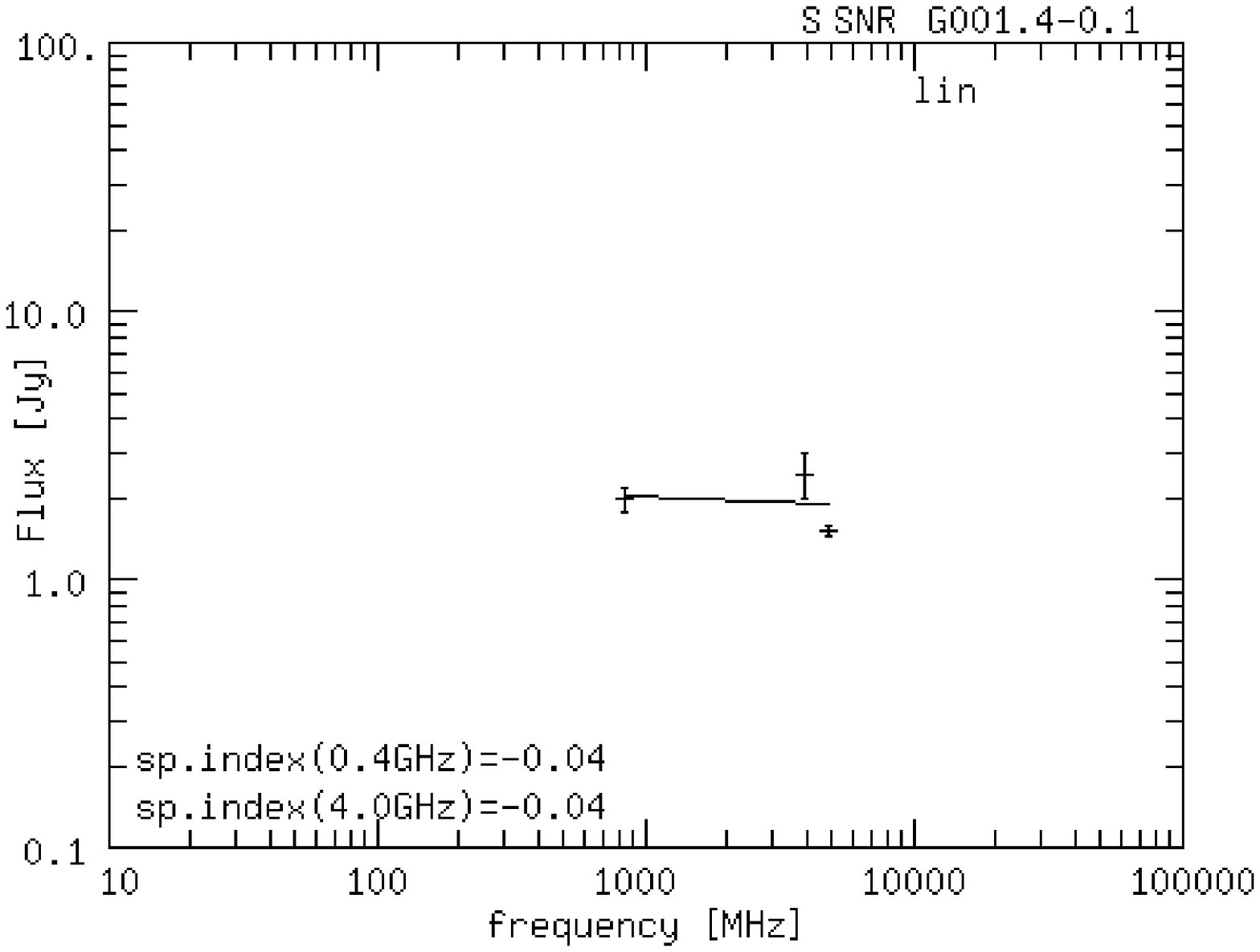,width=7.4cm,angle=0}}}\end{figure}
\begin{figure}\centerline{\vbox{\psfig{figure=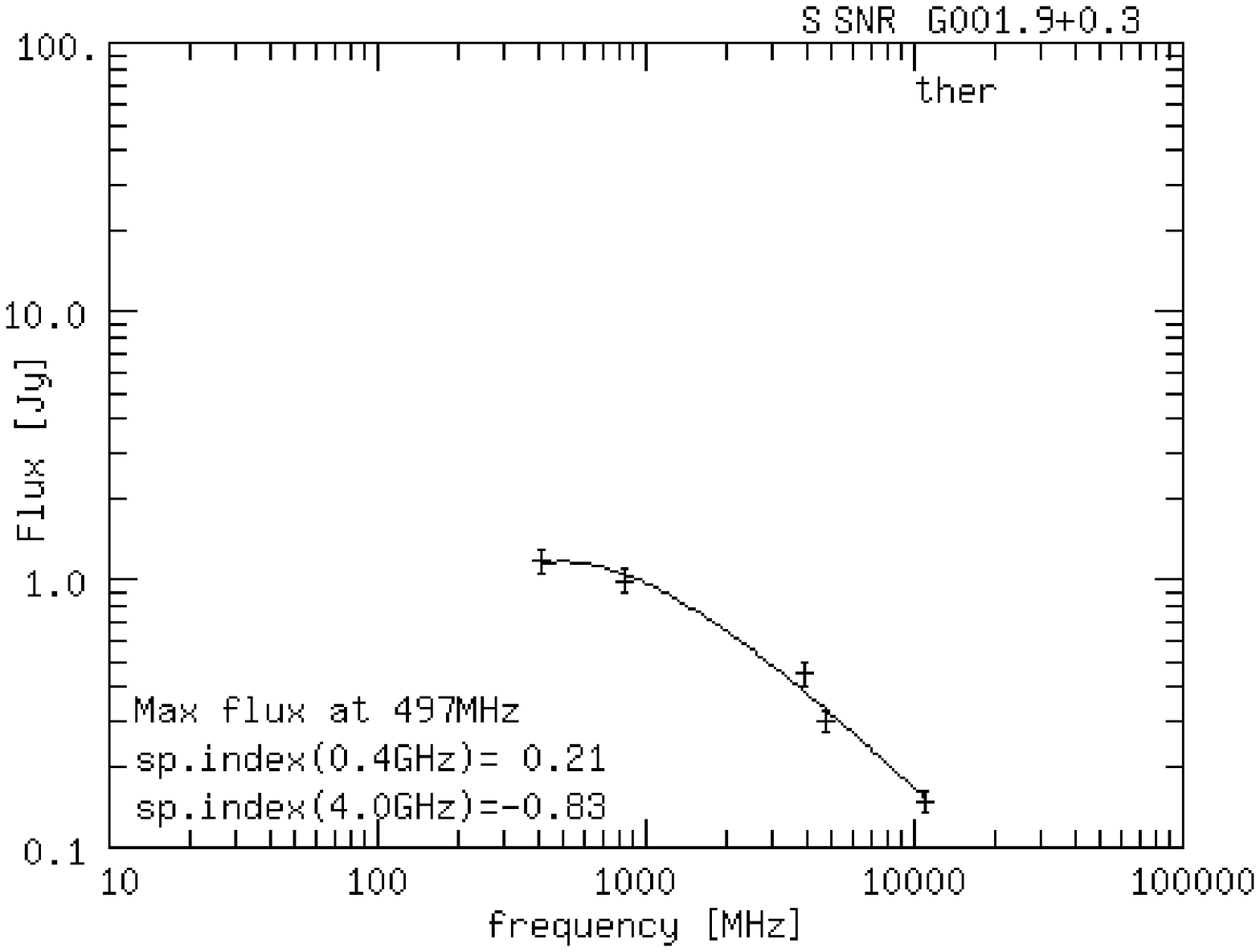,width=7.4cm,angle=0}}}\end{figure}
\begin{figure}\centerline{\vbox{\psfig{figure=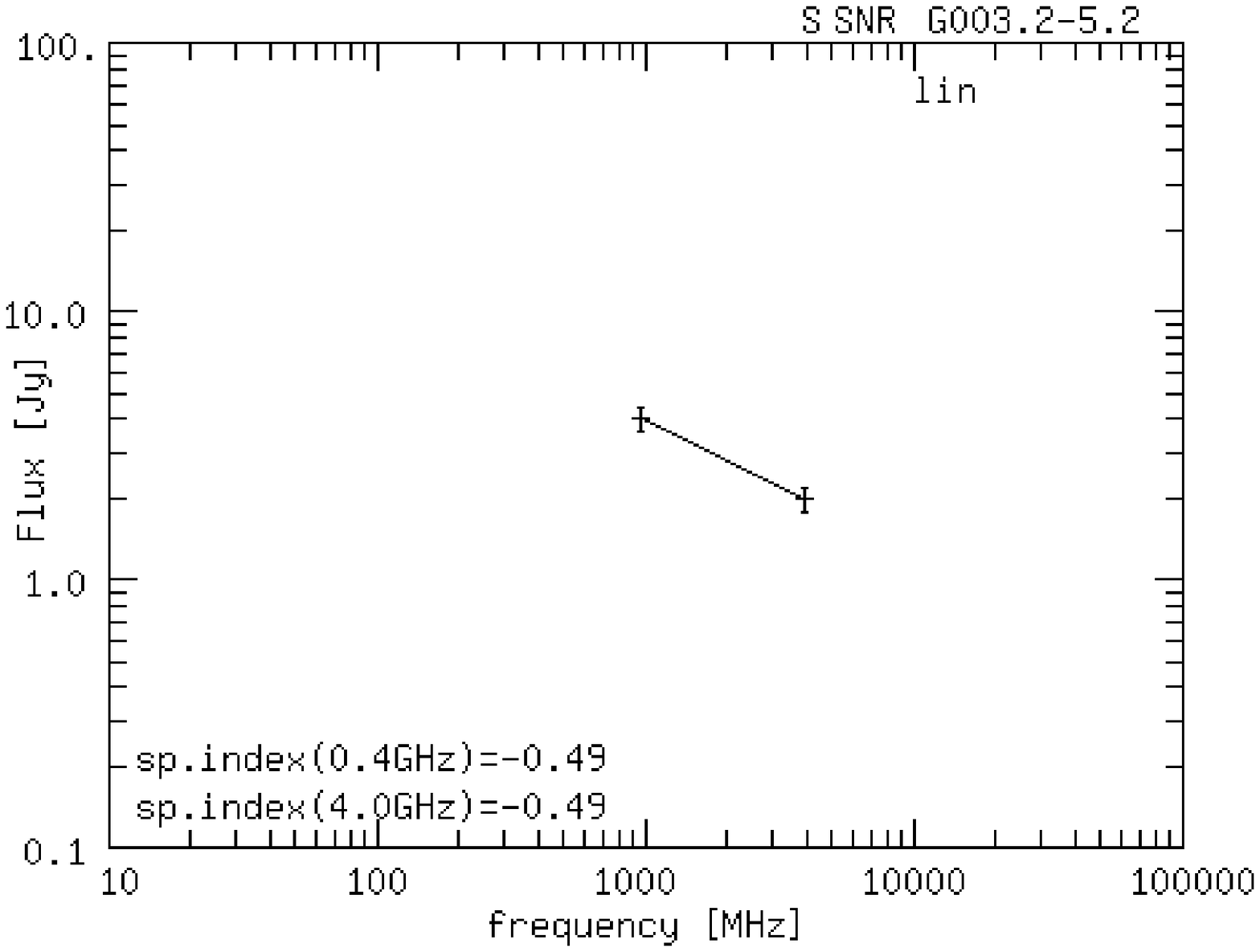,width=7.4cm,angle=0}}}\end{figure}
\begin{figure}\centerline{\vbox{\psfig{figure=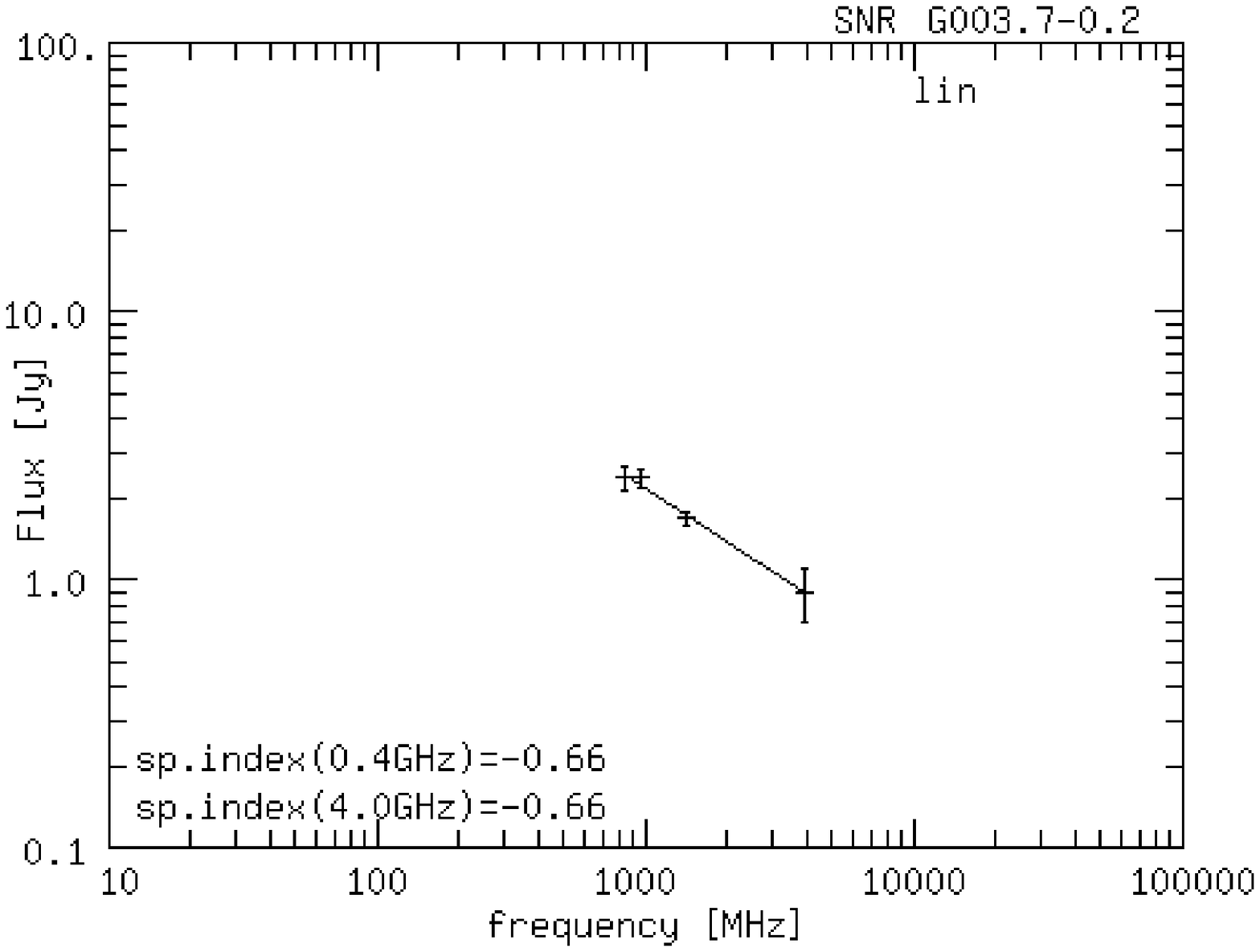,width=7.4cm,angle=0}}}\end{figure}
\begin{figure}\centerline{\vbox{\psfig{figure=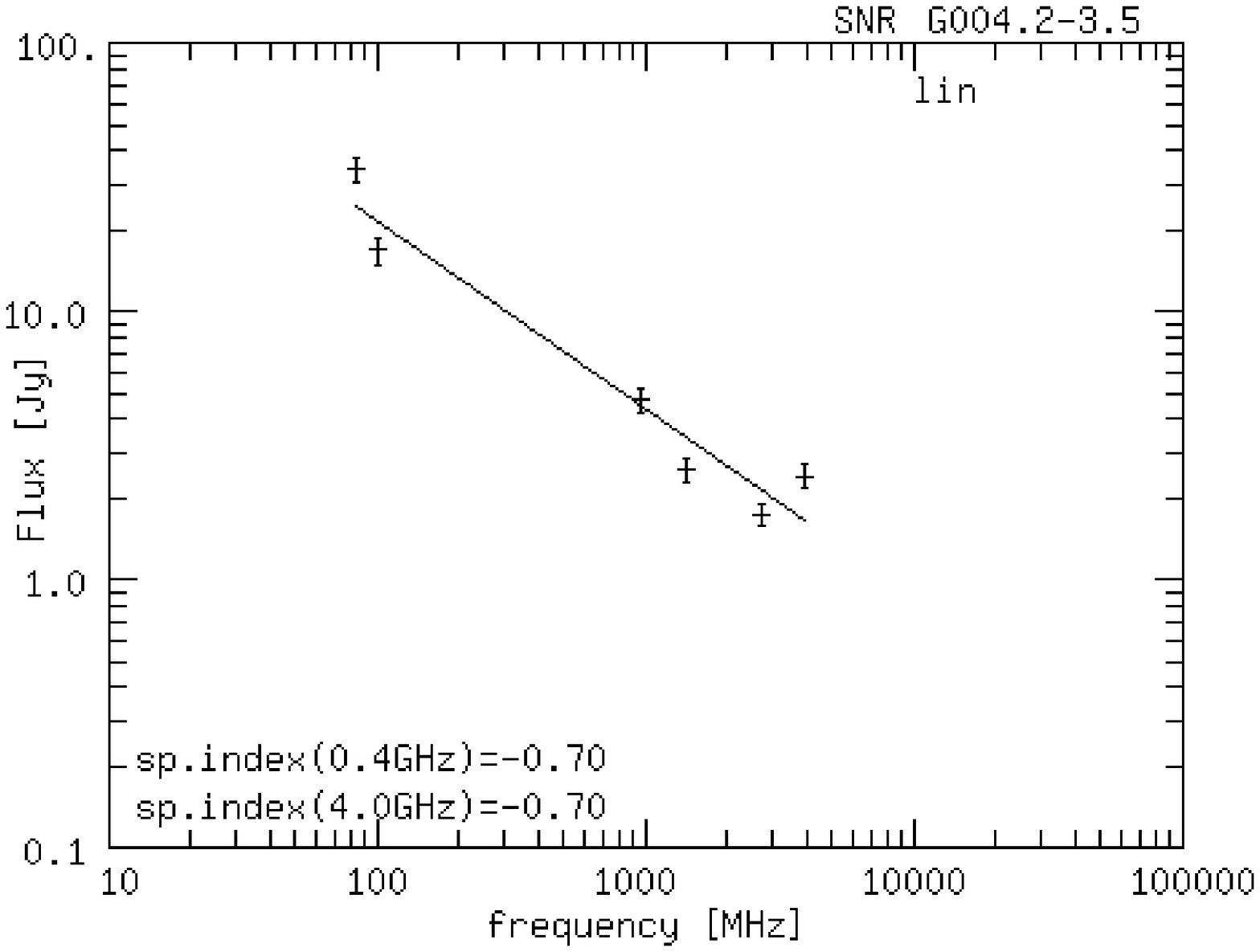,width=7.4cm,angle=0}}}\end{figure}\clearpage
\begin{figure}\centerline{\vbox{\psfig{figure=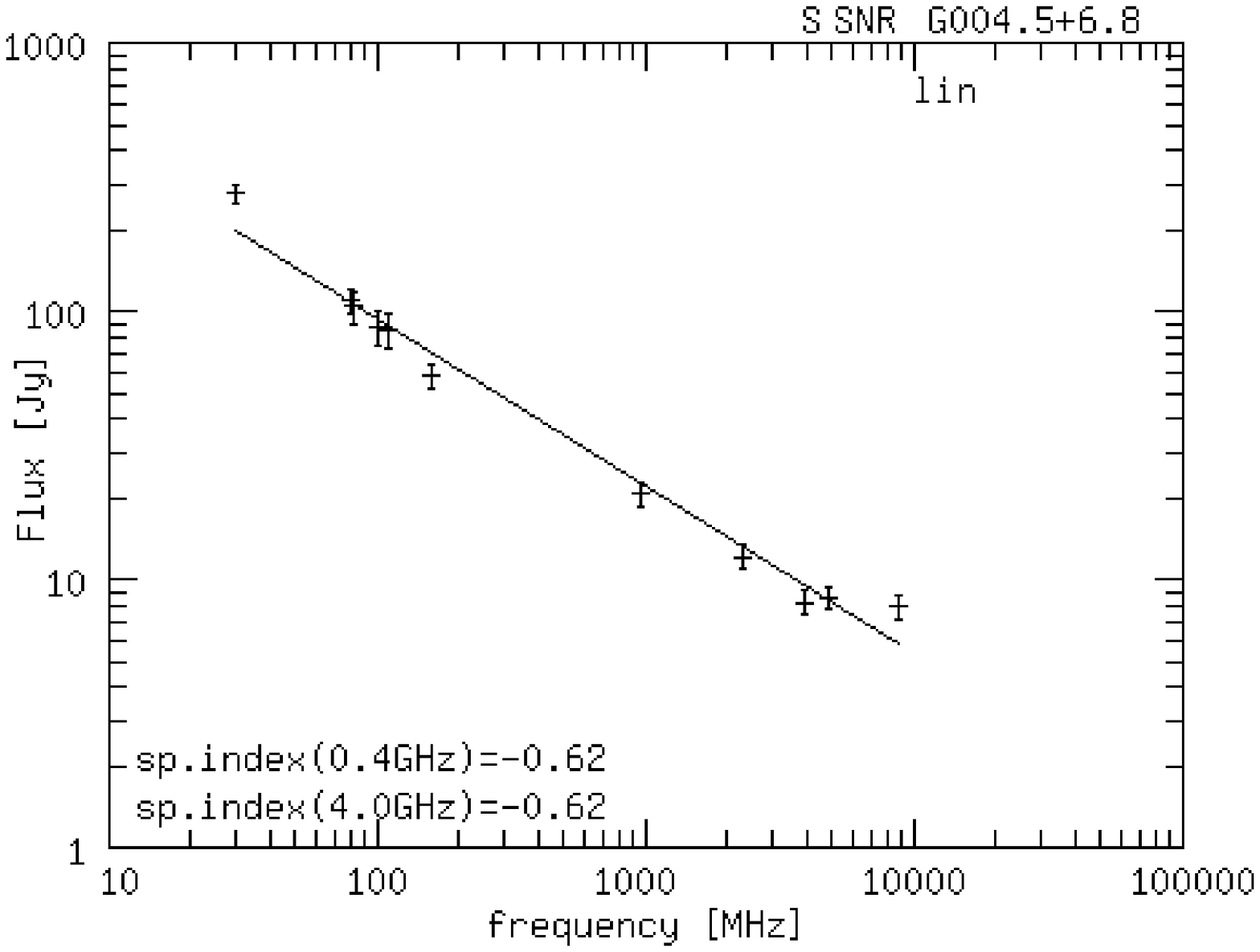,width=7.4cm,angle=0}}}\end{figure}
\begin{figure}\centerline{\vbox{\psfig{figure=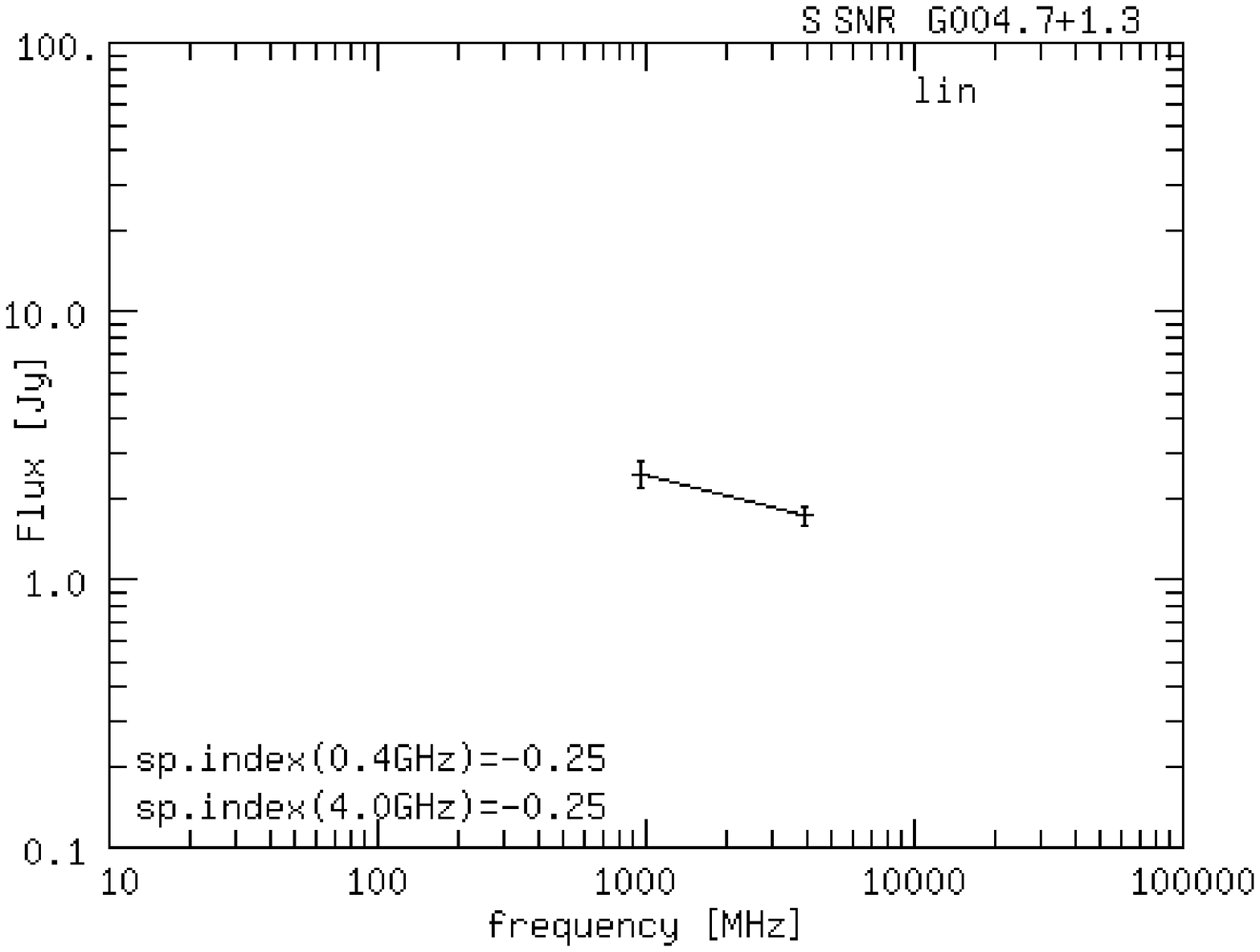,width=7.4cm,angle=0}}}\end{figure}
\begin{figure}\centerline{\vbox{\psfig{figure=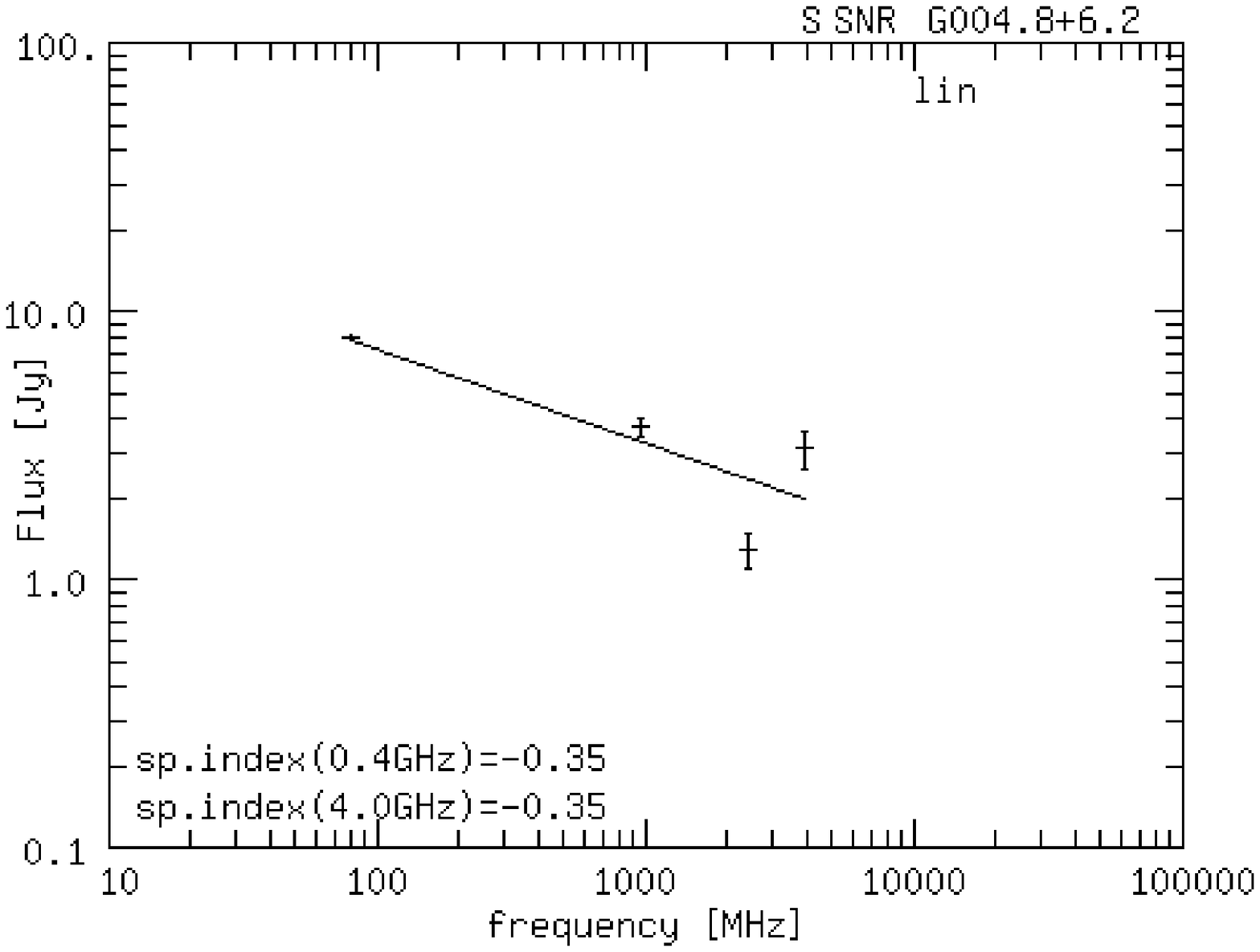,width=7.4cm,angle=0}}}\end{figure}
\begin{figure}\centerline{\vbox{\psfig{figure=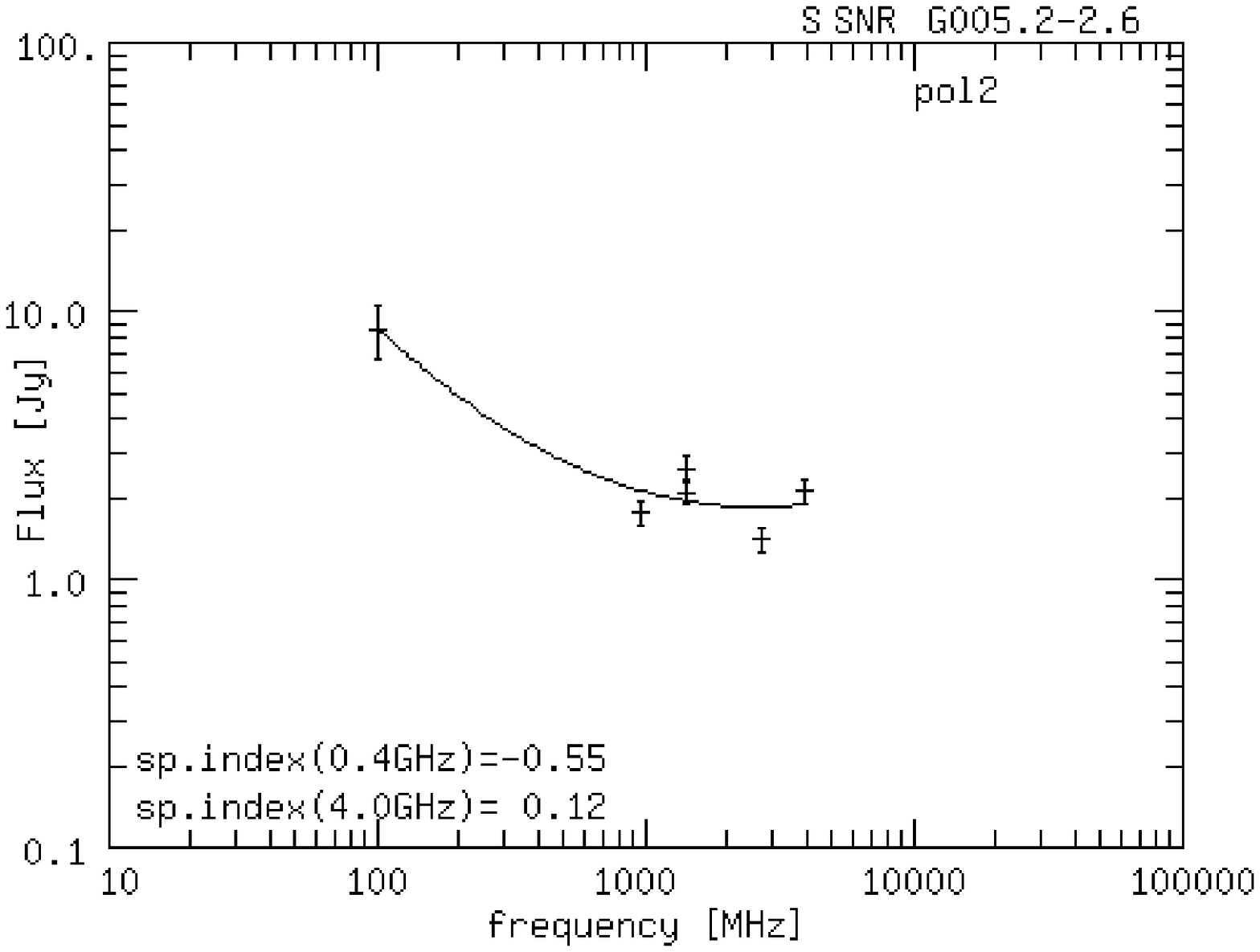,width=7.4cm,angle=0}}}\end{figure}
\begin{figure}\centerline{\vbox{\psfig{figure=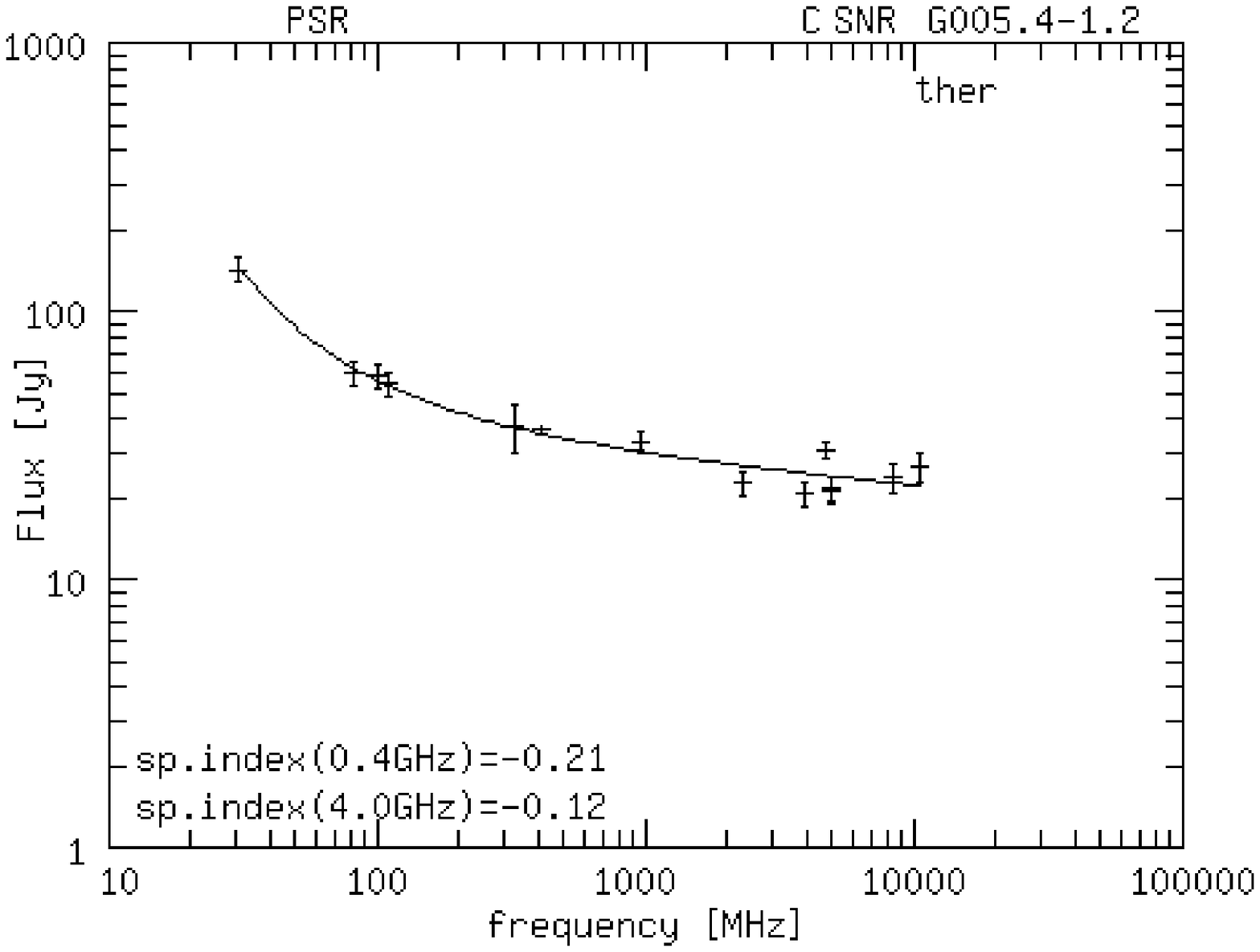,width=7.4cm,angle=0}}}\end{figure}
\begin{figure}\centerline{\vbox{\psfig{figure=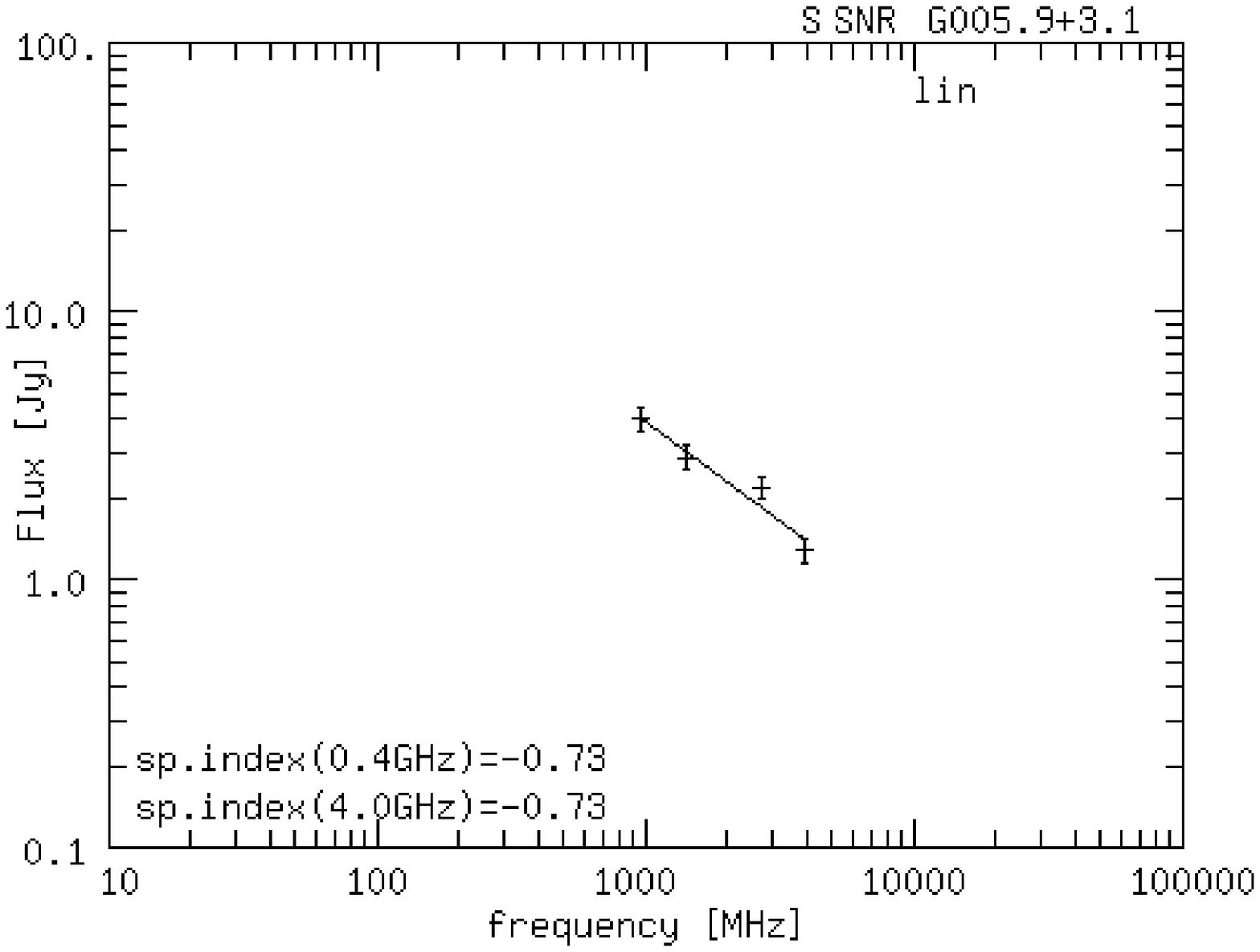,width=7.4cm,angle=0}}}\end{figure}
\begin{figure}\centerline{\vbox{\psfig{figure=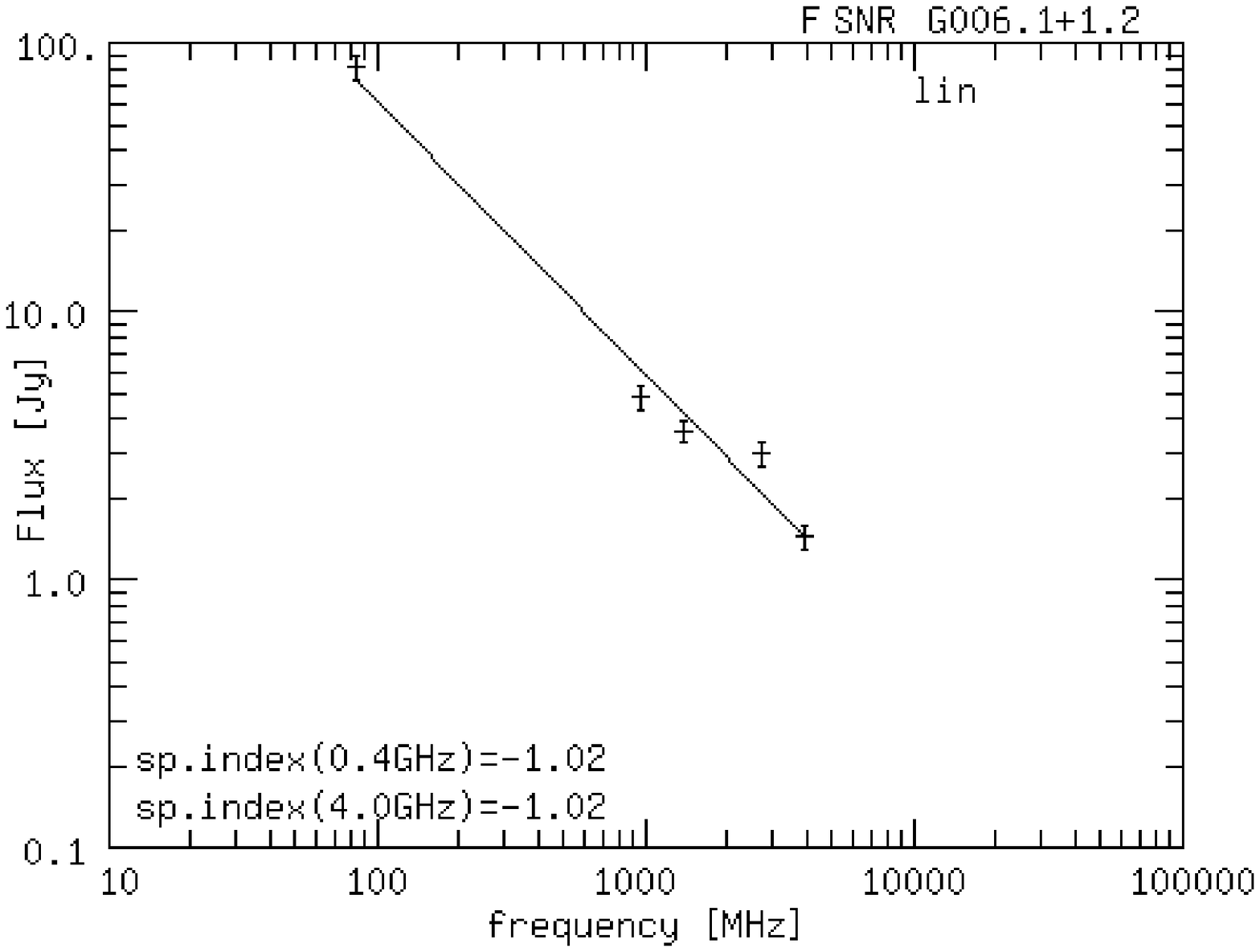,width=7.4cm,angle=0}}}\end{figure}
\begin{figure}\centerline{\vbox{\psfig{figure=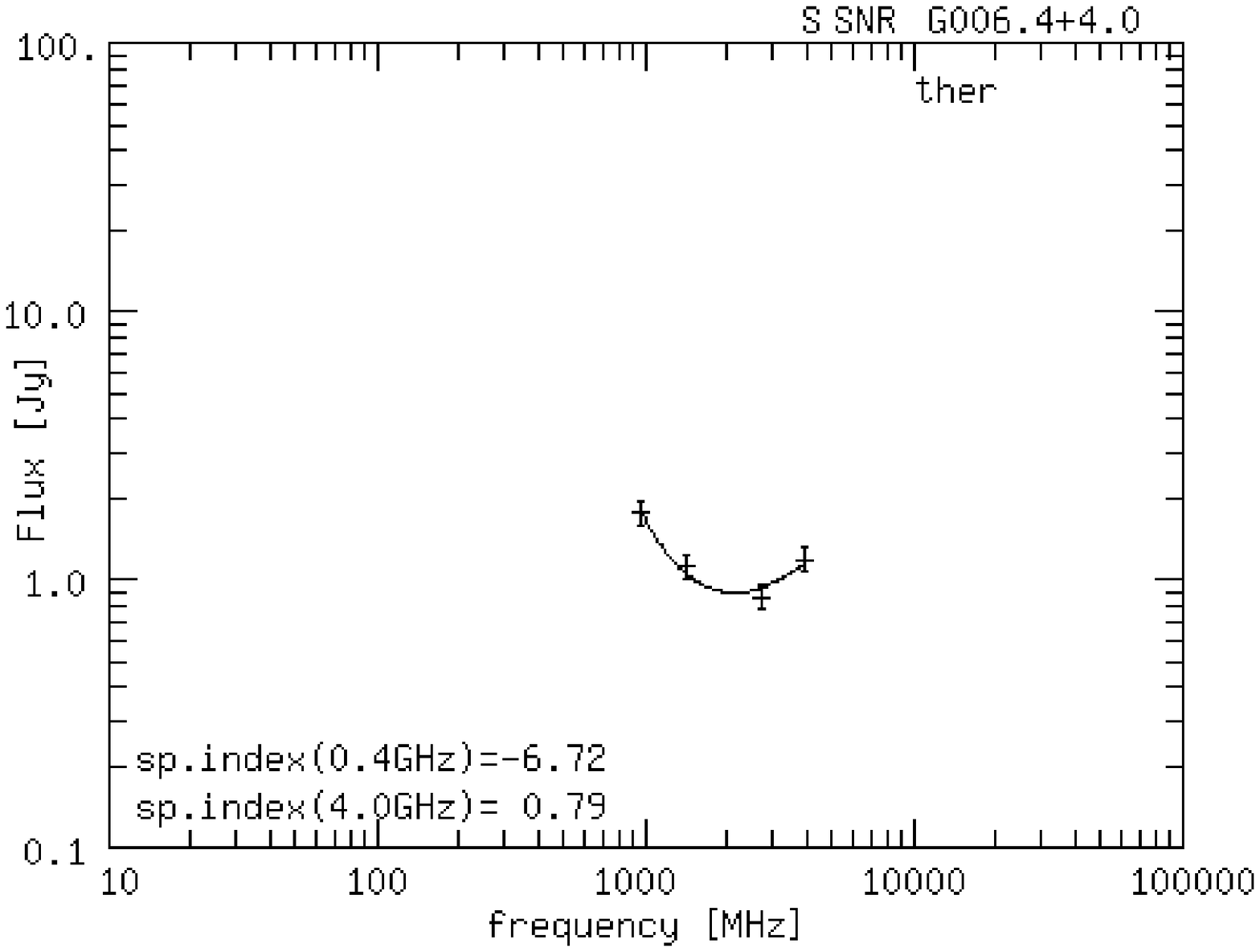,width=7.4cm,angle=0}}}\end{figure}\clearpage
\begin{figure}\centerline{\vbox{\psfig{figure=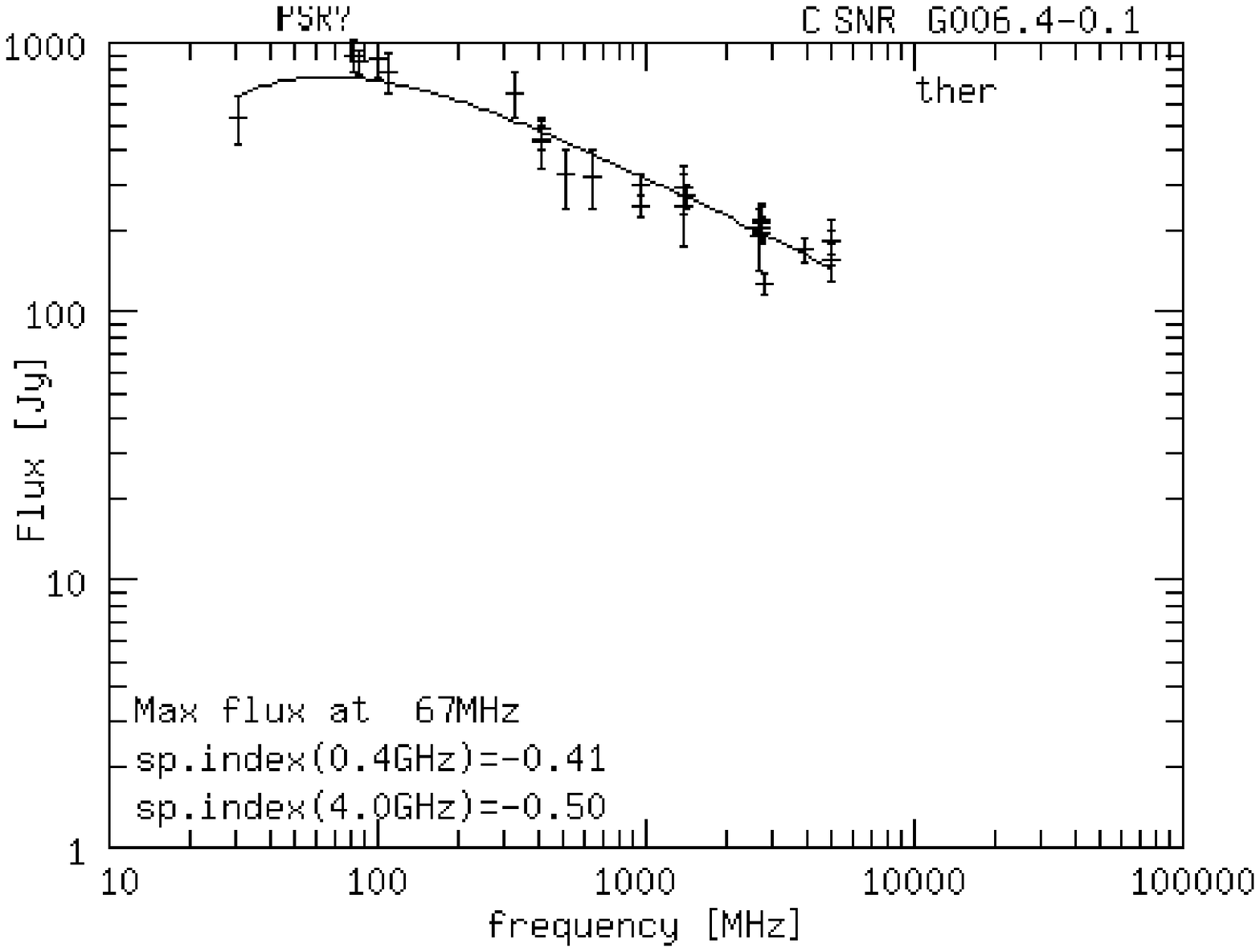,width=7.4cm,angle=0}}}\end{figure}
\begin{figure}\centerline{\vbox{\psfig{figure=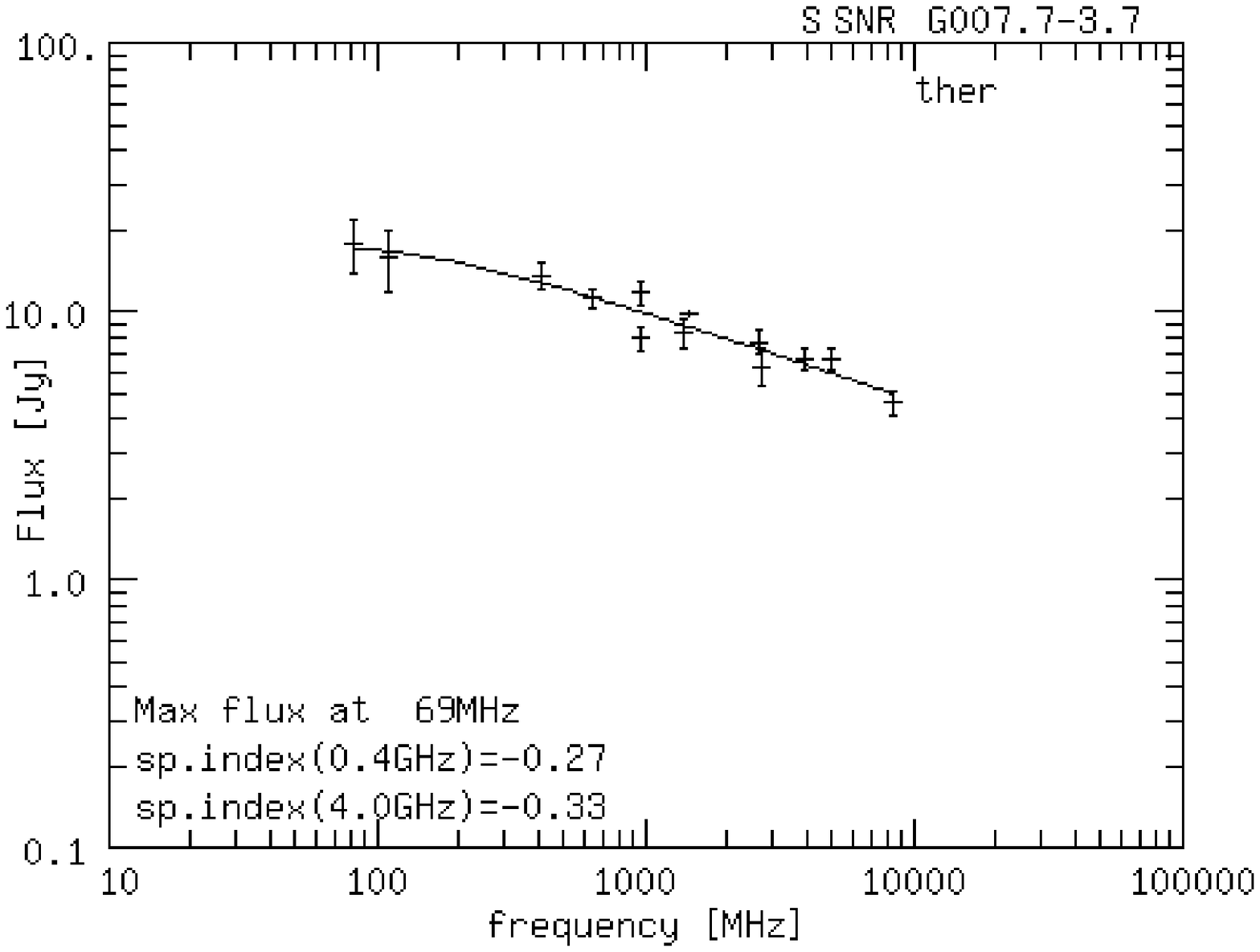,width=7.4cm,angle=0}}}\end{figure}
\begin{figure}\centerline{\vbox{\psfig{figure=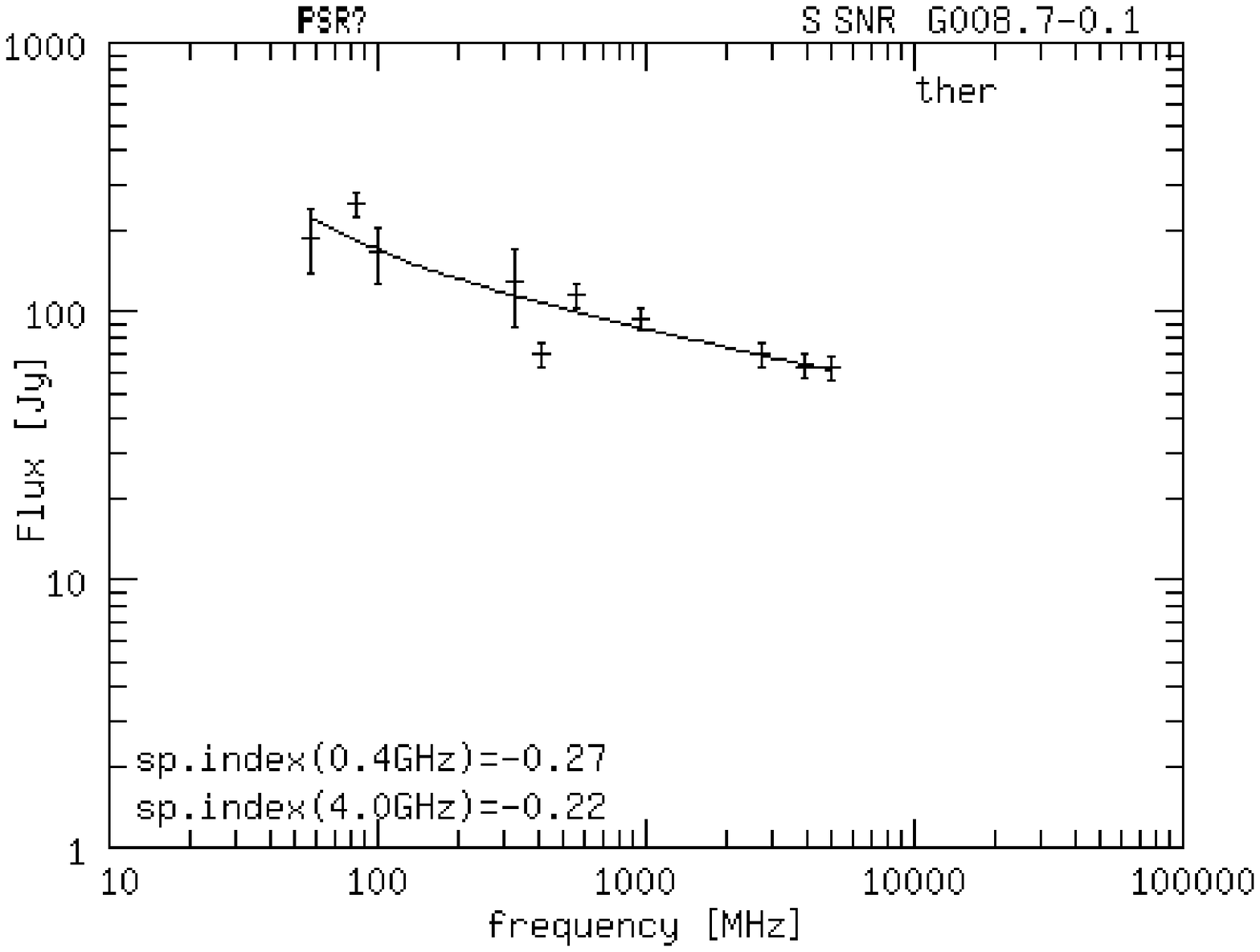,width=7.4cm,angle=0}}}\end{figure}
\begin{figure}\centerline{\vbox{\psfig{figure=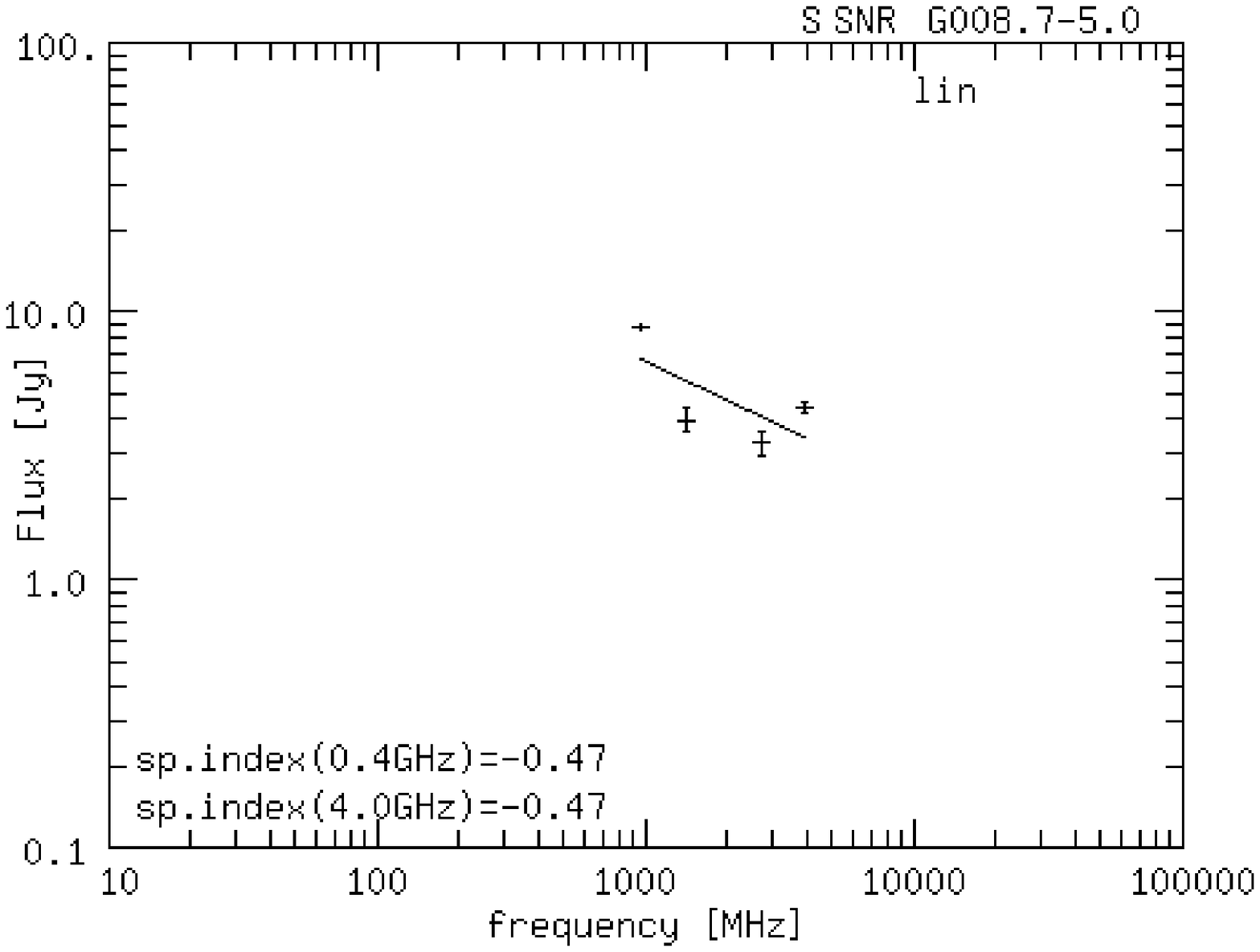,width=7.4cm,angle=0}}}\end{figure}
\begin{figure}\centerline{\vbox{\psfig{figure=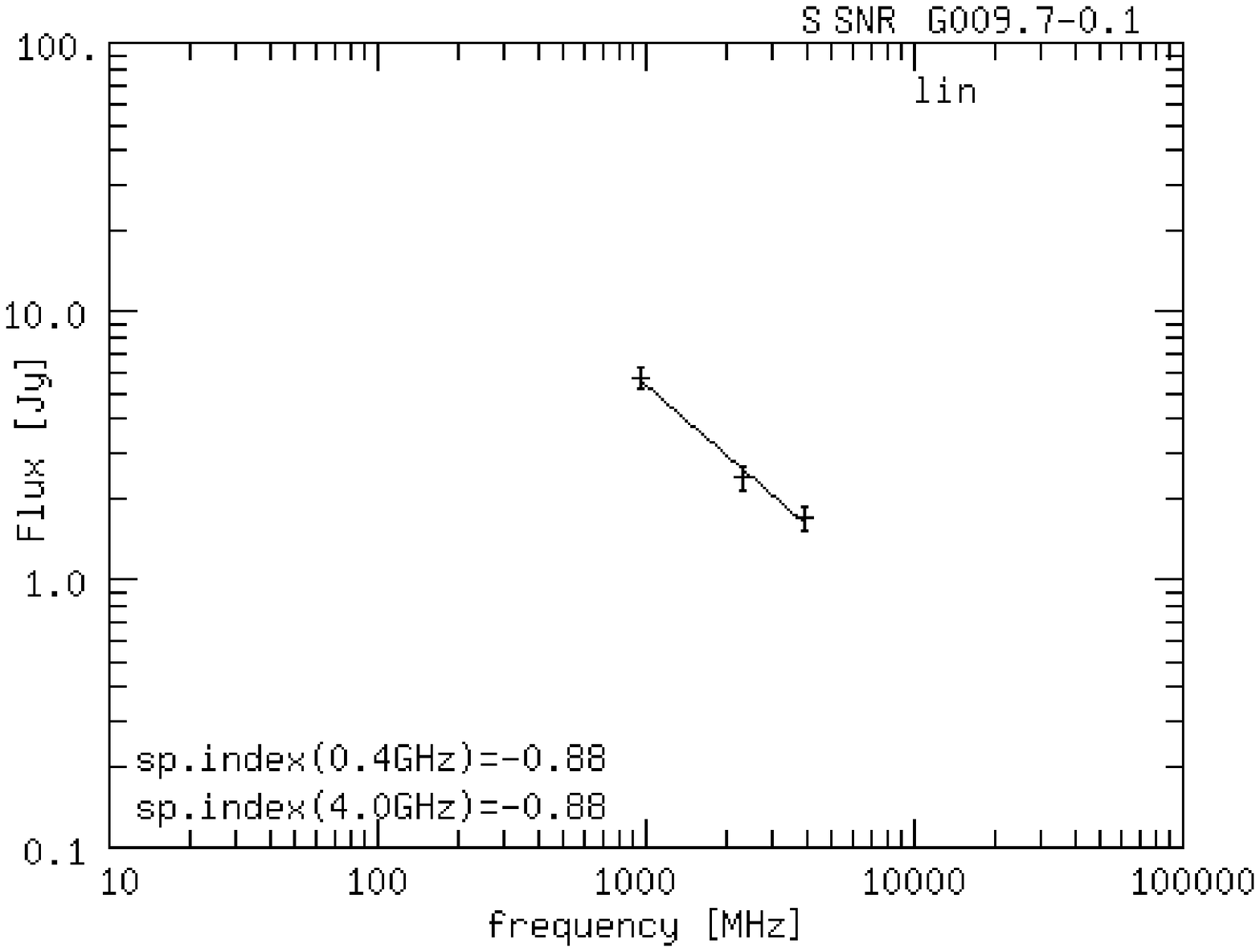,width=7.4cm,angle=0}}}\end{figure}
\begin{figure}\centerline{\vbox{\psfig{figure=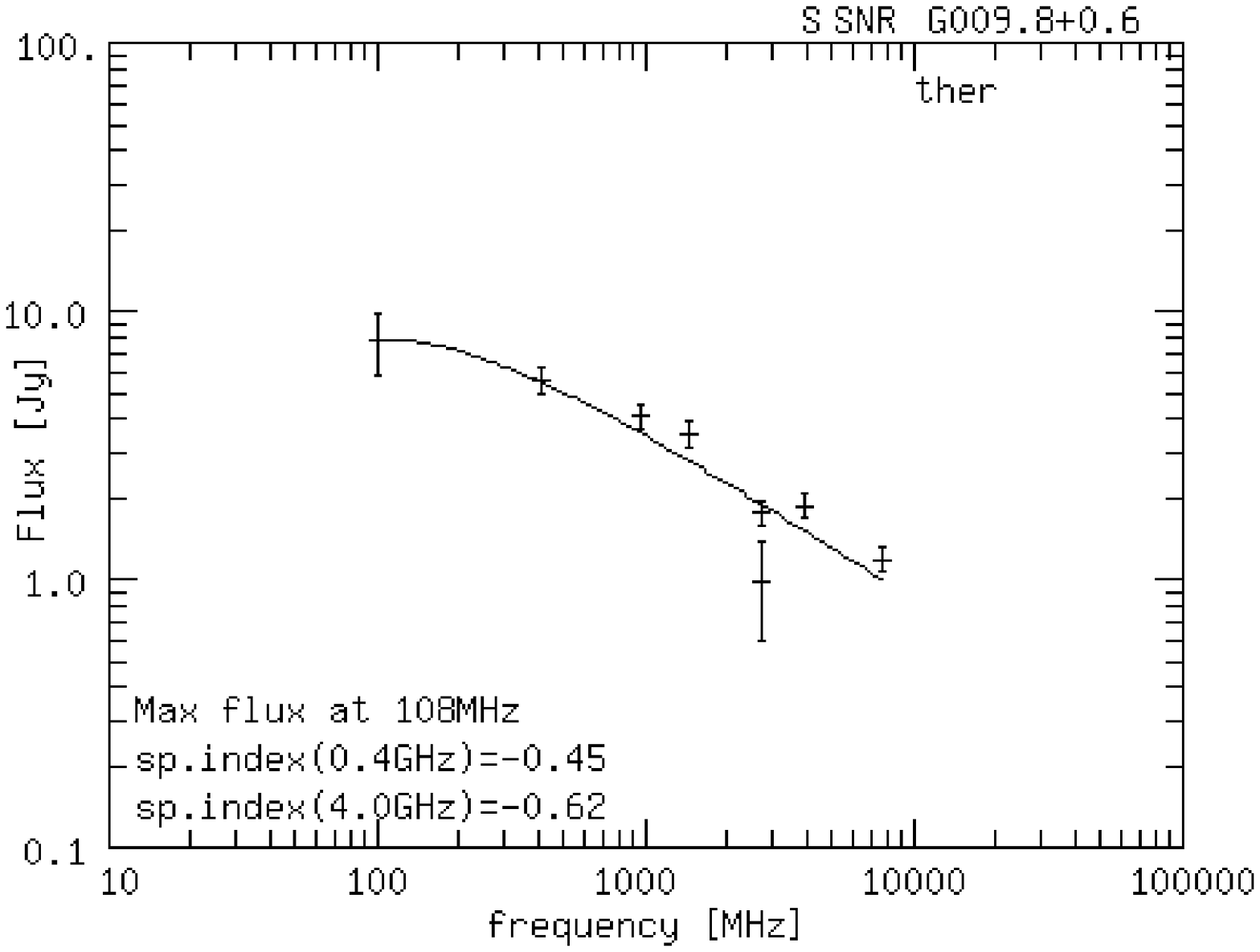,width=7.4cm,angle=0}}}\end{figure}
\begin{figure}\centerline{\vbox{\psfig{figure=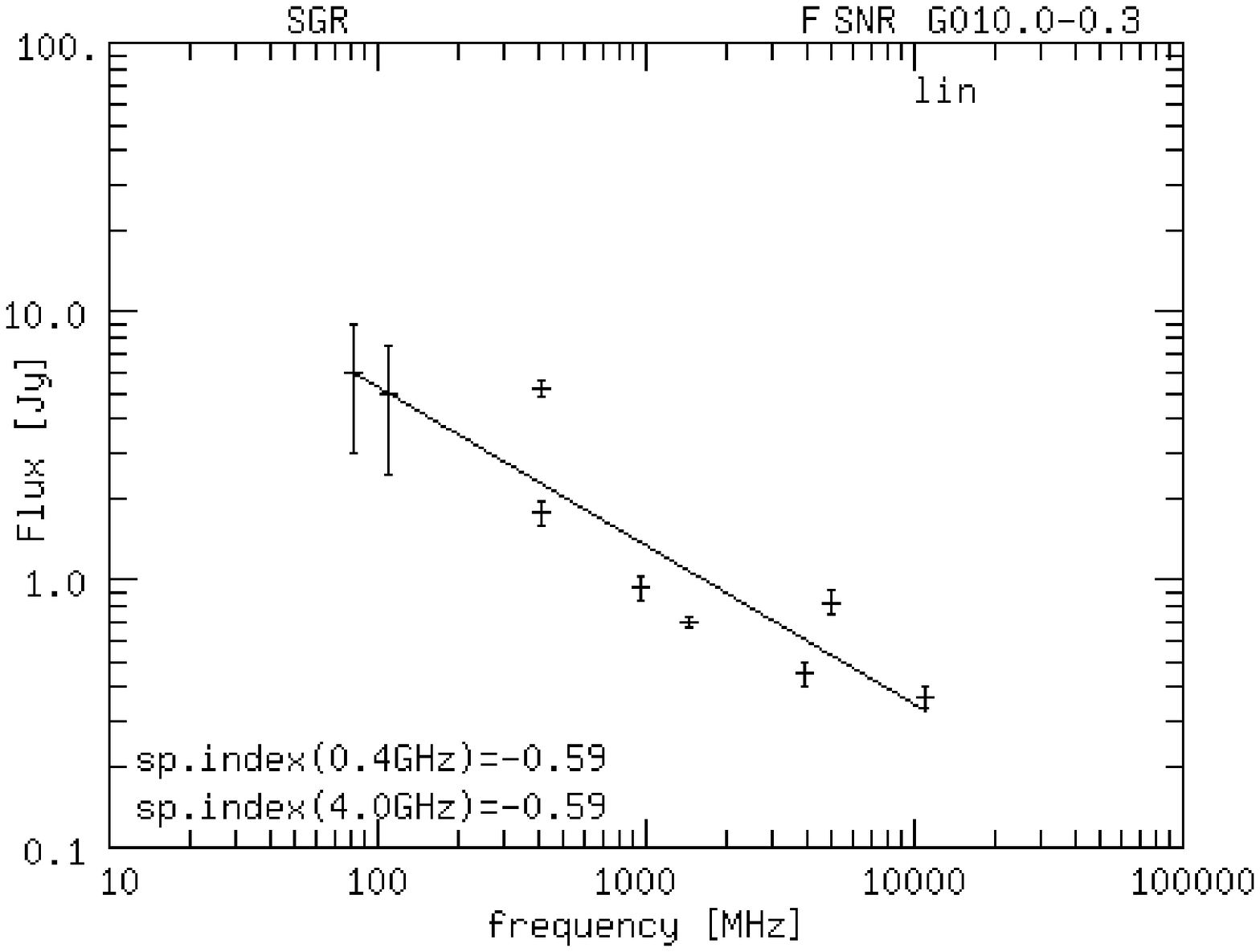,width=7.4cm,angle=0}}}\end{figure}
\begin{figure}\centerline{\vbox{\psfig{figure=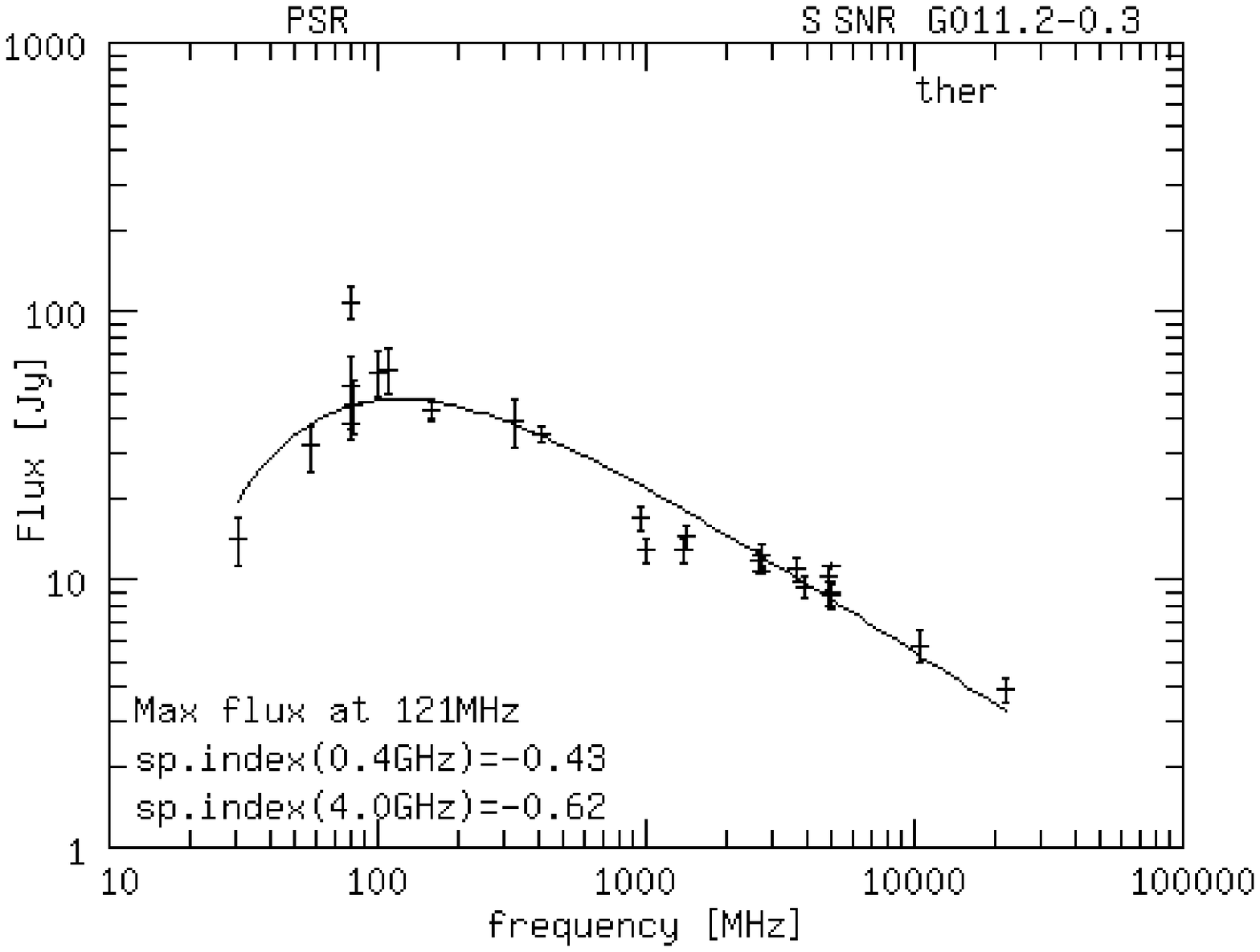,width=7.4cm,angle=0}}}\end{figure}\clearpage
\begin{figure}\centerline{\vbox{\psfig{figure=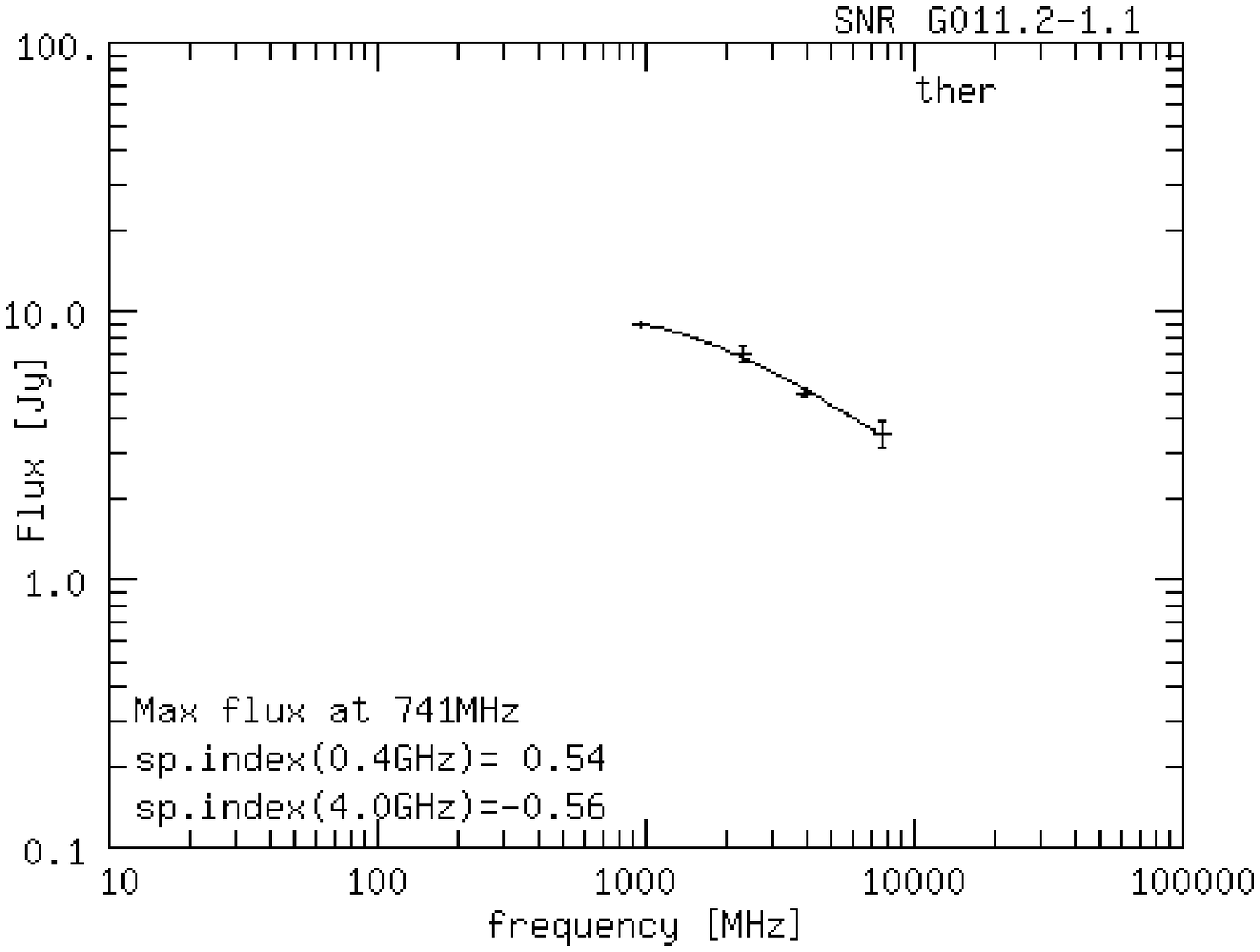,width=7.4cm,angle=0}}}\end{figure}
\begin{figure}\centerline{\vbox{\psfig{figure=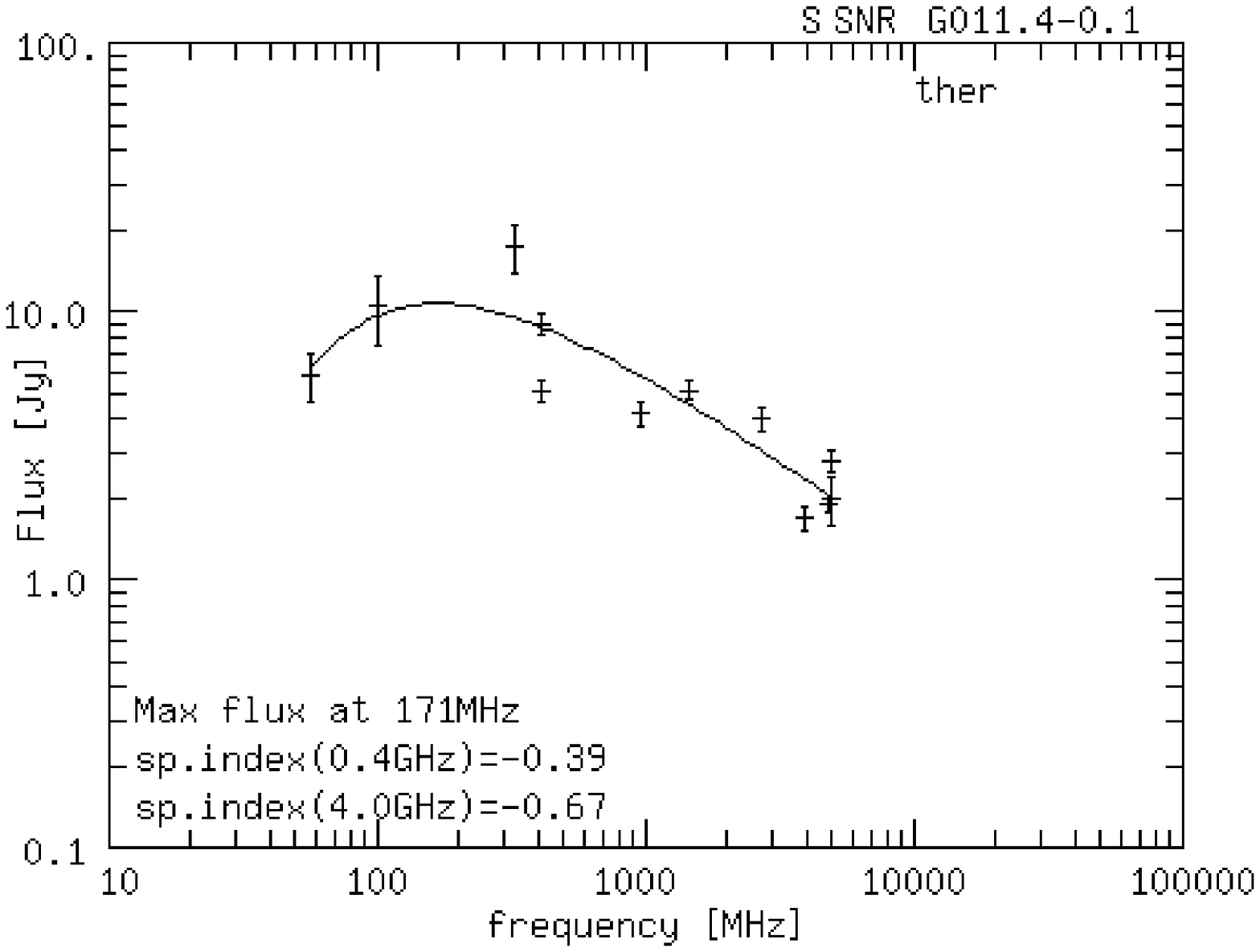,width=7.4cm,angle=0}}}\end{figure}
\begin{figure}\centerline{\vbox{\psfig{figure=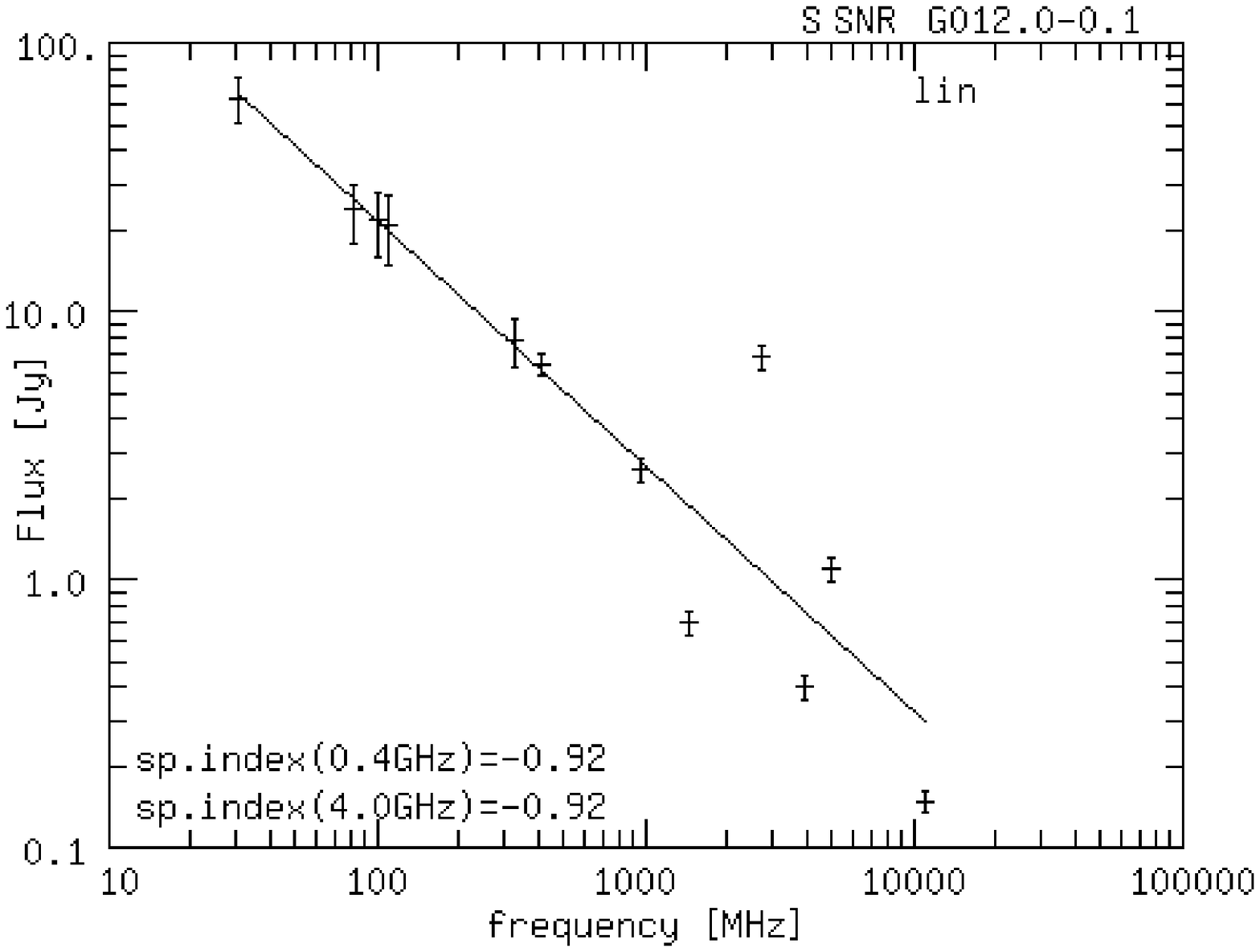,width=7.4cm,angle=0}}}\end{figure}
\begin{figure}\centerline{\vbox{\psfig{figure=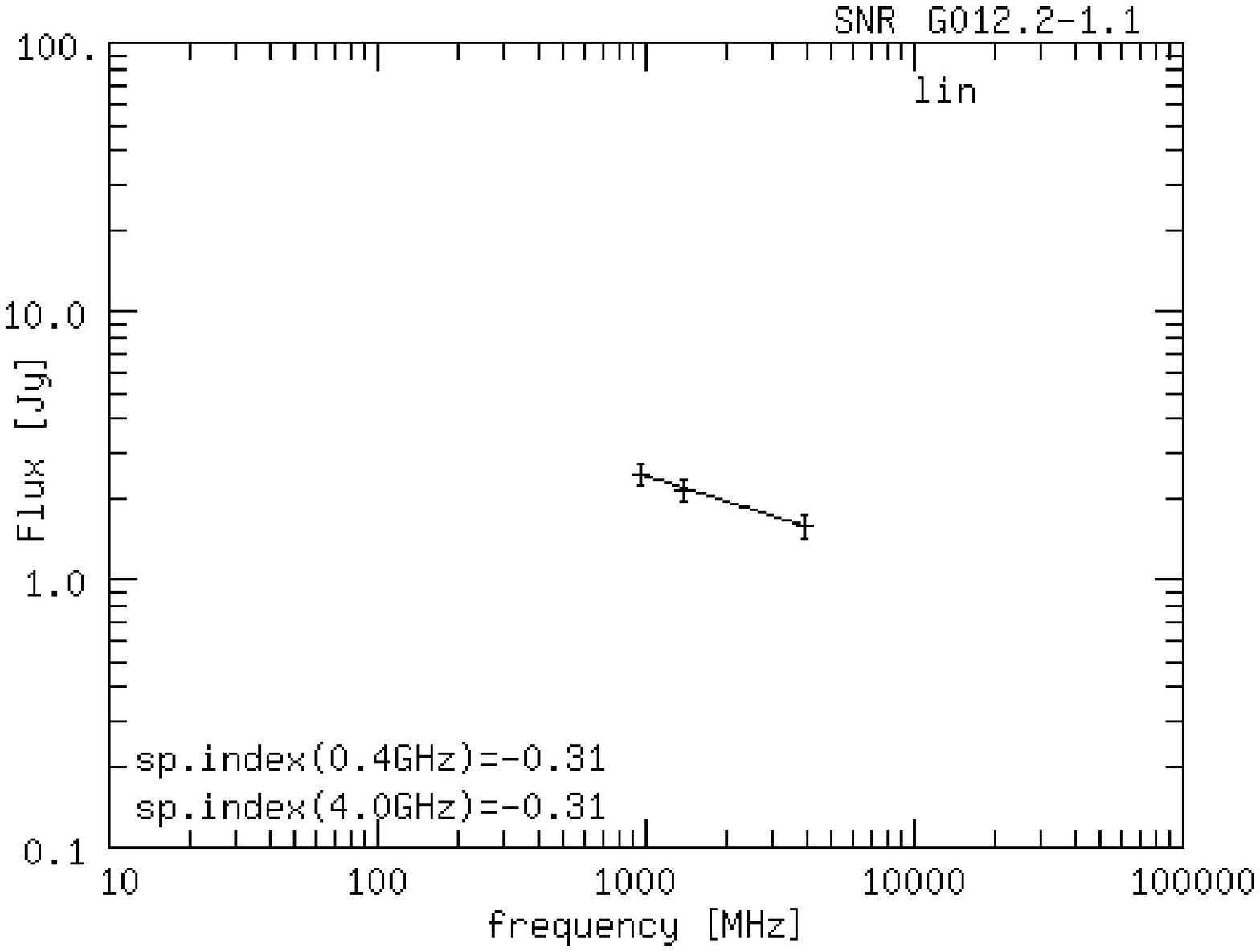,width=7.4cm,angle=0}}}\end{figure}
\begin{figure}\centerline{\vbox{\psfig{figure=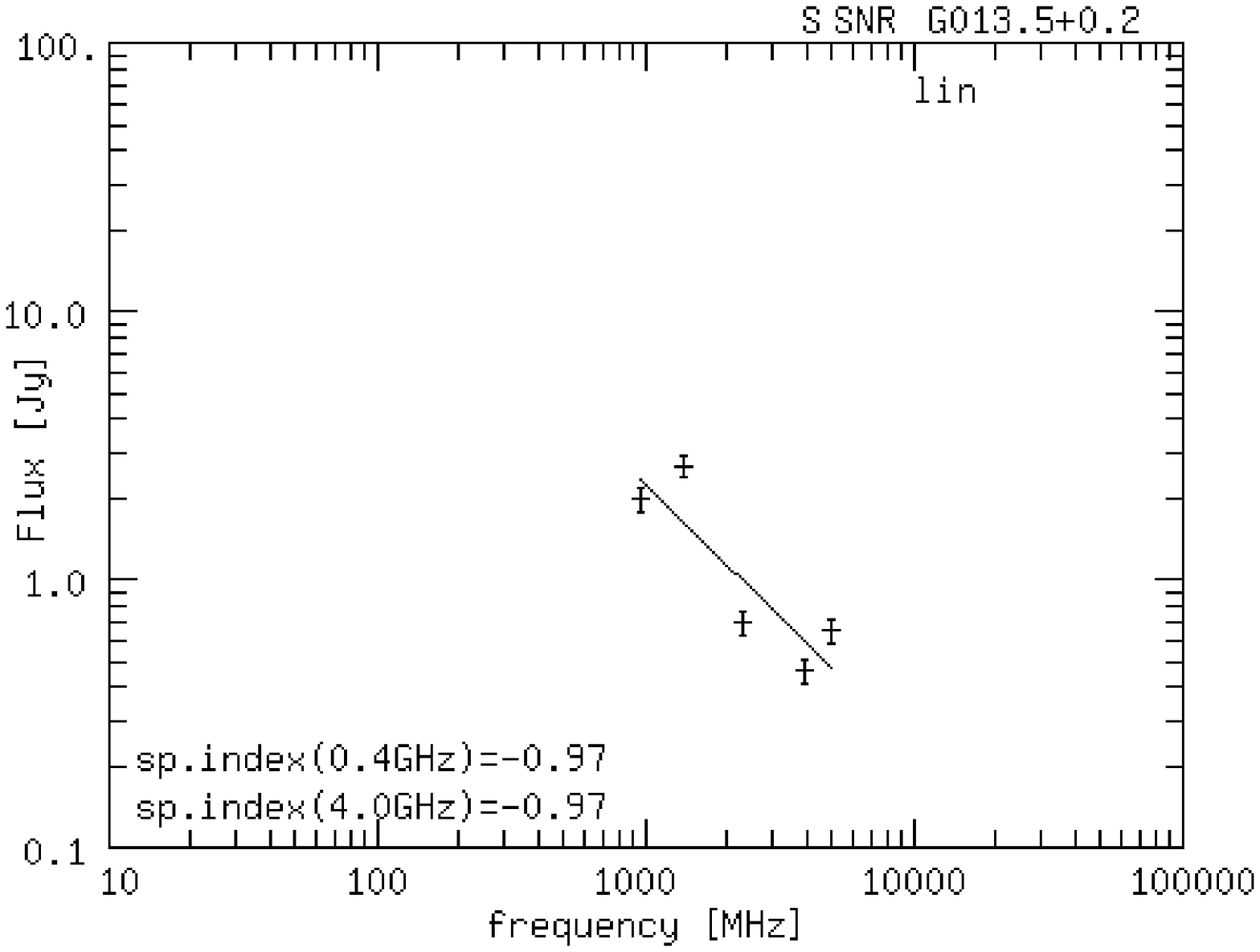,width=7.4cm,angle=0}}}\end{figure}
\begin{figure}\centerline{\vbox{\psfig{figure=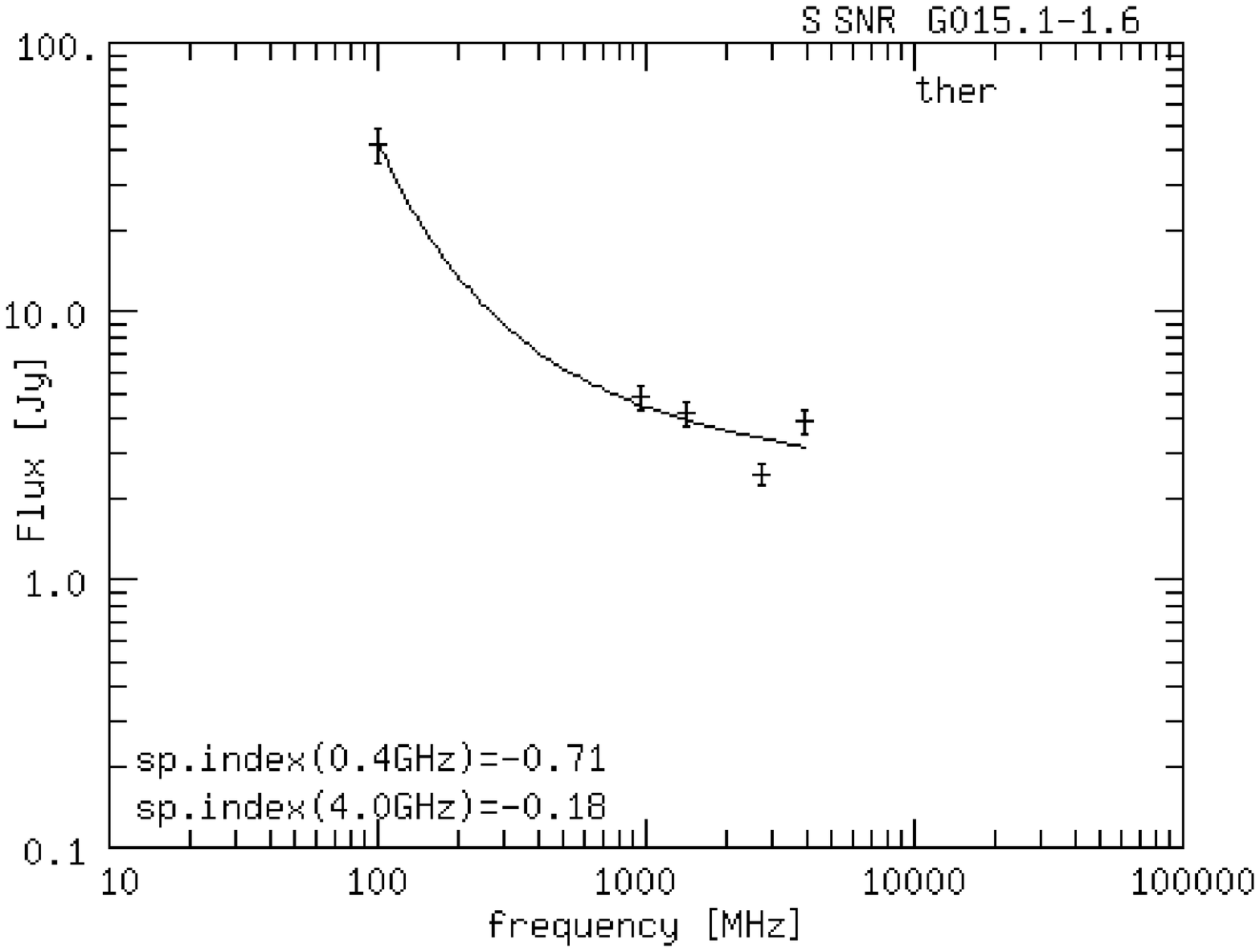,width=7.4cm,angle=0}}}\end{figure}
\begin{figure}\centerline{\vbox{\psfig{figure=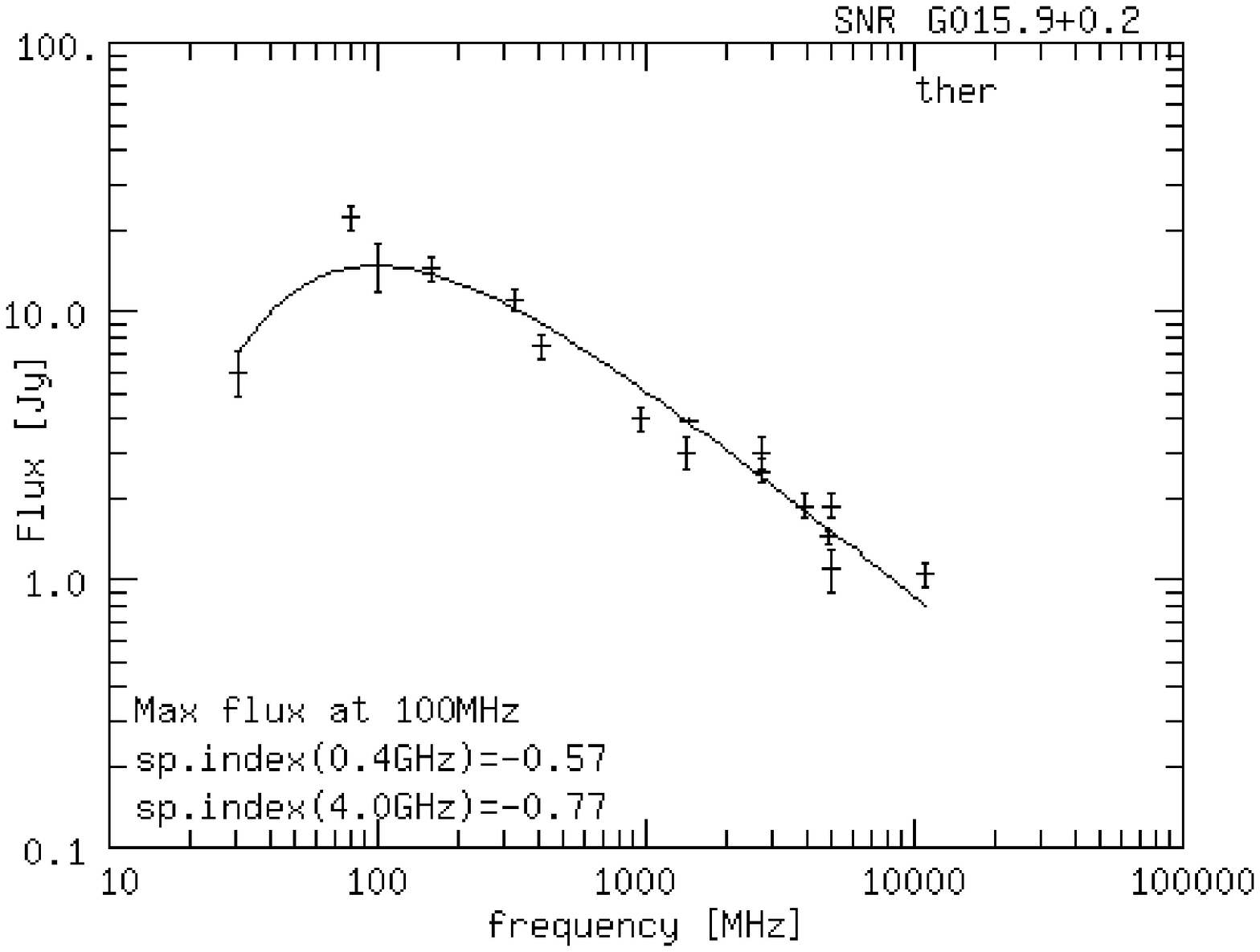,width=7.4cm,angle=0}}}\end{figure}
\begin{figure}\centerline{\vbox{\psfig{figure=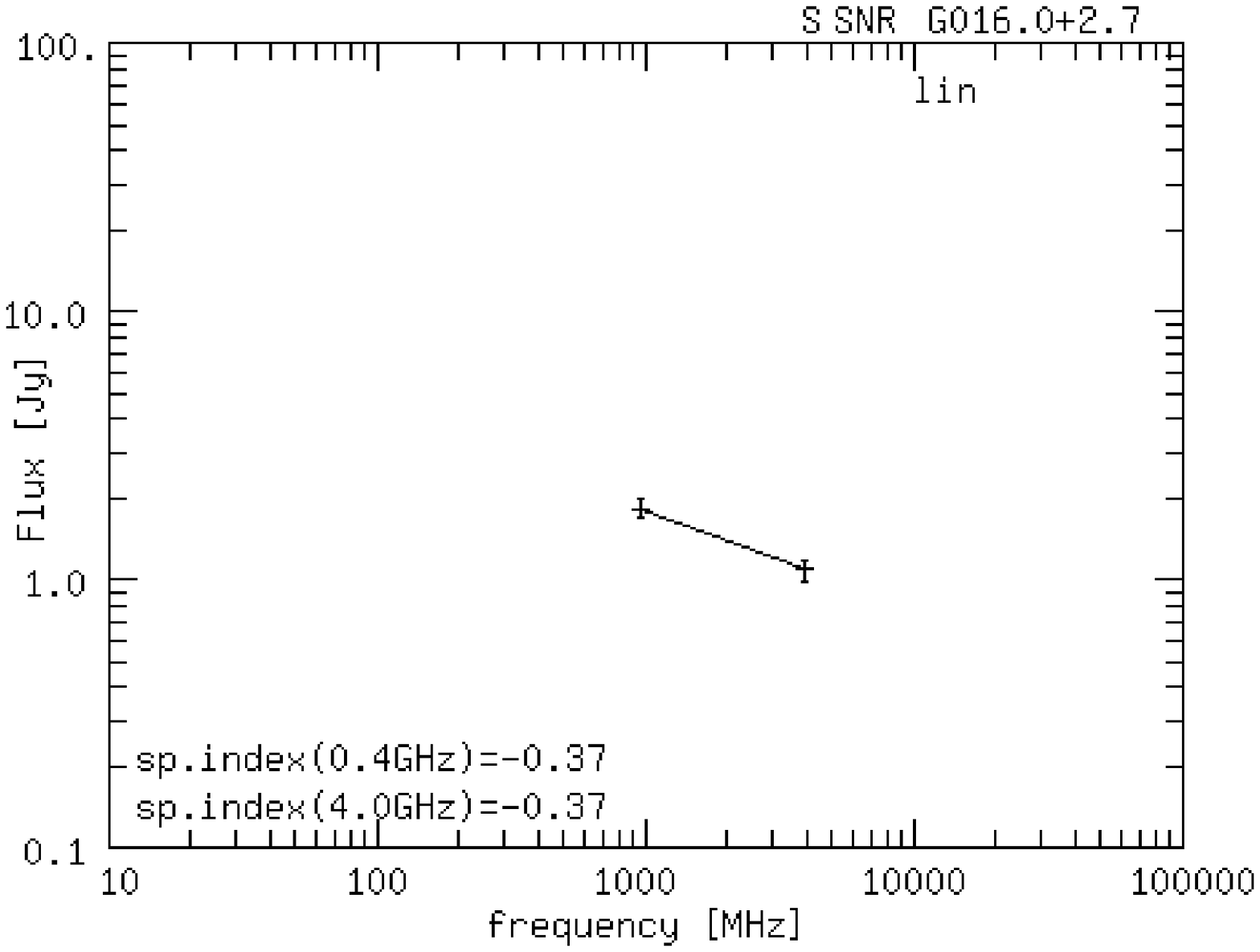,width=7.4cm,angle=0}}}\end{figure}\clearpage
\begin{figure}\centerline{\vbox{\psfig{figure=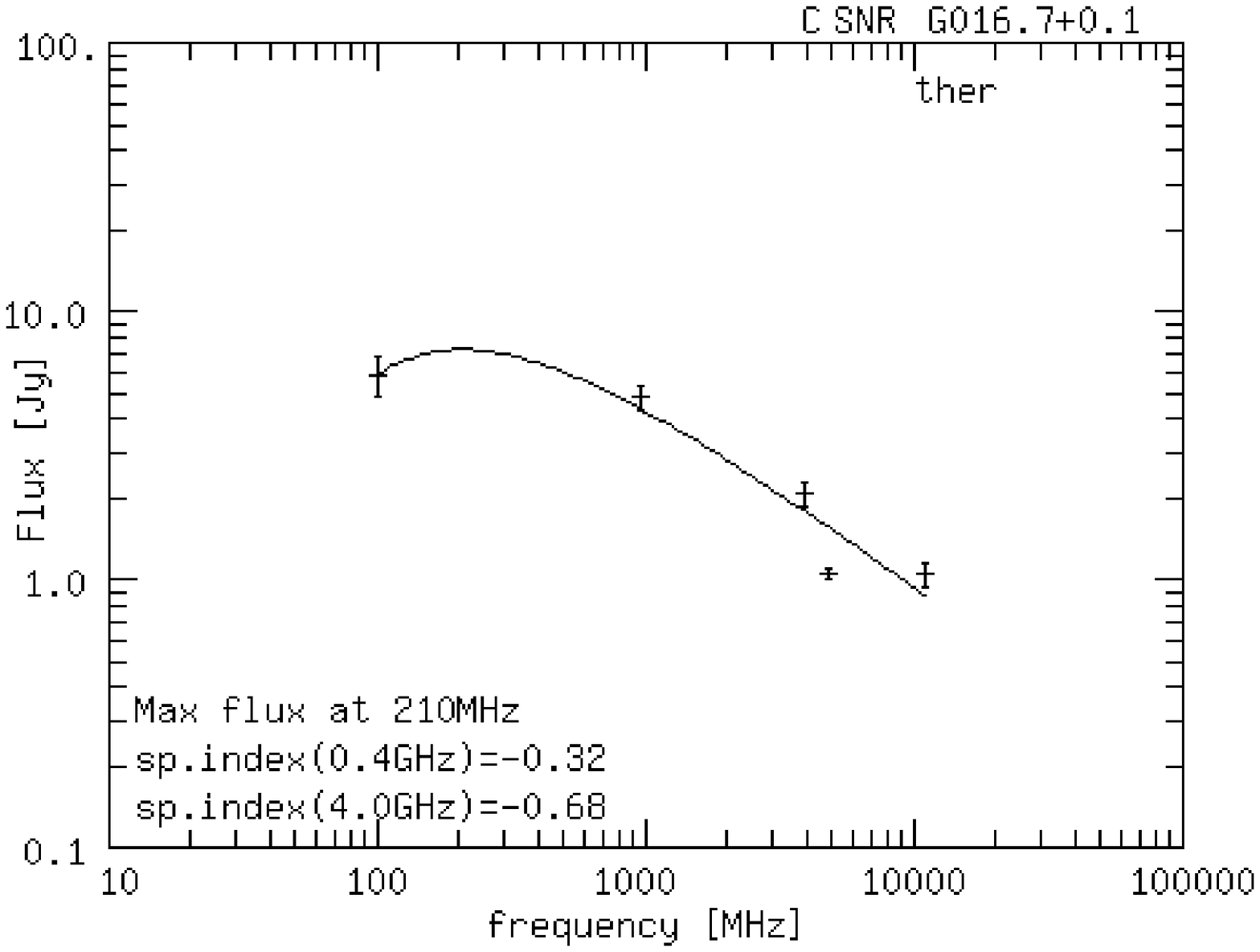,width=7.4cm,angle=0}}}\end{figure}
\begin{figure}\centerline{\vbox{\psfig{figure=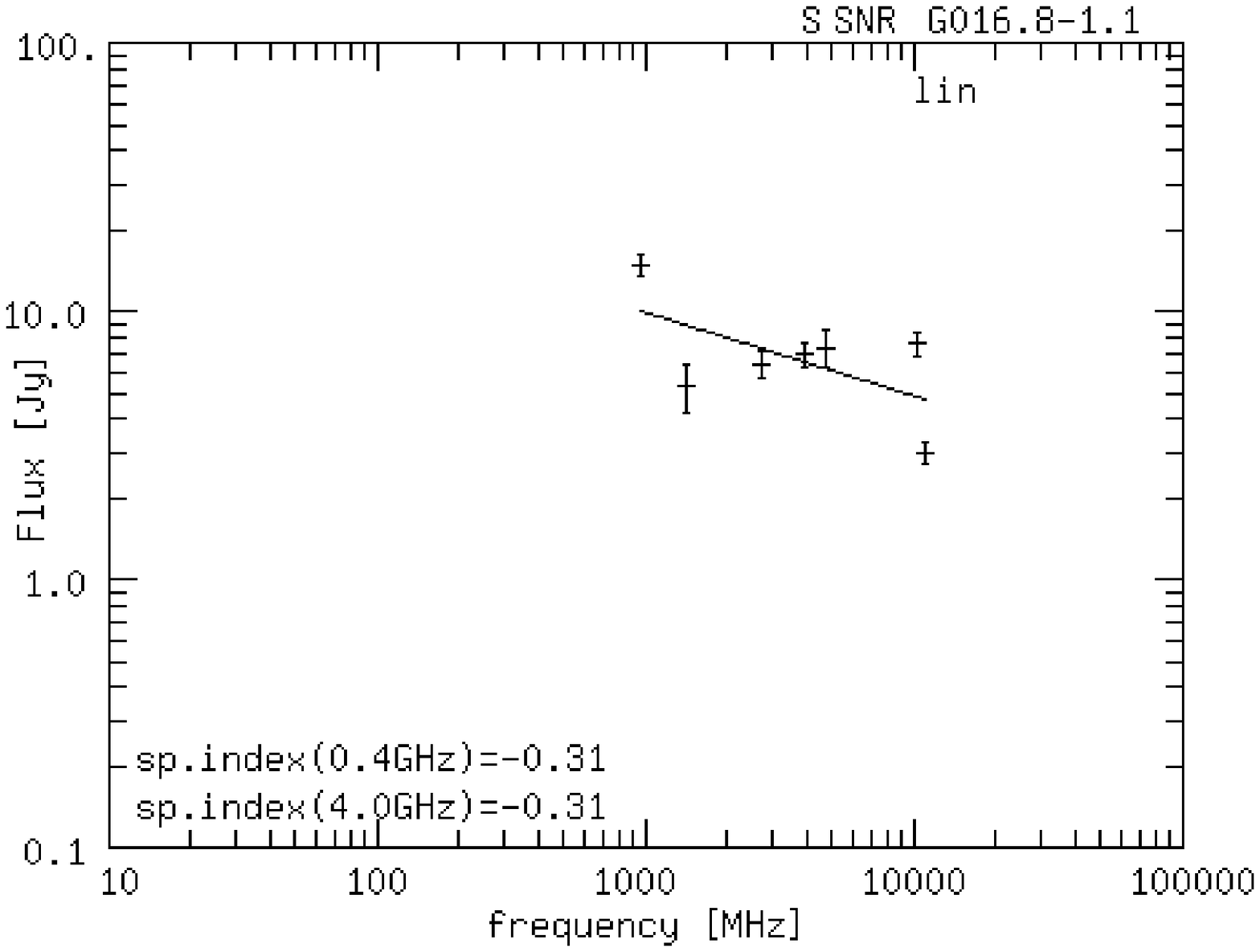,width=7.4cm,angle=0}}}\end{figure}
\begin{figure}\centerline{\vbox{\psfig{figure=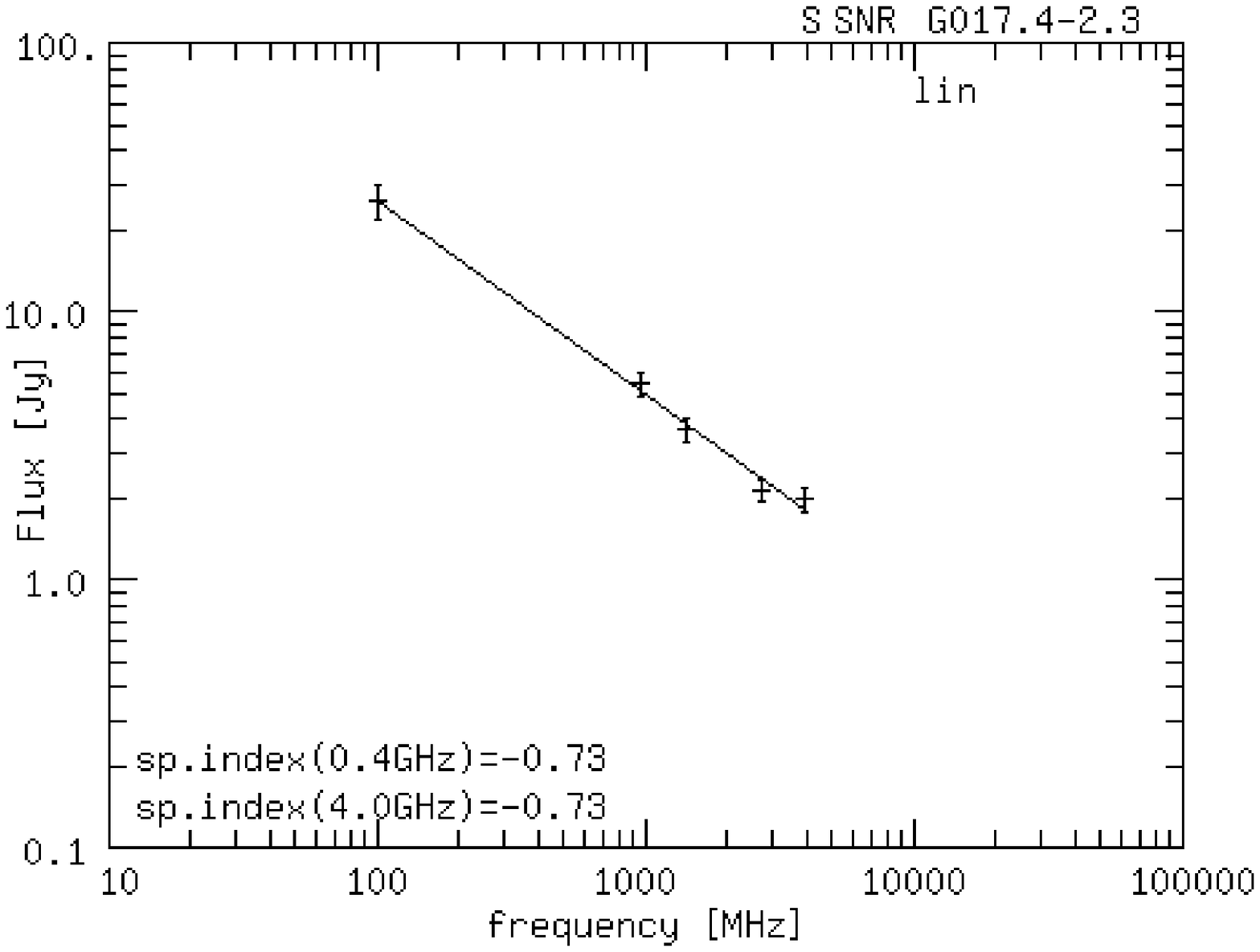,width=7.4cm,angle=0}}}\end{figure}
\begin{figure}\centerline{\vbox{\psfig{figure=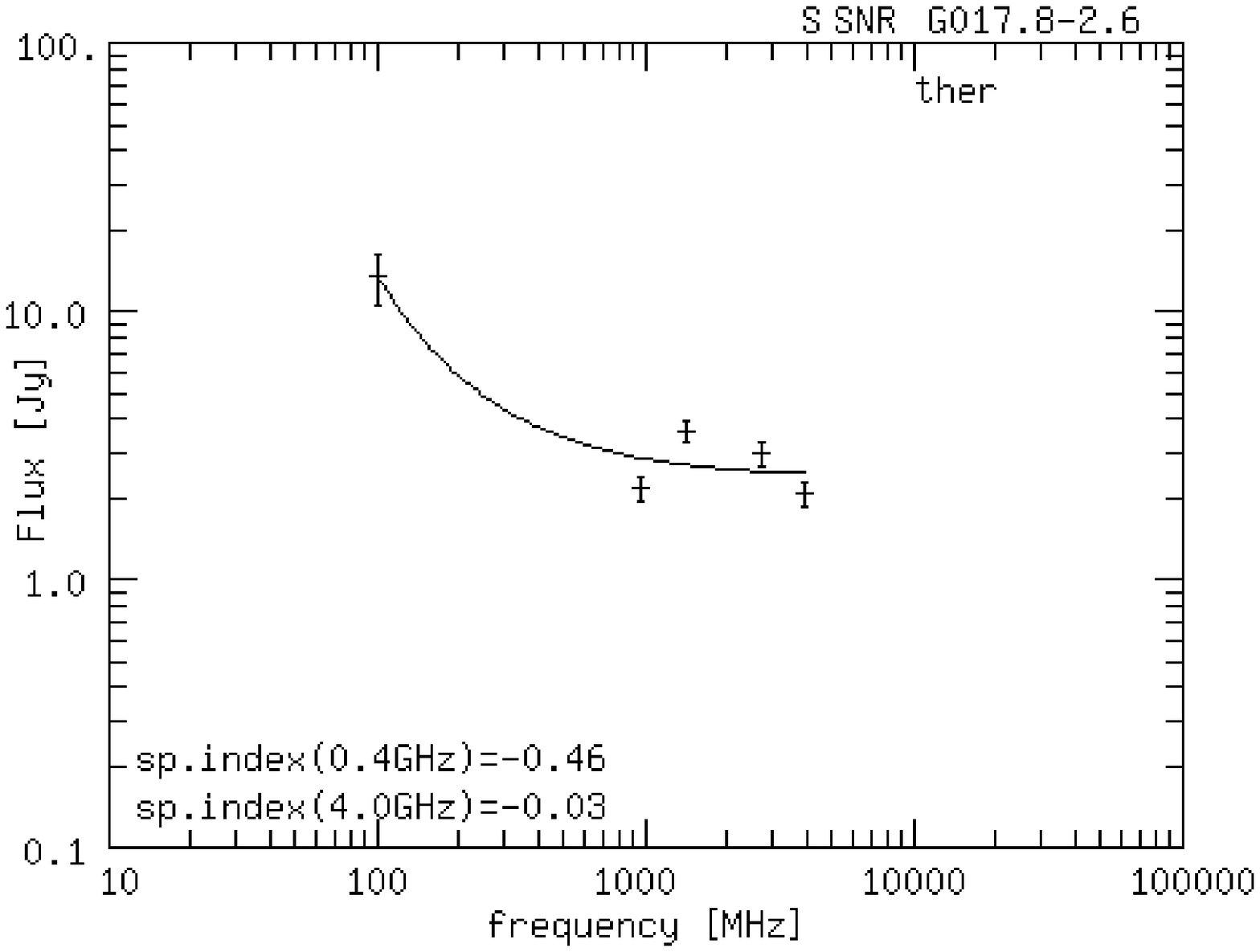,width=7.4cm,angle=0}}}\end{figure}
\begin{figure}\centerline{\vbox{\psfig{figure=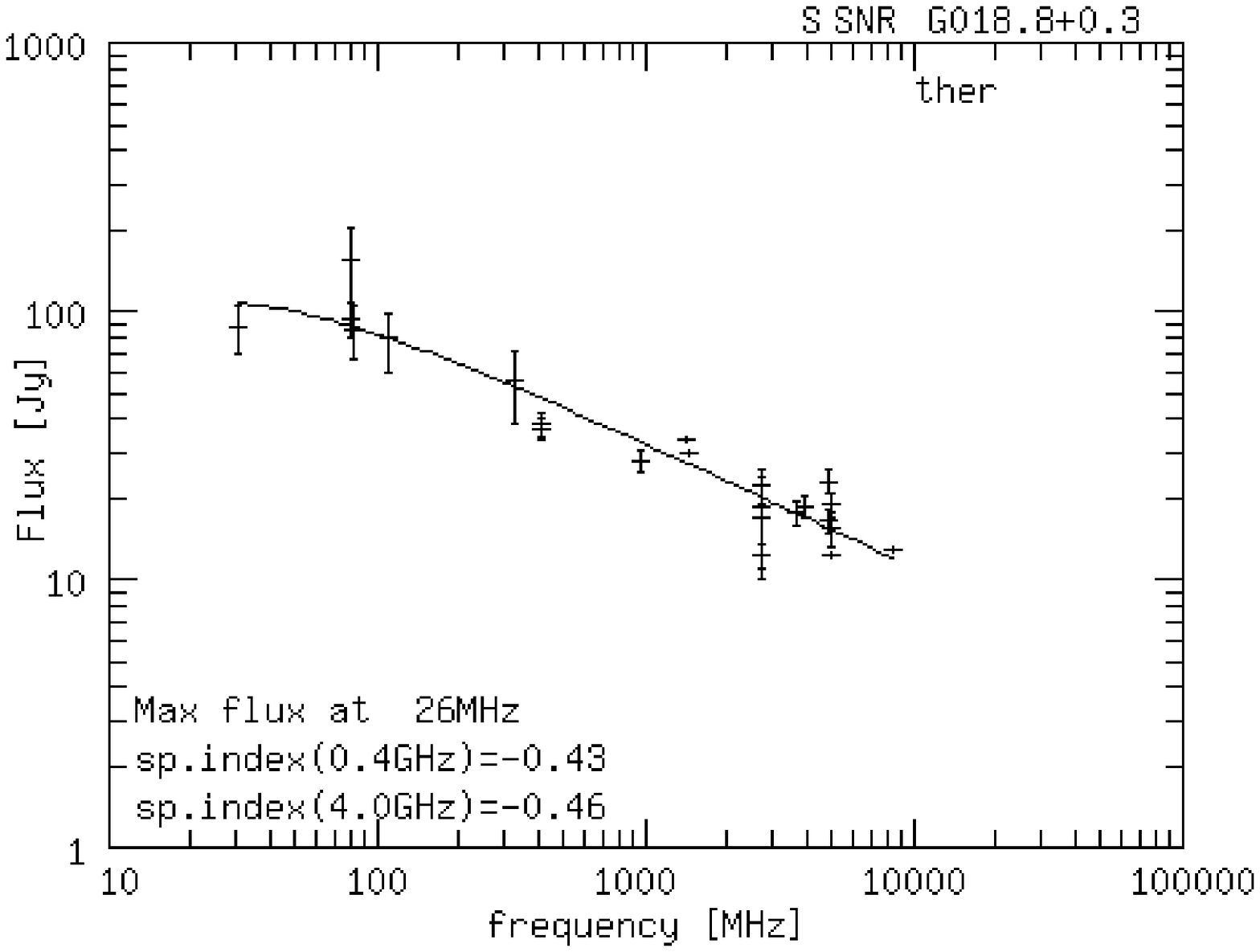,width=7.4cm,angle=0}}}\end{figure}
\begin{figure}\centerline{\vbox{\psfig{figure=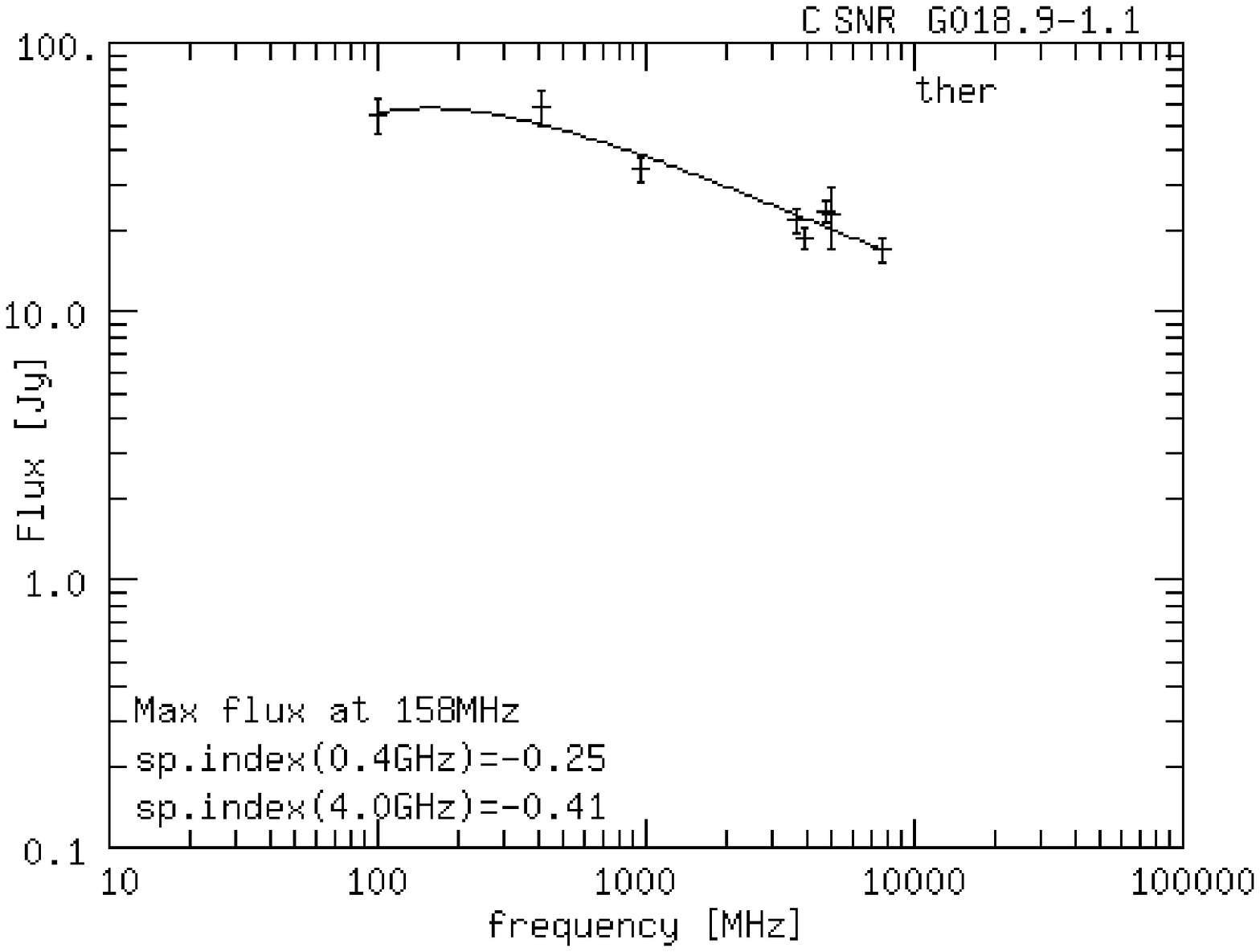,width=7.4cm,angle=0}}}\end{figure}
\begin{figure}\centerline{\vbox{\psfig{figure=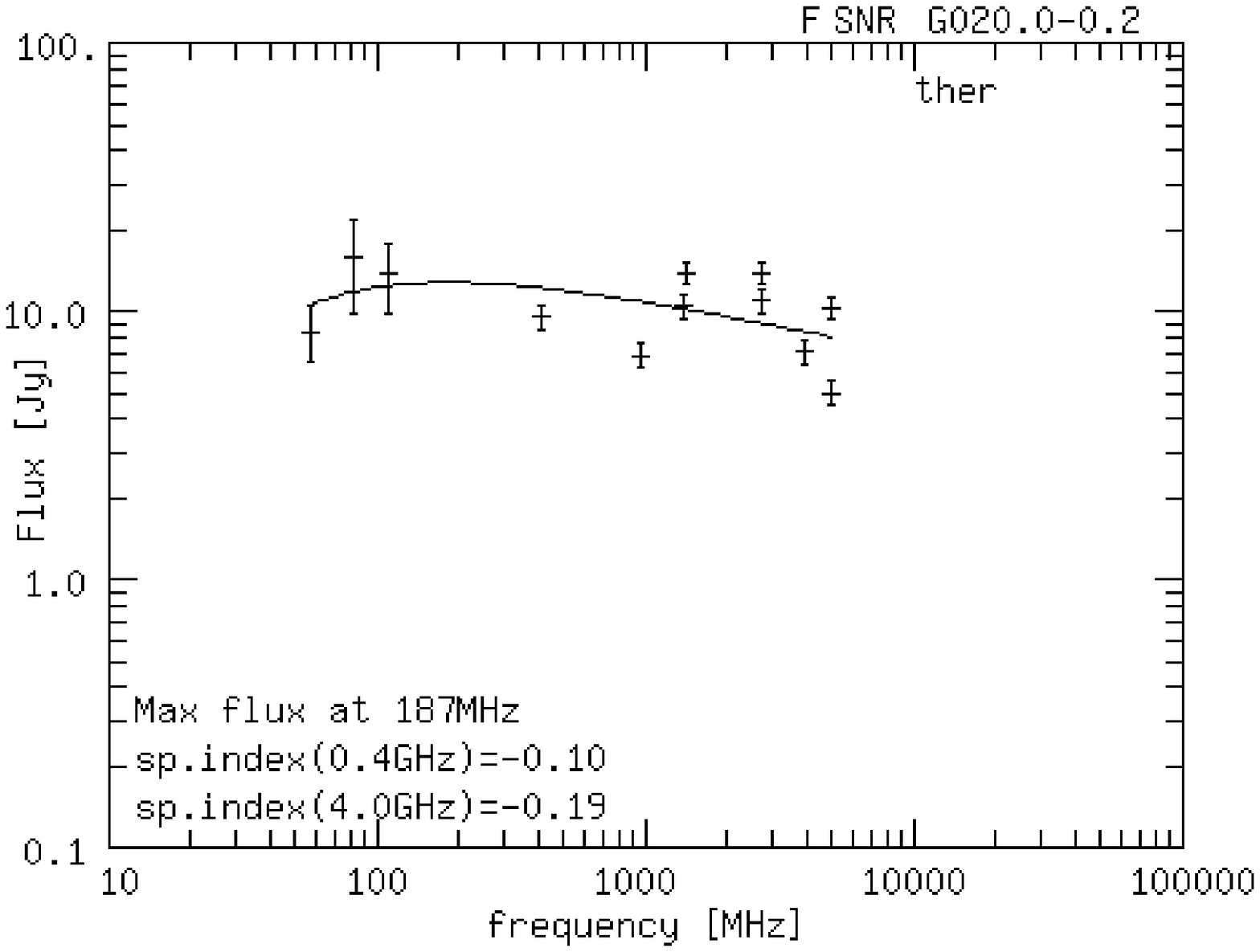,width=7.4cm,angle=0}}}\end{figure}
\begin{figure}\centerline{\vbox{\psfig{figure=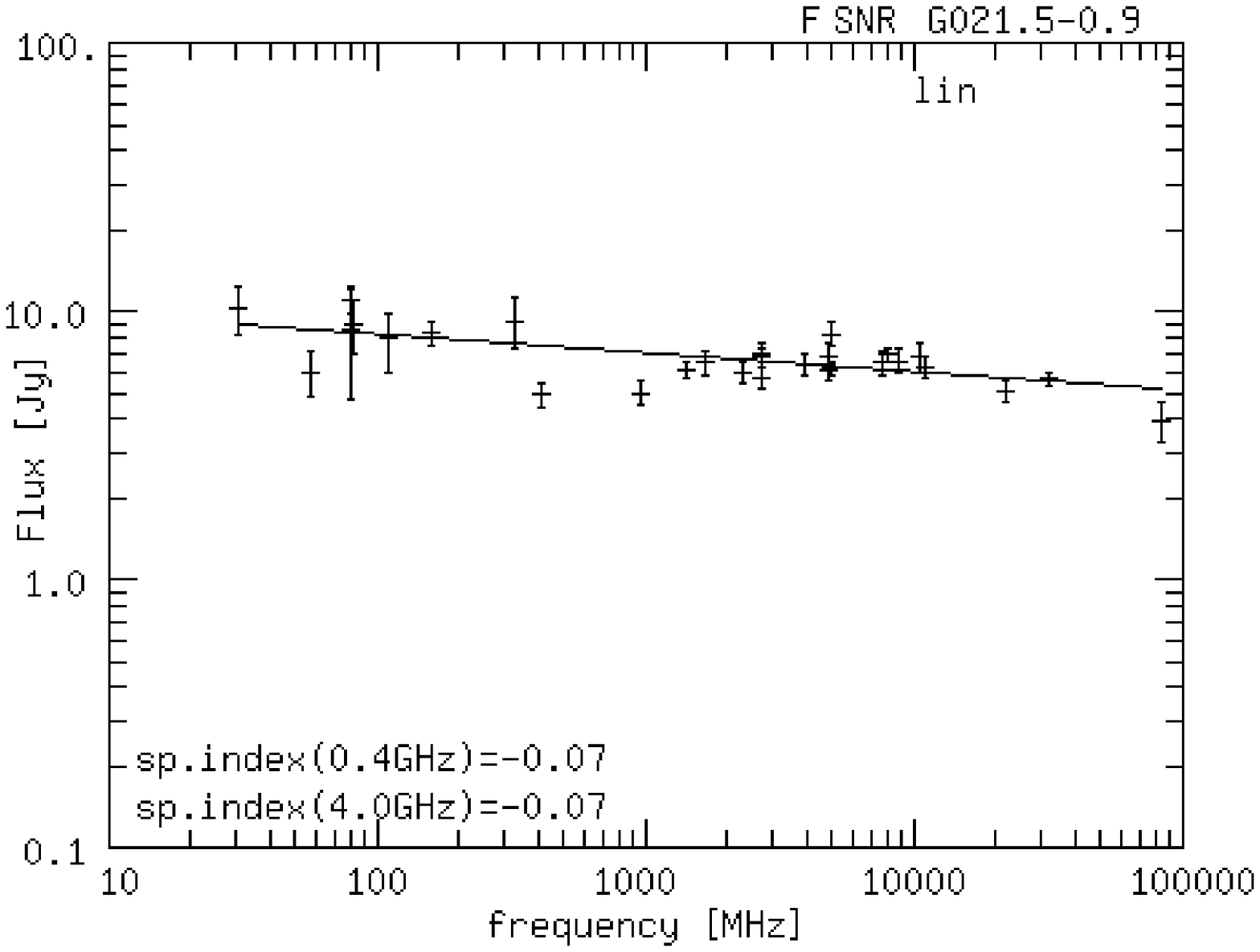,width=7.4cm,angle=0}}}\end{figure}\clearpage
\begin{figure}\centerline{\vbox{\psfig{figure=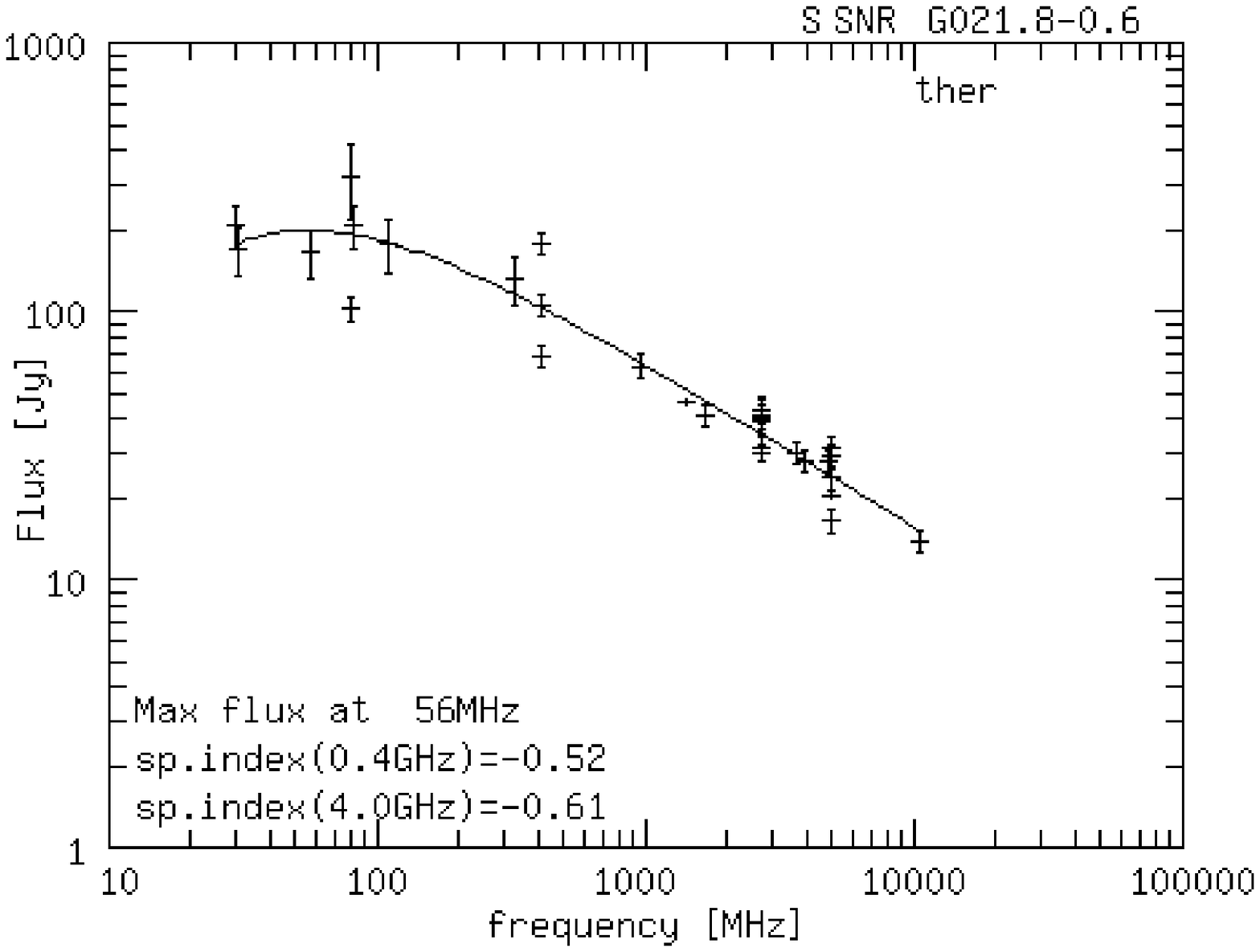,width=7.4cm,angle=0}}}\end{figure}
\begin{figure}\centerline{\vbox{\psfig{figure=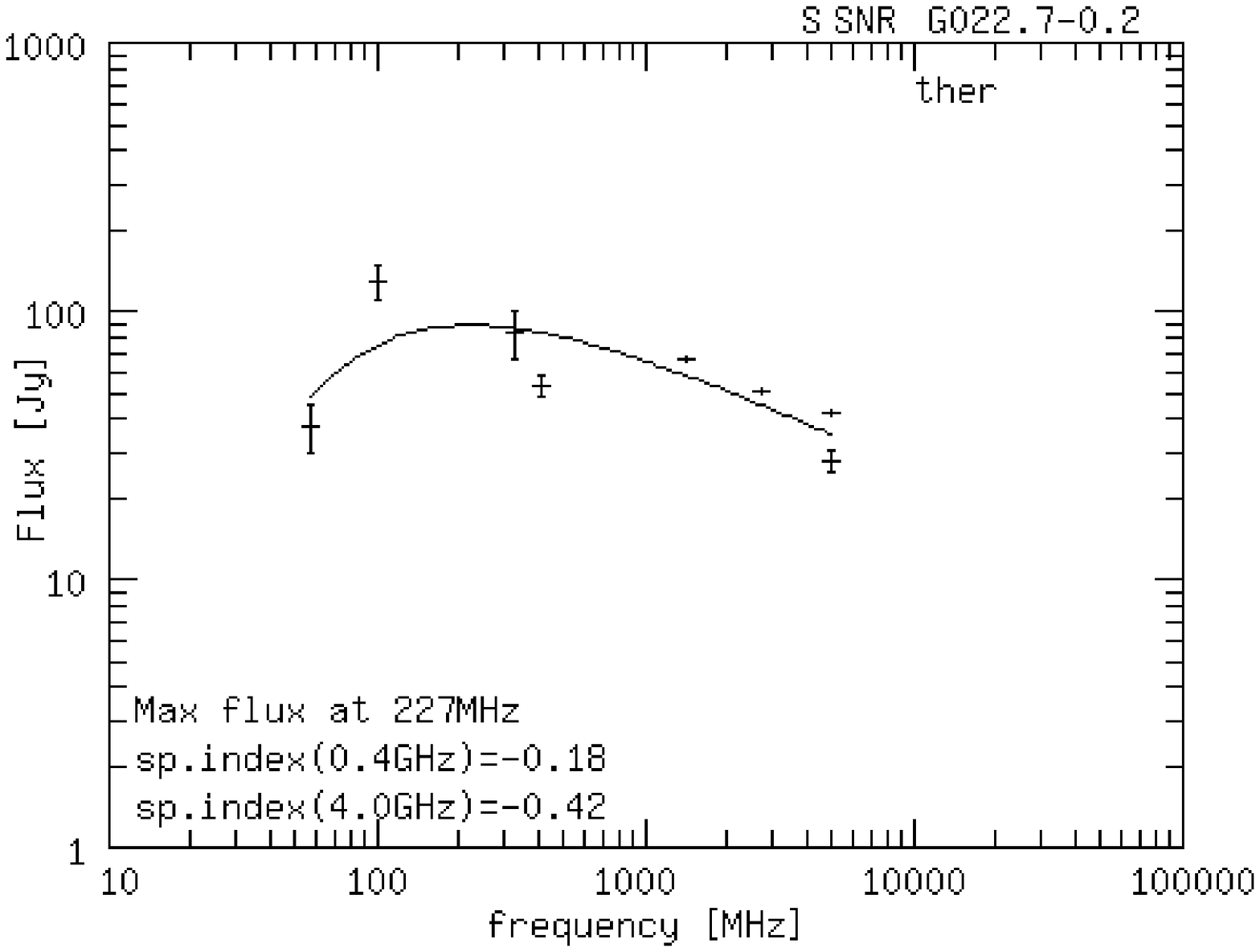,width=7.4cm,angle=0}}}\end{figure}
\begin{figure}\centerline{\vbox{\psfig{figure=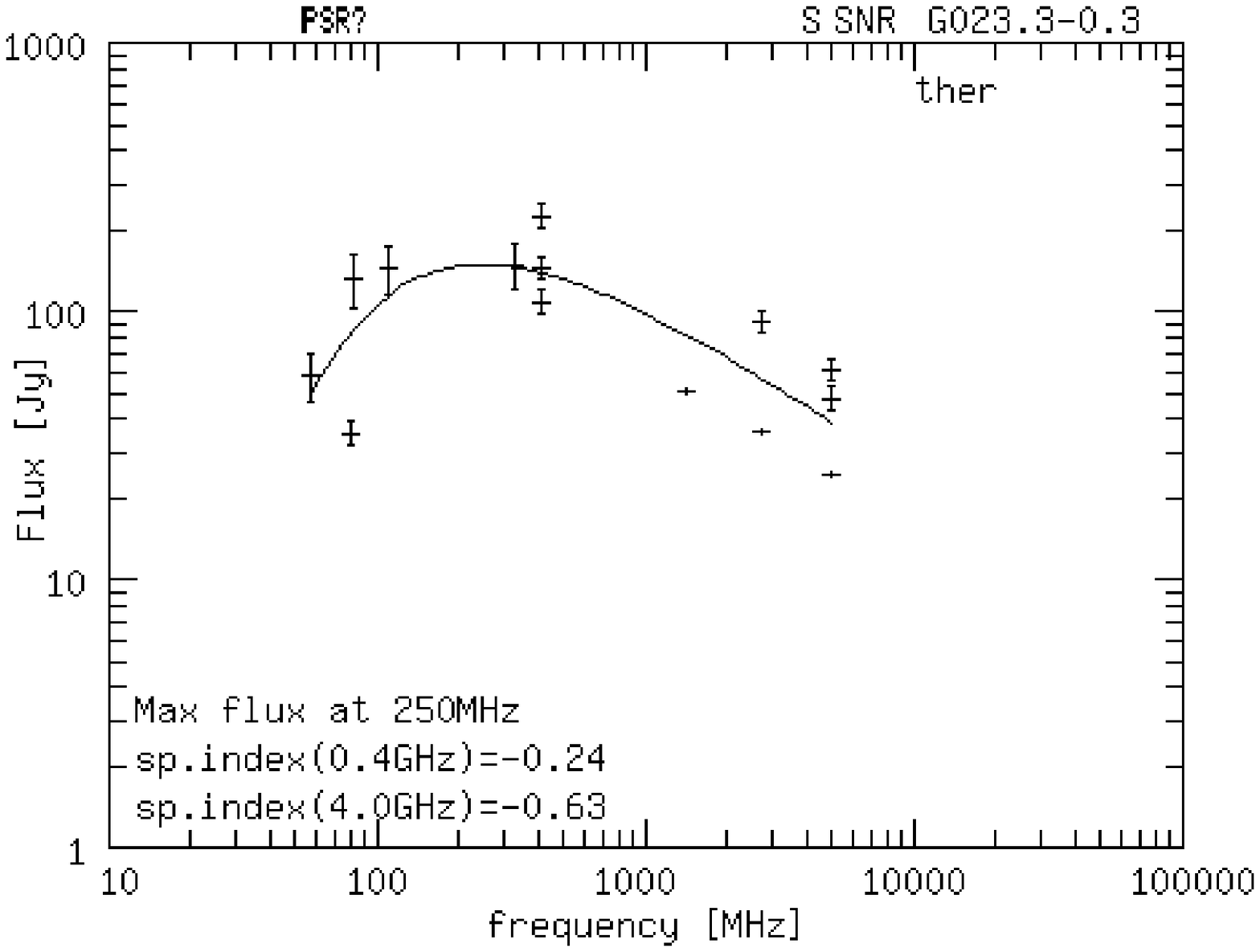,width=7.4cm,angle=0}}}\end{figure}
\begin{figure}\centerline{\vbox{\psfig{figure=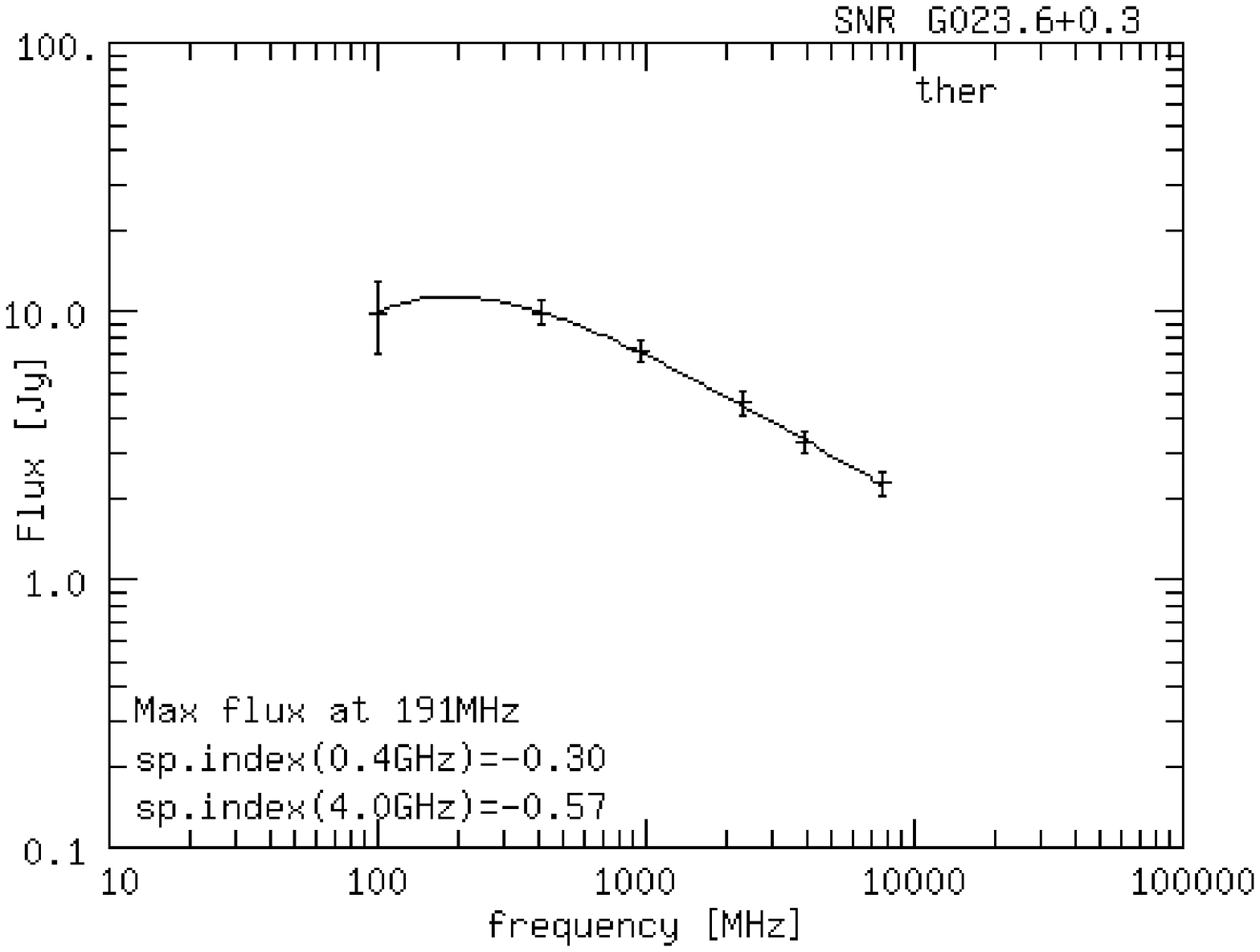,width=7.4cm,angle=0}}}\end{figure}
\begin{figure}\centerline{\vbox{\psfig{figure=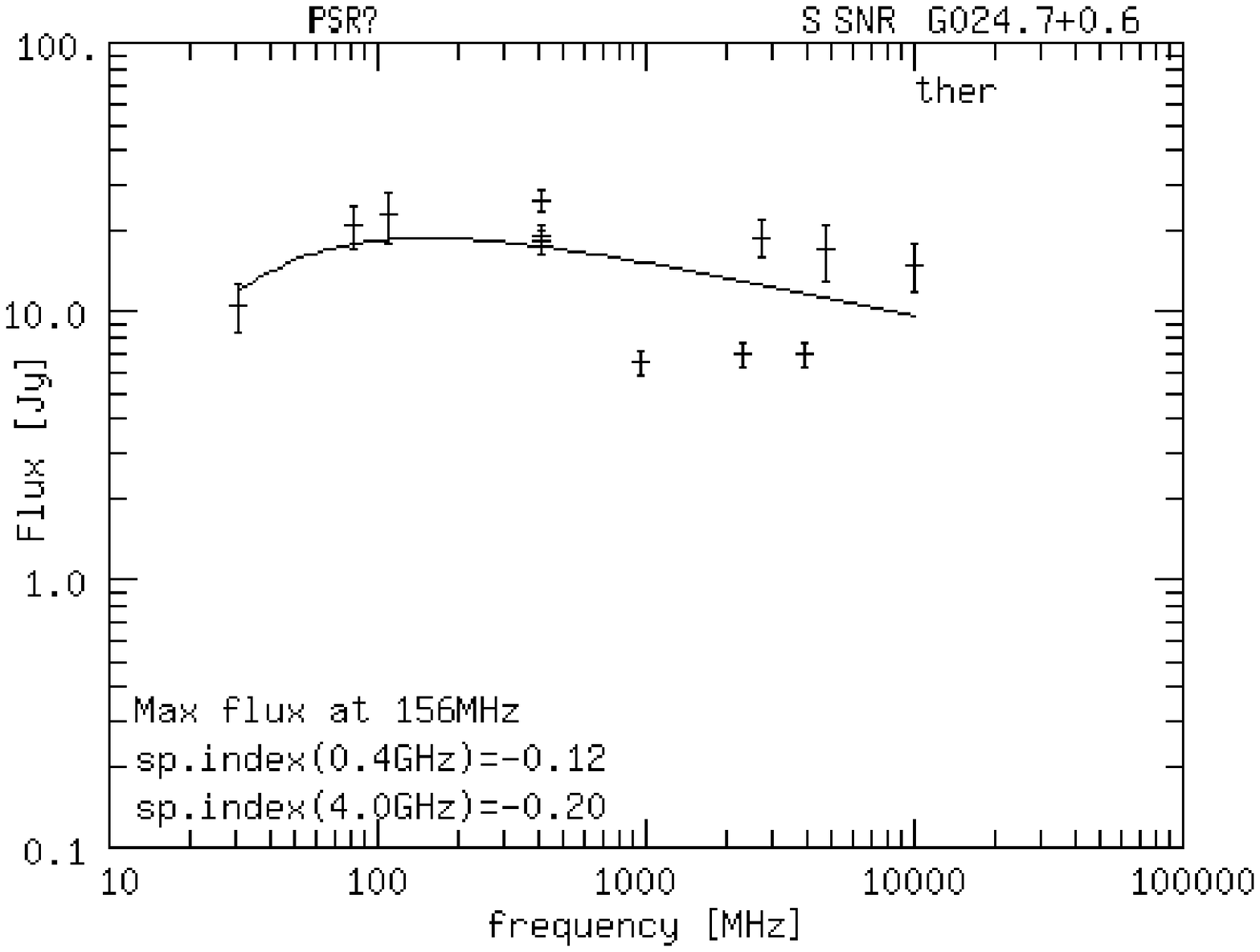,width=7.4cm,angle=0}}}\end{figure}
\begin{figure}\centerline{\vbox{\psfig{figure=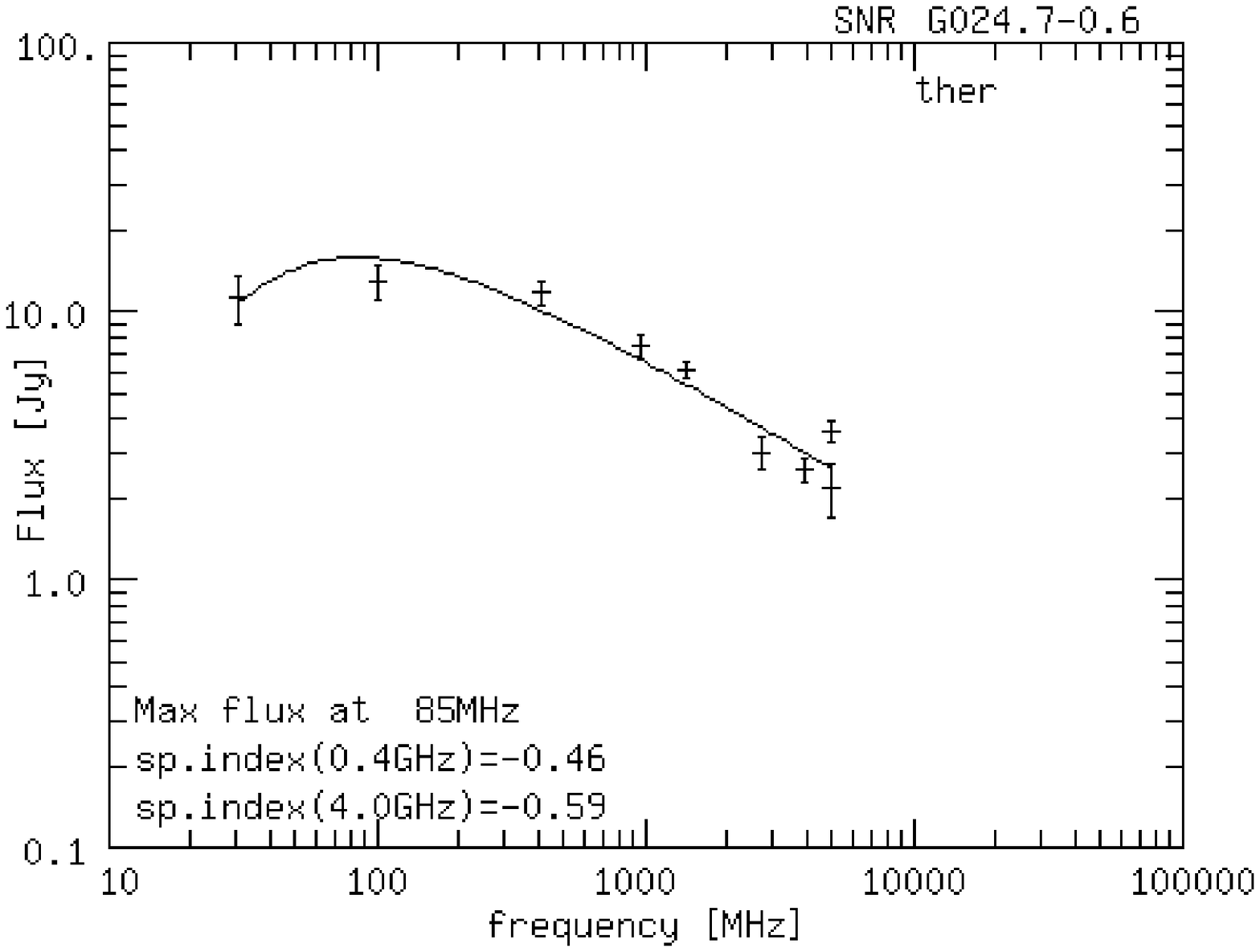,width=7.4cm,angle=0}}}\end{figure}
\begin{figure}\centerline{\vbox{\psfig{figure=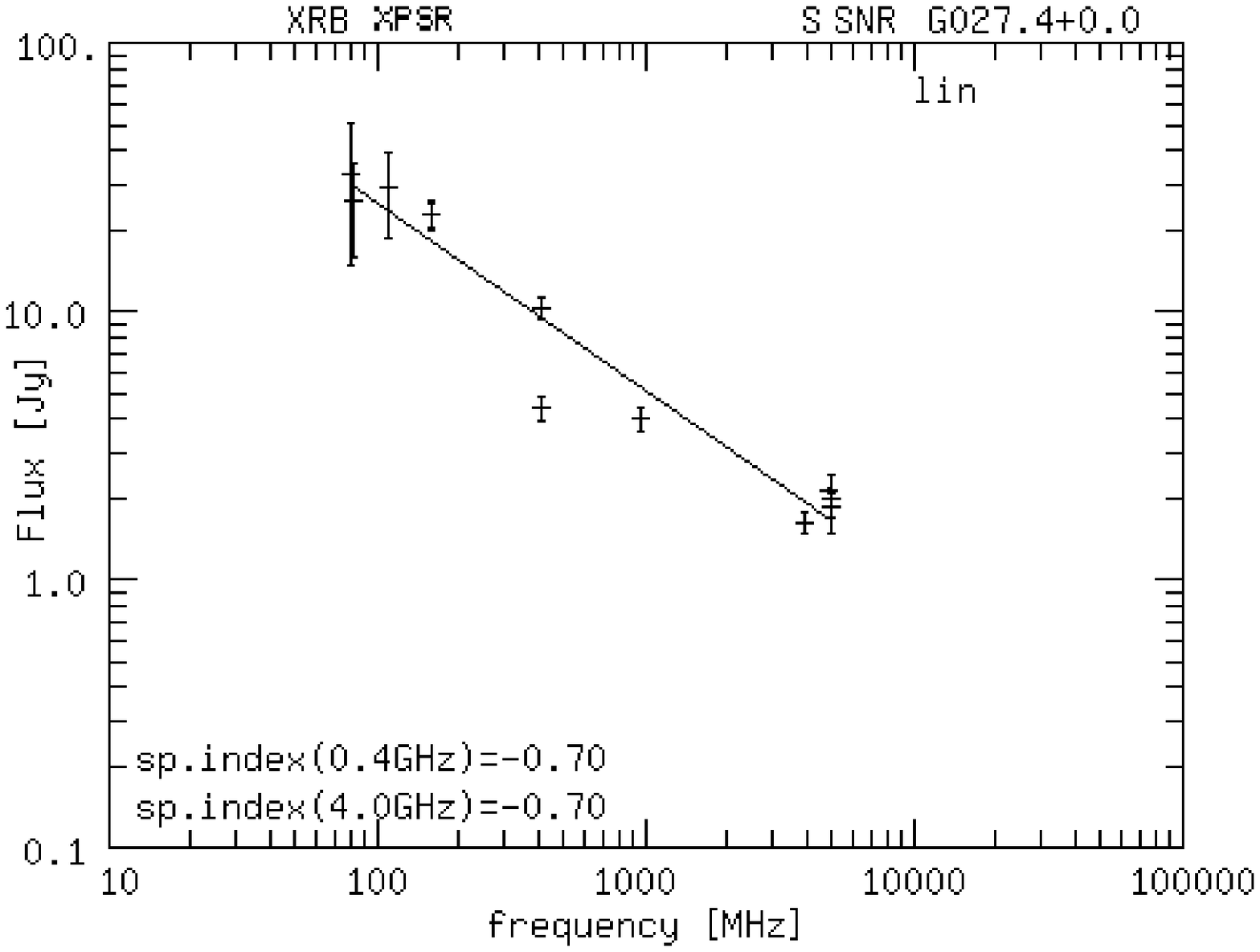,width=7.4cm,angle=0}}}\end{figure}
\begin{figure}\centerline{\vbox{\psfig{figure=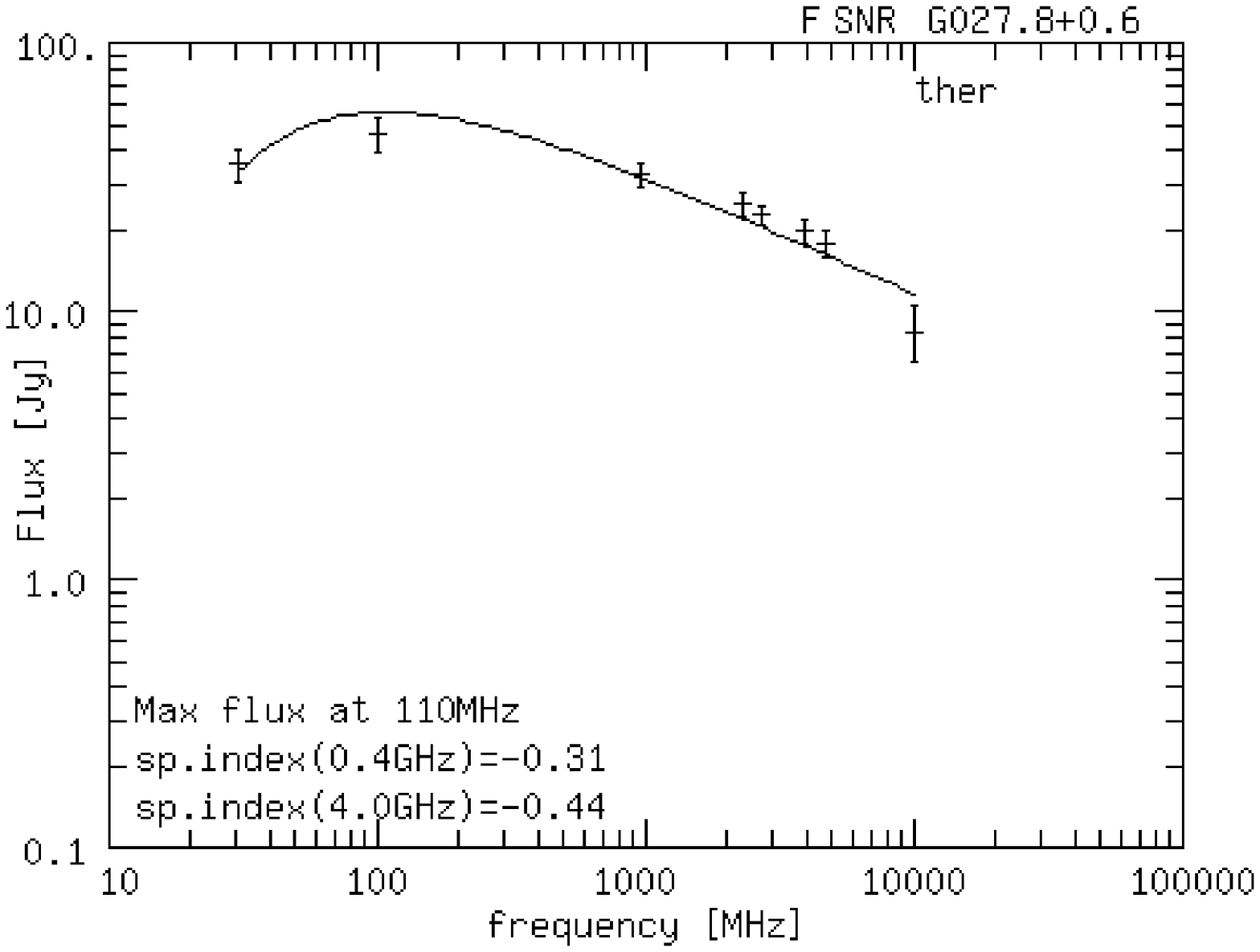,width=7.4cm,angle=0}}}\end{figure}\clearpage
\begin{figure}\centerline{\vbox{\psfig{figure=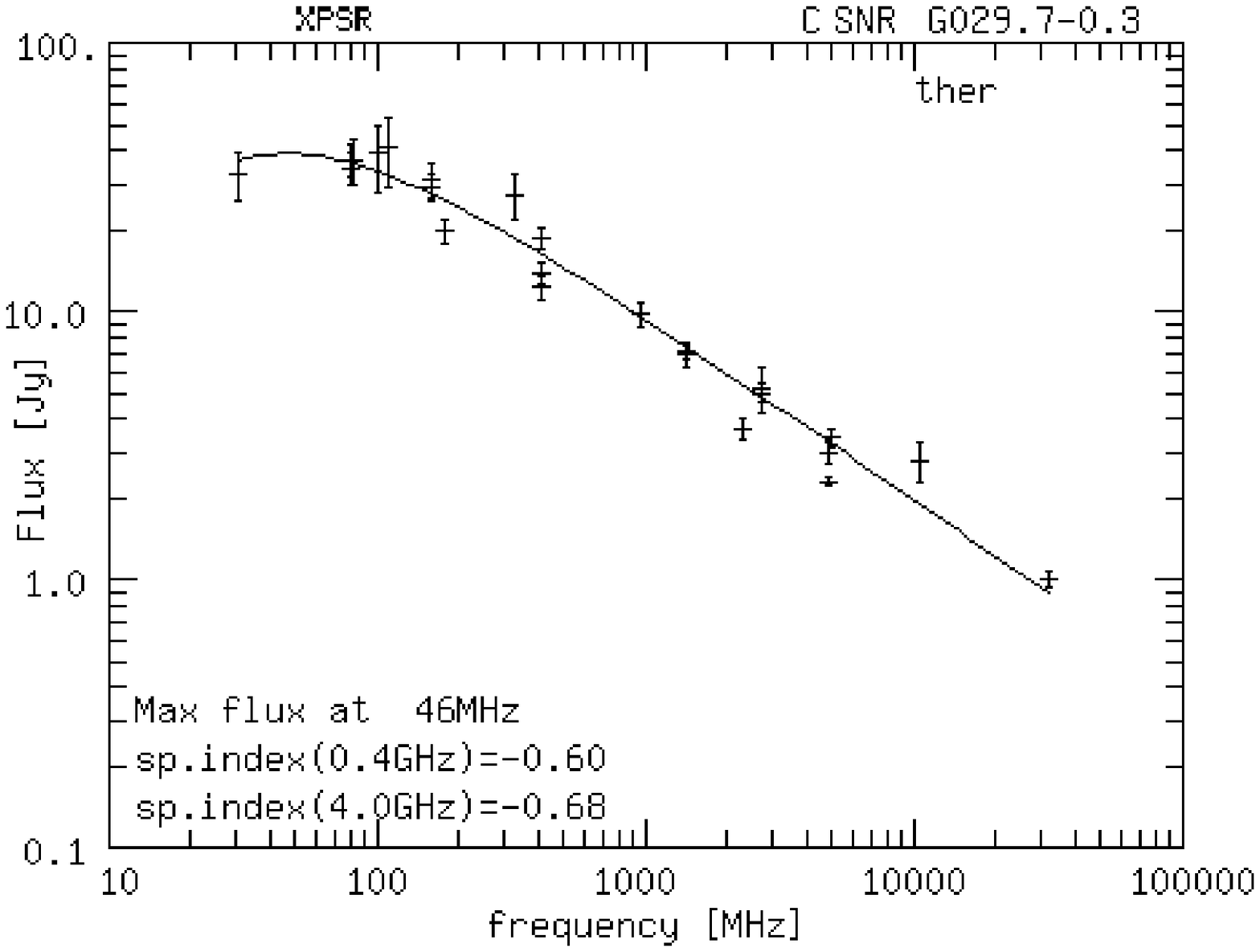,width=7.4cm,angle=0}}}\end{figure}
\begin{figure}\centerline{\vbox{\psfig{figure=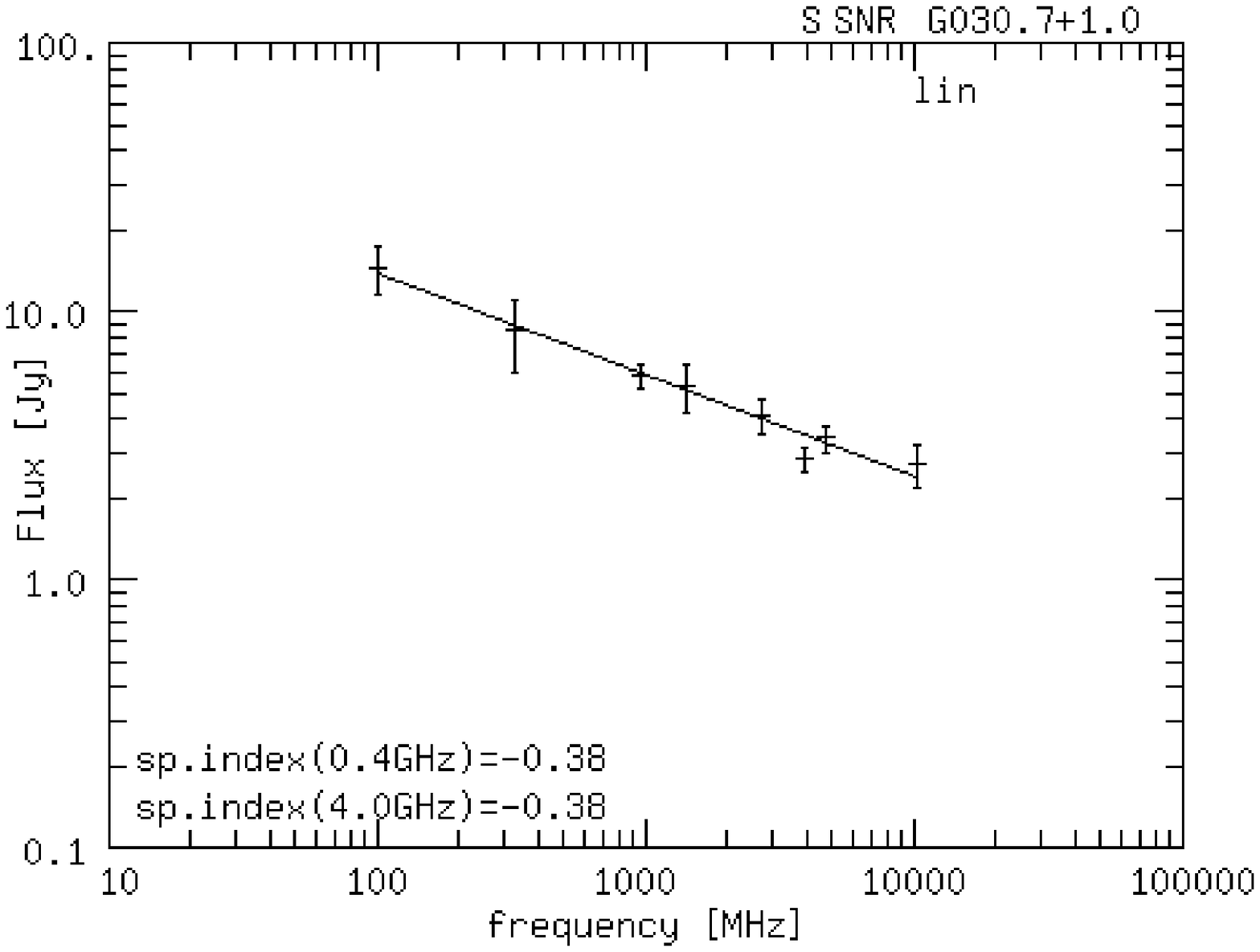,width=7.4cm,angle=0}}}\end{figure}
\begin{figure}\centerline{\vbox{\psfig{figure=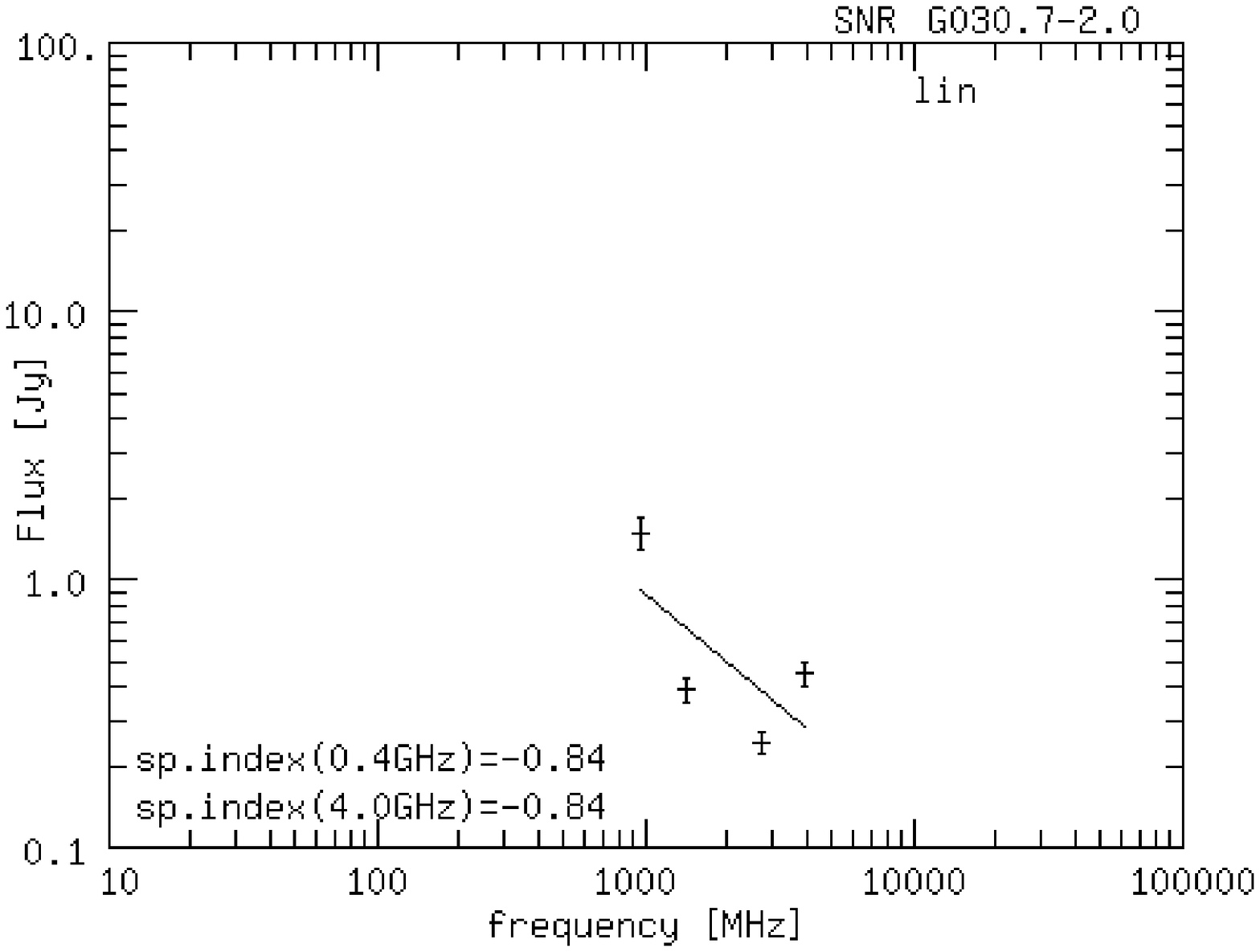,width=7.4cm,angle=0}}}\end{figure}
\begin{figure}\centerline{\vbox{\psfig{figure=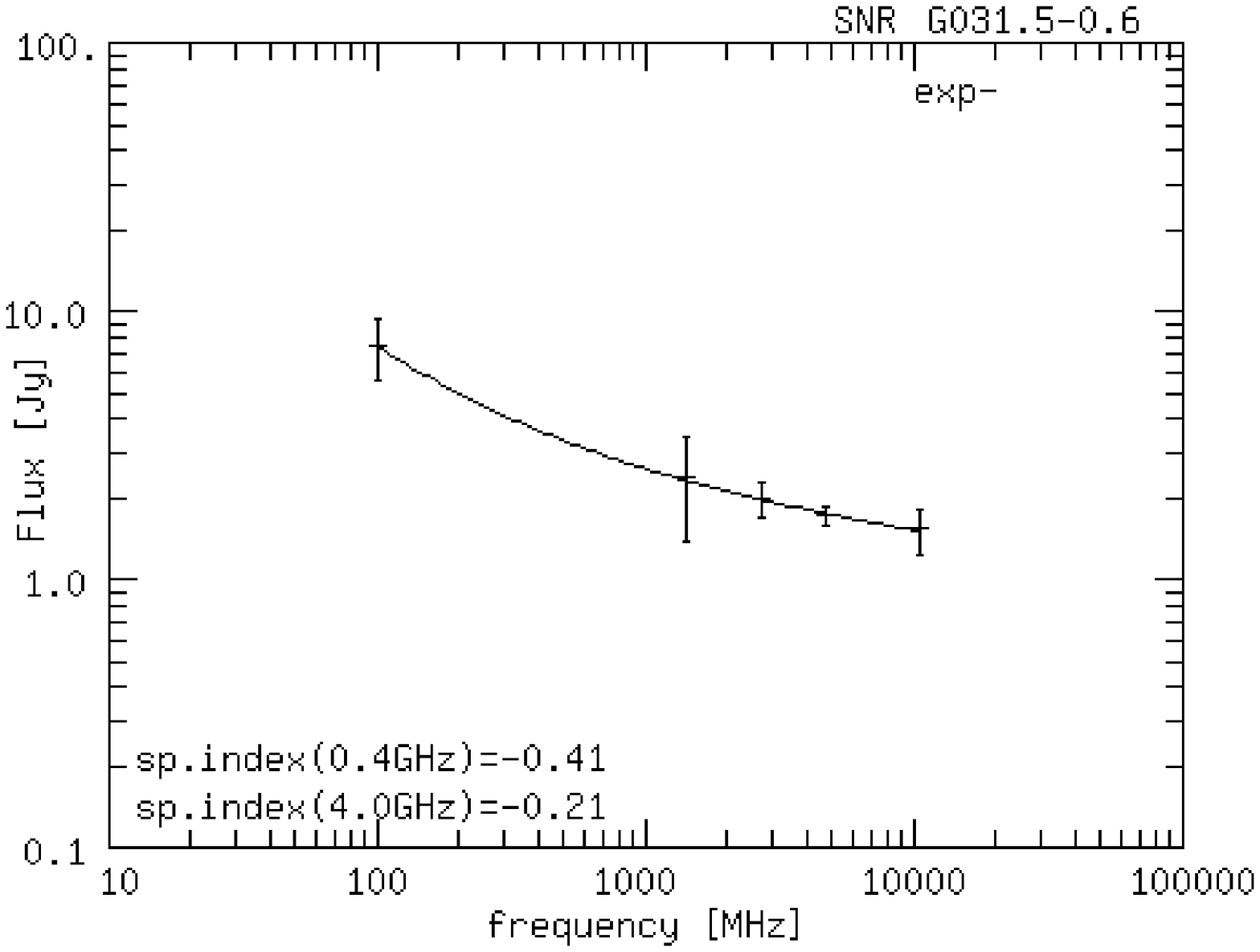,width=7.4cm,angle=0}}}\end{figure}
\begin{figure}\centerline{\vbox{\psfig{figure=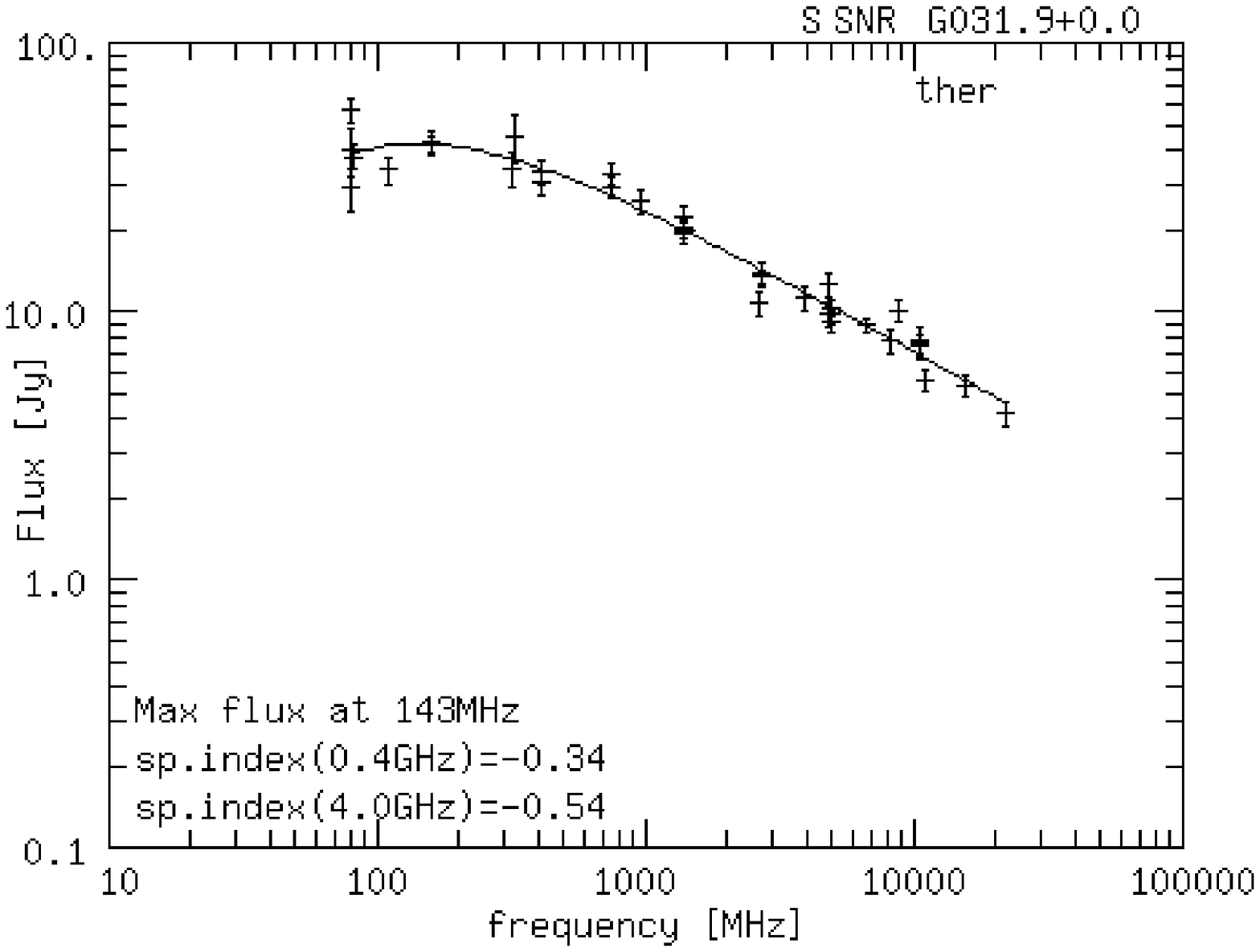,width=7.4cm,angle=0}}}\end{figure}
\begin{figure}\centerline{\vbox{\psfig{figure=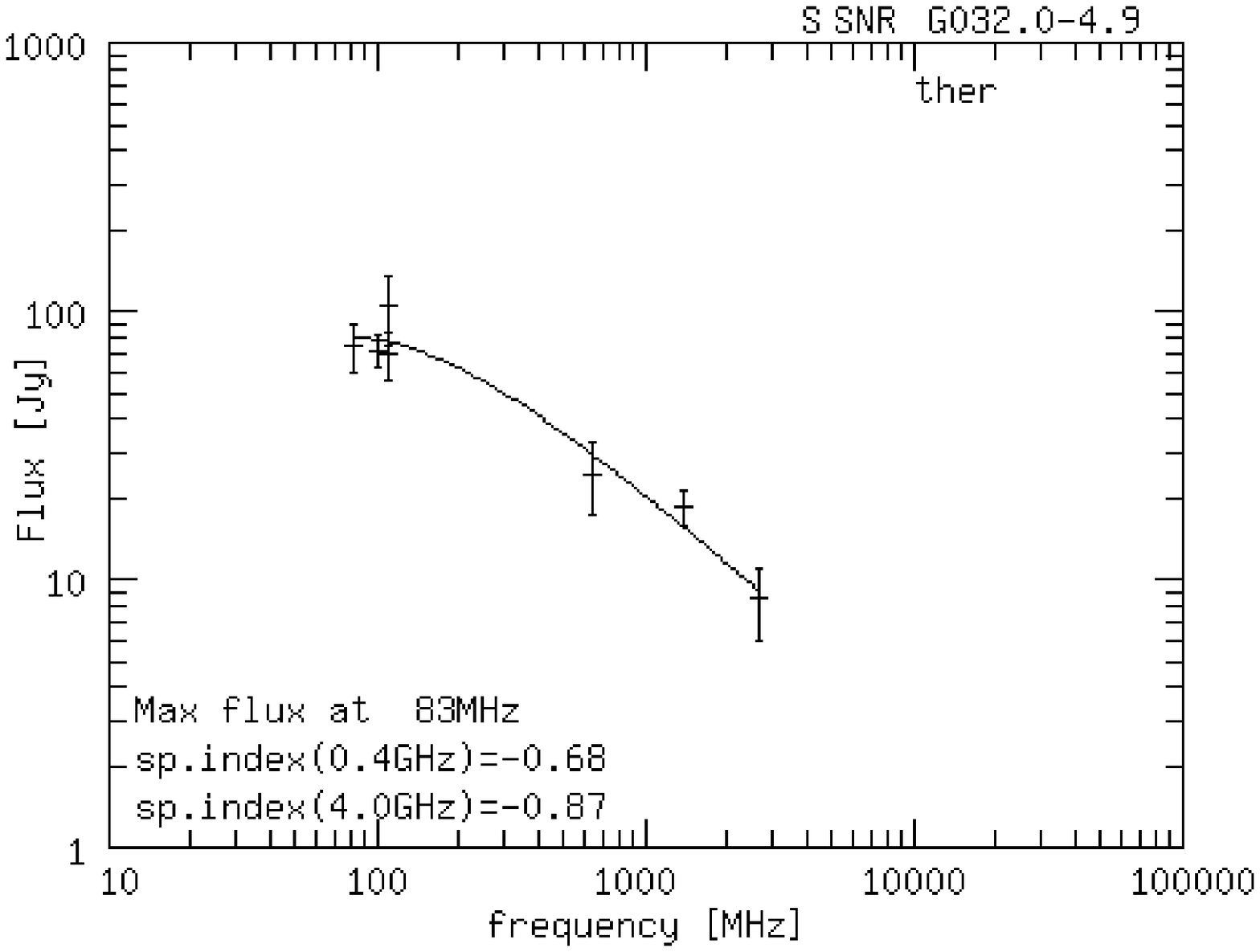,width=7.4cm,angle=0}}}\end{figure}
\begin{figure}\centerline{\vbox{\psfig{figure=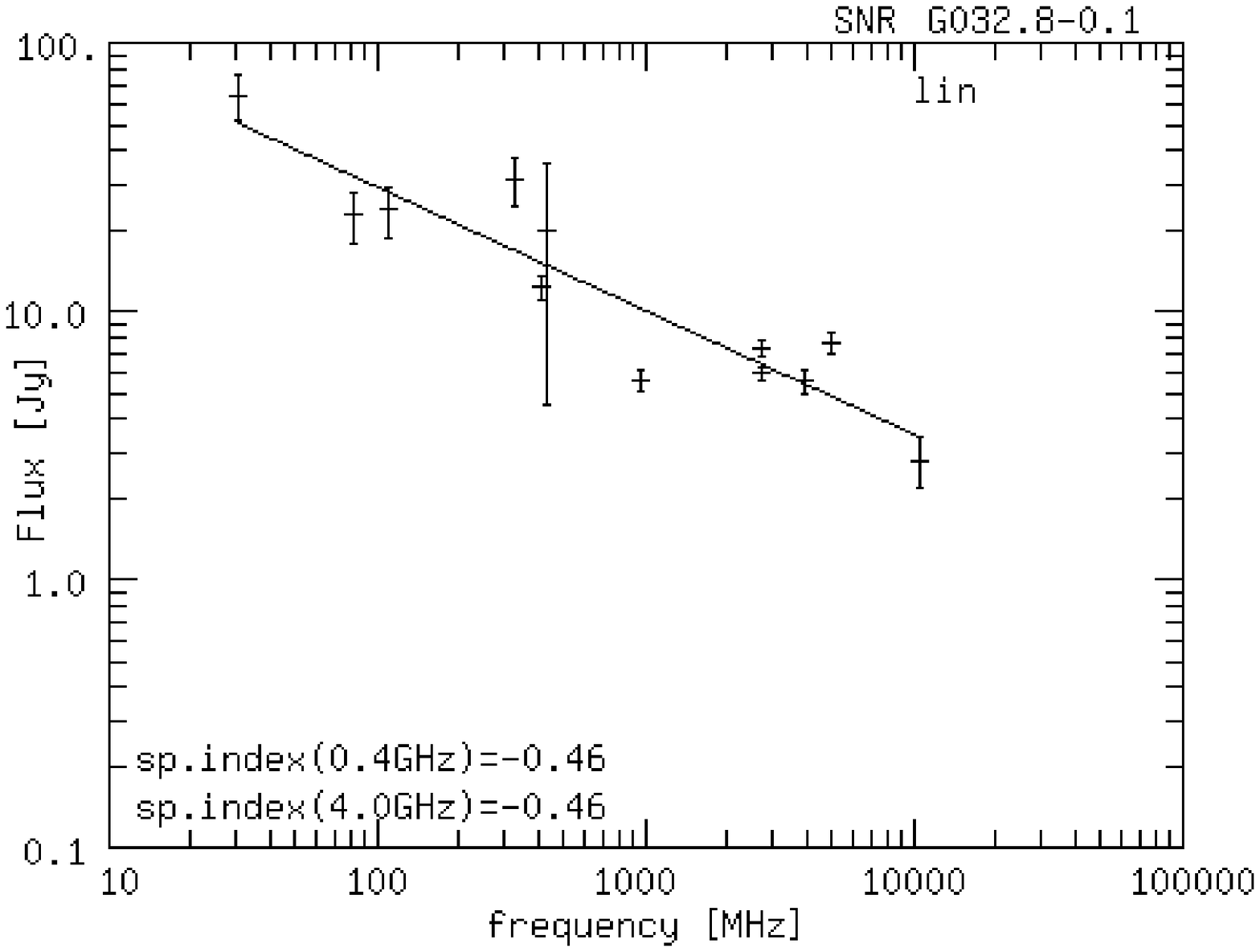,width=7.4cm,angle=0}}}\end{figure}
\begin{figure}\centerline{\vbox{\psfig{figure=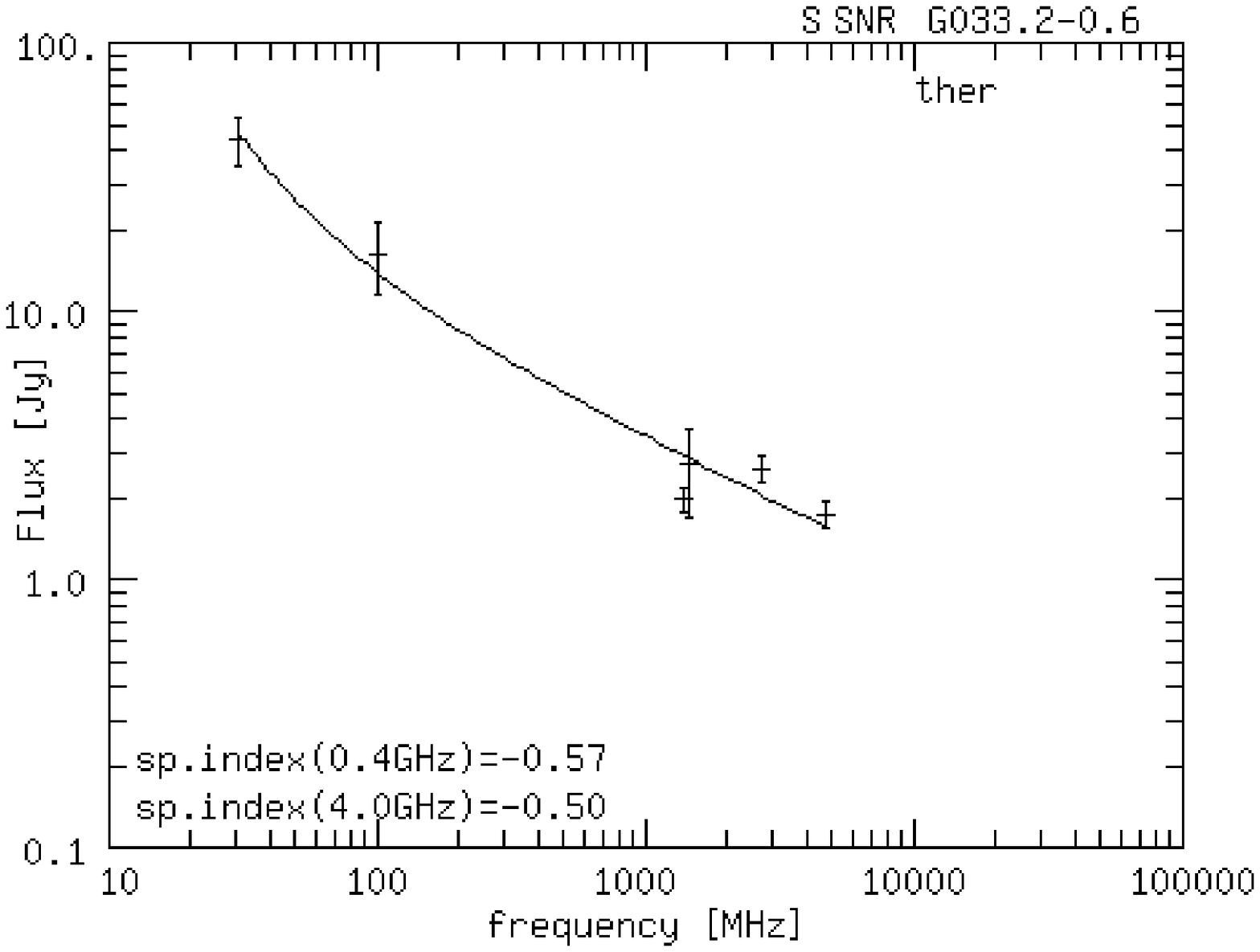,width=7.4cm,angle=0}}}\end{figure}\clearpage
\begin{figure}\centerline{\vbox{\psfig{figure=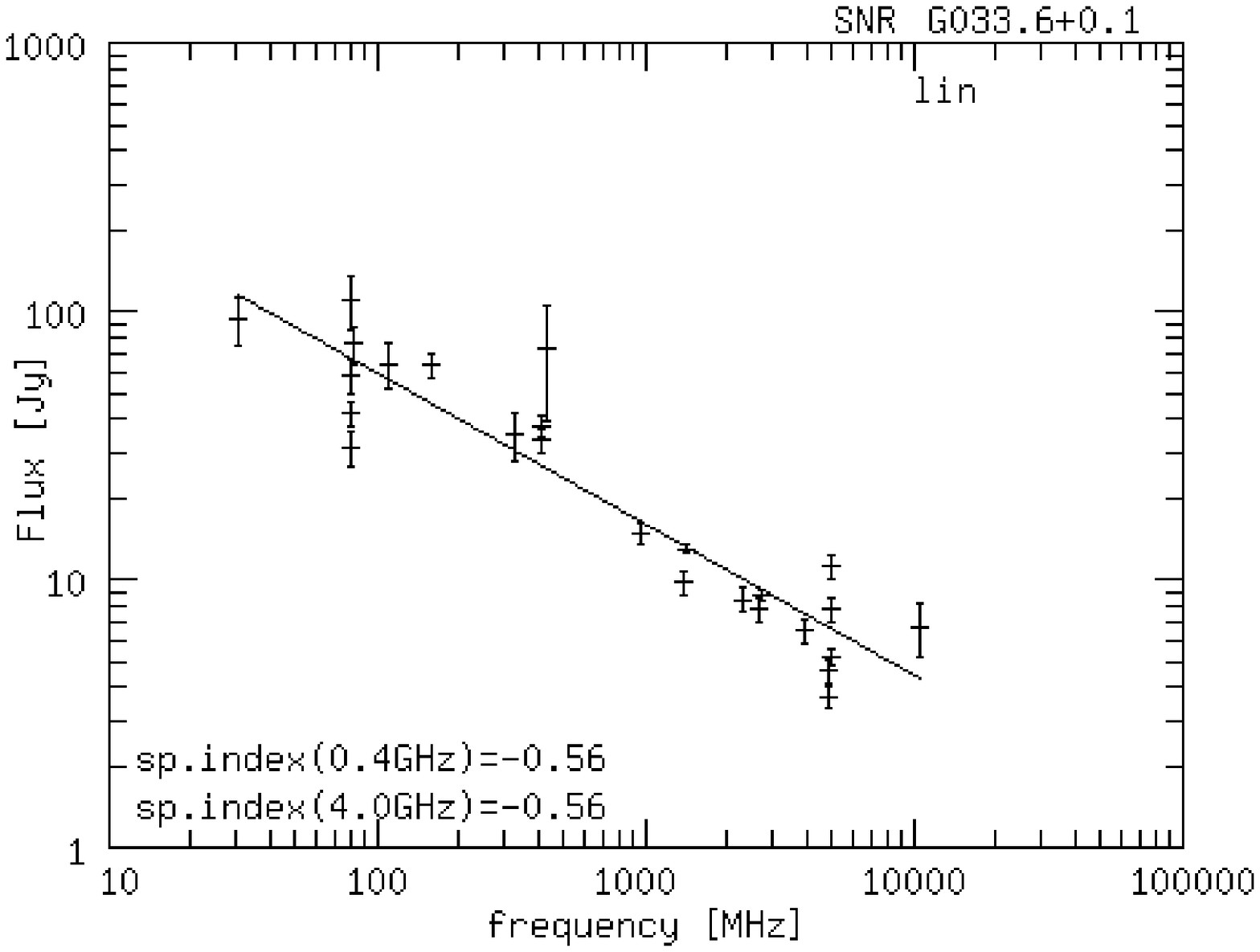,width=7.4cm,angle=0}}}\end{figure}
\begin{figure}\centerline{\vbox{\psfig{figure=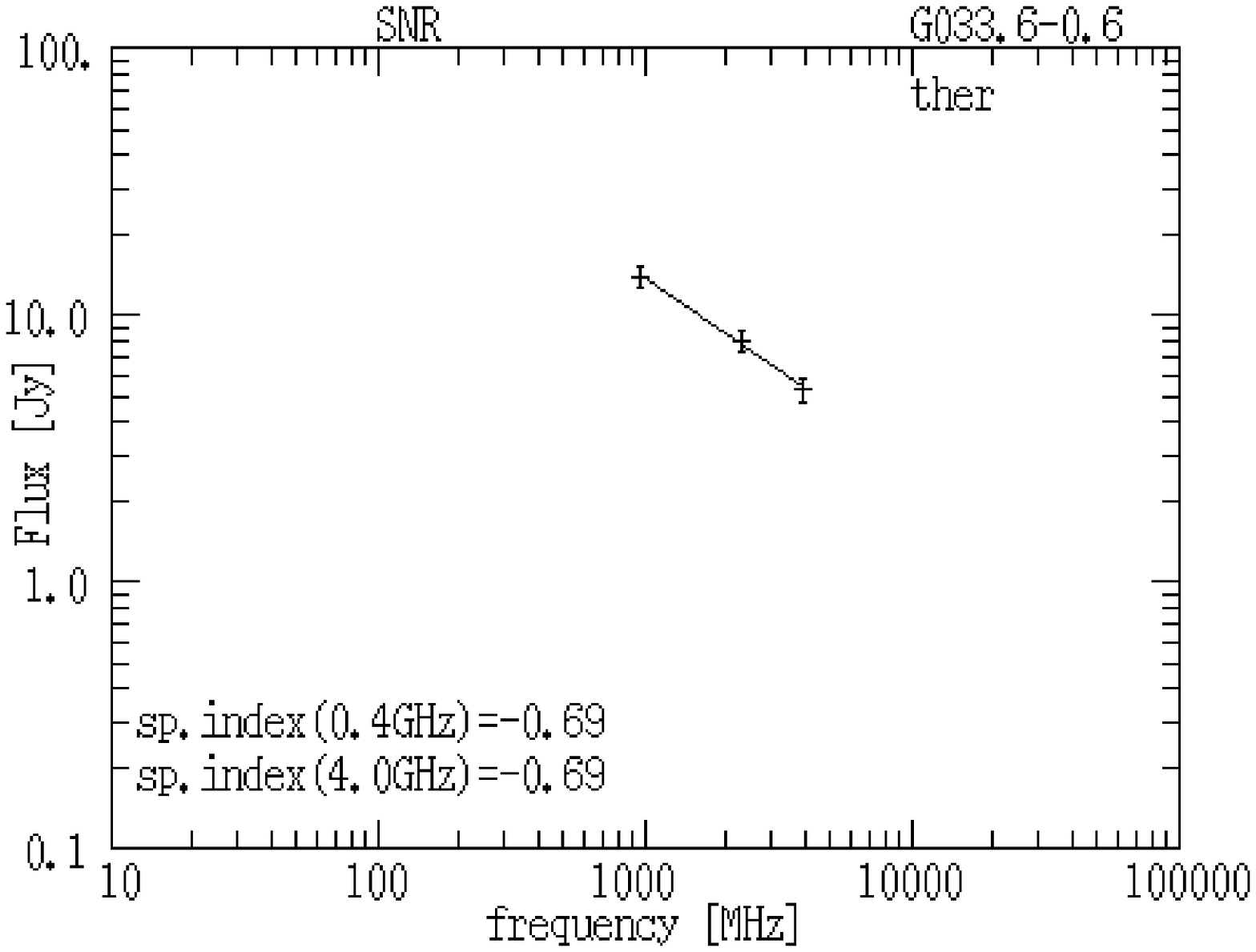,width=7.4cm,angle=0}}}\end{figure}
\begin{figure}\centerline{\vbox{\psfig{figure=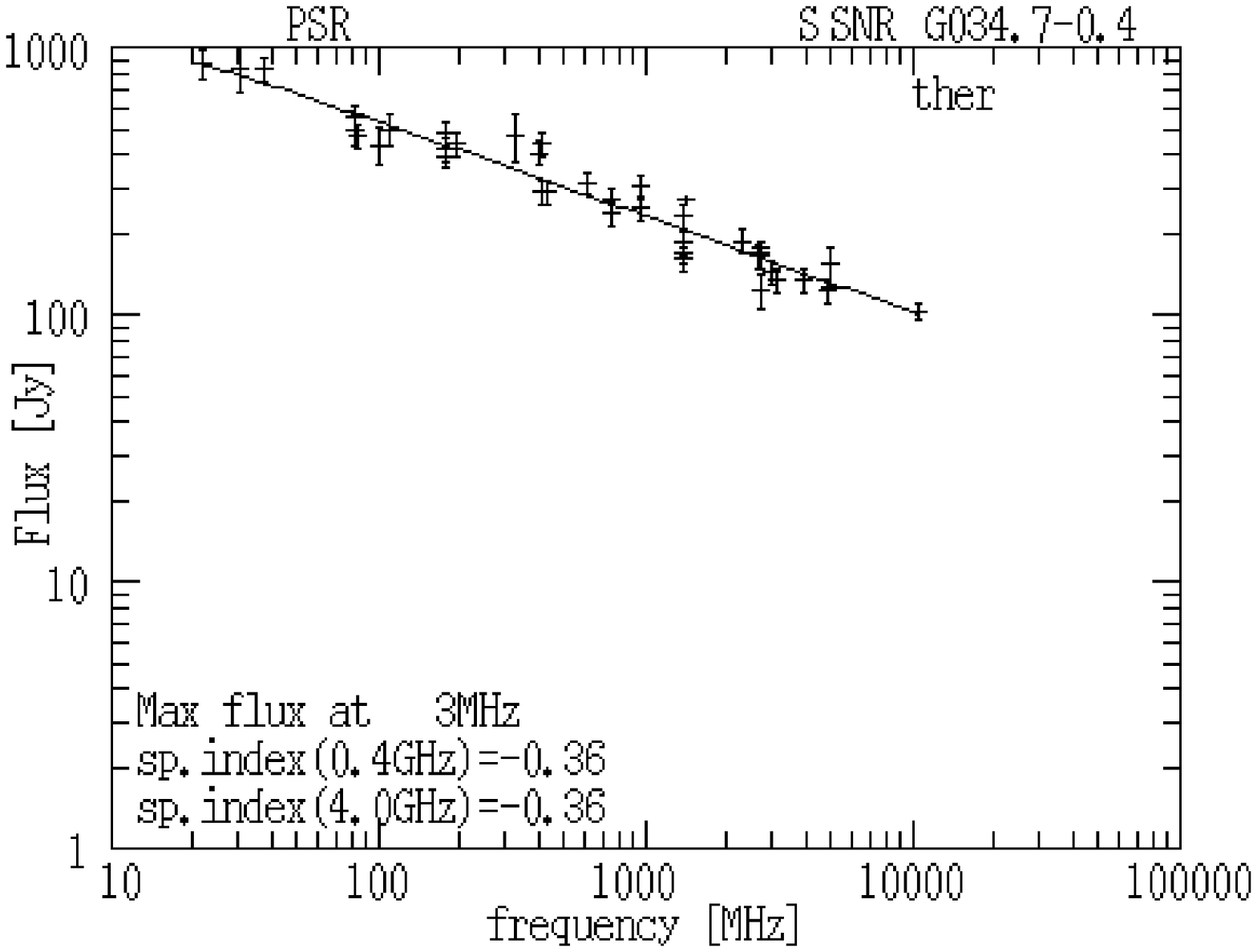,width=7.4cm,angle=0}}}\end{figure}
\begin{figure}\centerline{\vbox{\psfig{figure=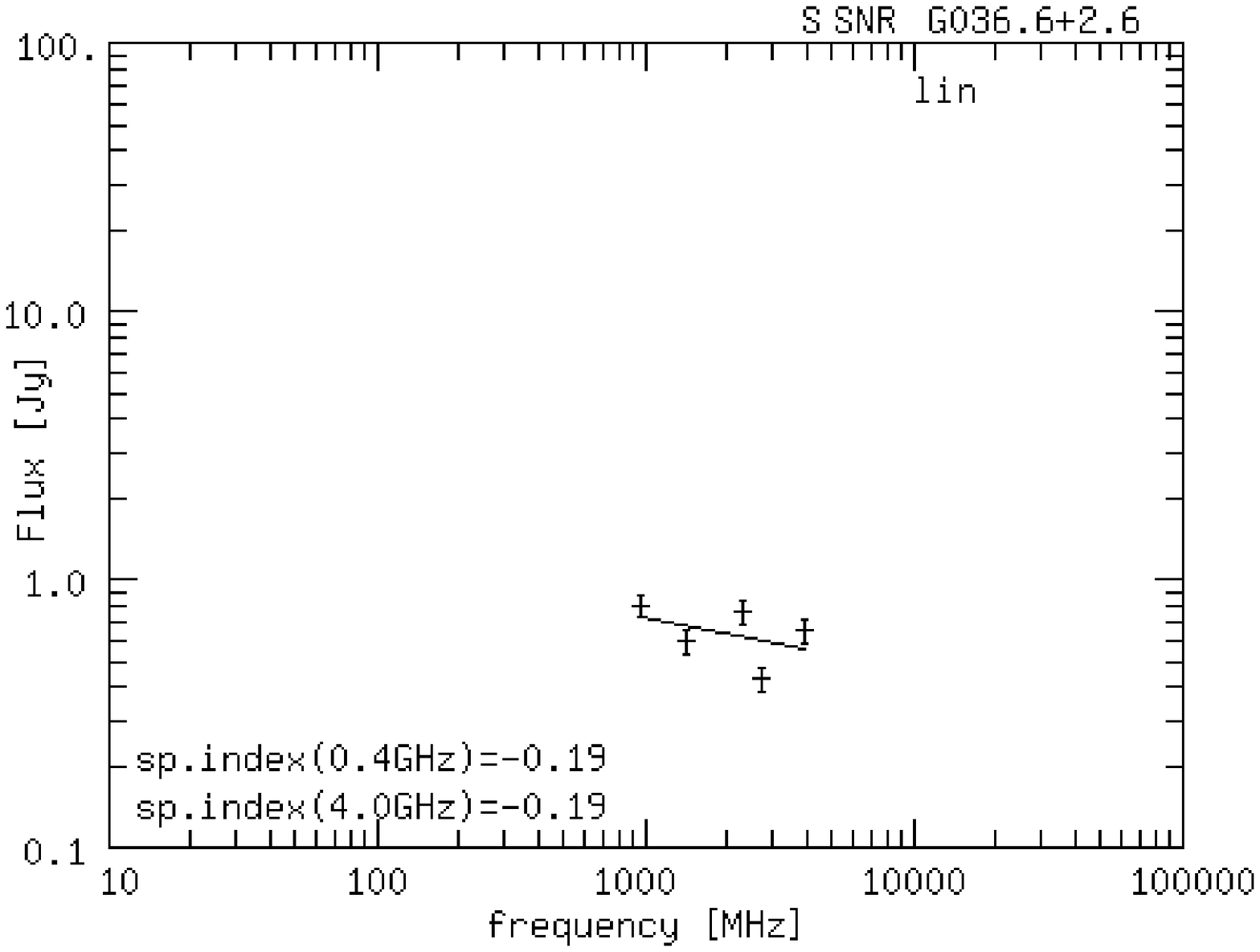,width=7.4cm,angle=0}}}\end{figure}
\begin{figure}\centerline{\vbox{\psfig{figure=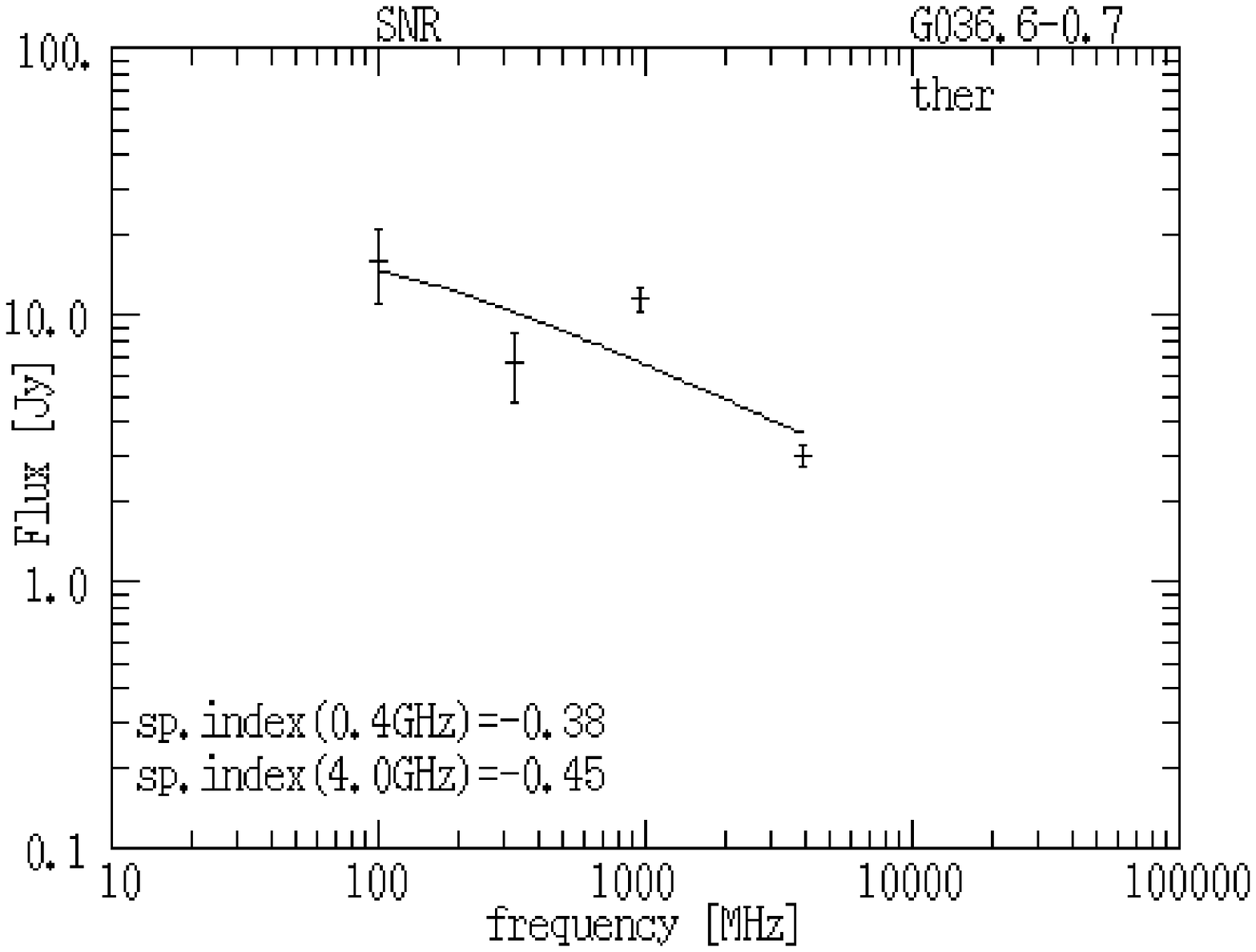,width=7.4cm,angle=0}}}\end{figure}
\begin{figure}\centerline{\vbox{\psfig{figure=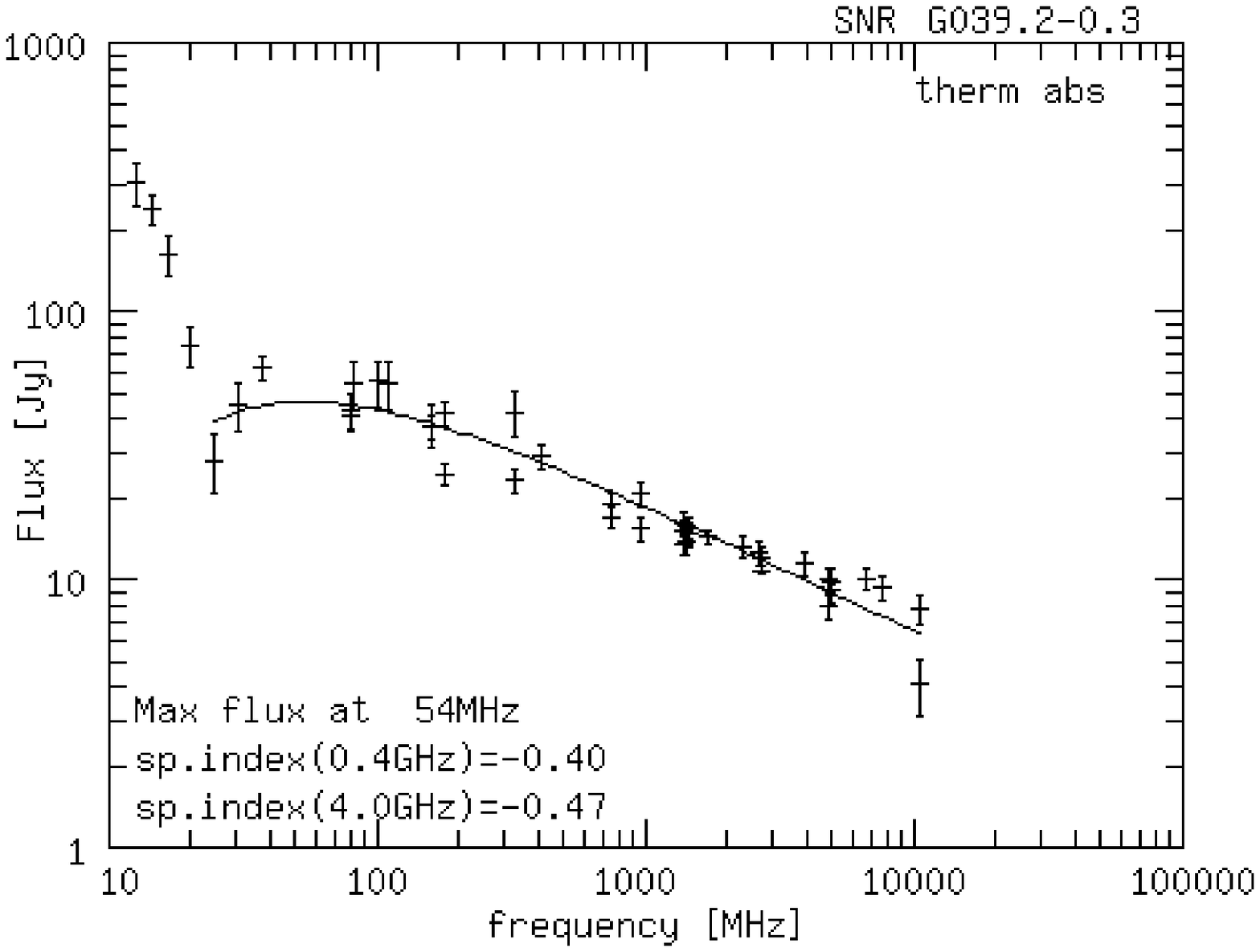,width=7.4cm,angle=0}}}\end{figure}
\begin{figure}\centerline{\vbox{\psfig{figure=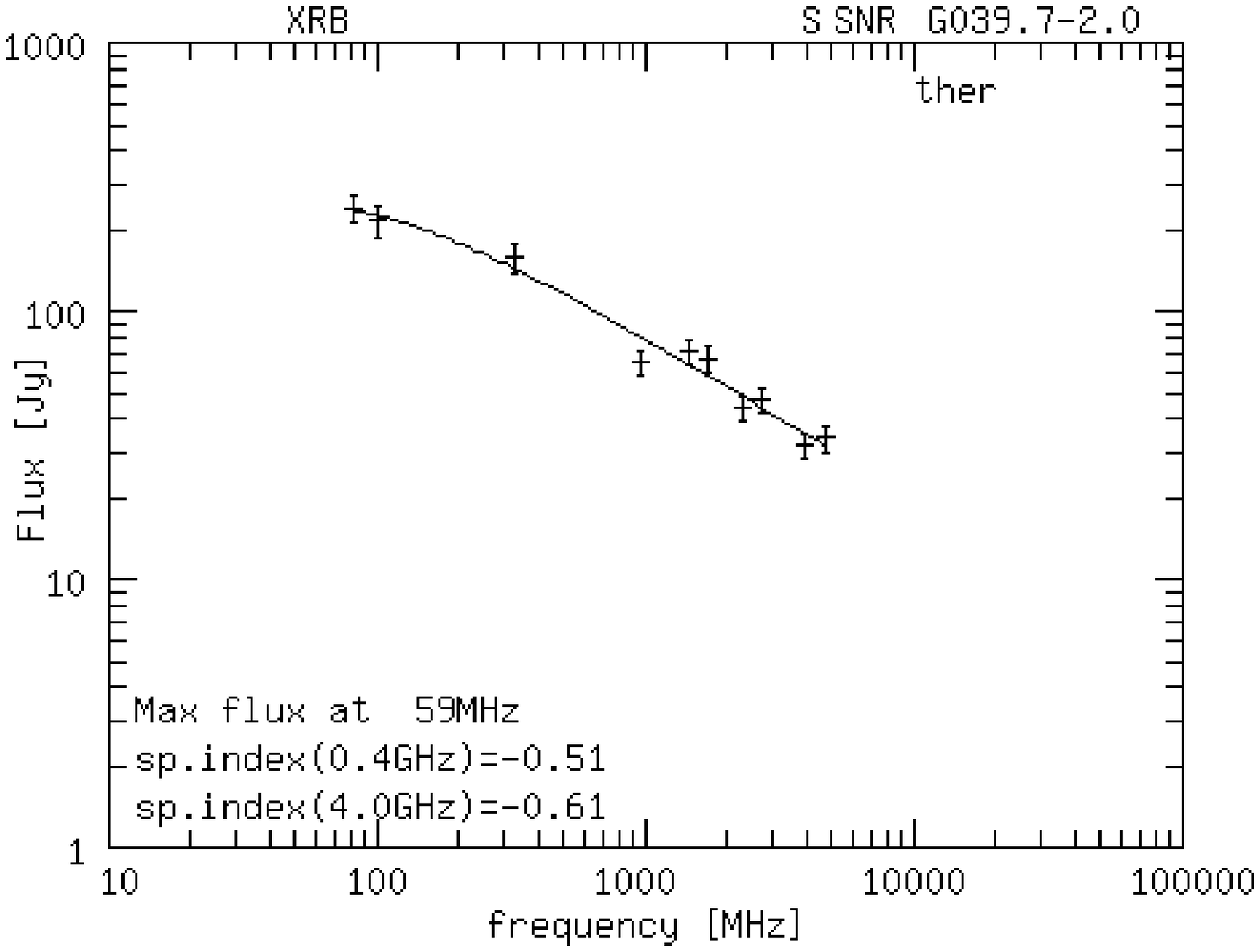,width=7.4cm,angle=0}}}\end{figure}
\begin{figure}\centerline{\vbox{\psfig{figure=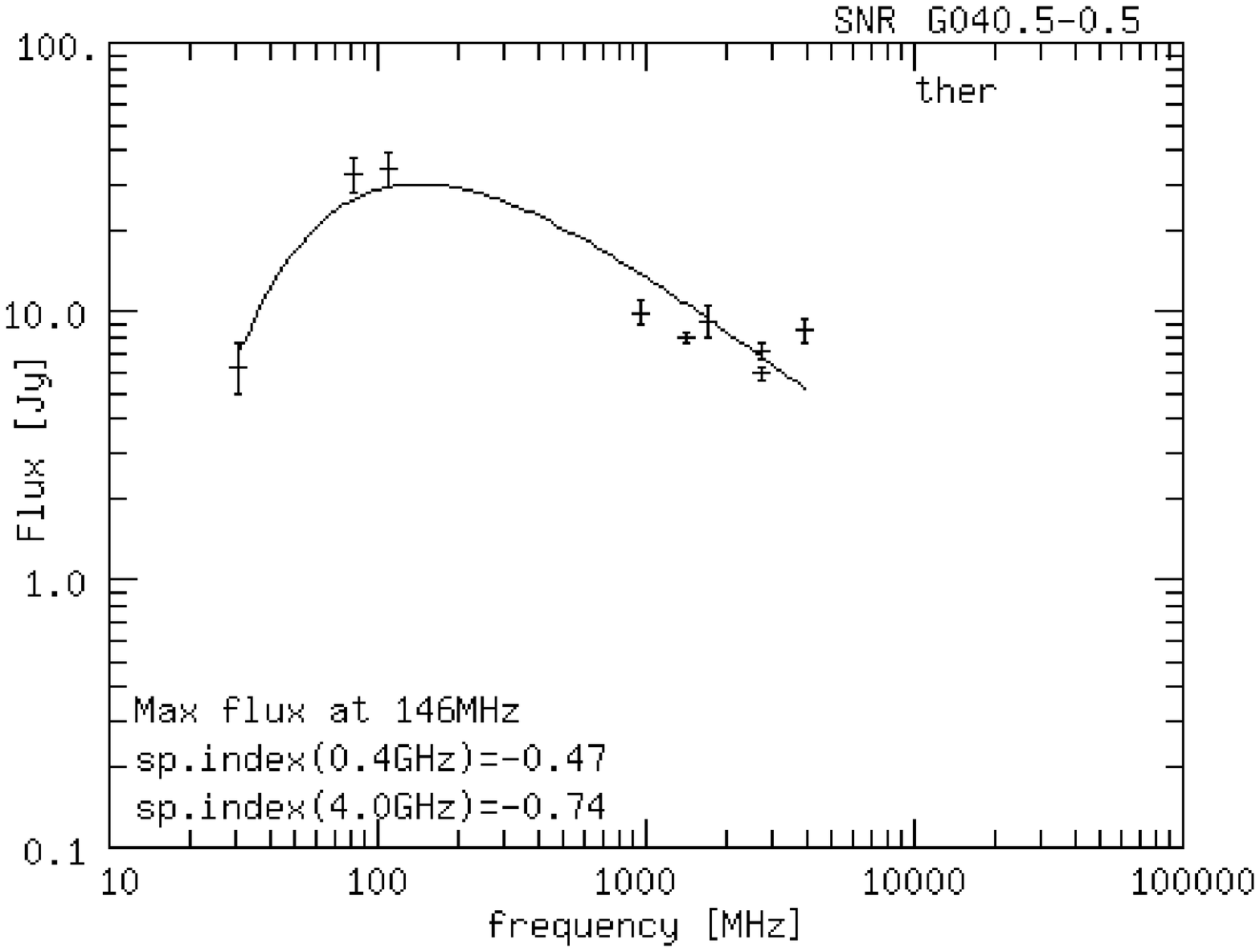,width=7.4cm,angle=0}}}\end{figure}\clearpage
\begin{figure}\centerline{\vbox{\psfig{figure=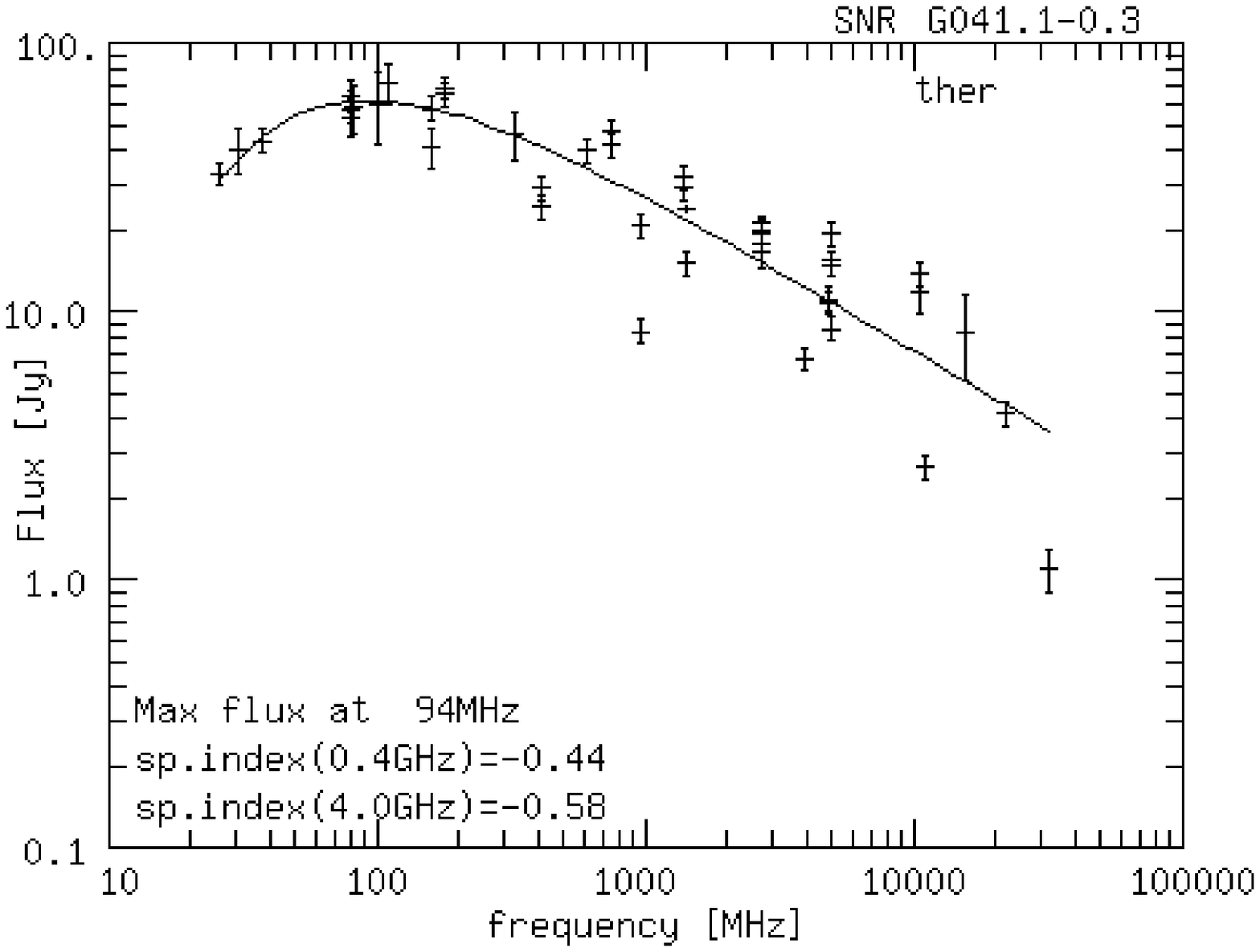,width=7.4cm,angle=0}}}\end{figure}
\begin{figure}\centerline{\vbox{\psfig{figure=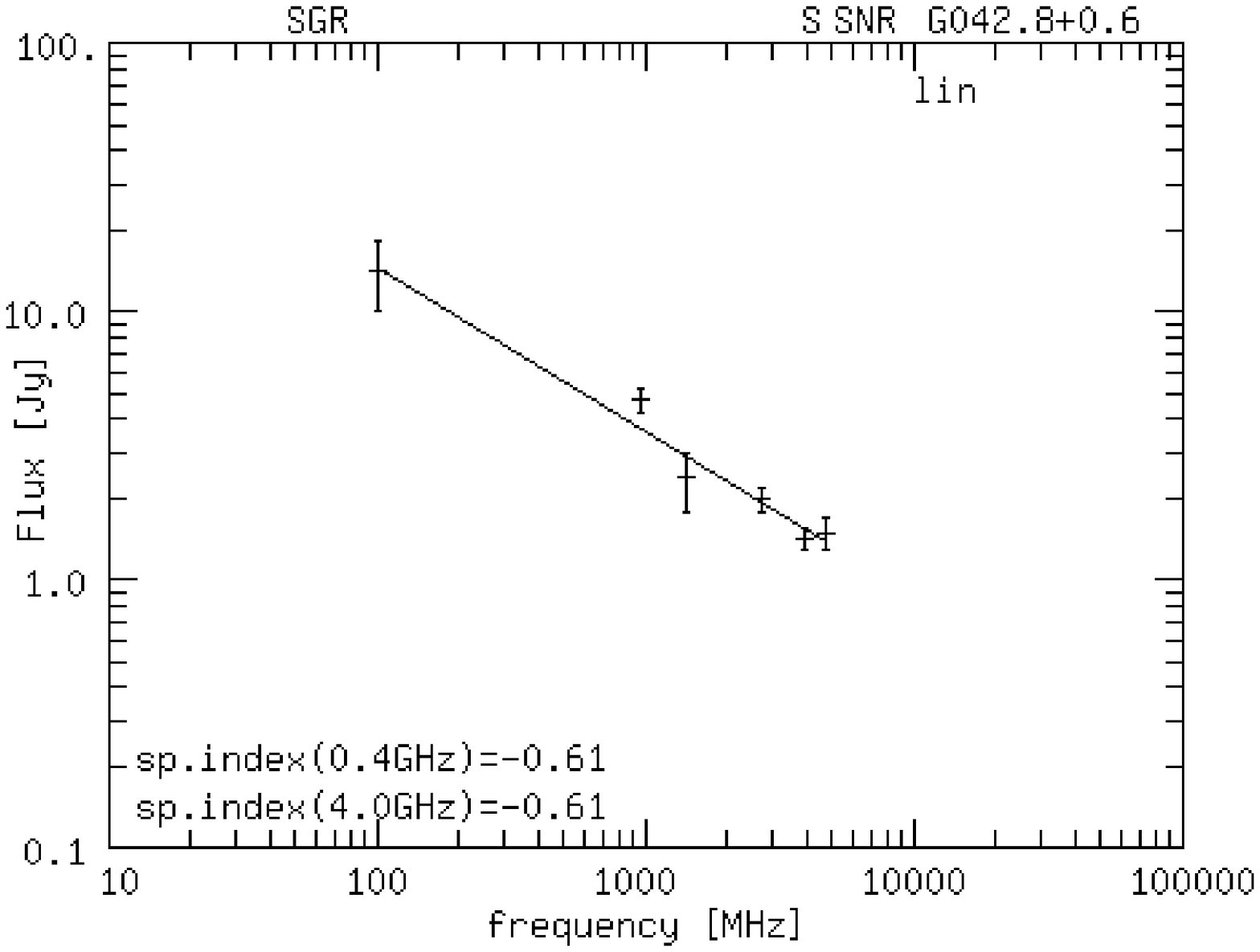,width=7.4cm,angle=0}}}\end{figure}
\begin{figure}\centerline{\vbox{\psfig{figure=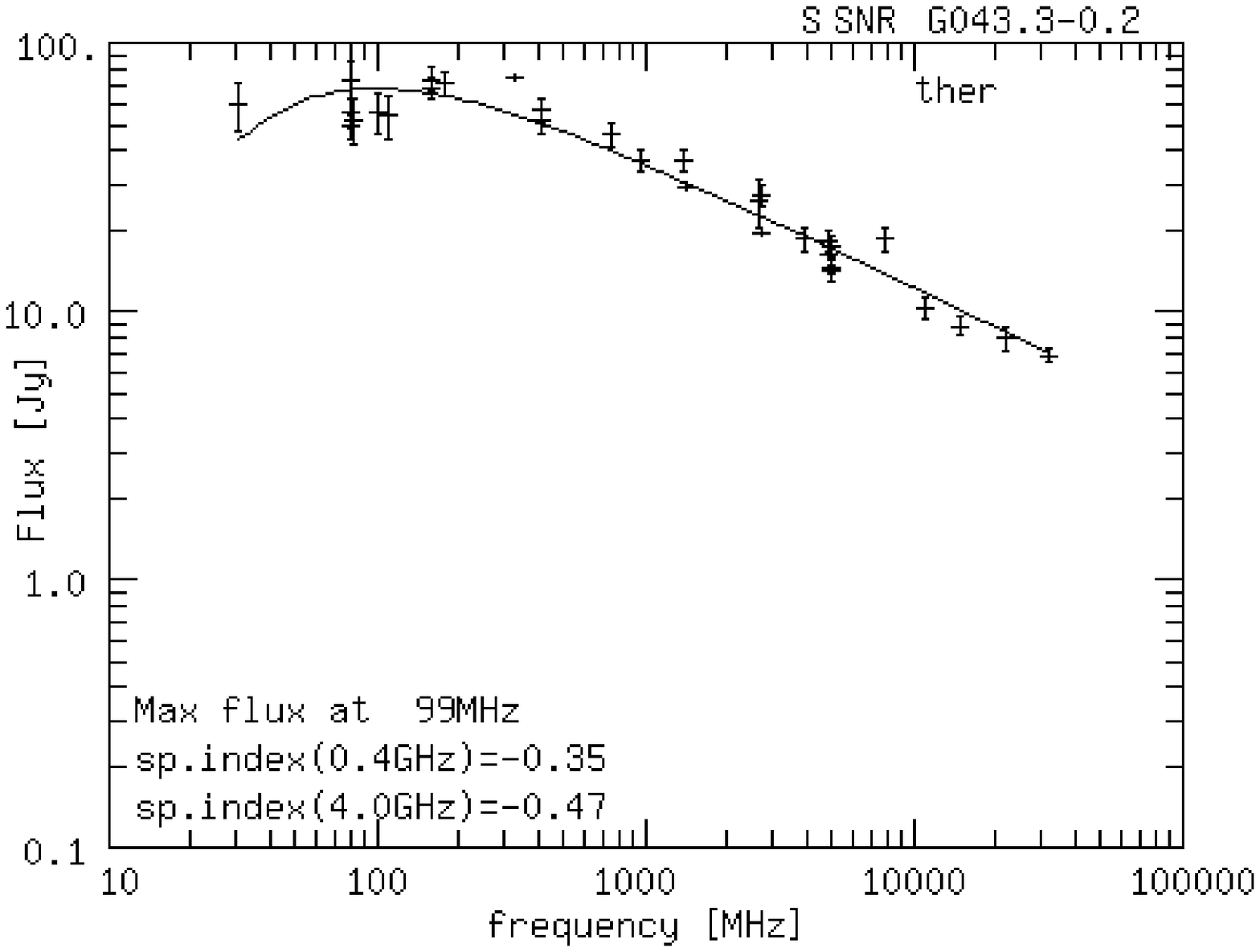,width=7.4cm,angle=0}}}\end{figure}
\begin{figure}\centerline{\vbox{\psfig{figure=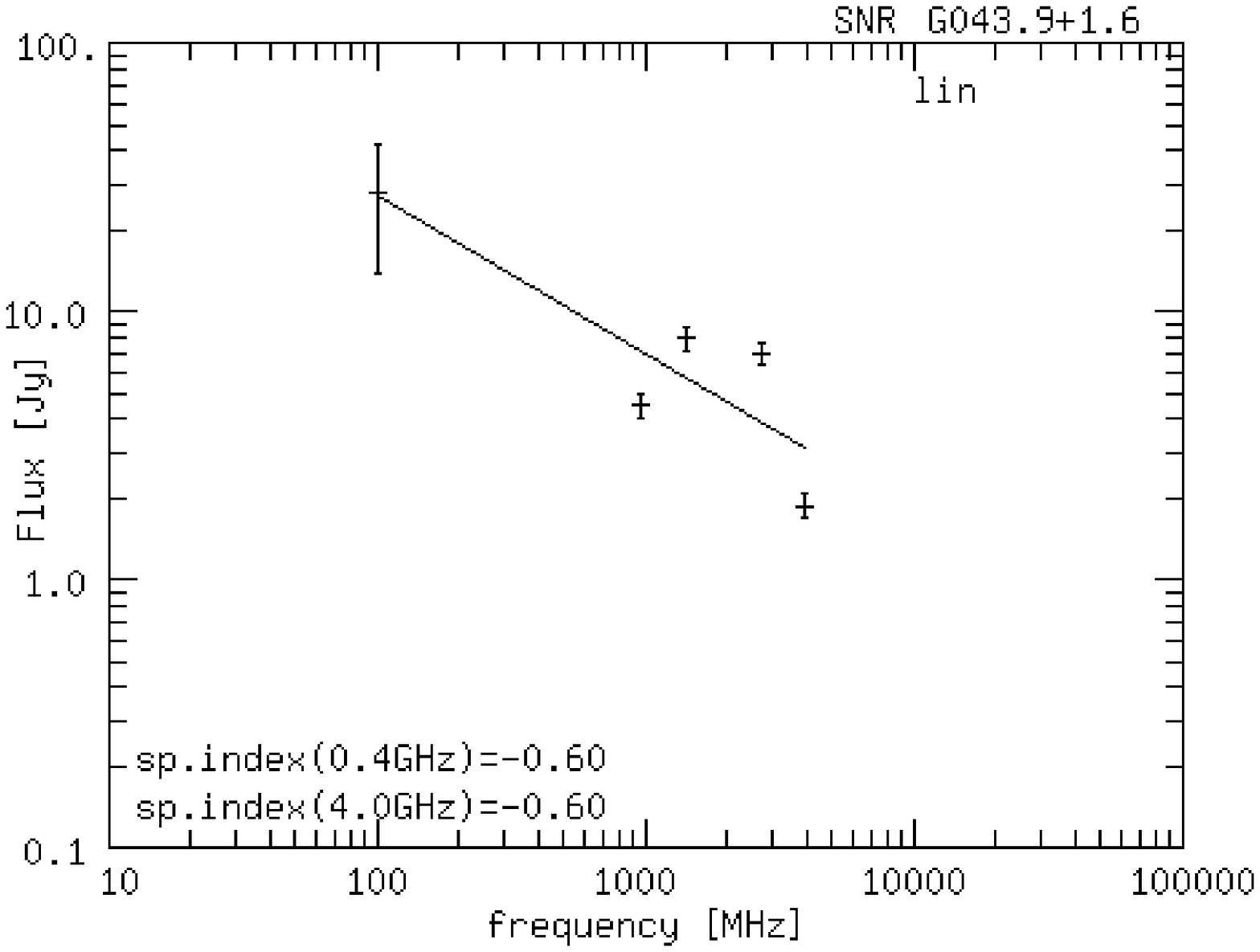,width=7.4cm,angle=0}}}\end{figure}
\begin{figure}\centerline{\vbox{\psfig{figure=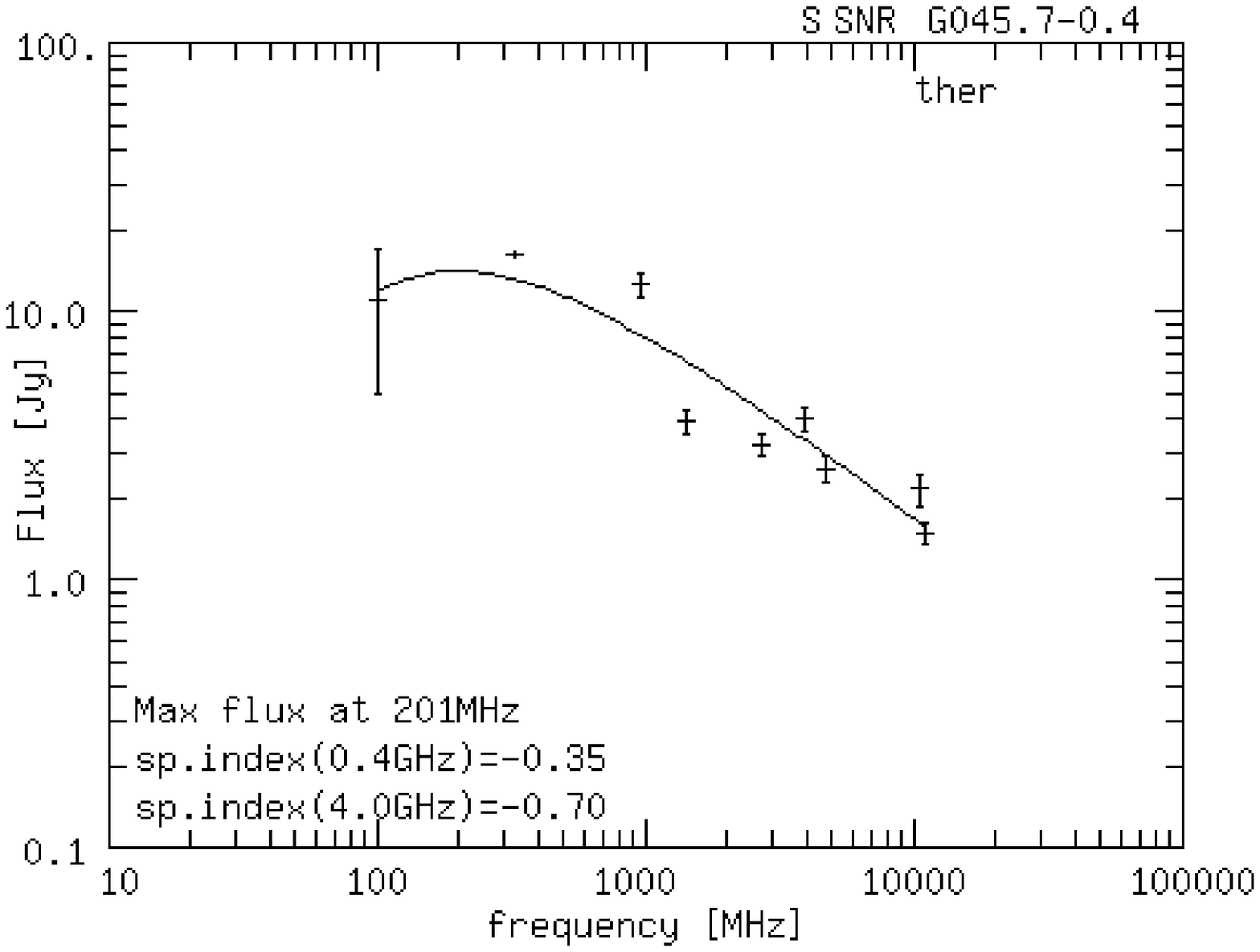,width=7.4cm,angle=0}}}\end{figure}
\begin{figure}\centerline{\vbox{\psfig{figure=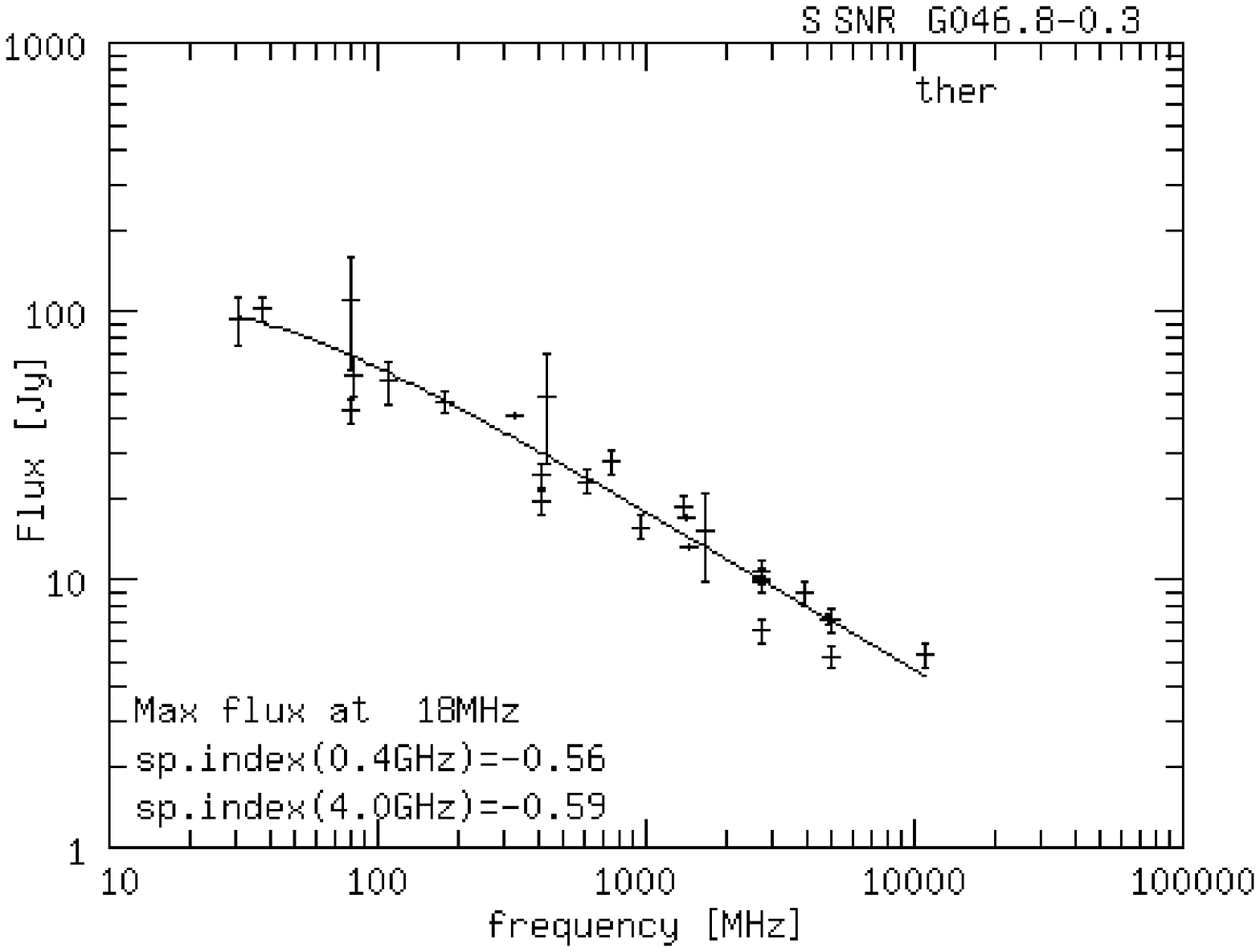,width=7.4cm,angle=0}}}\end{figure}
\begin{figure}\centerline{\vbox{\psfig{figure=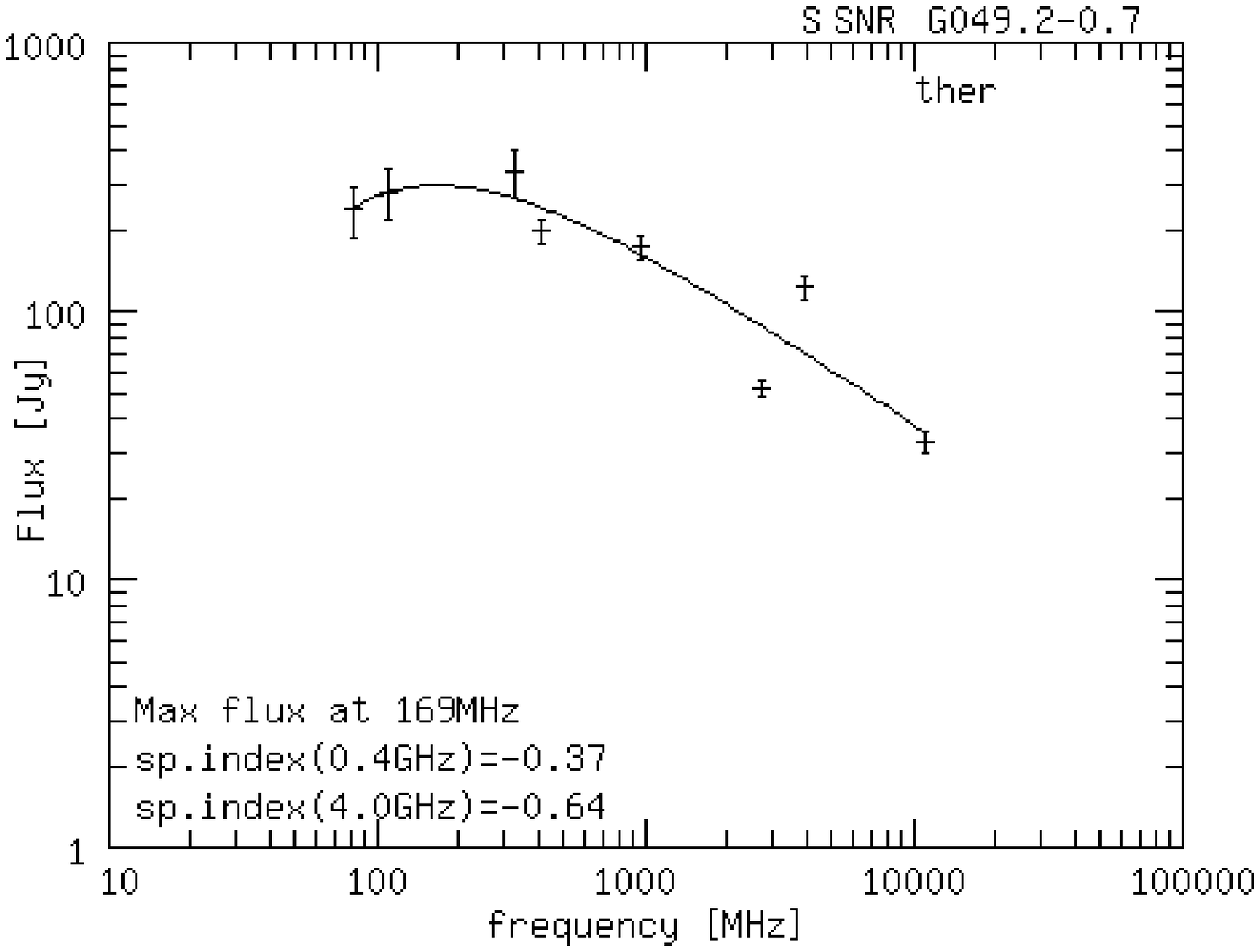,width=7.4cm,angle=0}}}\end{figure}
\begin{figure}\centerline{\vbox{\psfig{figure=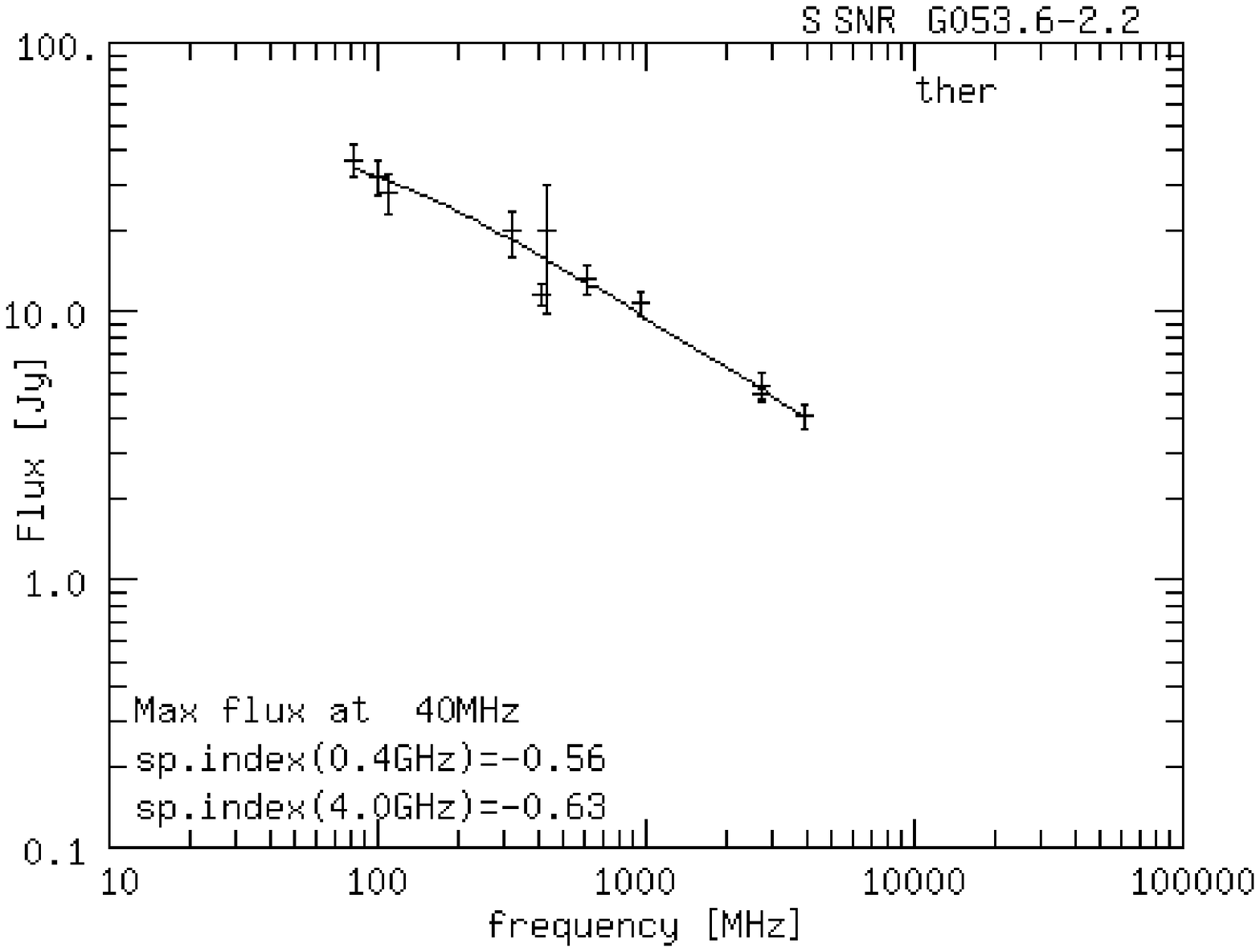,width=7.4cm,angle=0}}}\end{figure}\clearpage
\begin{figure}\centerline{\vbox{\psfig{figure=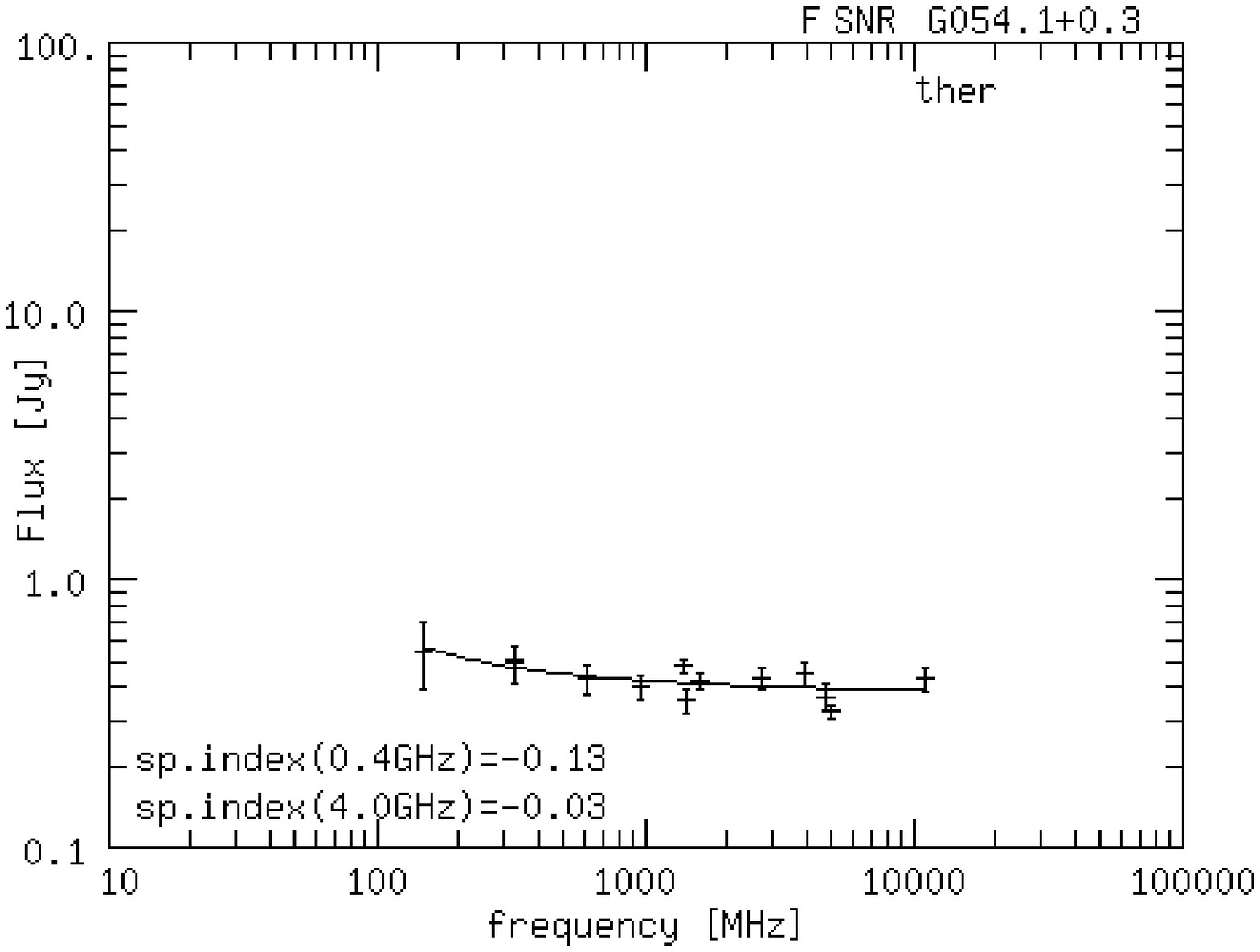,width=7.4cm,angle=0}}}\end{figure}
\begin{figure}\centerline{\vbox{\psfig{figure=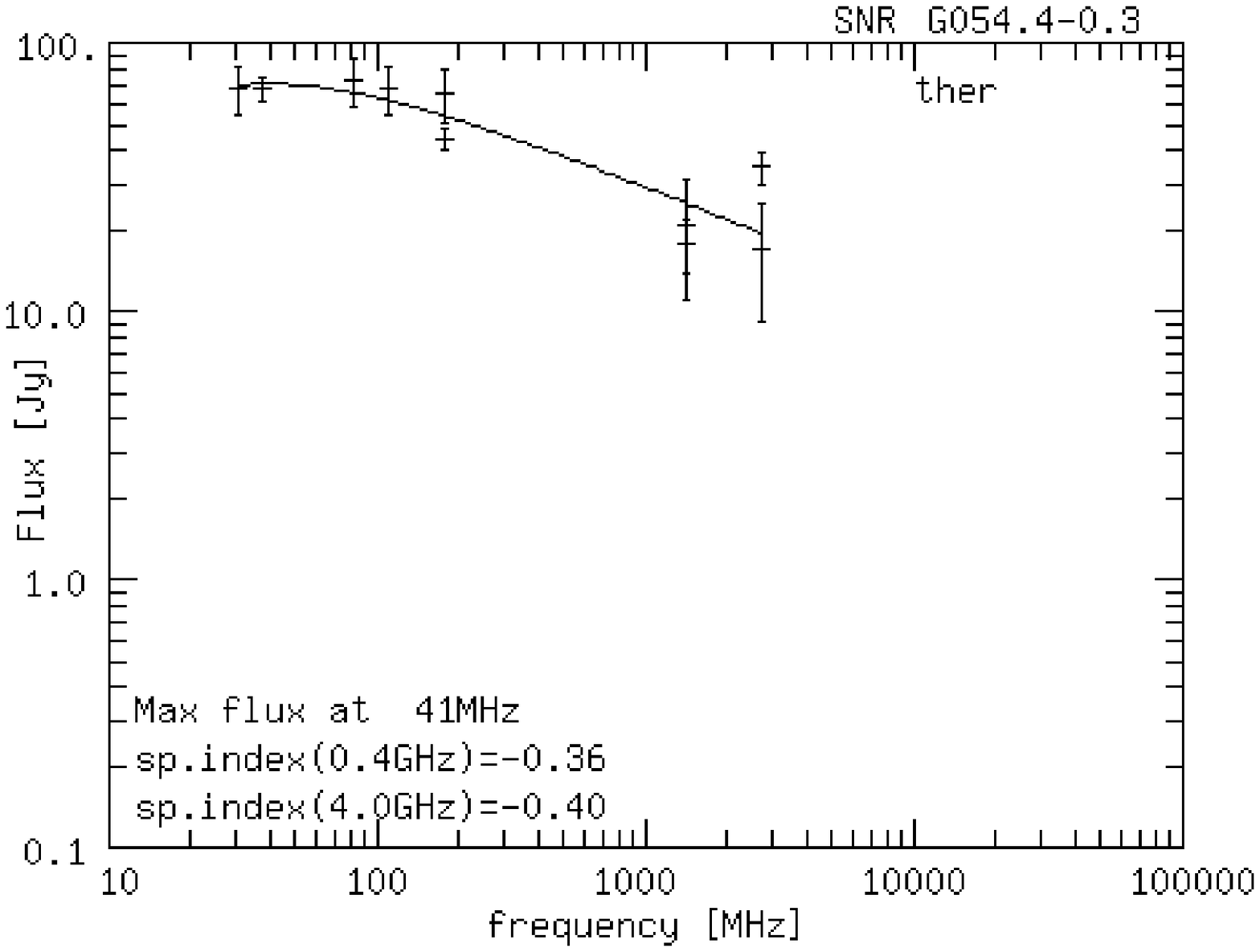,width=7.4cm,angle=0}}}\end{figure}
\begin{figure}\centerline{\vbox{\psfig{figure=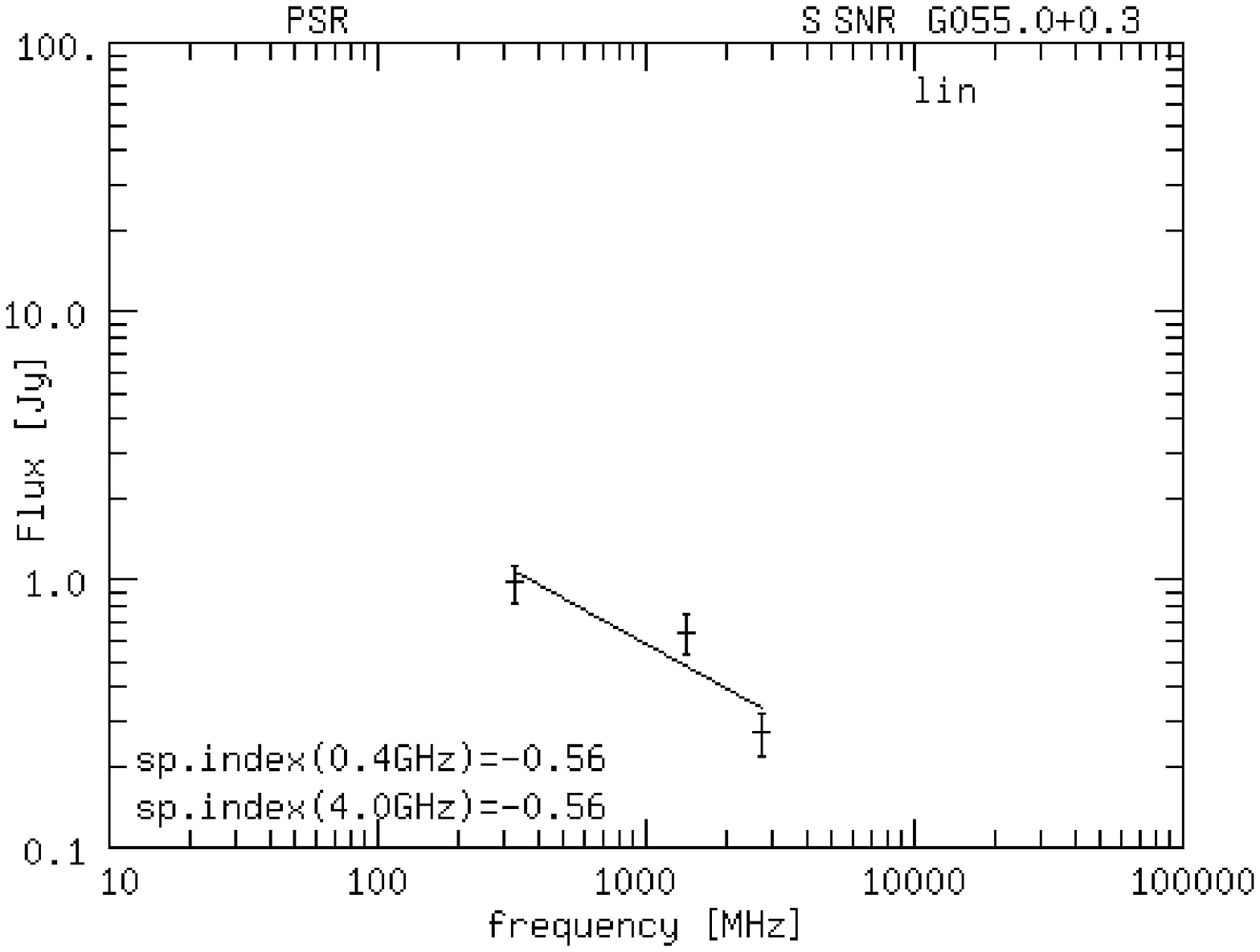,width=7.4cm,angle=0}}}\end{figure}
\begin{figure}\centerline{\vbox{\psfig{figure=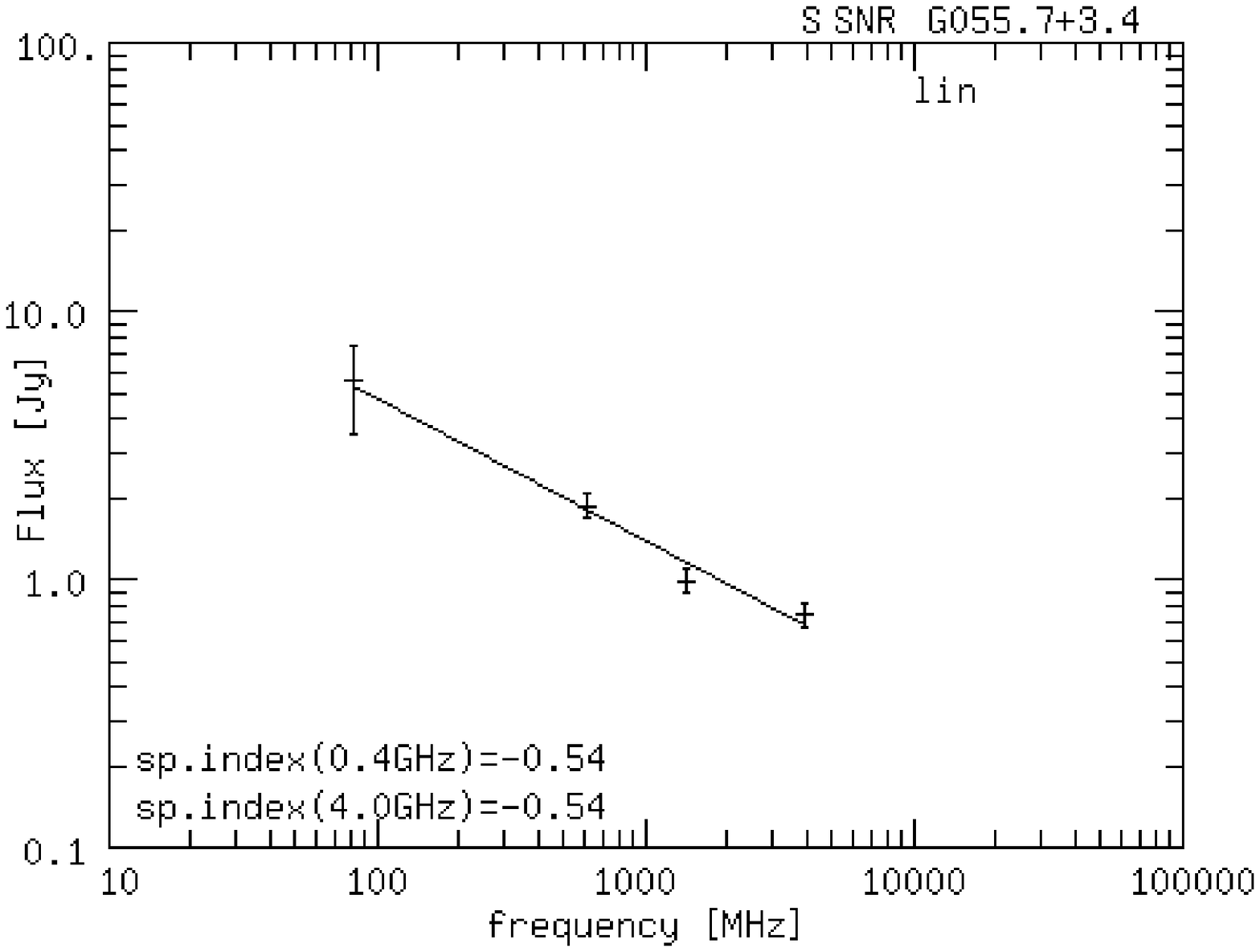,width=7.4cm,angle=0}}}\end{figure}
\begin{figure}\centerline{\vbox{\psfig{figure=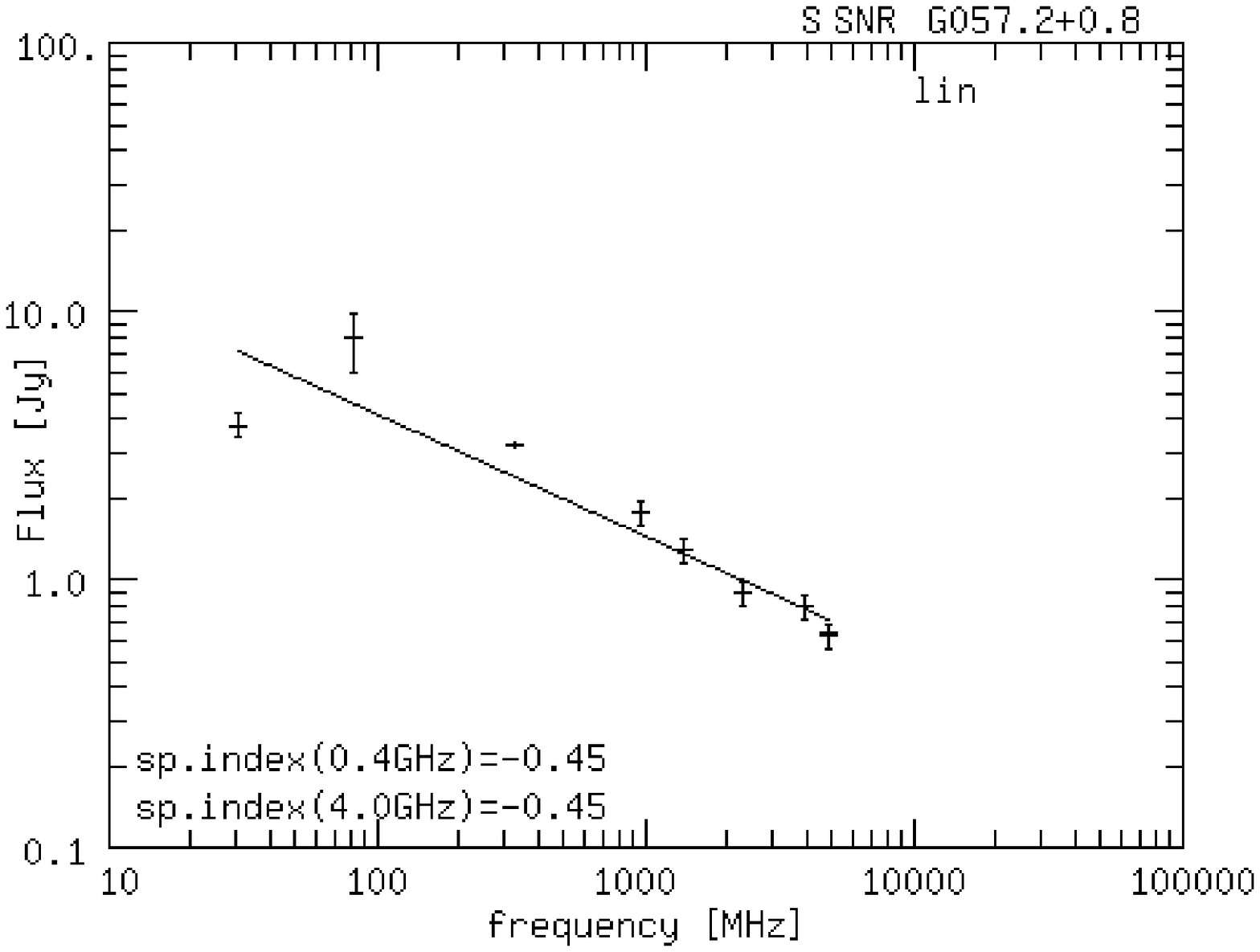,width=7.4cm,angle=0}}}\end{figure}
\begin{figure}\centerline{\vbox{\psfig{figure=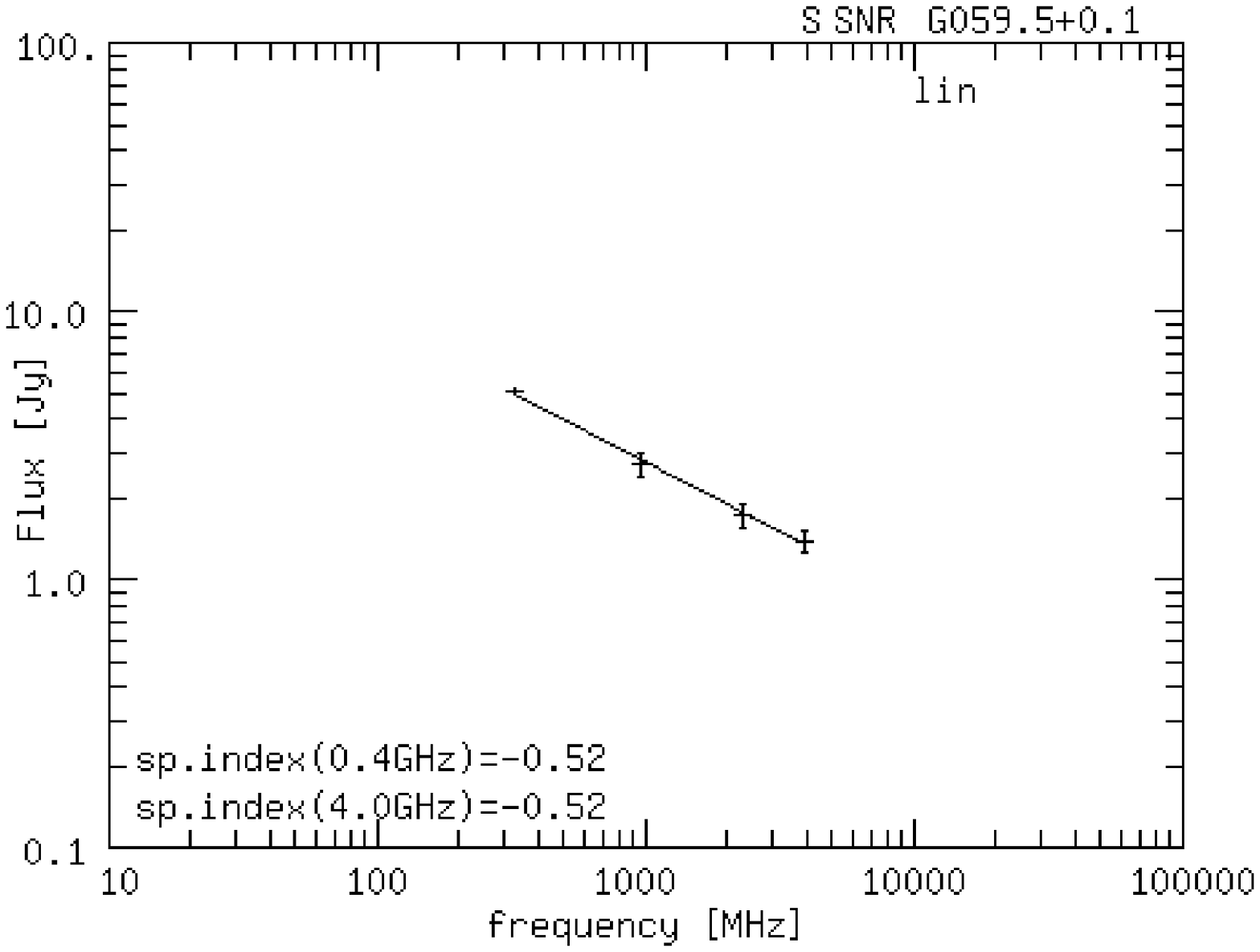,width=7.4cm,angle=0}}}\end{figure}
\begin{figure}\centerline{\vbox{\psfig{figure=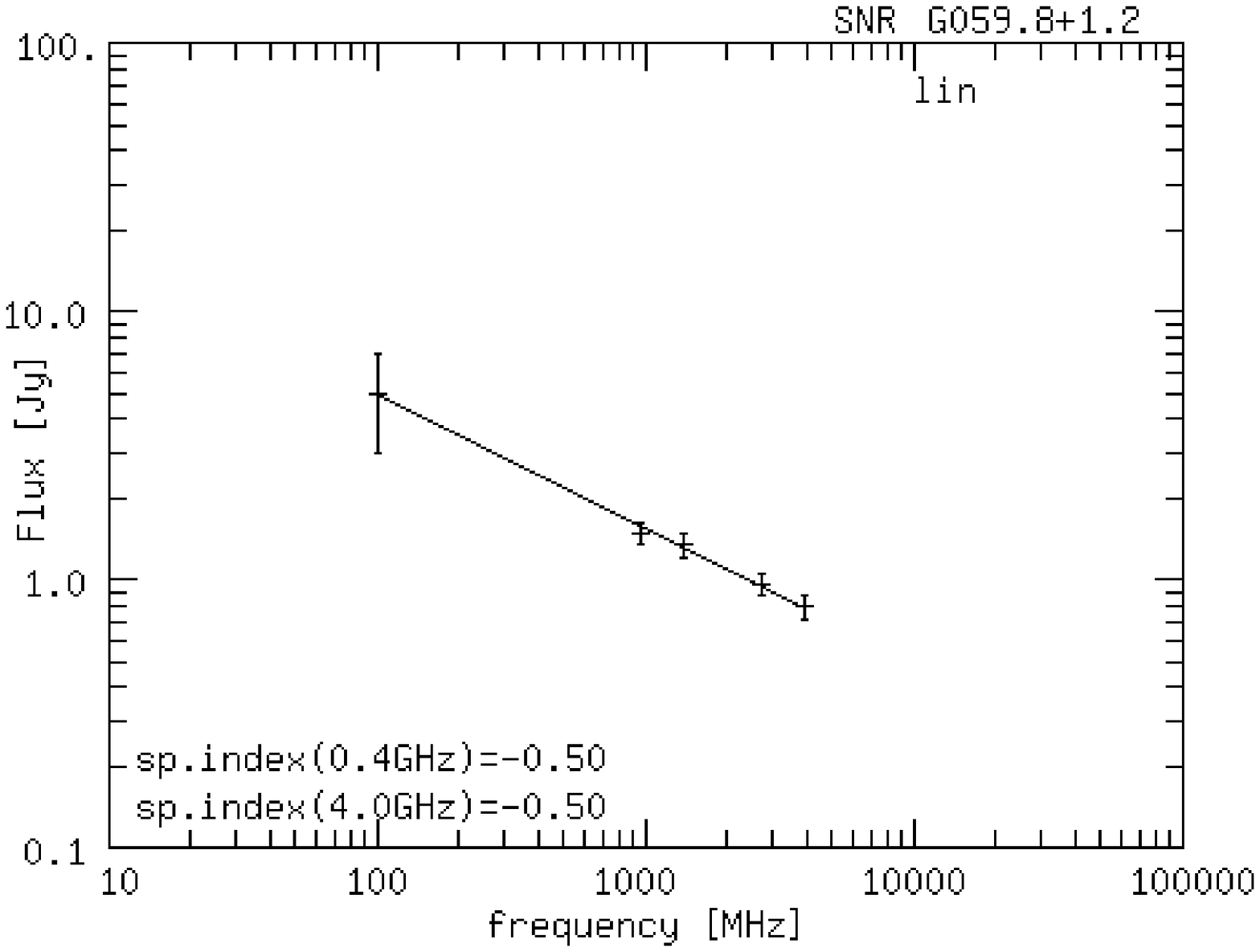,width=7.4cm,angle=0}}}\end{figure}
\begin{figure}\centerline{\vbox{\psfig{figure=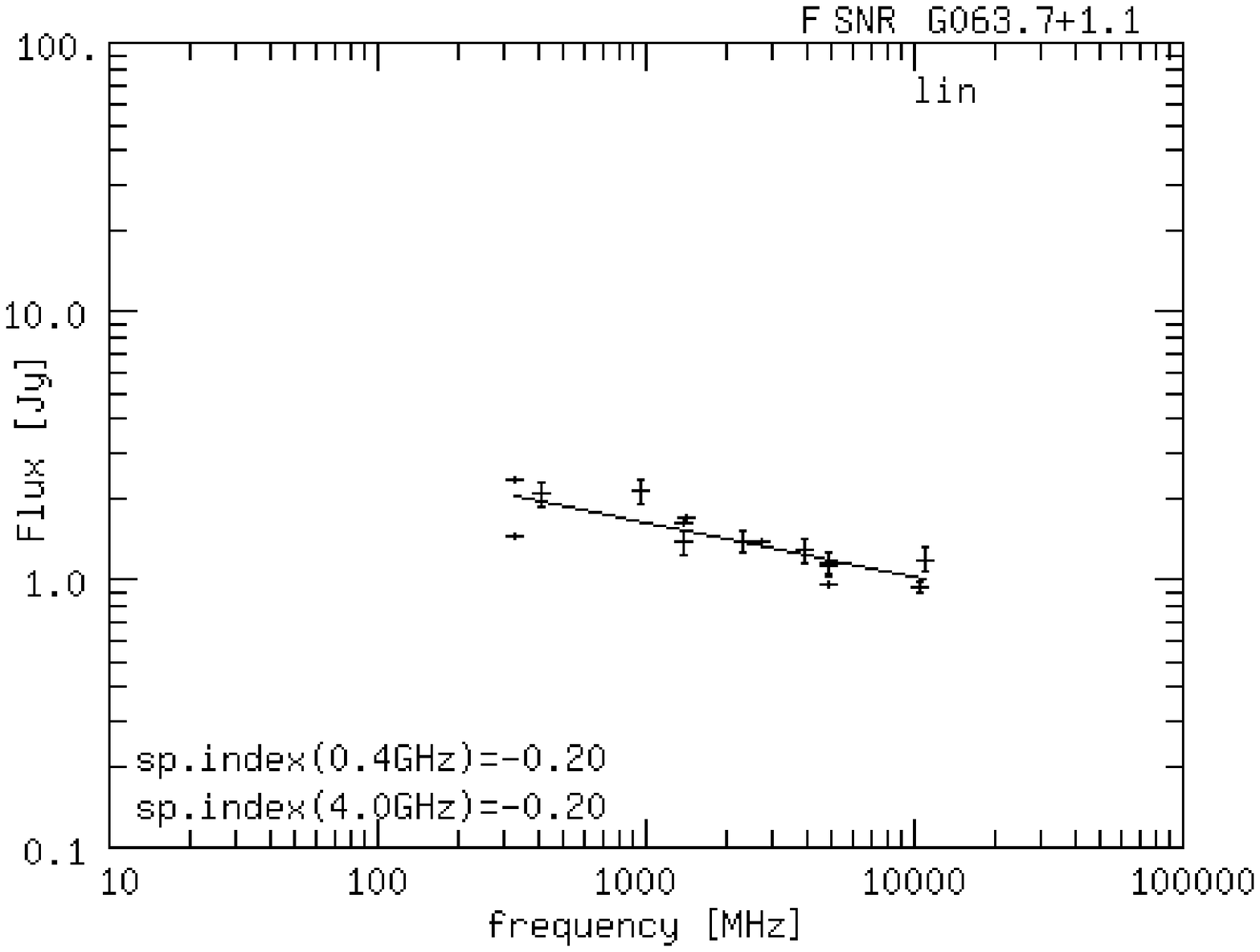,width=7.4cm,angle=0}}}\end{figure}\clearpage
\begin{figure}\centerline{\vbox{\psfig{figure=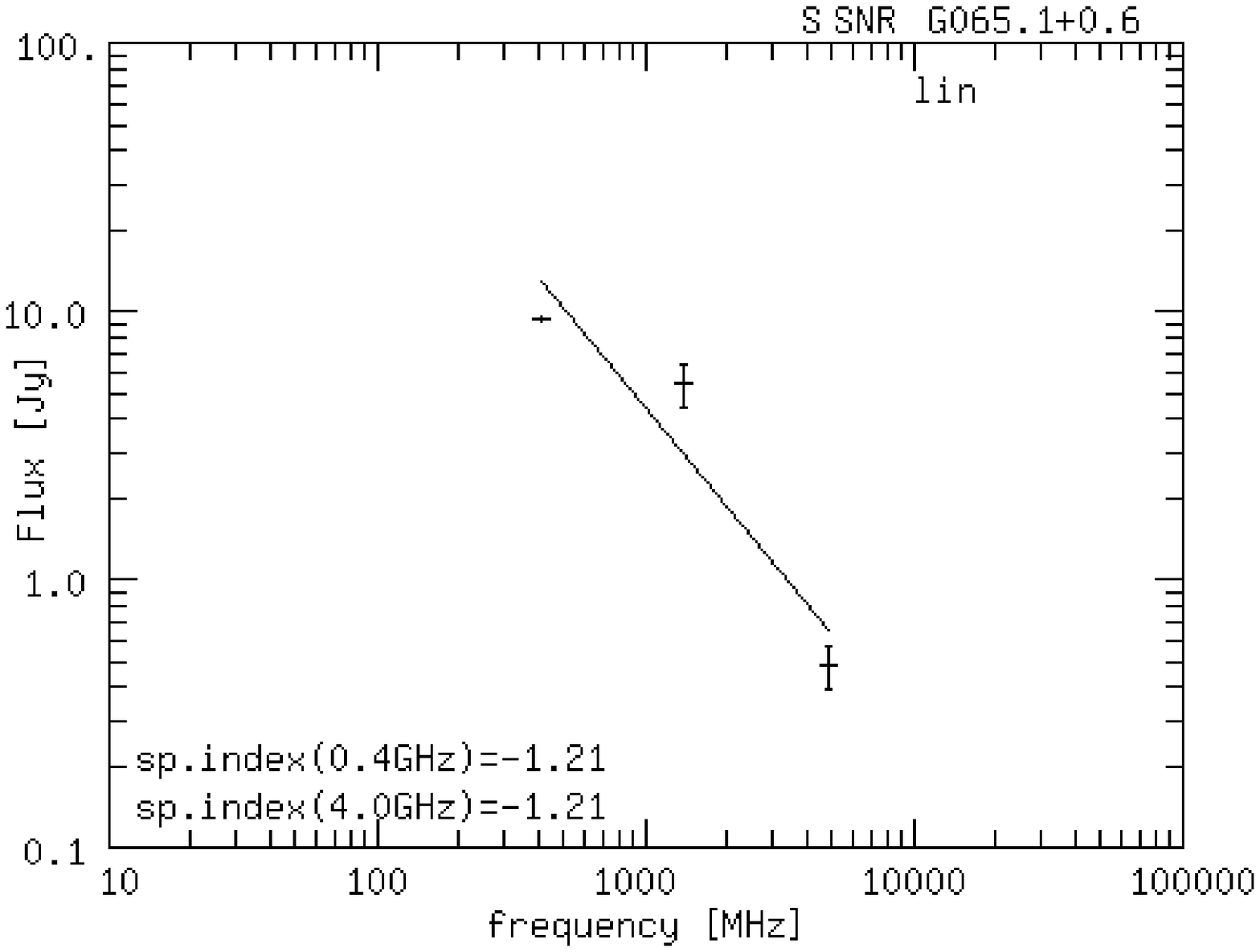,width=7.4cm,angle=0}}}\end{figure}
\begin{figure}\centerline{\vbox{\psfig{figure=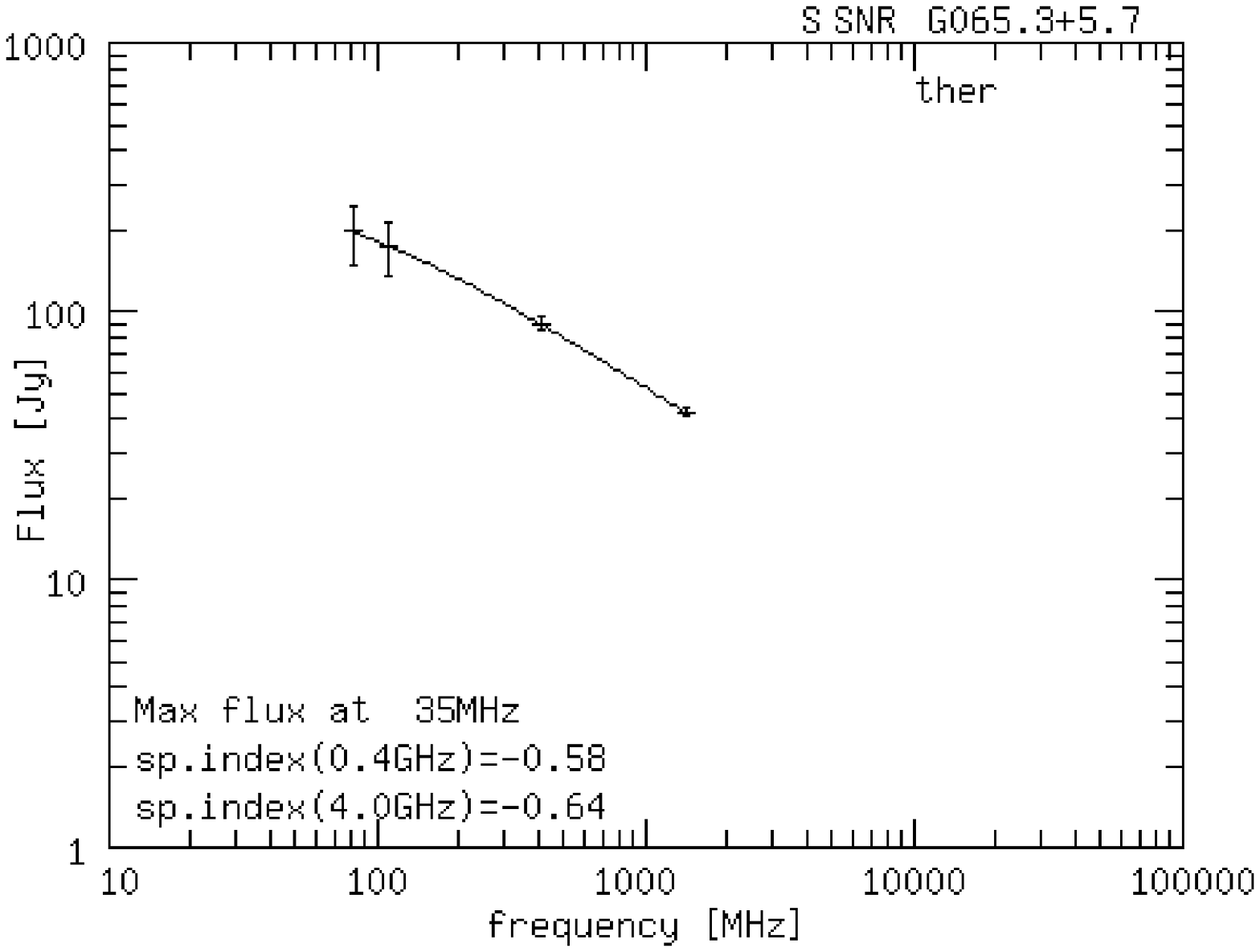,width=7.4cm,angle=0}}}\end{figure}
\begin{figure}\centerline{\vbox{\psfig{figure=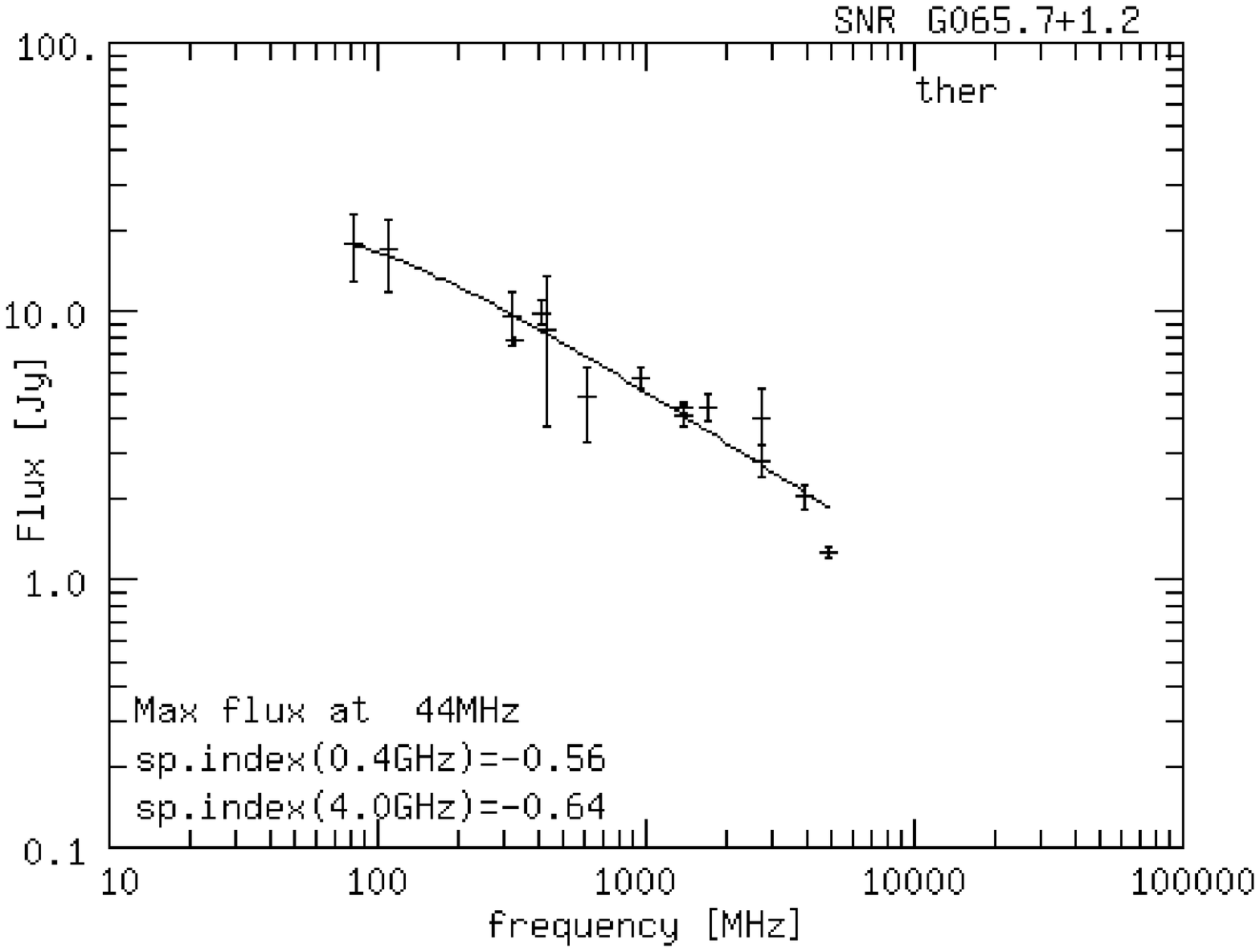,width=7.4cm,angle=0}}}\end{figure}
\begin{figure}\centerline{\vbox{\psfig{figure=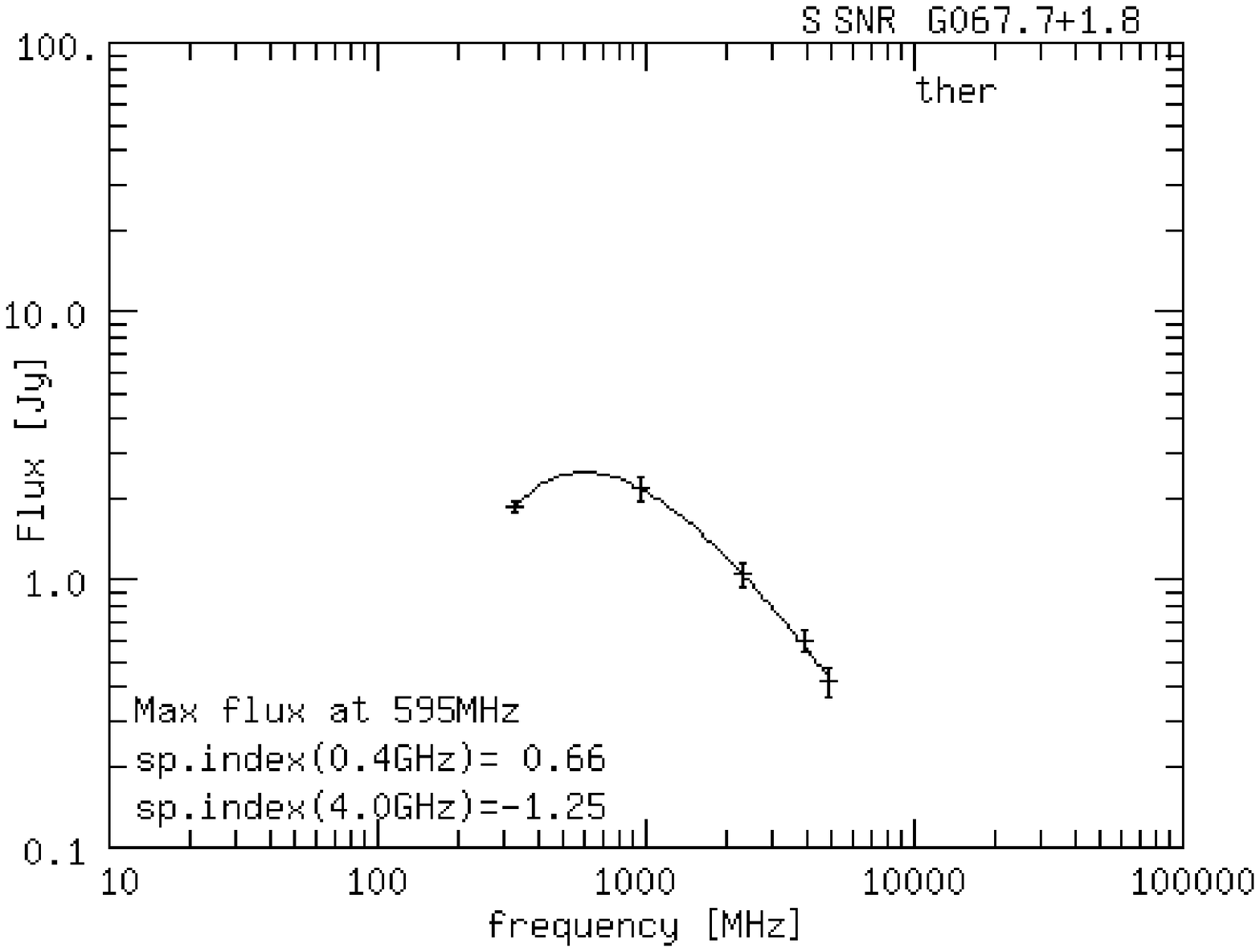,width=7.4cm,angle=0}}}\end{figure}
\begin{figure}\centerline{\vbox{\psfig{figure=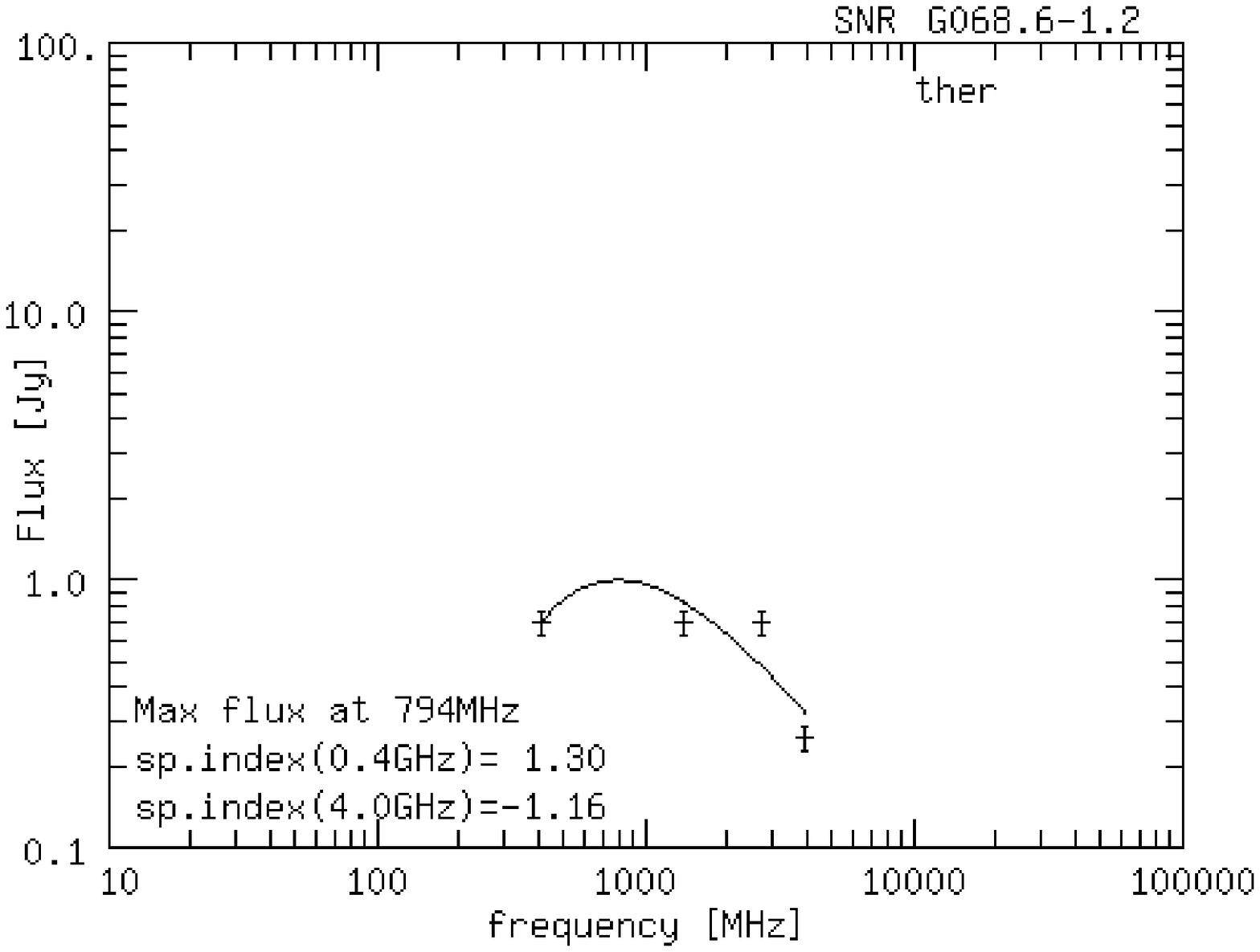,width=7.4cm,angle=0}}}\end{figure}
\begin{figure}\centerline{\vbox{\psfig{figure=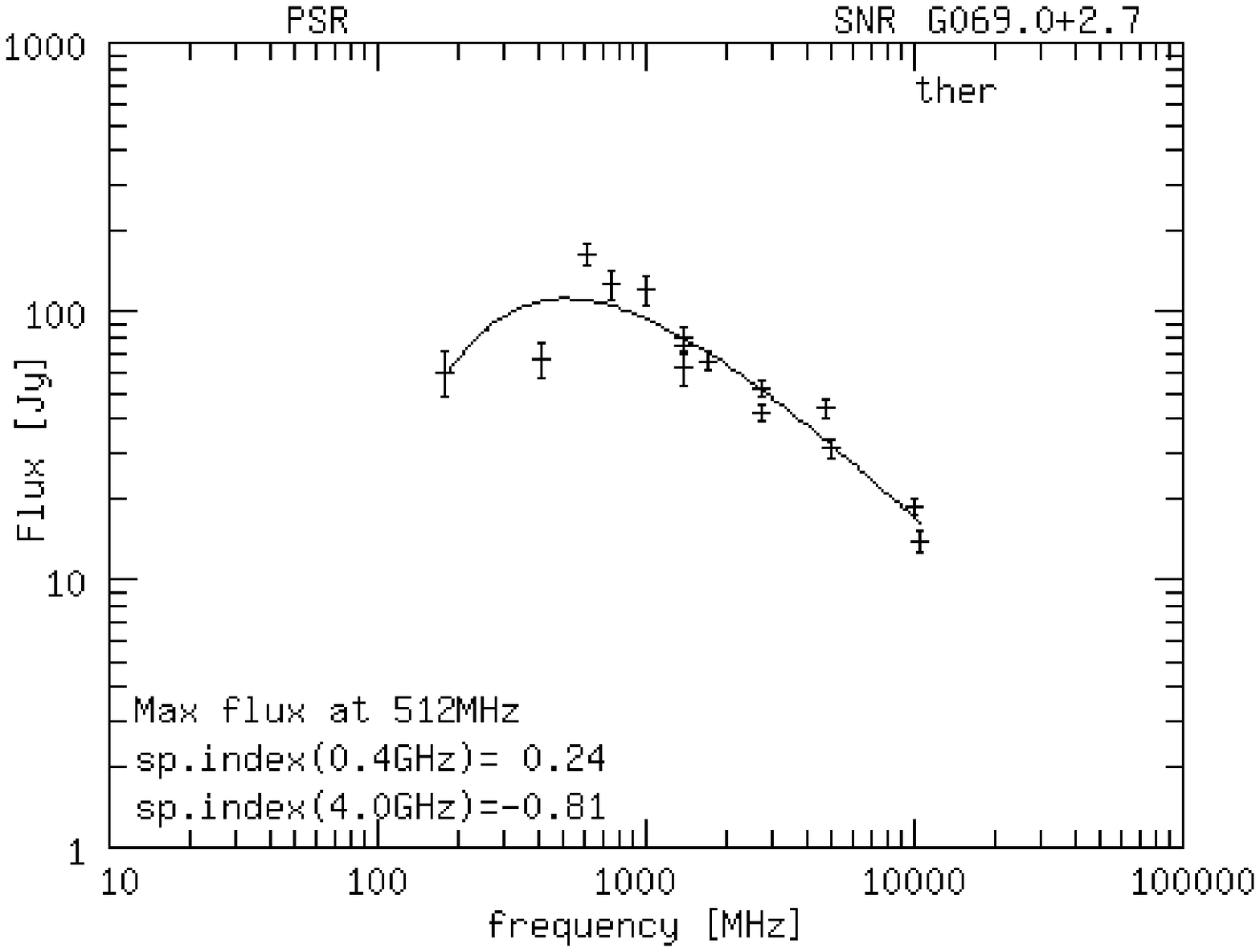,width=7.4cm,angle=0}}}\end{figure}
\begin{figure}\centerline{\vbox{\psfig{figure=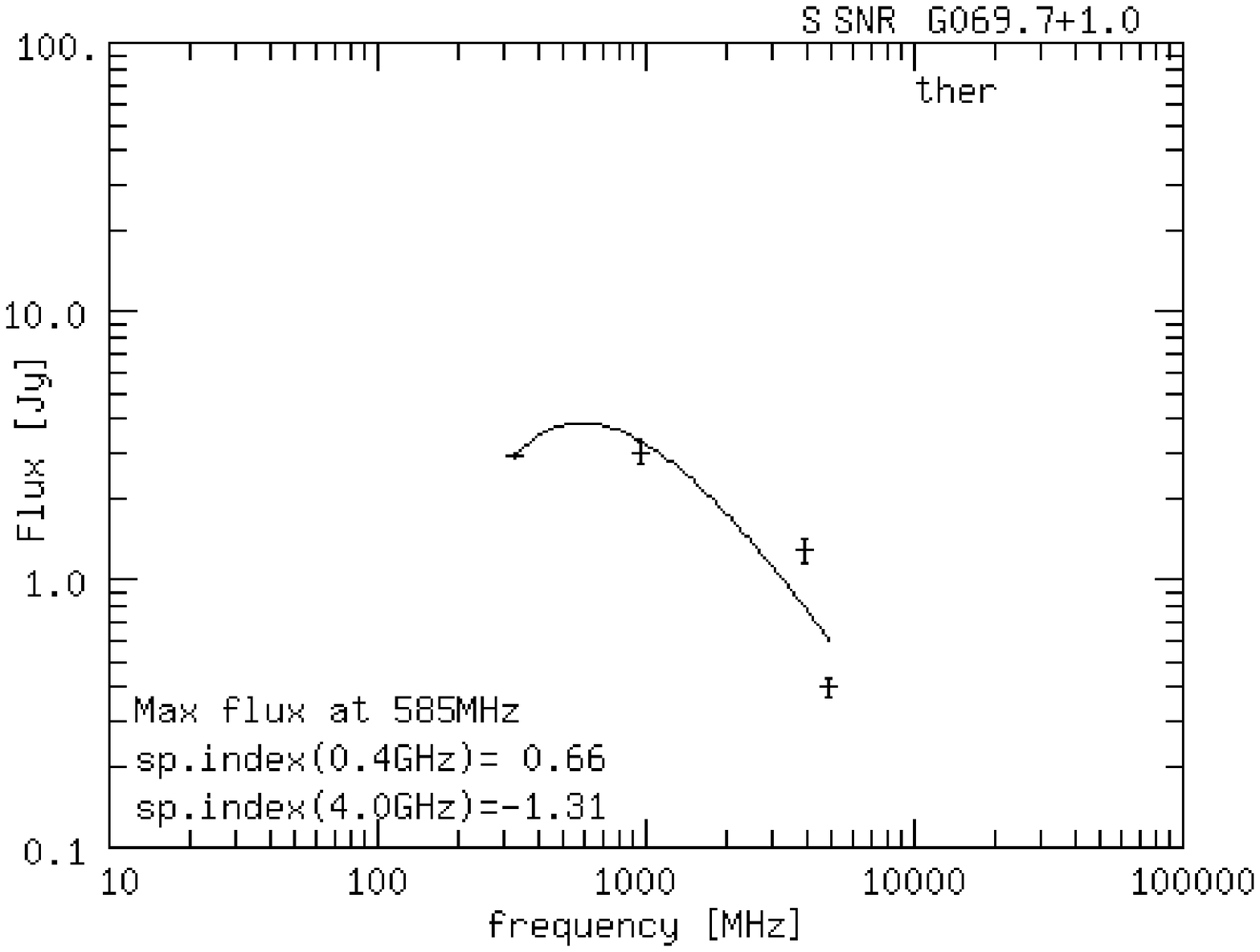,width=7.4cm,angle=0}}}\end{figure}
\begin{figure}\centerline{\vbox{\psfig{figure=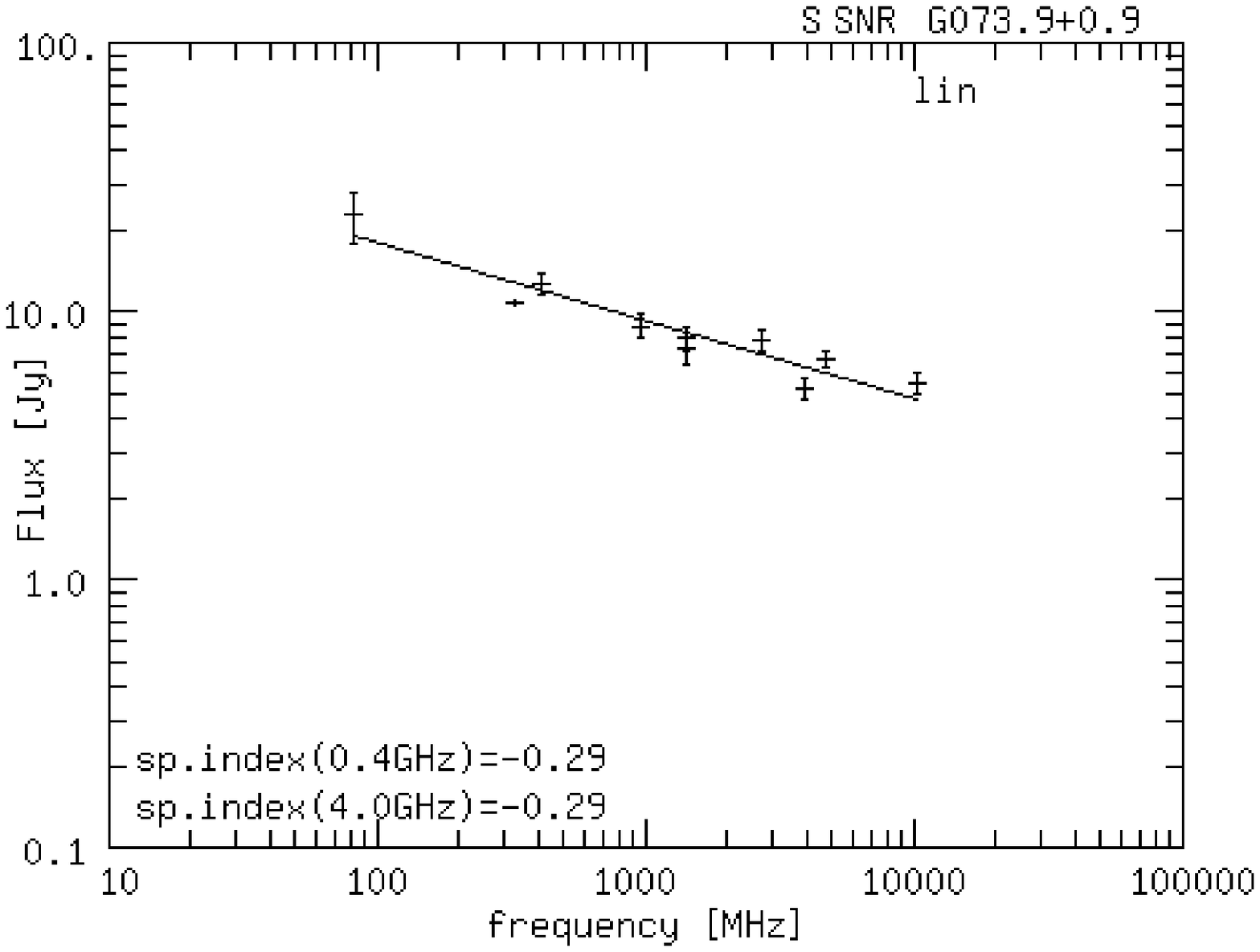,width=7.4cm,angle=0}}}\end{figure}\clearpage
\begin{figure}\centerline{\vbox{\psfig{figure=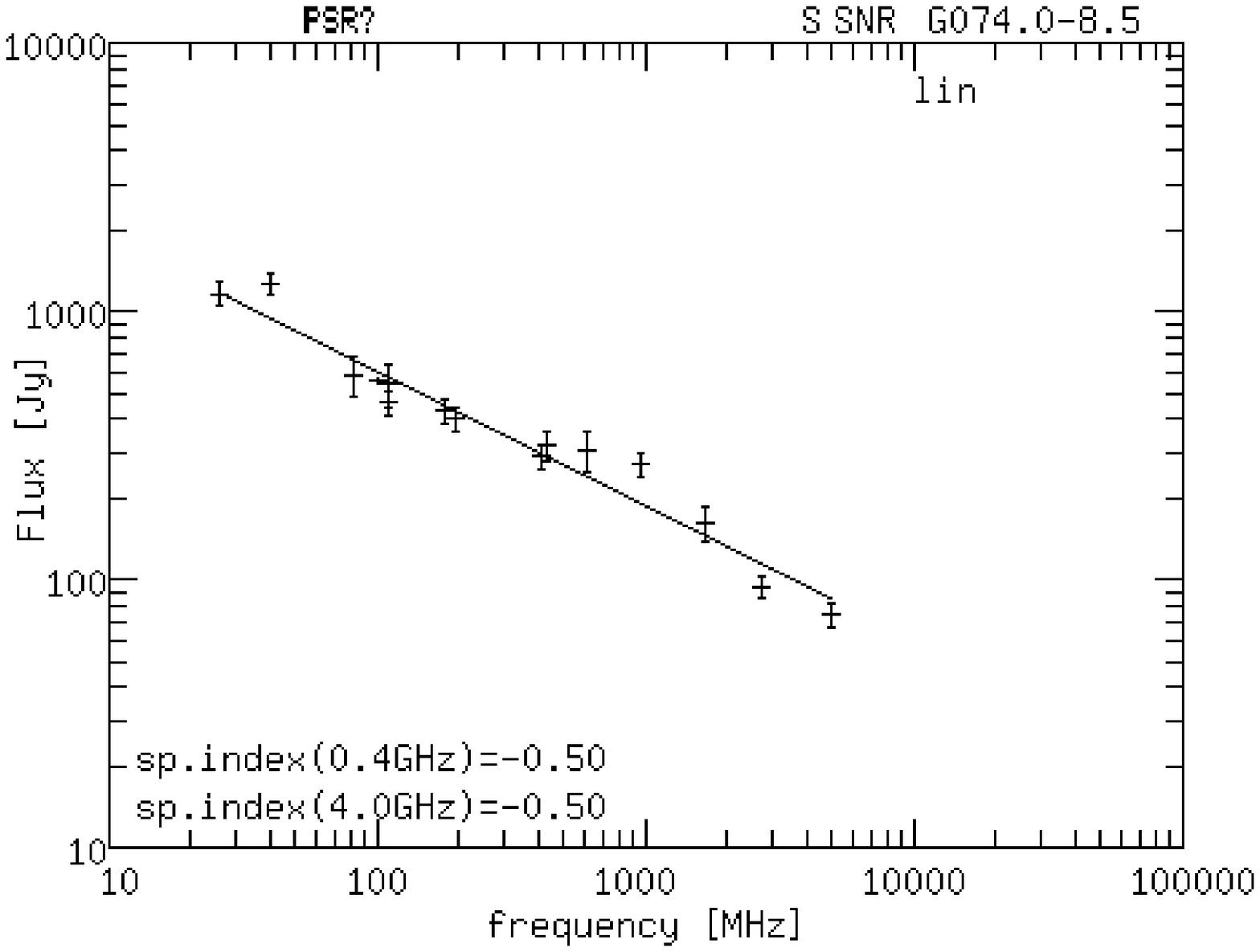,width=7.4cm,angle=0}}}\end{figure}
\begin{figure}\centerline{\vbox{\psfig{figure=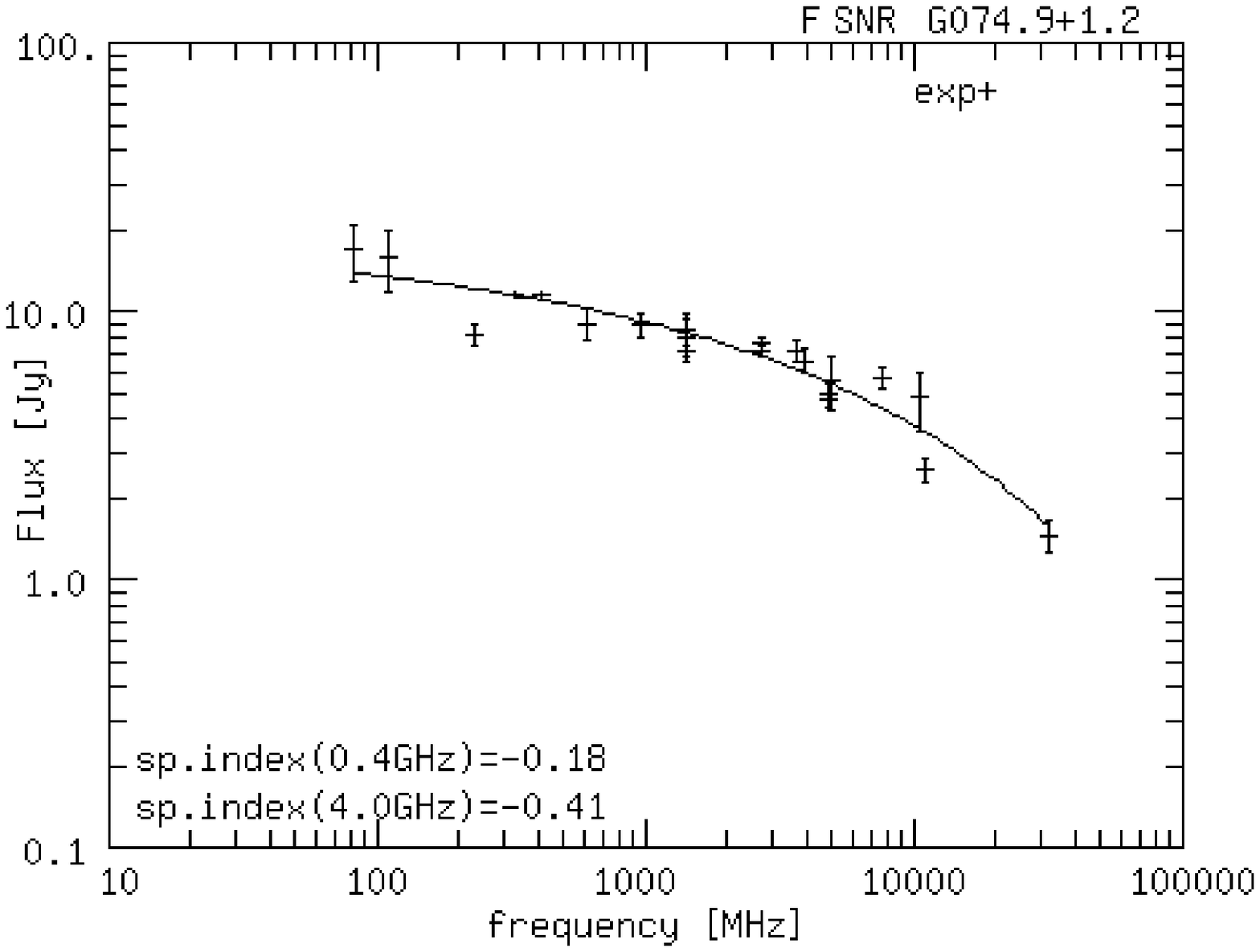,width=7.4cm,angle=0}}}\end{figure}
\begin{figure}\centerline{\vbox{\psfig{figure=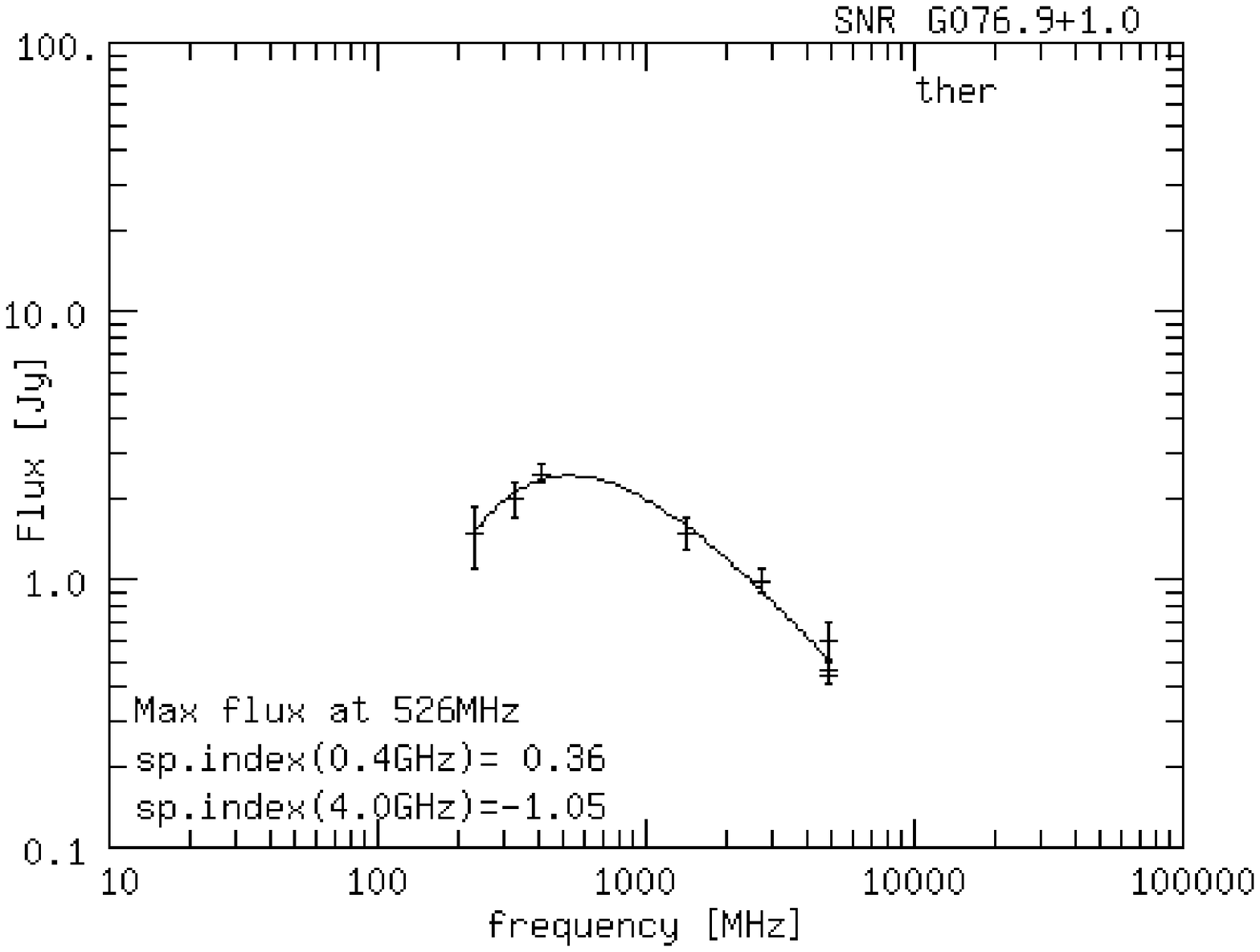,width=7.4cm,angle=0}}}\end{figure}
\begin{figure}\centerline{\vbox{\psfig{figure=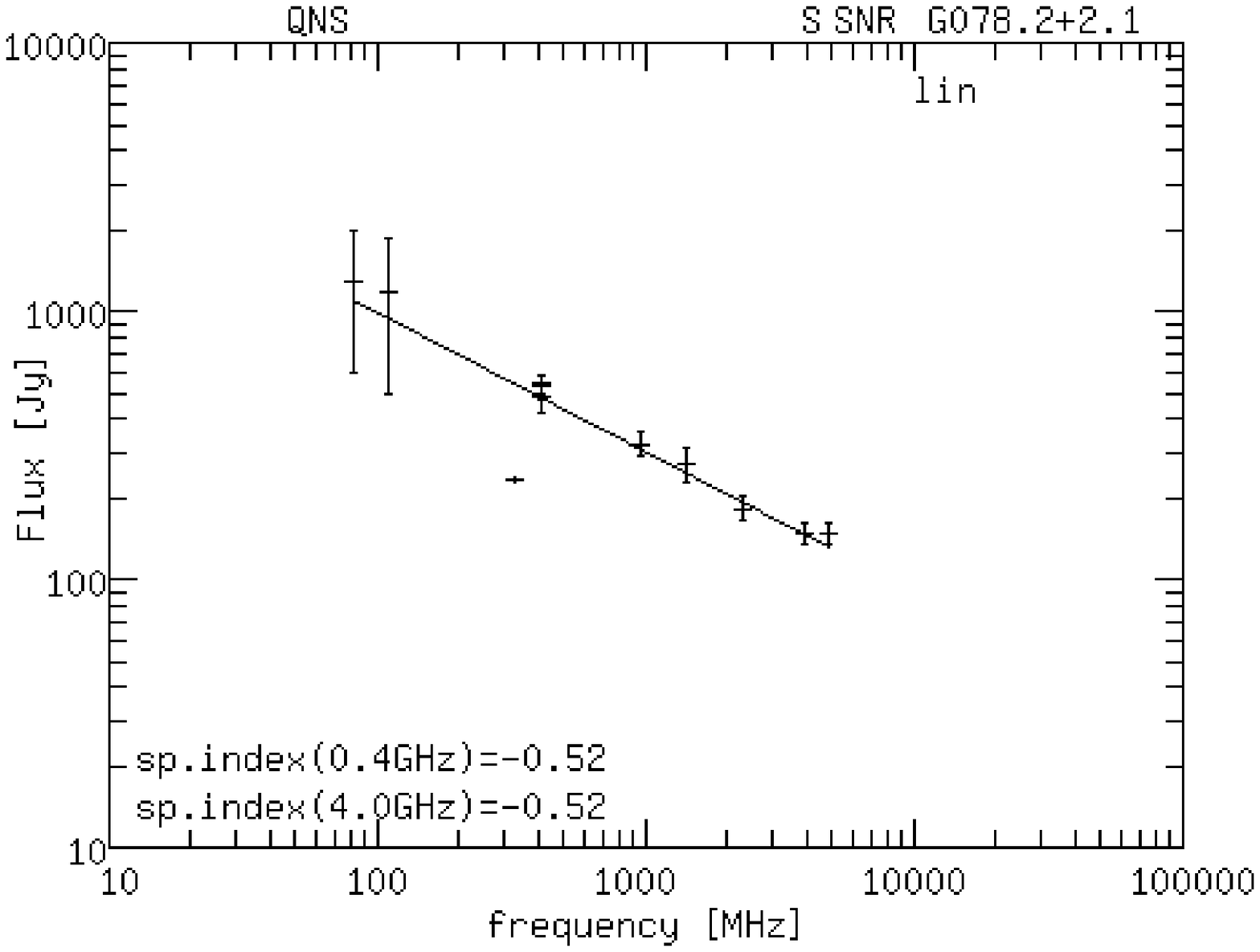,width=7.4cm,angle=0}}}\end{figure}
\begin{figure}\centerline{\vbox{\psfig{figure=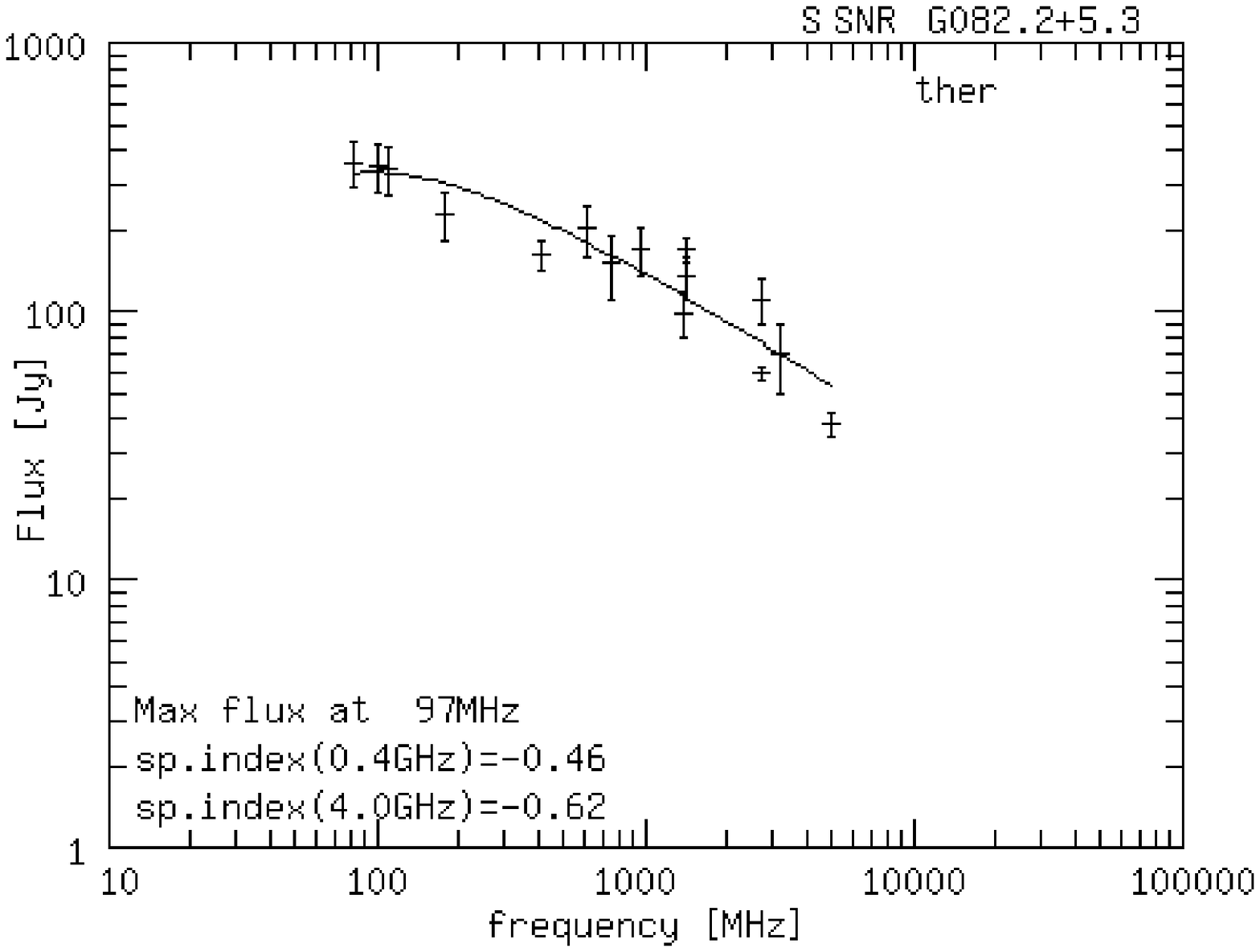,width=7.4cm,angle=0}}}\end{figure}
\begin{figure}\centerline{\vbox{\psfig{figure=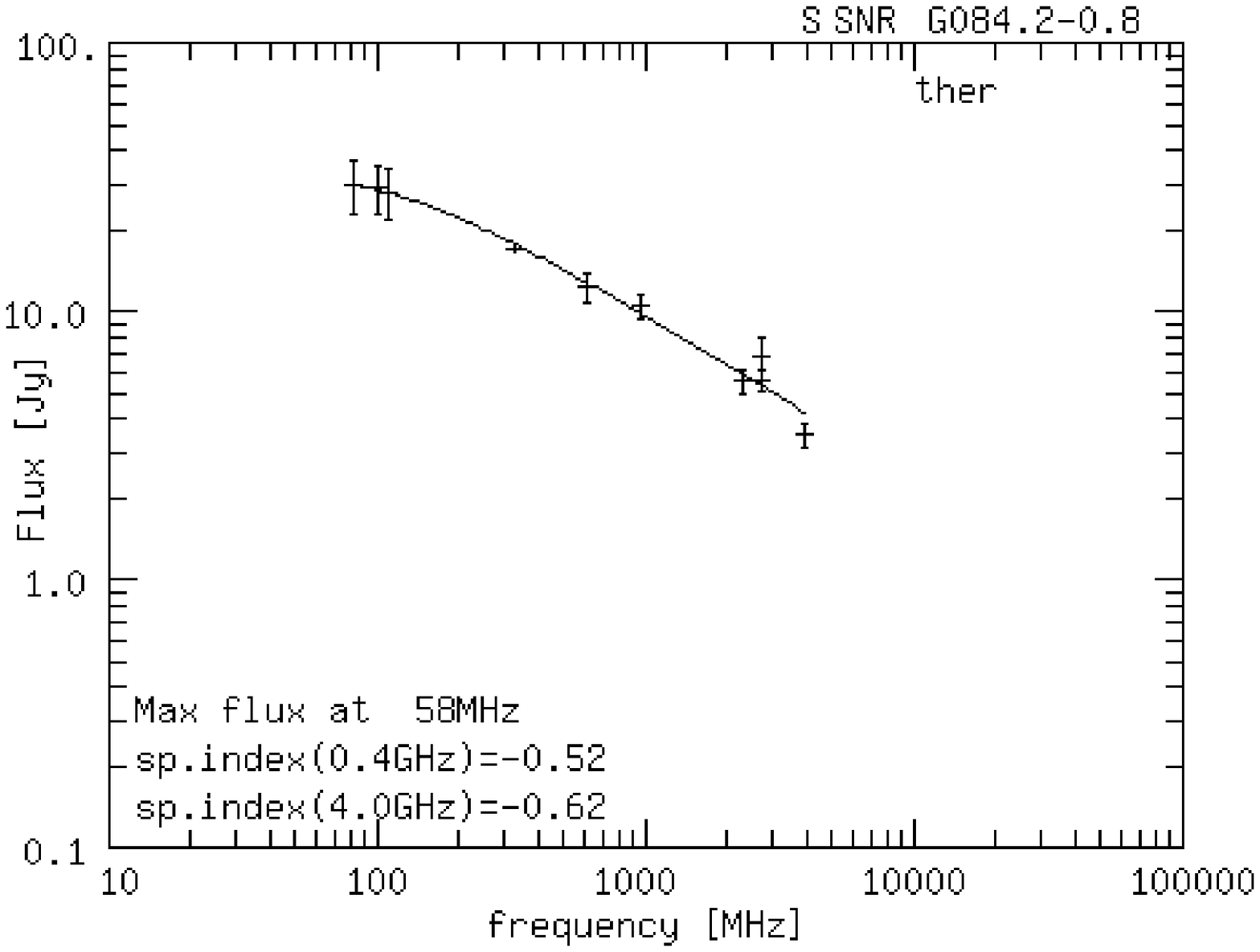,width=7.4cm,angle=0}}}\end{figure}
\begin{figure}\centerline{\vbox{\psfig{figure=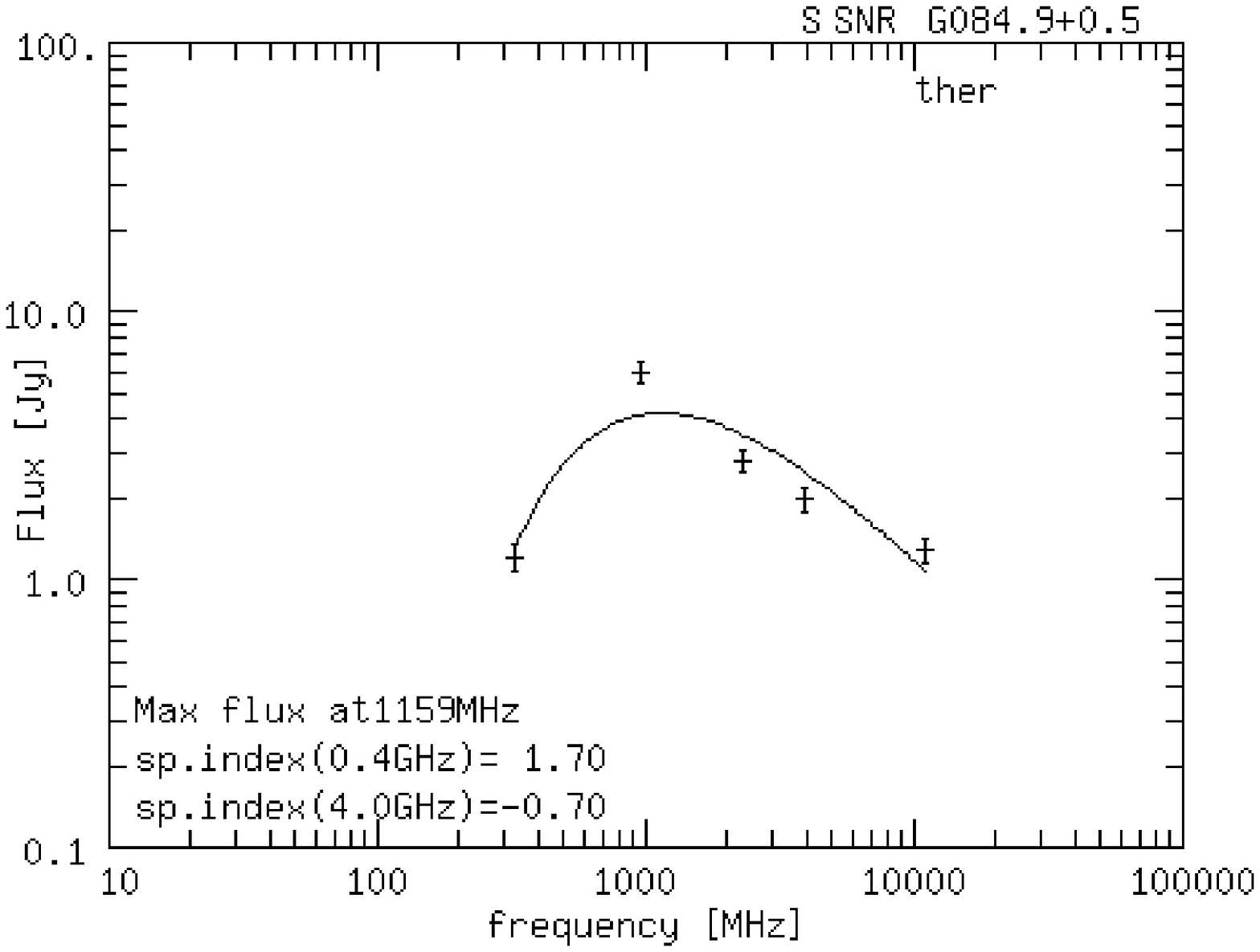,width=7.4cm,angle=0}}}\end{figure}
\begin{figure}\centerline{\vbox{\psfig{figure=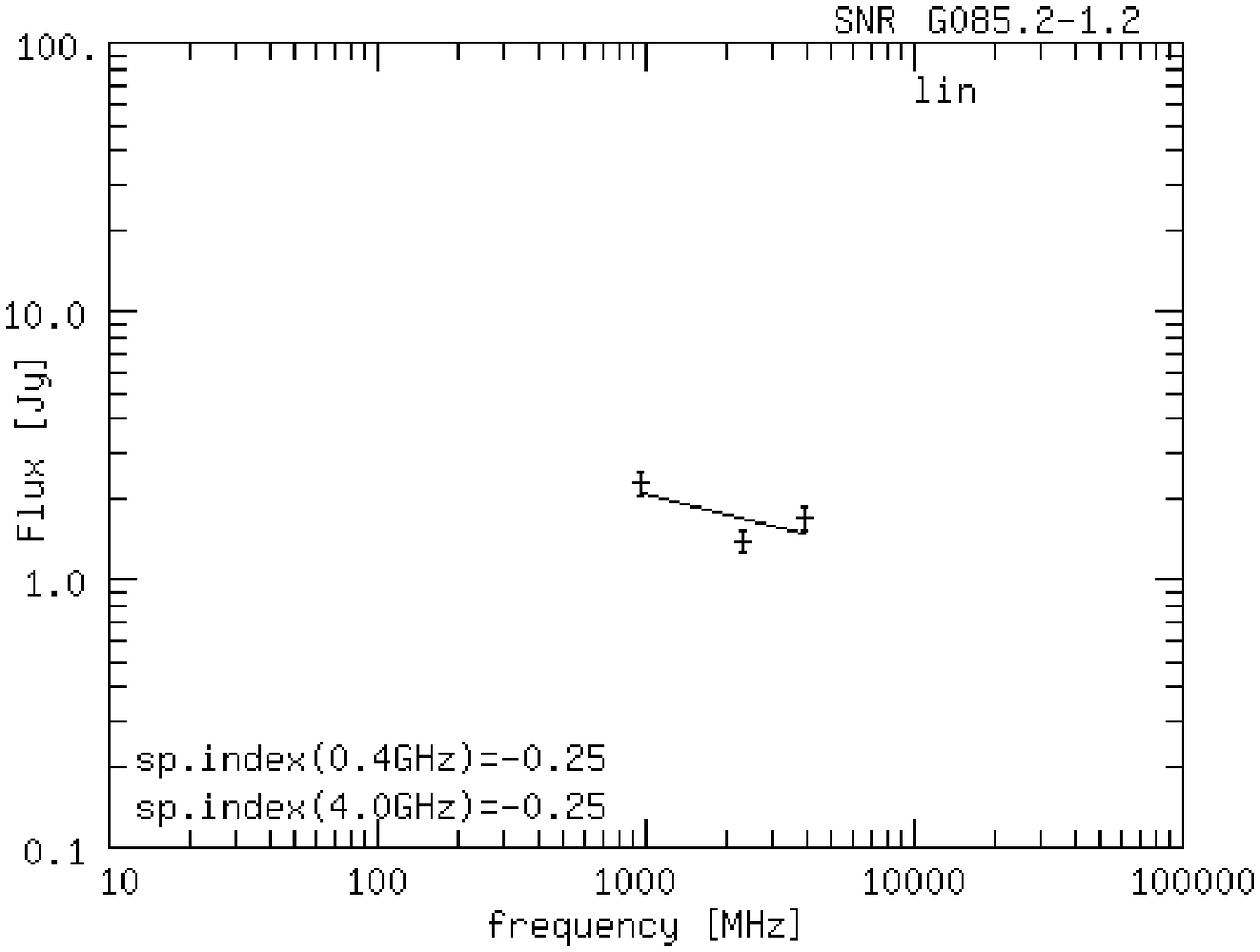,width=7.4cm,angle=0}}}\end{figure}\clearpage
\begin{figure}\centerline{\vbox{\psfig{figure=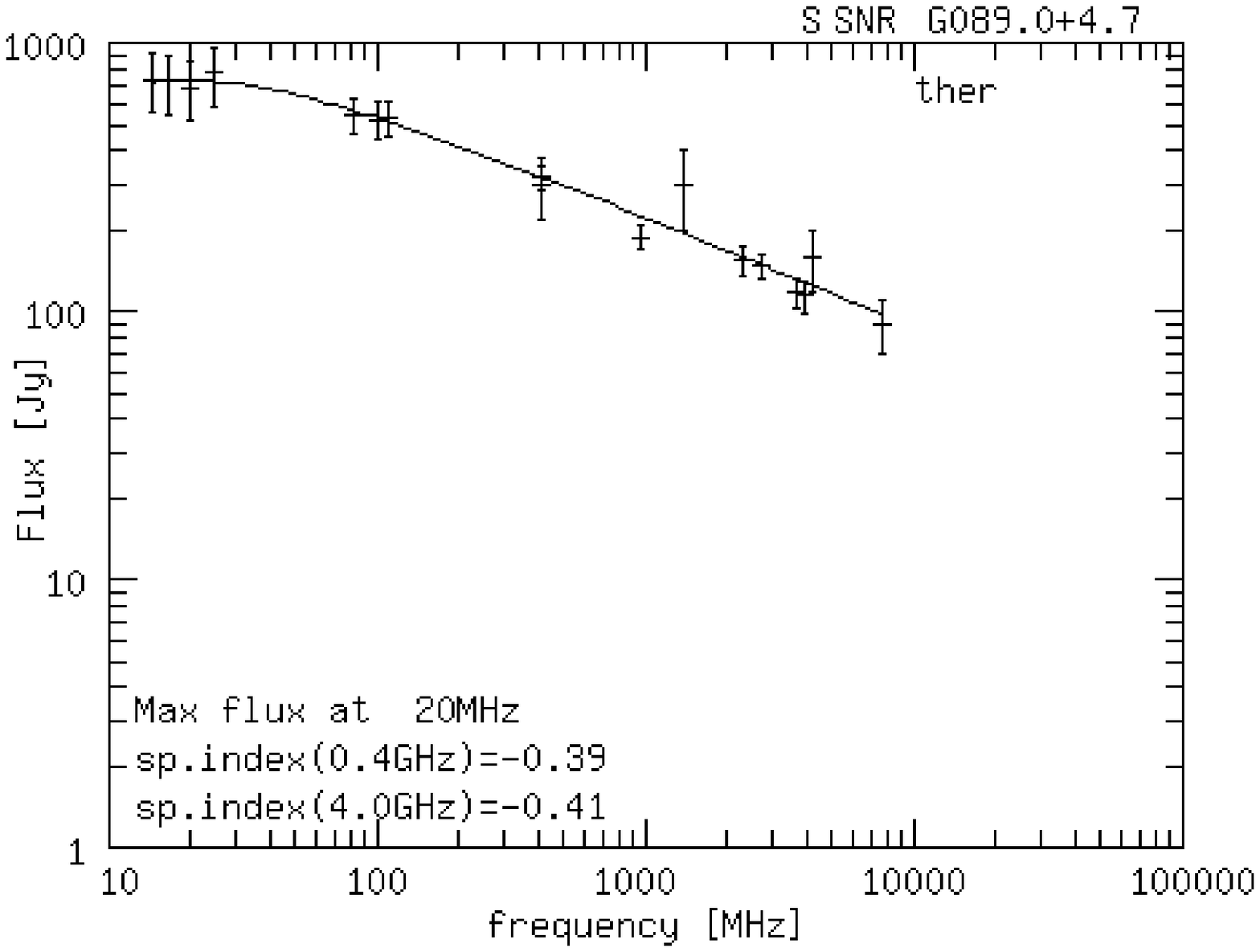,width=7.4cm,angle=0}}}\end{figure}
\begin{figure}\centerline{\vbox{\psfig{figure=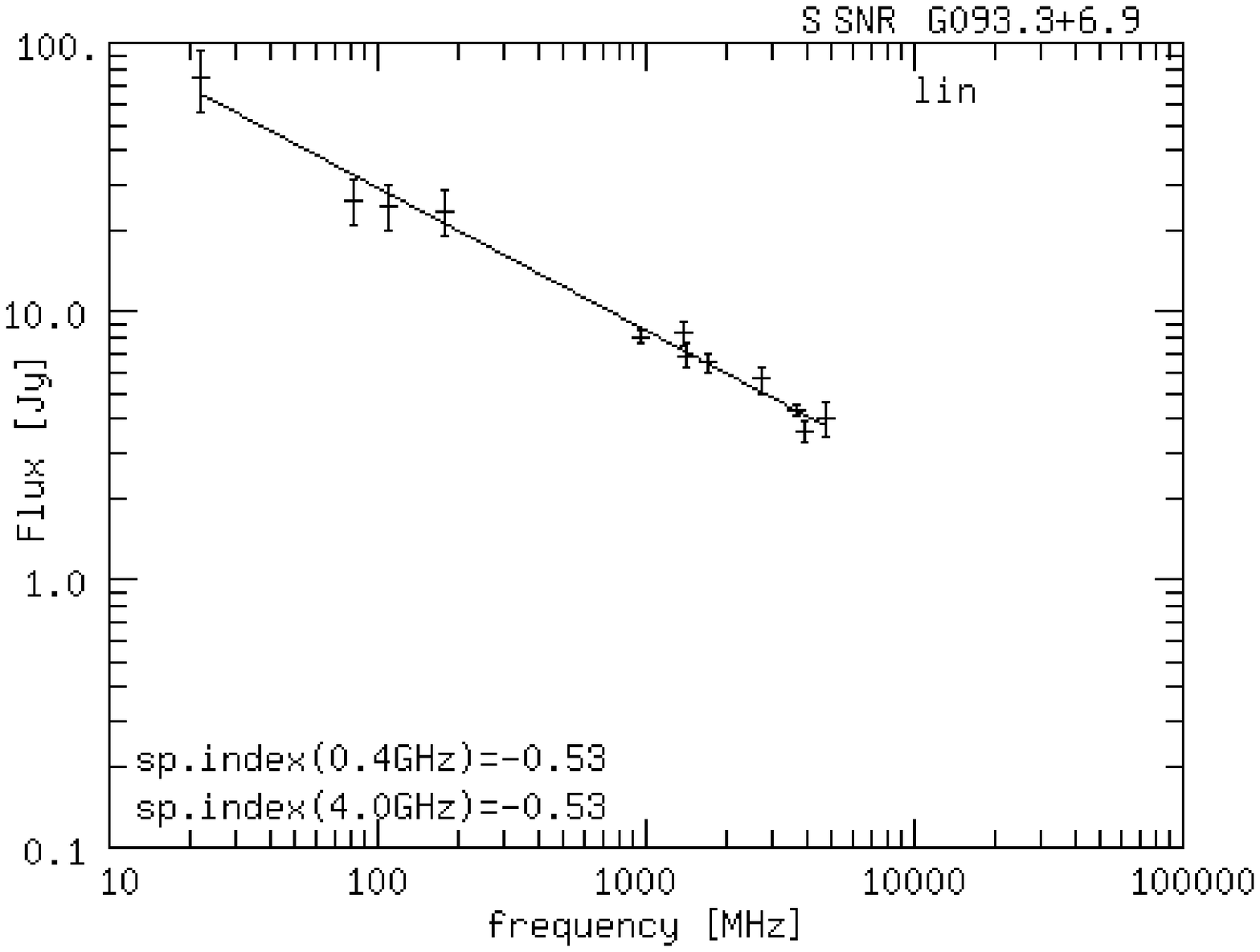,width=7.4cm,angle=0}}}\end{figure}
\begin{figure}\centerline{\vbox{\psfig{figure=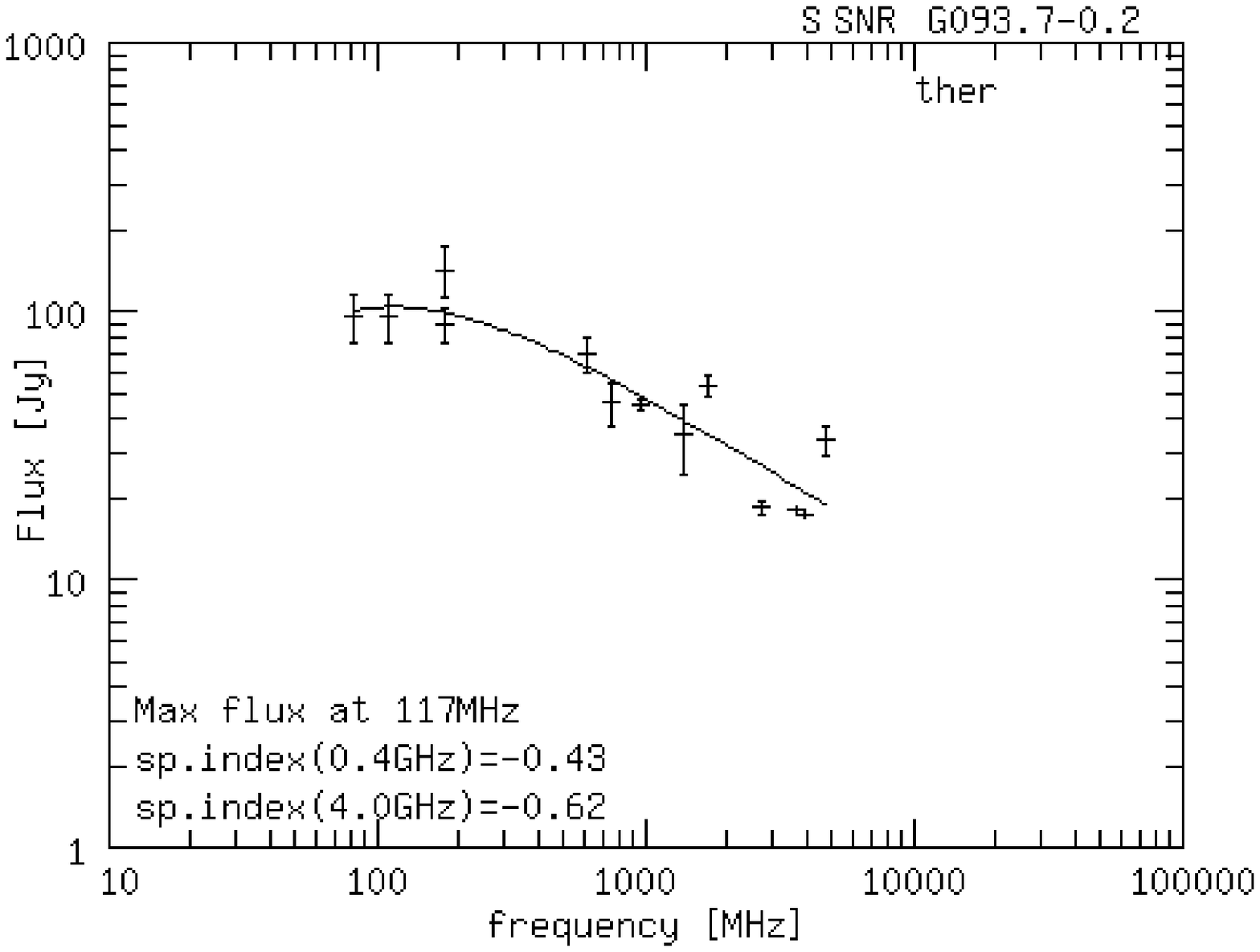,width=7.4cm,angle=0}}}\end{figure}
\begin{figure}\centerline{\vbox{\psfig{figure=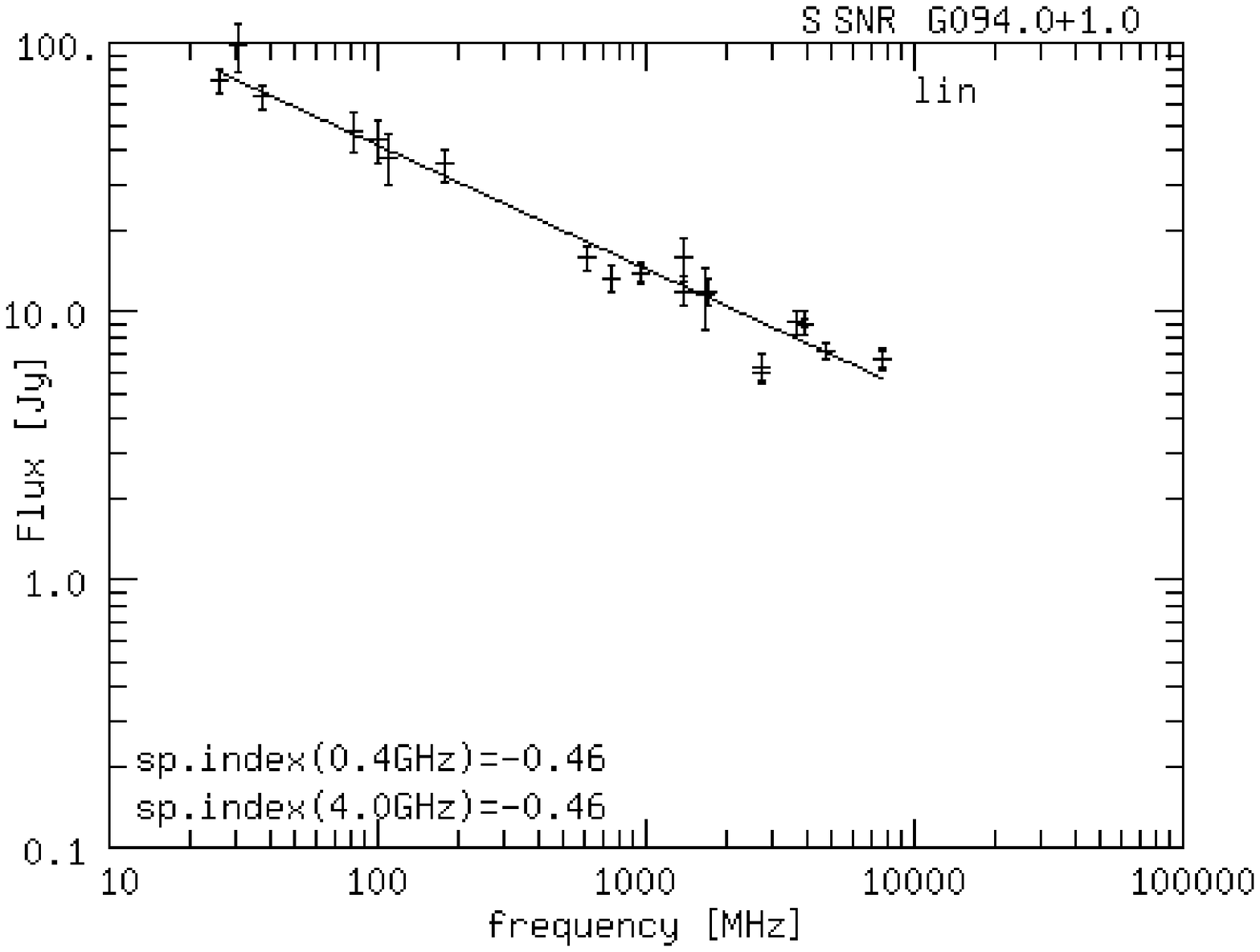,width=7.4cm,angle=0}}}\end{figure}
\begin{figure}\centerline{\vbox{\psfig{figure=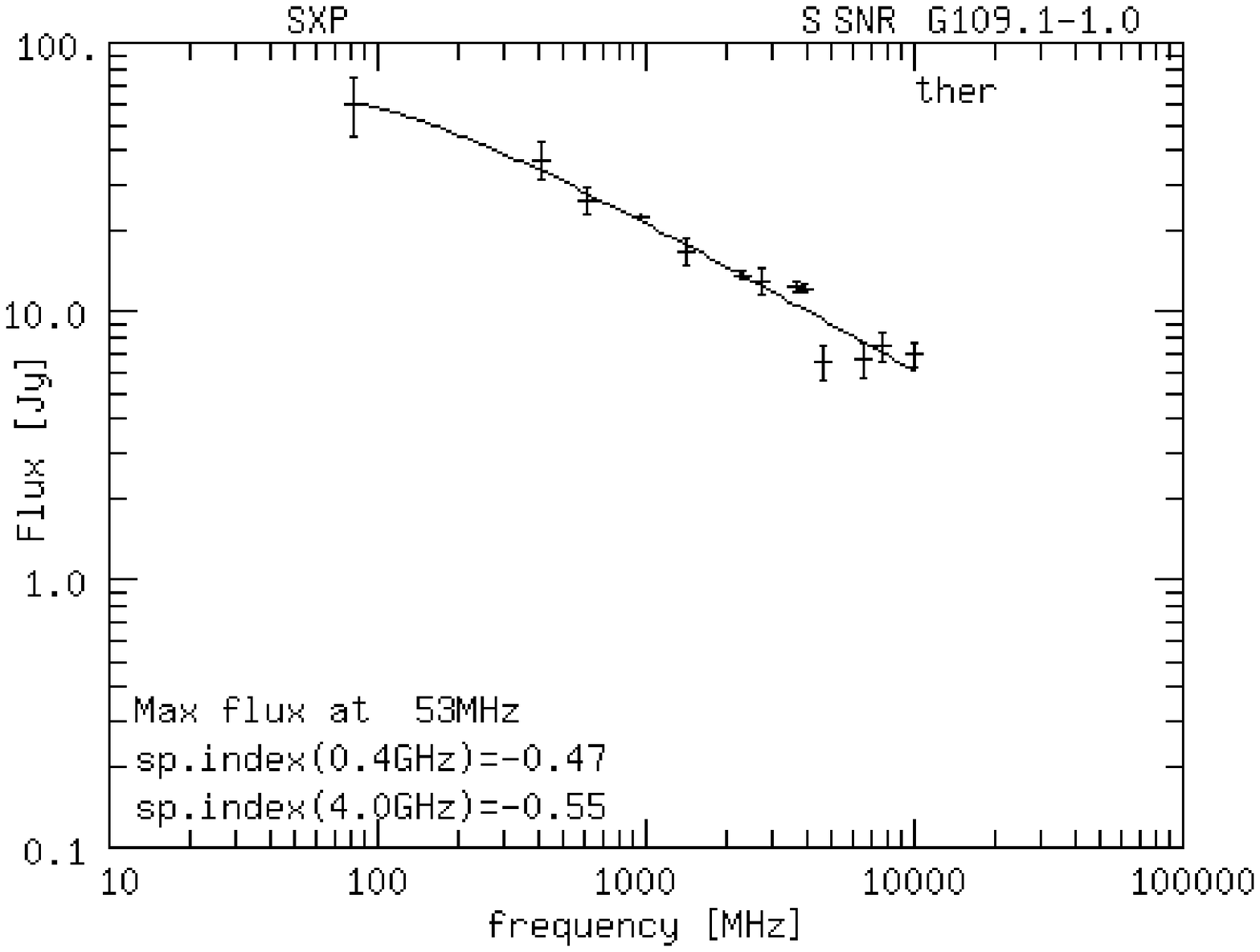,width=7.4cm,angle=0}}}\end{figure}
\begin{figure}\centerline{\vbox{\psfig{figure=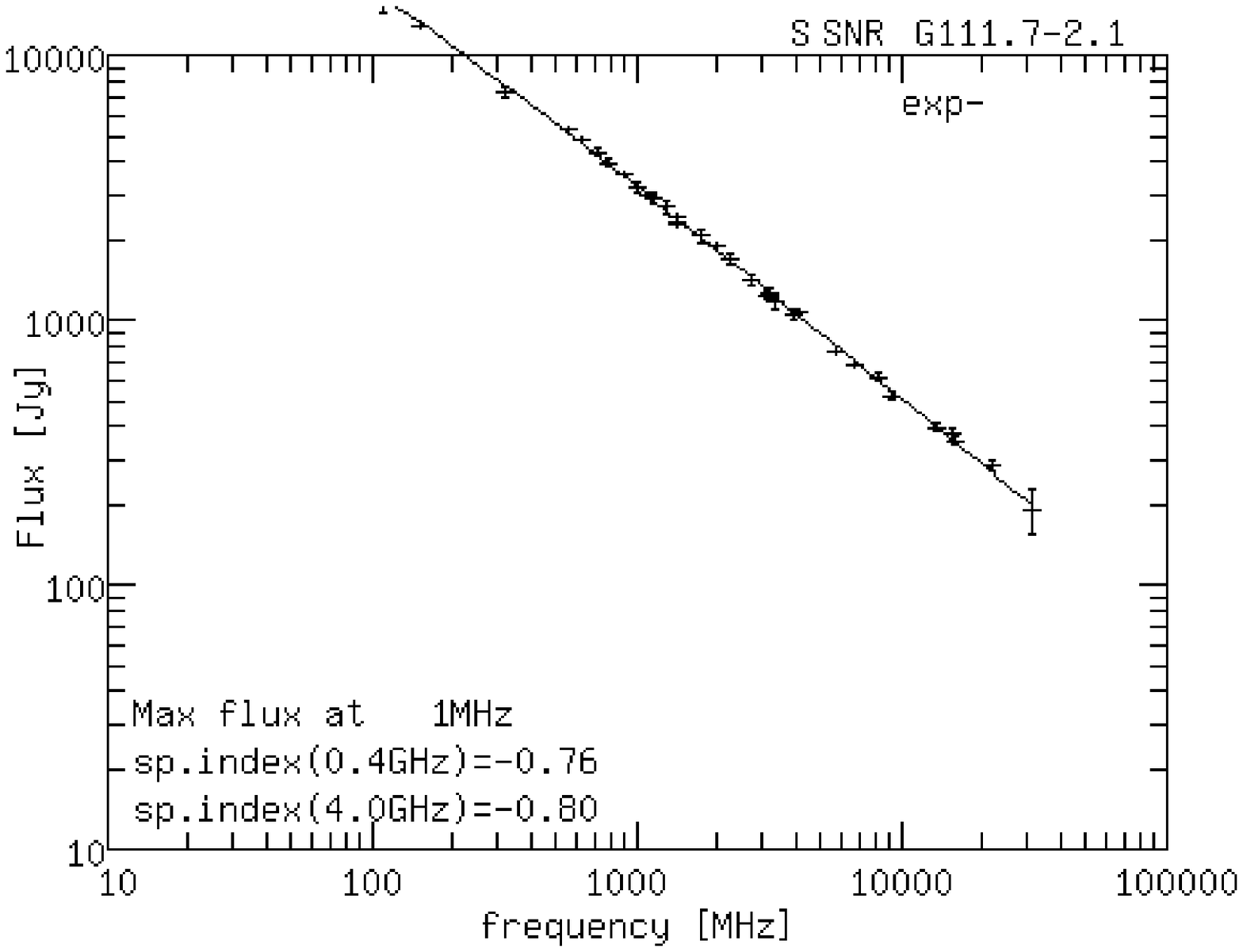,width=7.4cm,angle=0}}}\end{figure}
\begin{figure}\centerline{\vbox{\psfig{figure=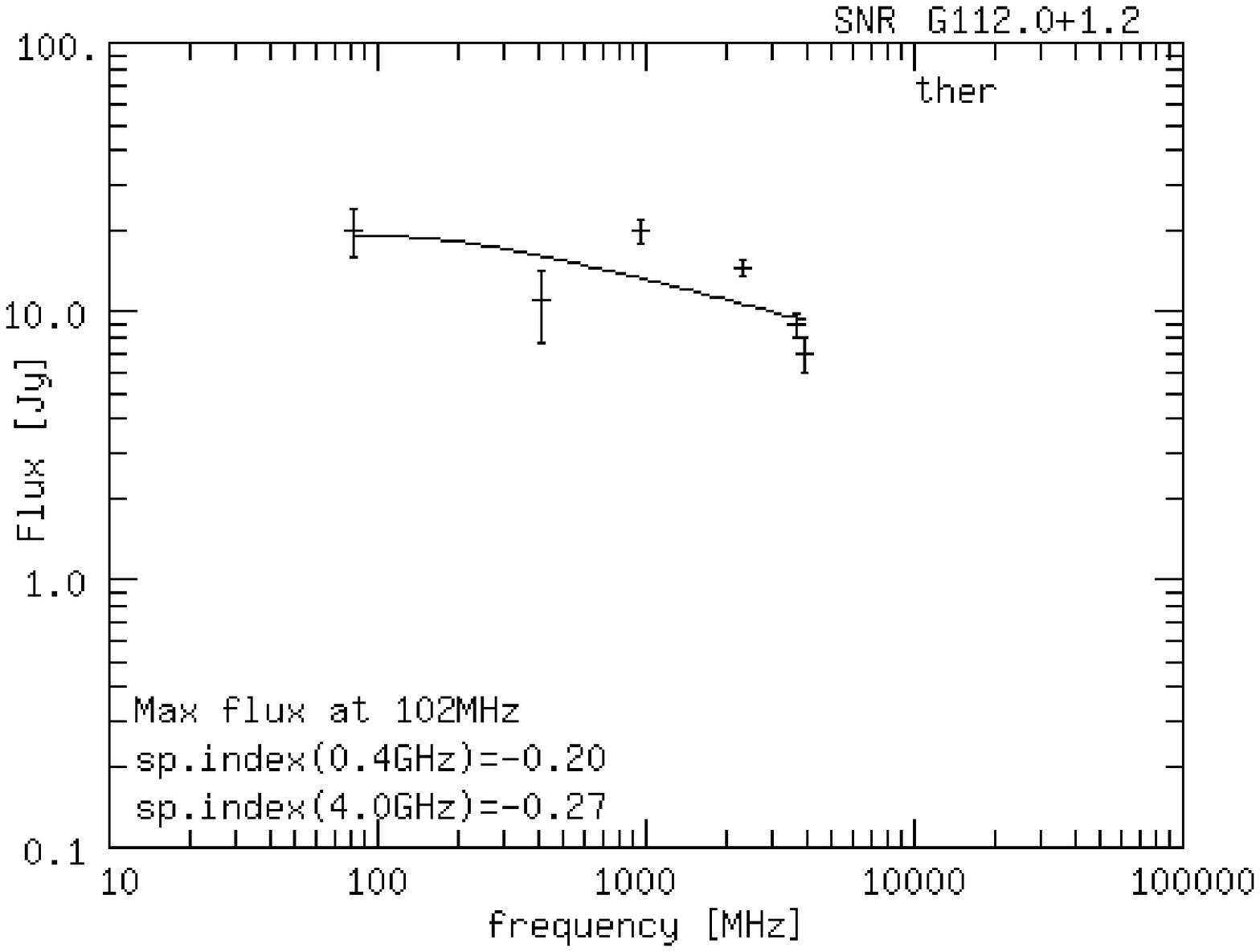,width=7.4cm,angle=0}}}\end{figure}
\begin{figure}\centerline{\vbox{\psfig{figure=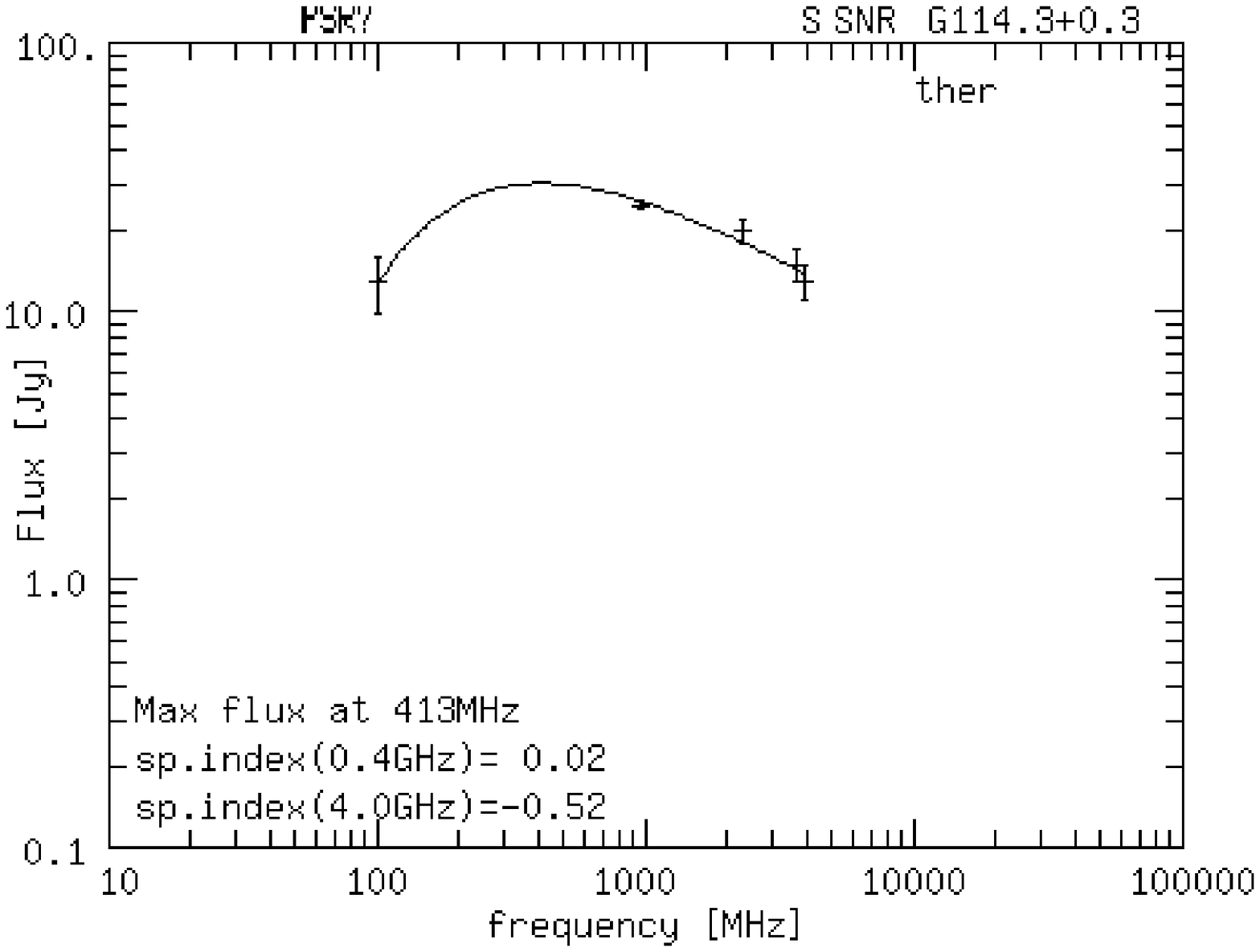,width=7.4cm,angle=0}}}\end{figure}\clearpage
\begin{figure}\centerline{\vbox{\psfig{figure=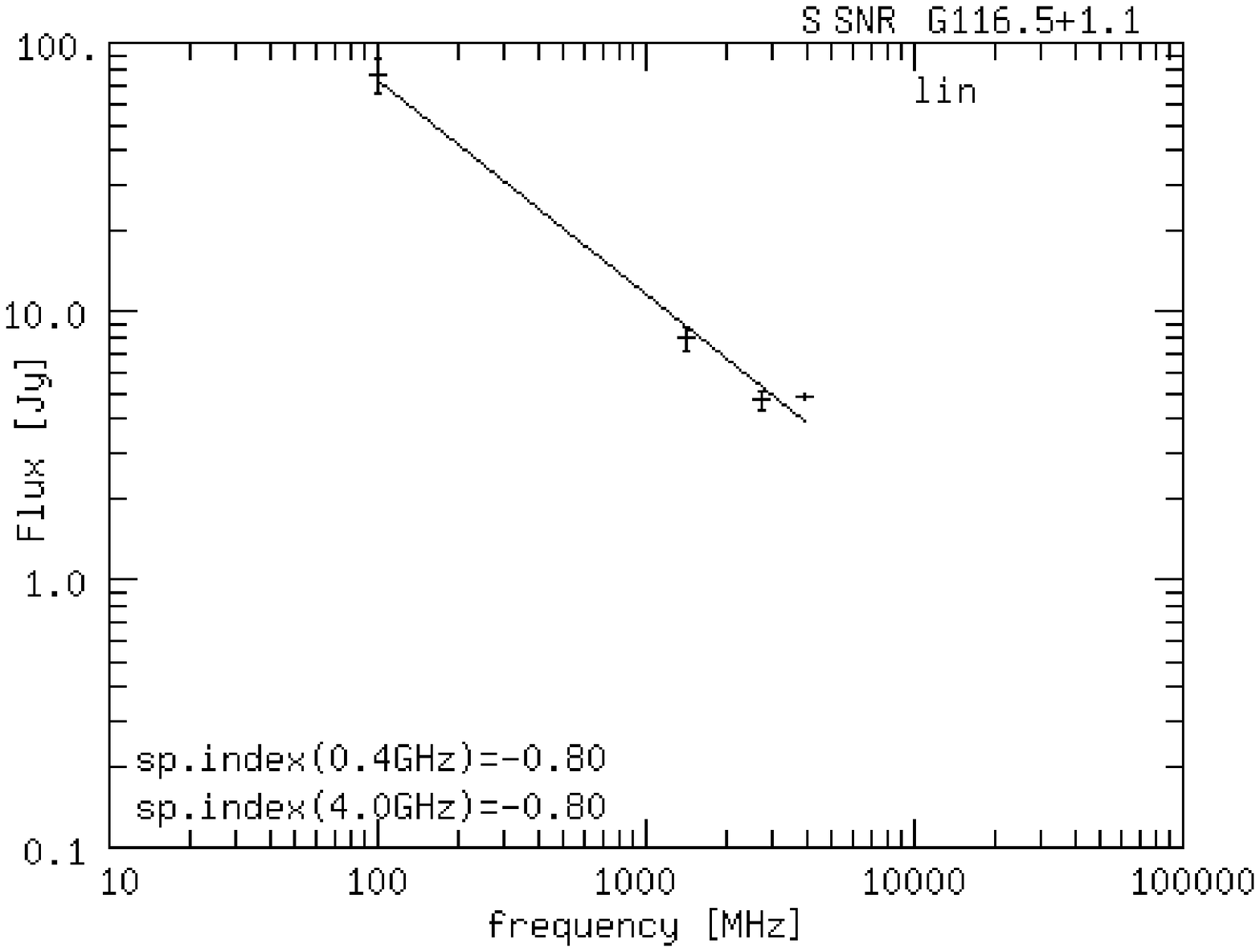,width=7.4cm,angle=0}}}\end{figure}
\begin{figure}\centerline{\vbox{\psfig{figure=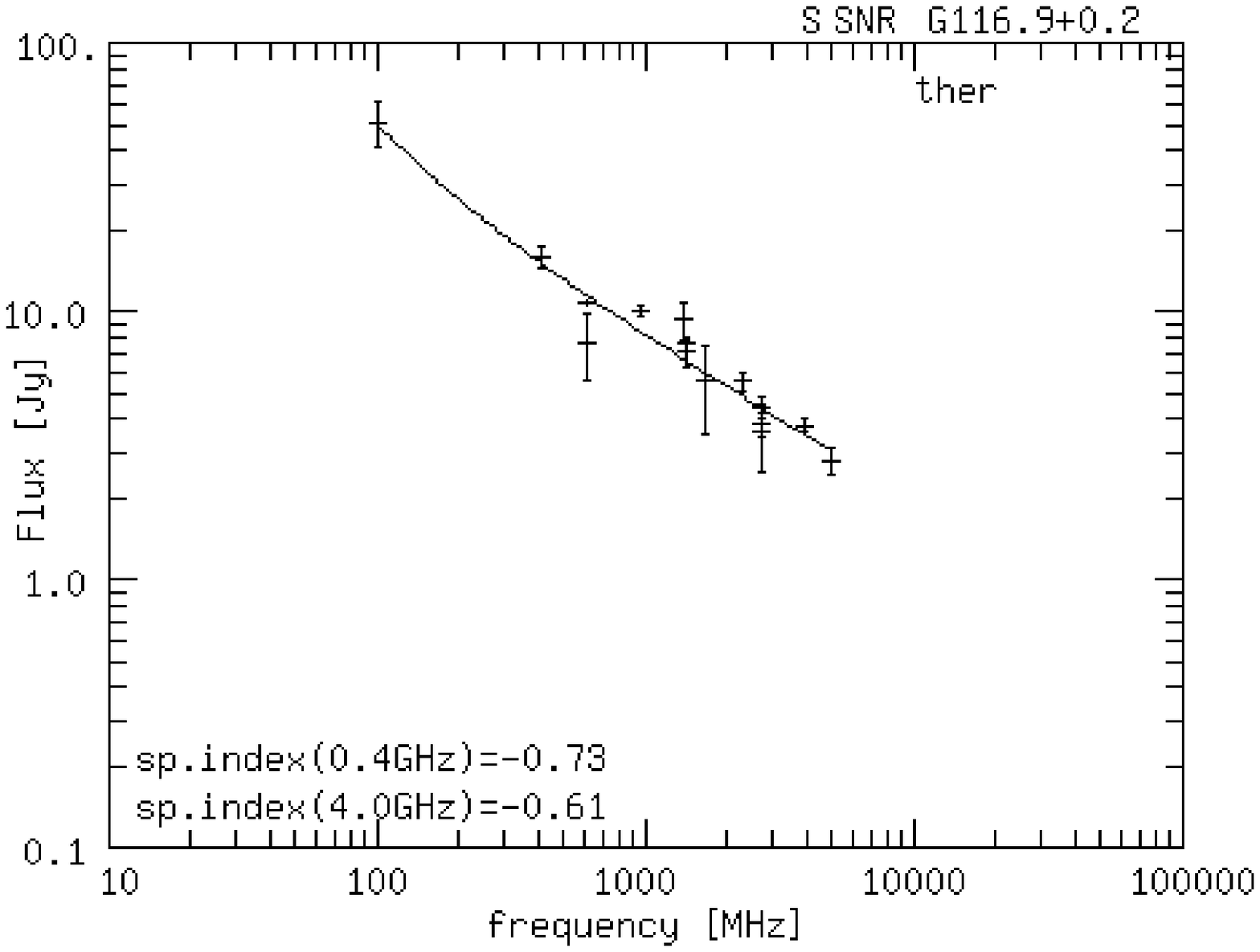,width=7.4cm,angle=0}}}\end{figure}
\begin{figure}\centerline{\vbox{\psfig{figure=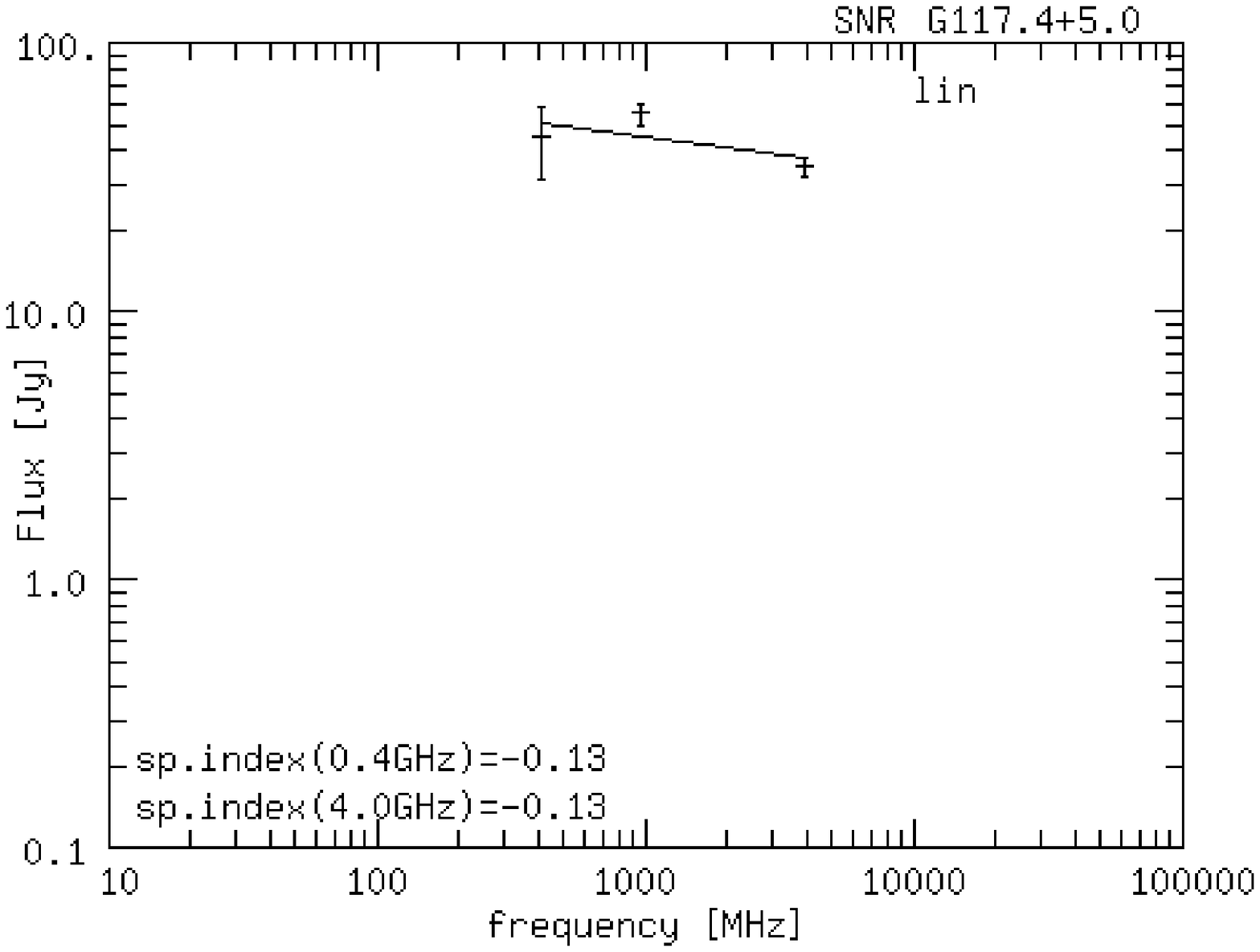,width=7.4cm,angle=0}}}\end{figure}
\begin{figure}\centerline{\vbox{\psfig{figure=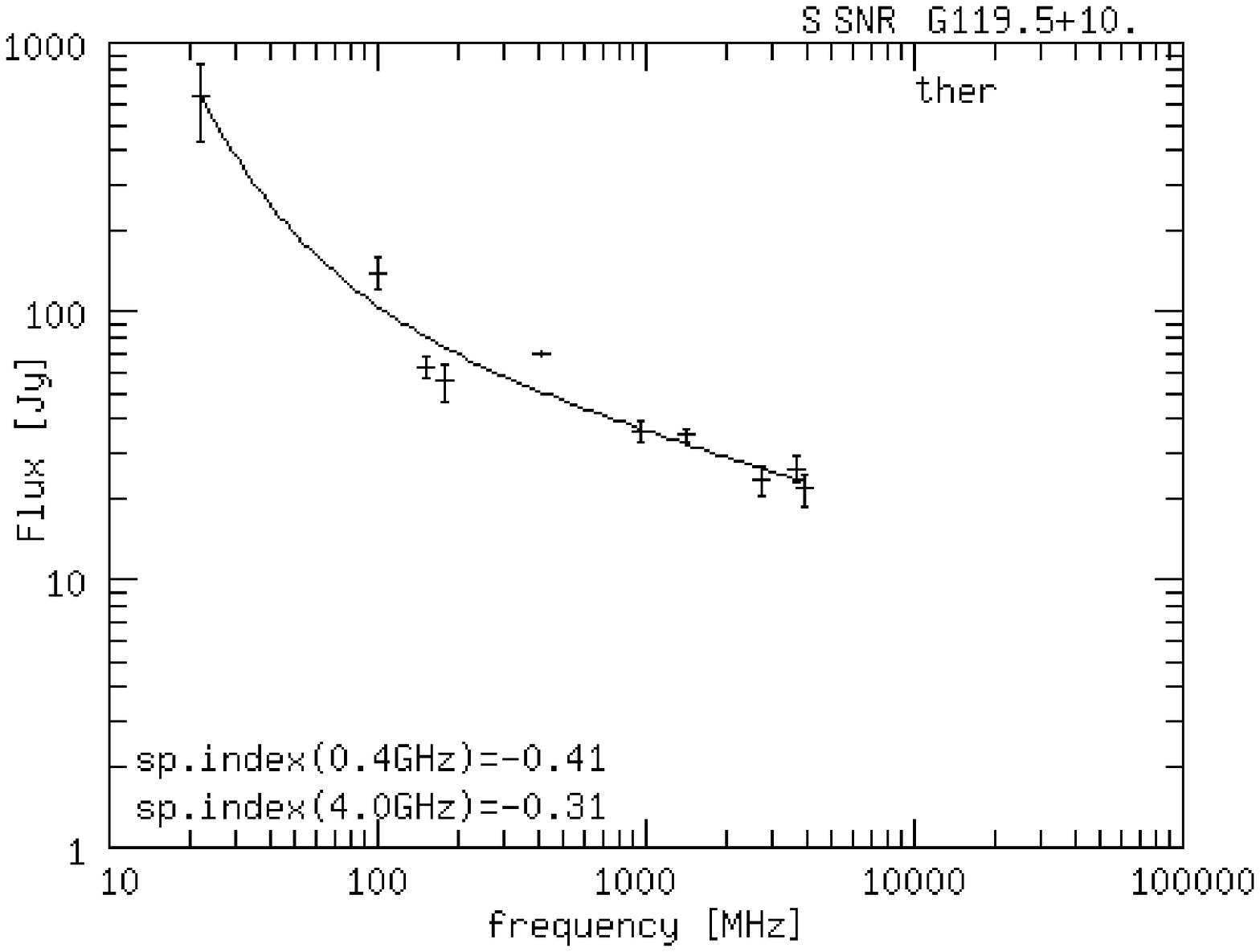,width=7.4cm,angle=0}}}\end{figure}
\begin{figure}\centerline{\vbox{\psfig{figure=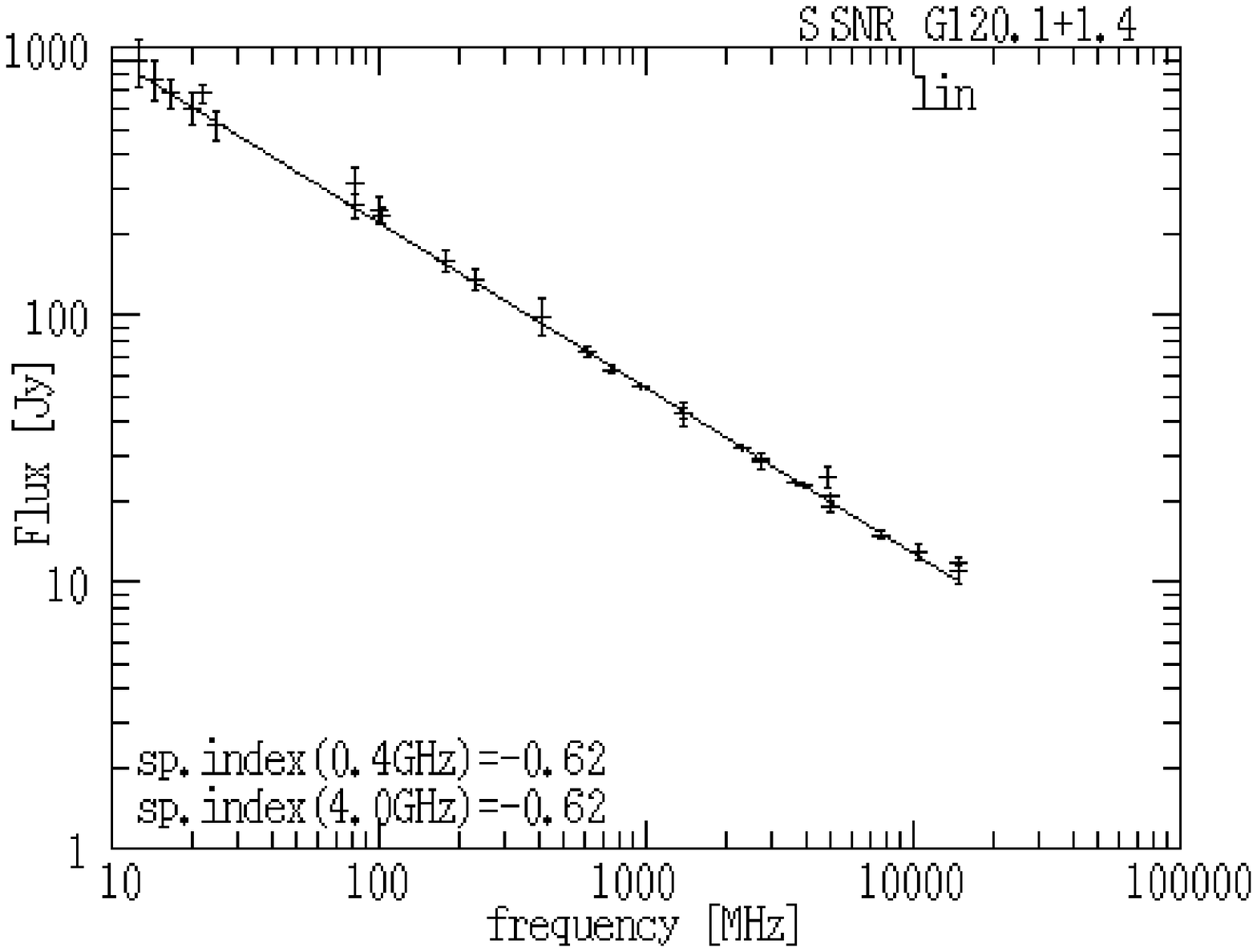,width=7.4cm,angle=0}}}\end{figure}
\begin{figure}\centerline{\vbox{\psfig{figure=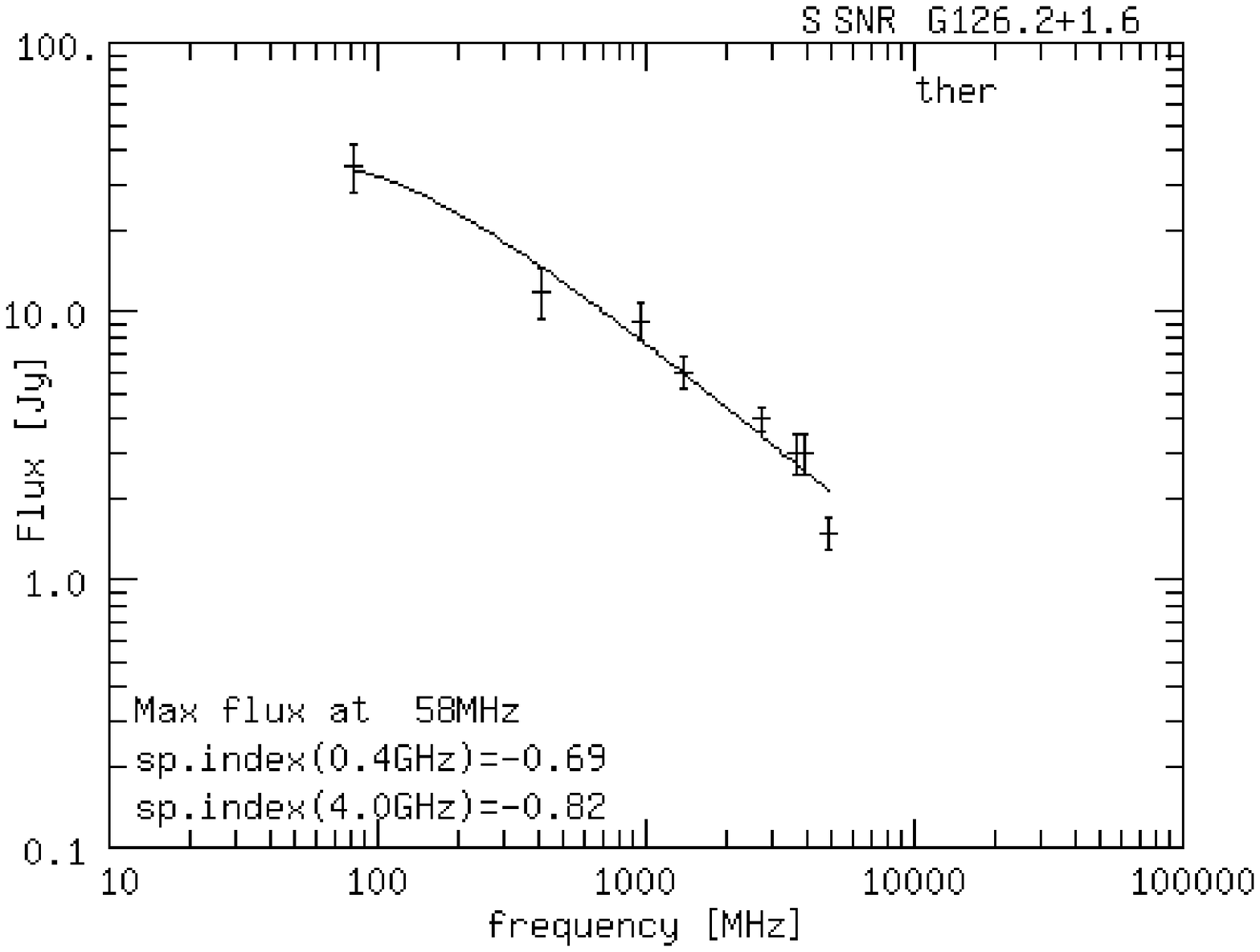,width=7.4cm,angle=0}}}\end{figure}
\begin{figure}\centerline{\vbox{\psfig{figure=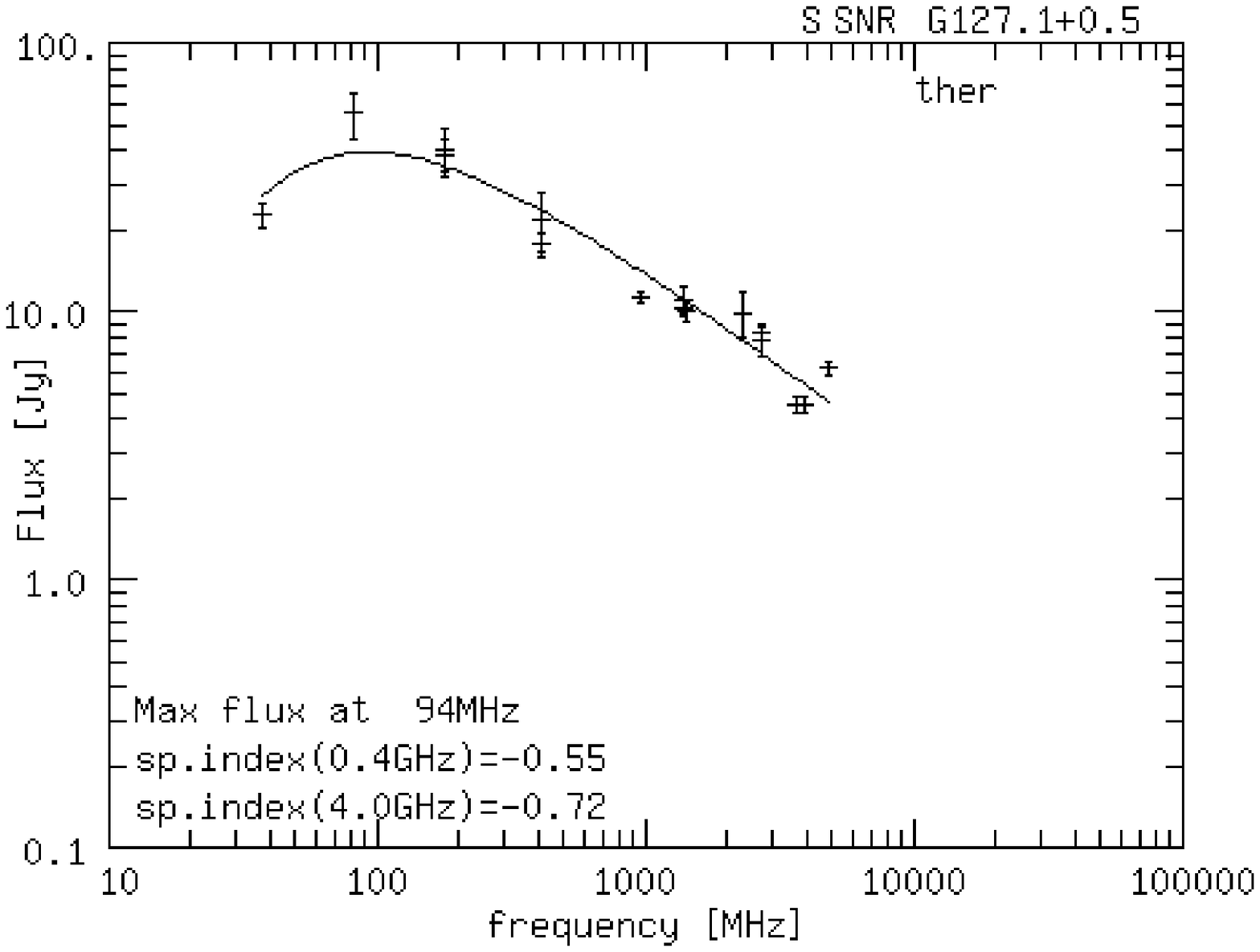,width=7.4cm,angle=0}}}\end{figure}
\begin{figure}\centerline{\vbox{\psfig{figure=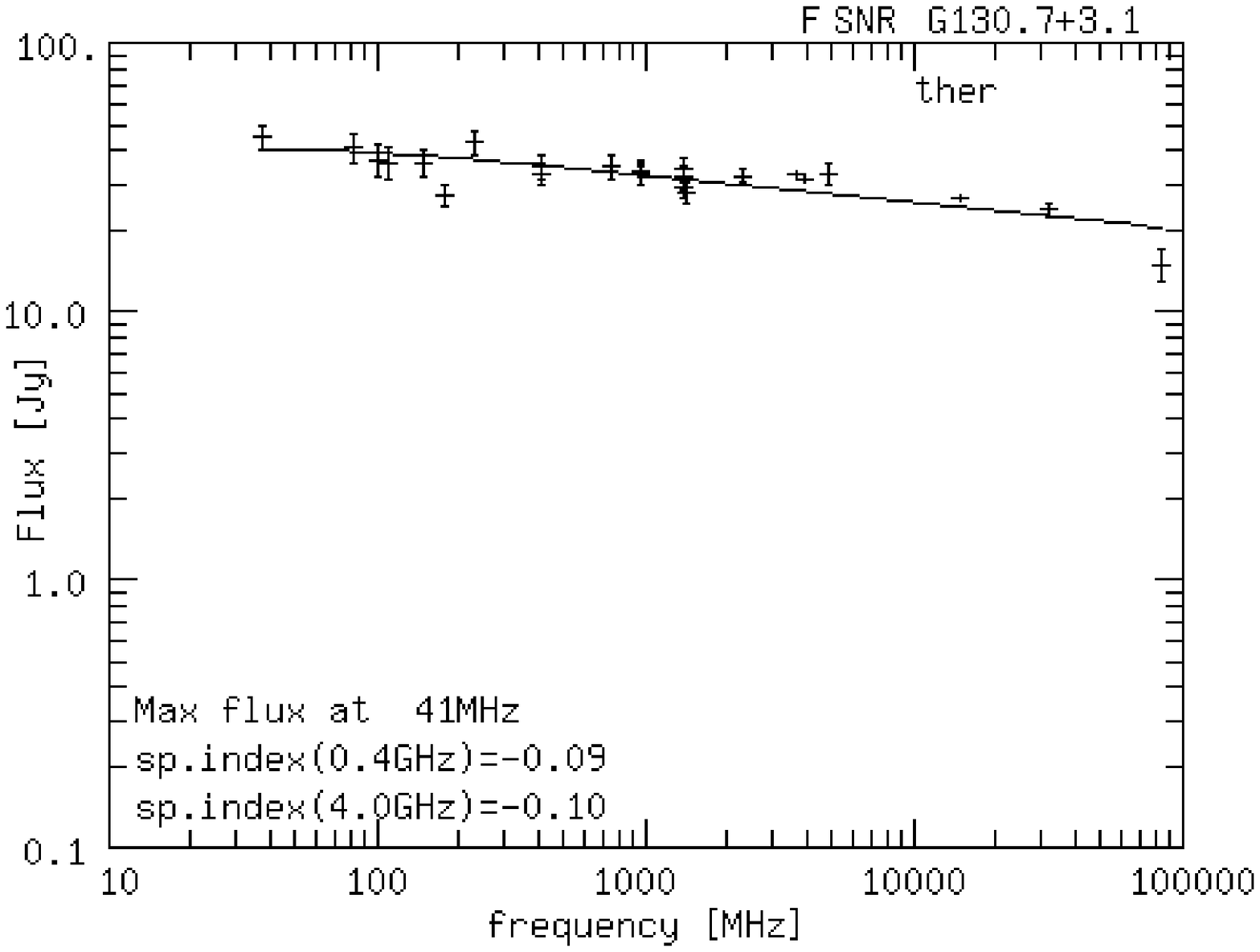,width=7.4cm,angle=0}}}\end{figure}\clearpage
\begin{figure}\centerline{\vbox{\psfig{figure=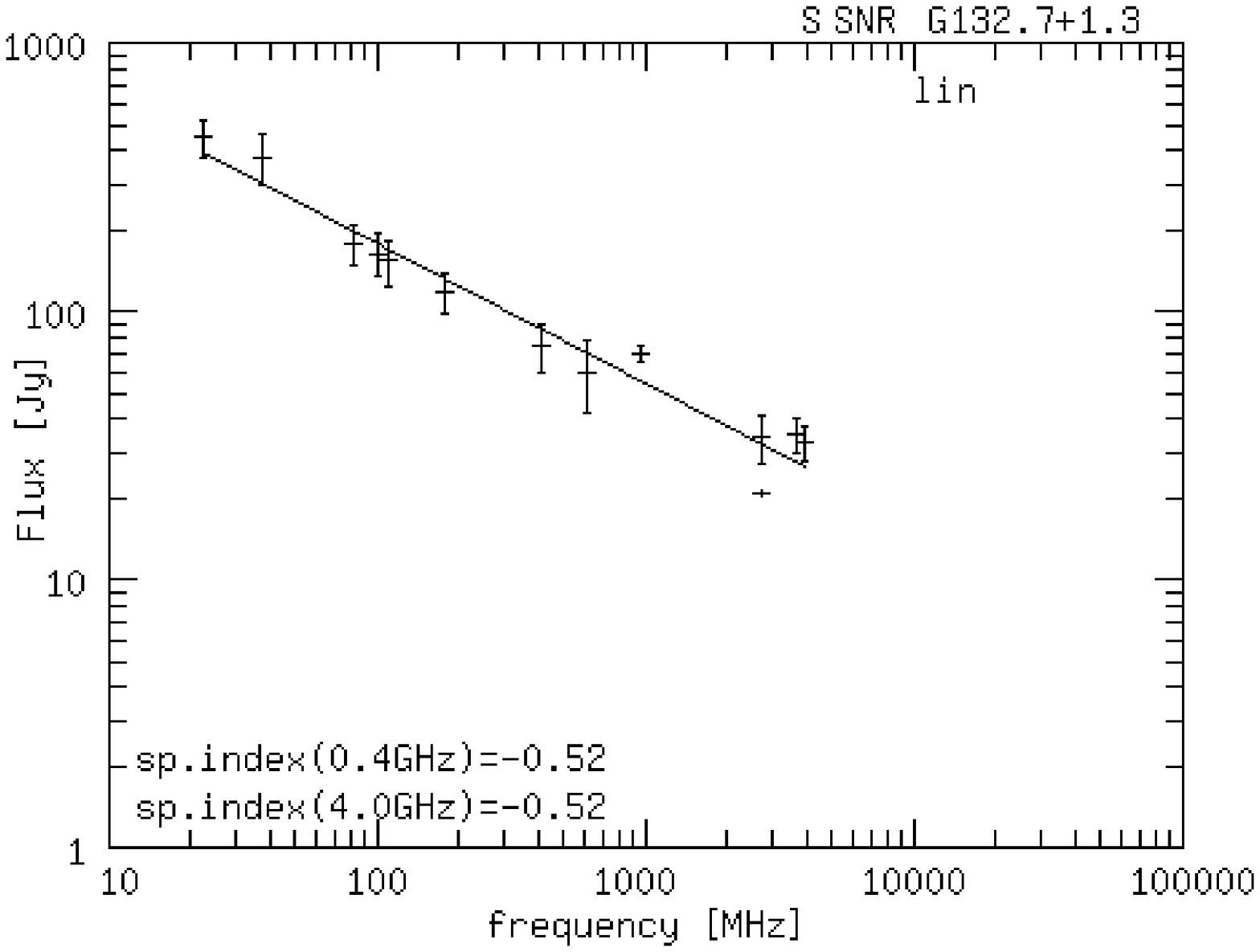,width=7.4cm,angle=0}}}\end{figure}
\begin{figure}\centerline{\vbox{\psfig{figure=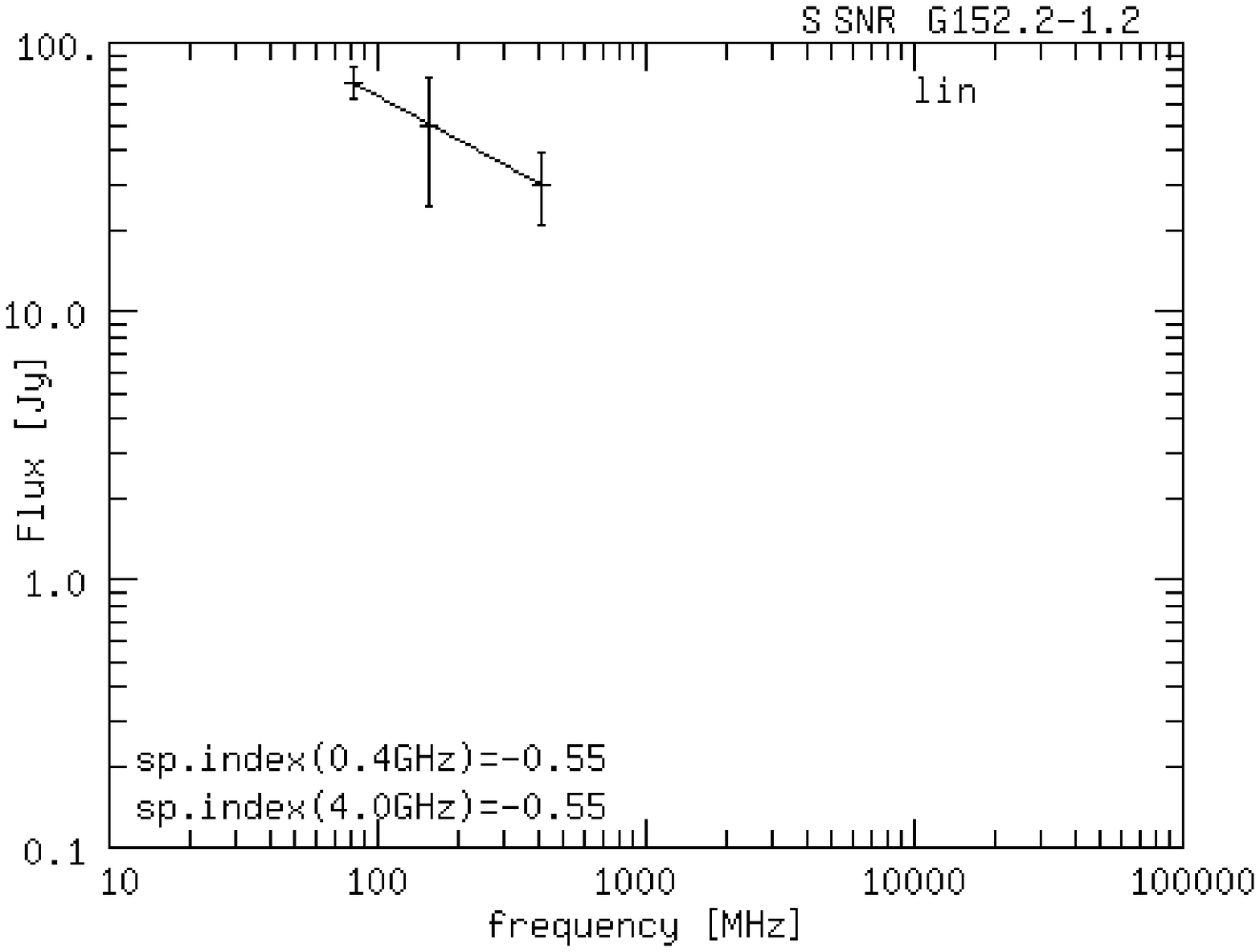,width=7.4cm,angle=0}}}\end{figure}
\begin{figure}\centerline{\vbox{\psfig{figure=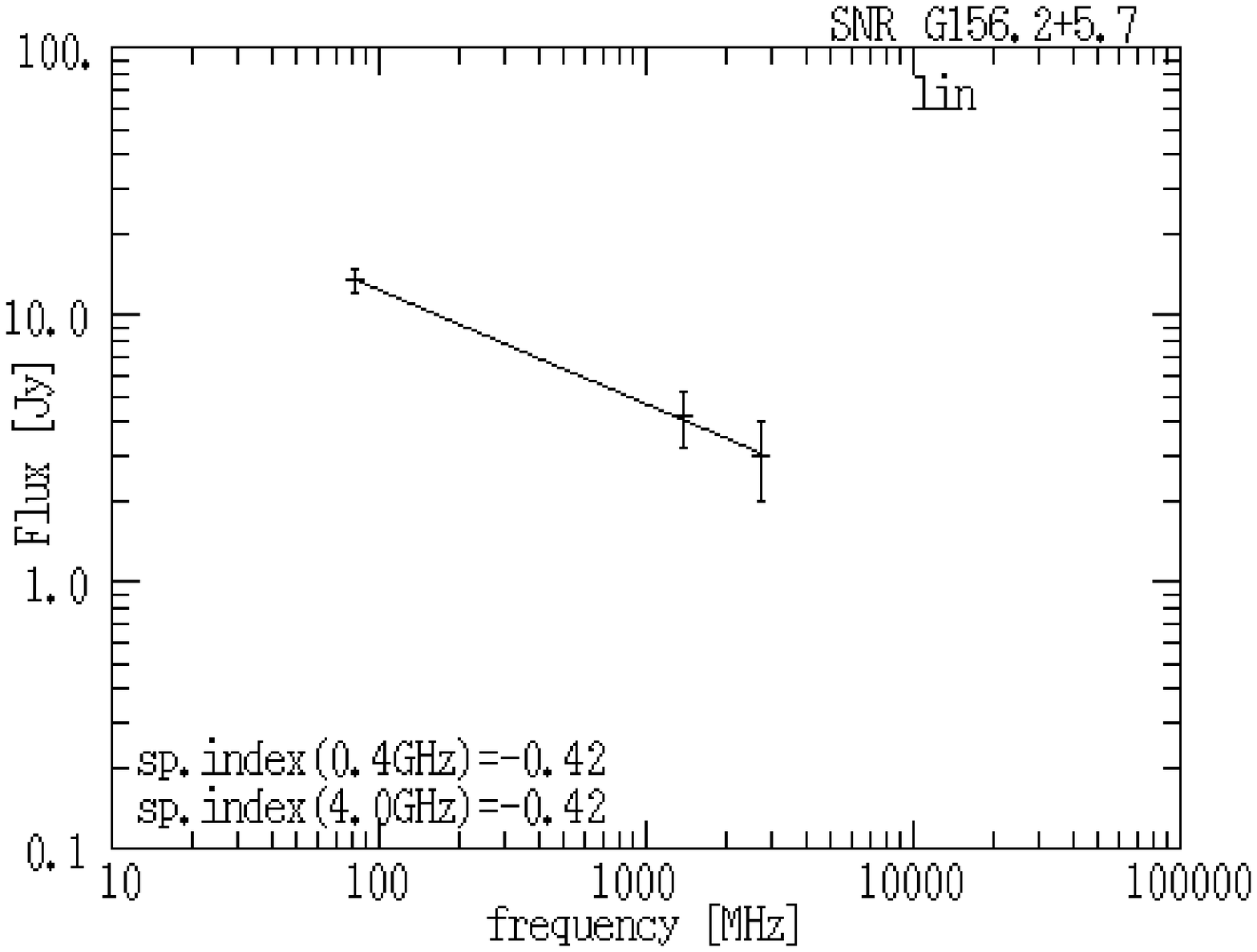,width=7.4cm,angle=0}}}\end{figure}
\begin{figure}\centerline{\vbox{\psfig{figure=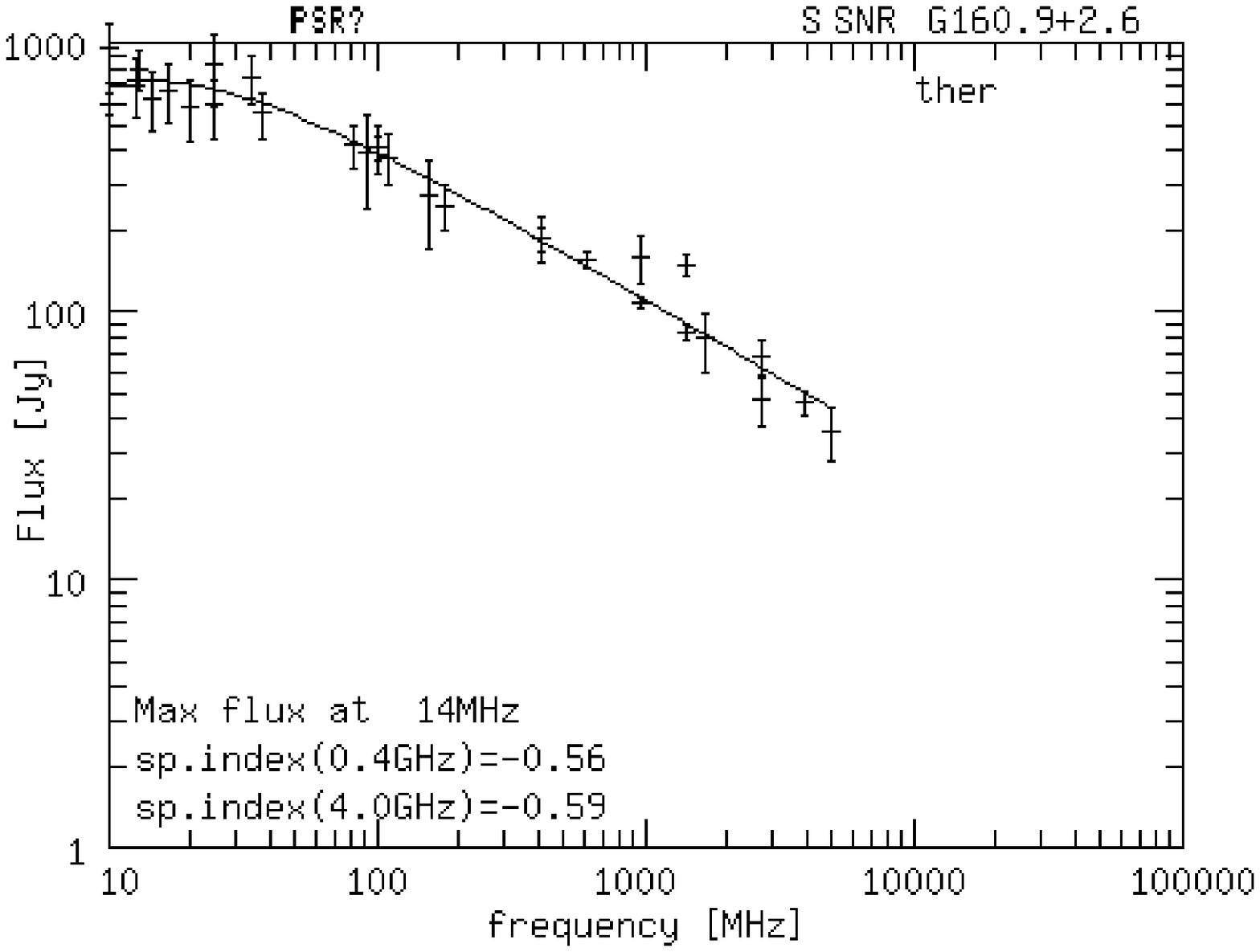,width=7.4cm,angle=0}}}\end{figure}
\begin{figure}\centerline{\vbox{\psfig{figure=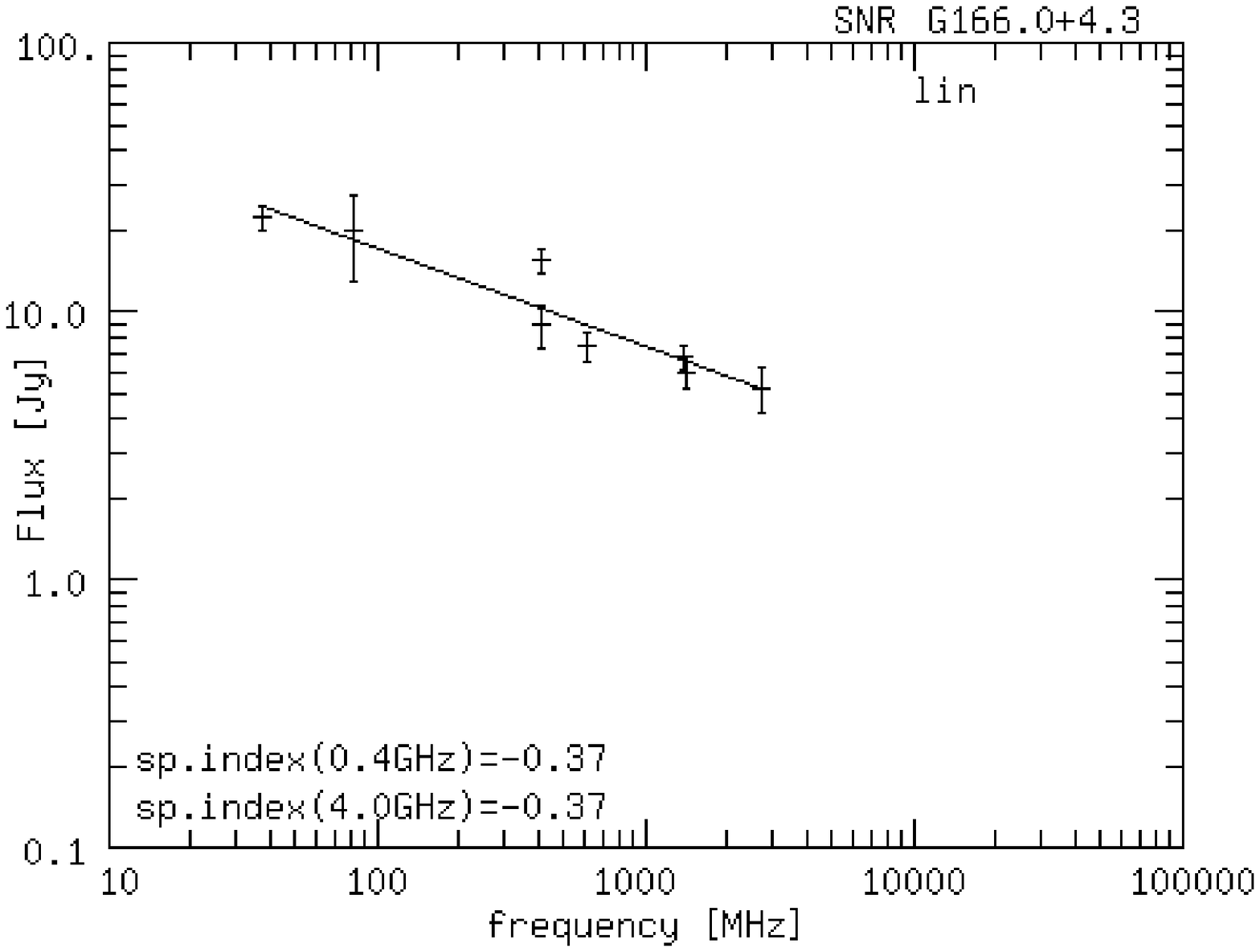,width=7.4cm,angle=0}}}\end{figure}
\begin{figure}\centerline{\vbox{\psfig{figure=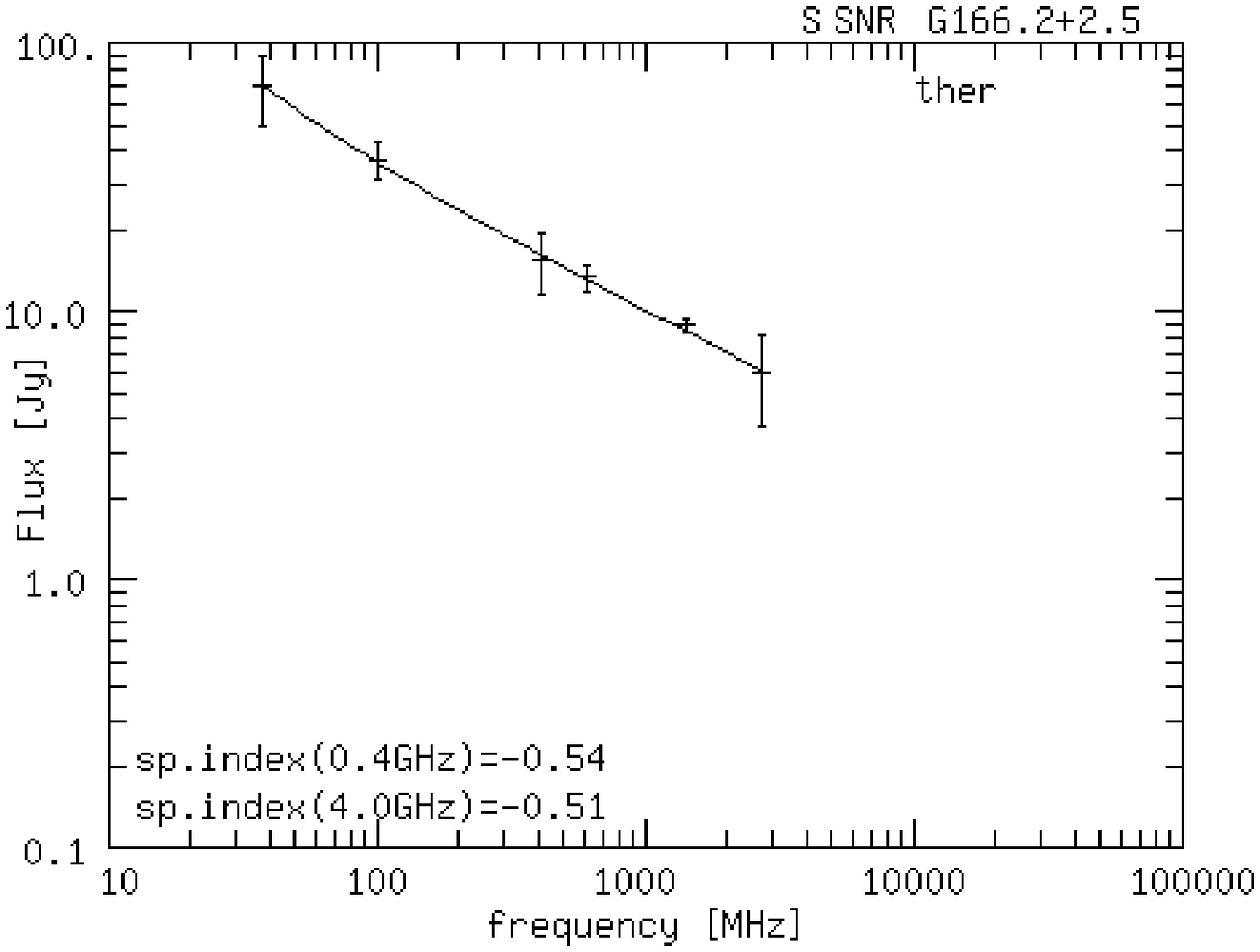,width=7.4cm,angle=0}}}\end{figure}
\begin{figure}\centerline{\vbox{\psfig{figure=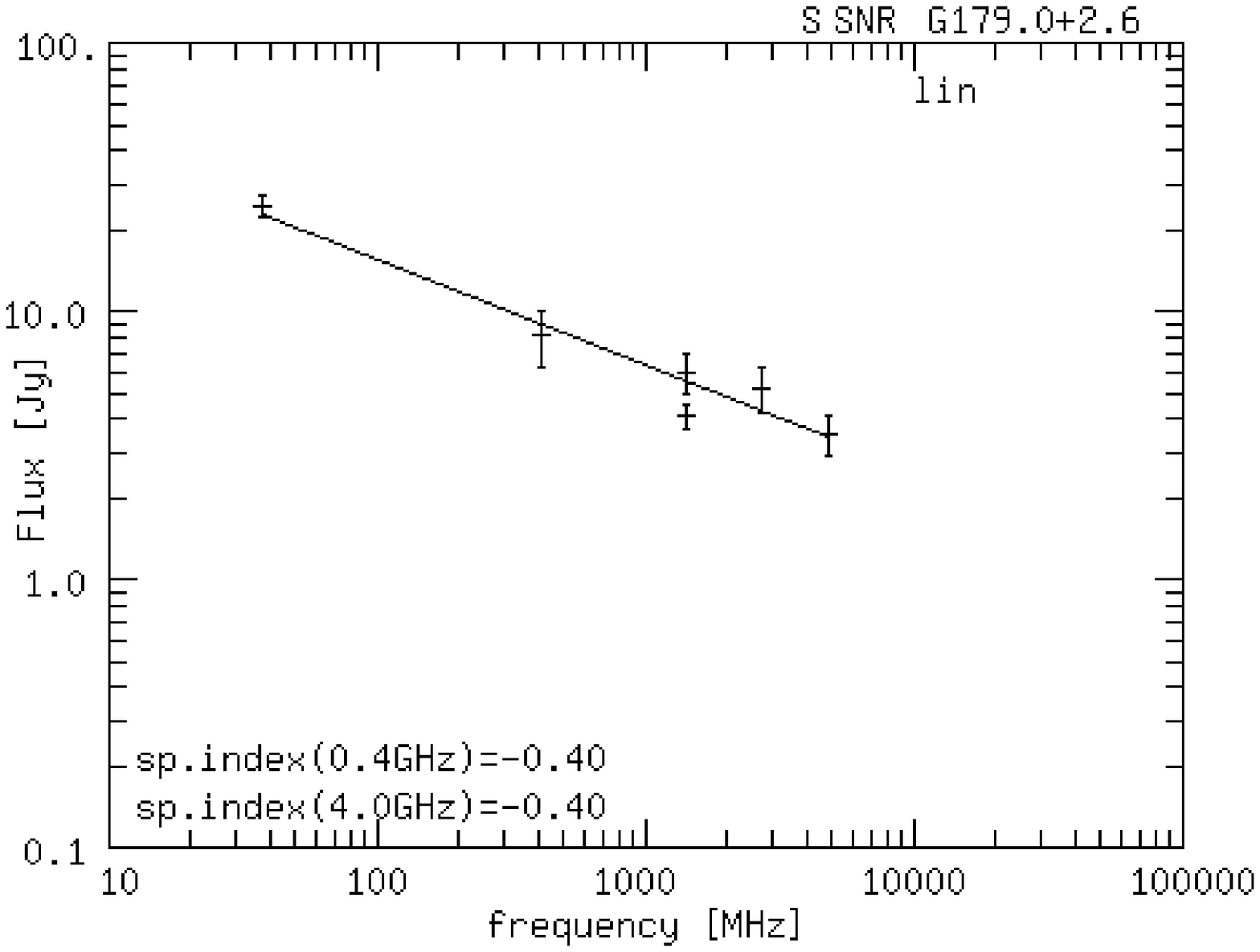,width=7.4cm,angle=0}}}\end{figure}
\begin{figure}\centerline{\vbox{\psfig{figure=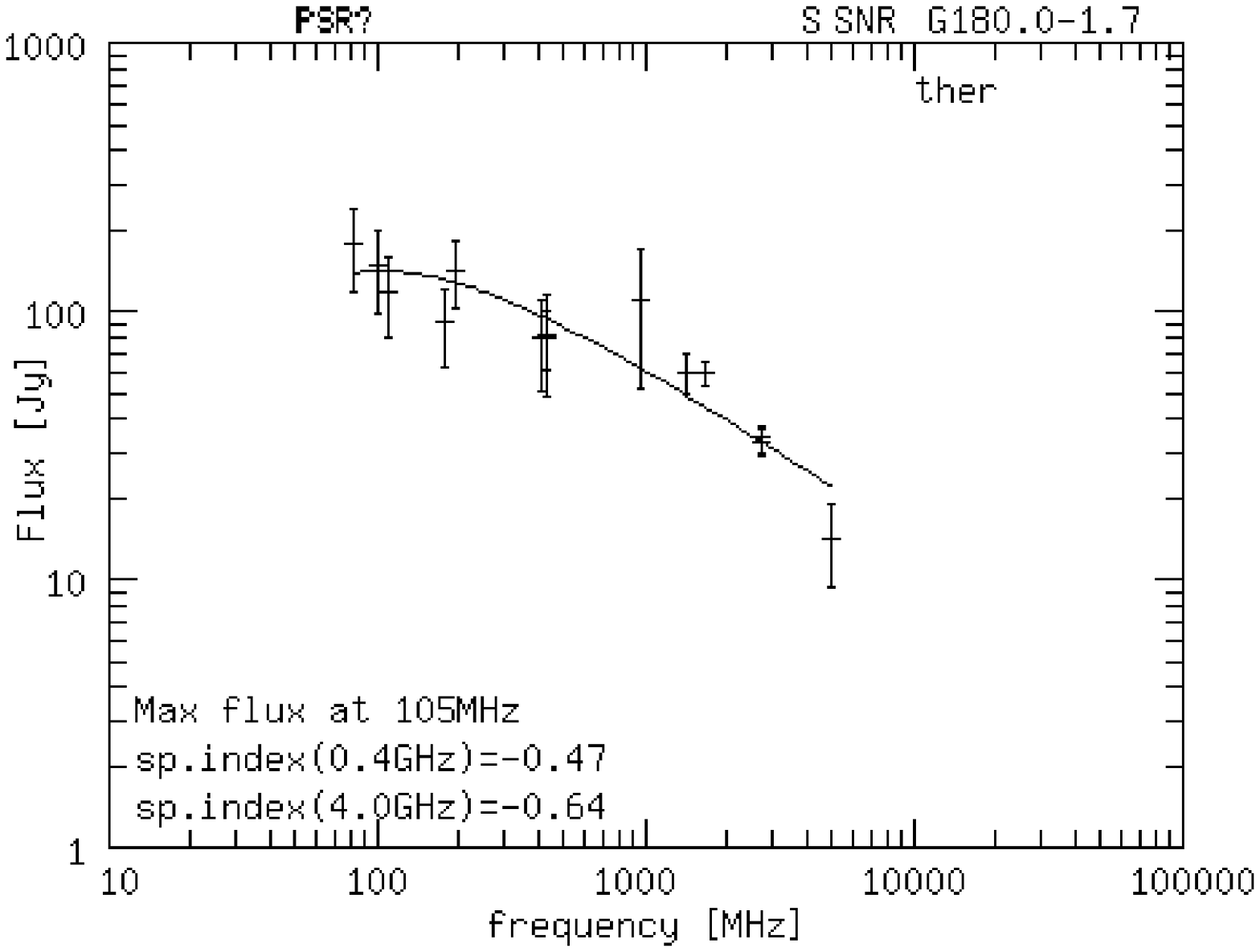,width=7.4cm,angle=0}}}\end{figure}\clearpage
\begin{figure}\centerline{\vbox{\psfig{figure=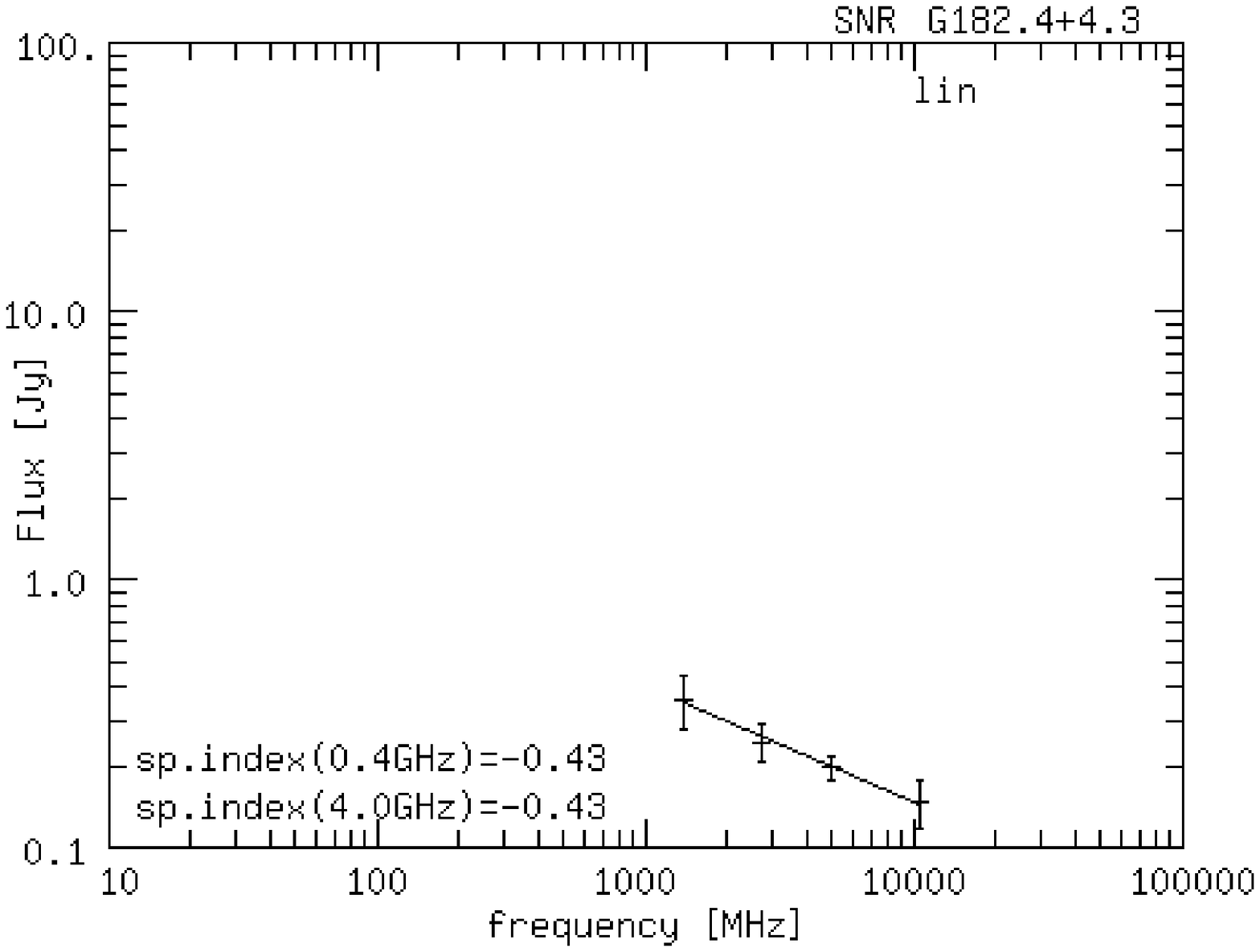,width=7.4cm,angle=0}}}\end{figure}
\begin{figure}\centerline{\vbox{\psfig{figure=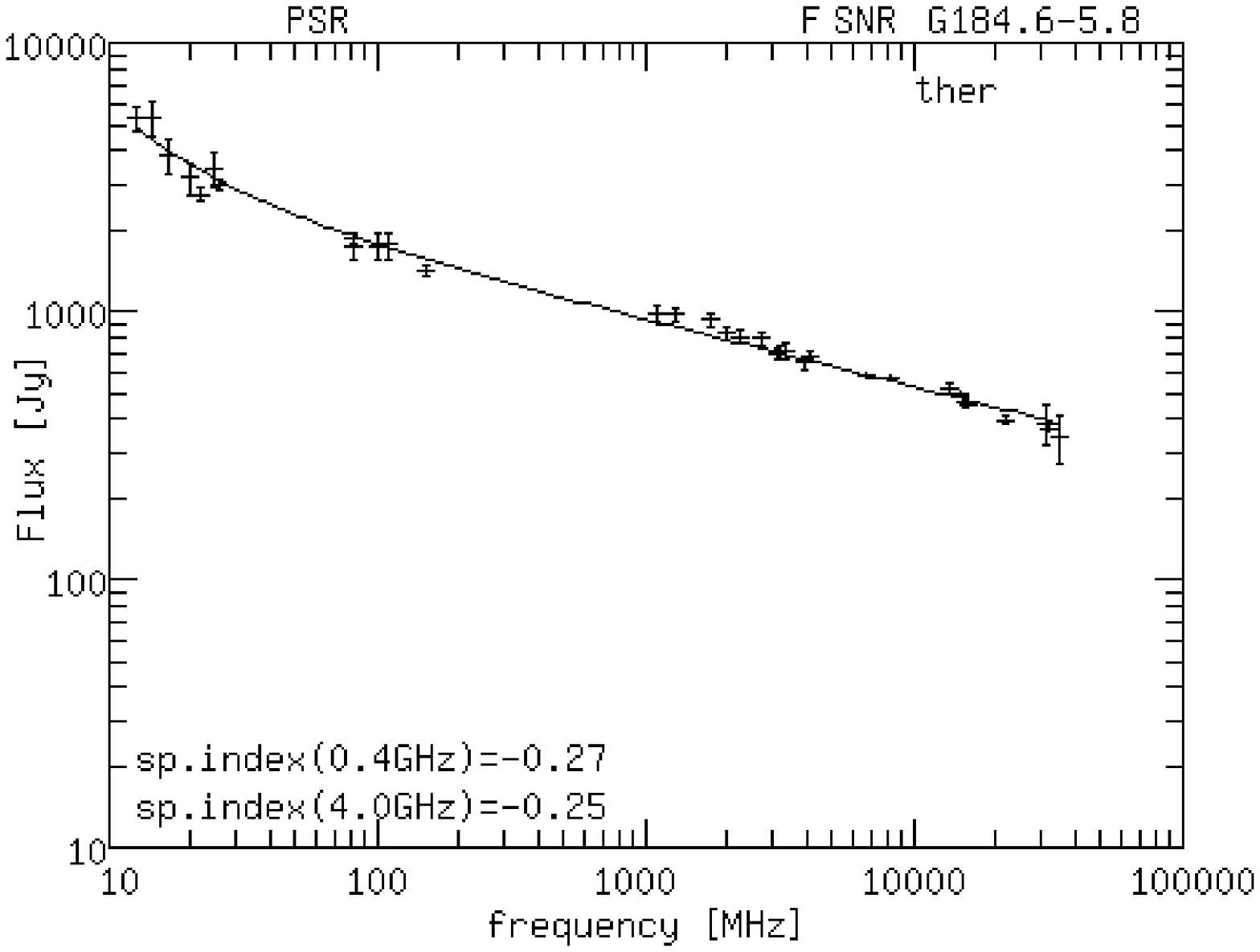,width=7.4cm,angle=0}}}\end{figure}
\begin{figure}\centerline{\vbox{\psfig{figure=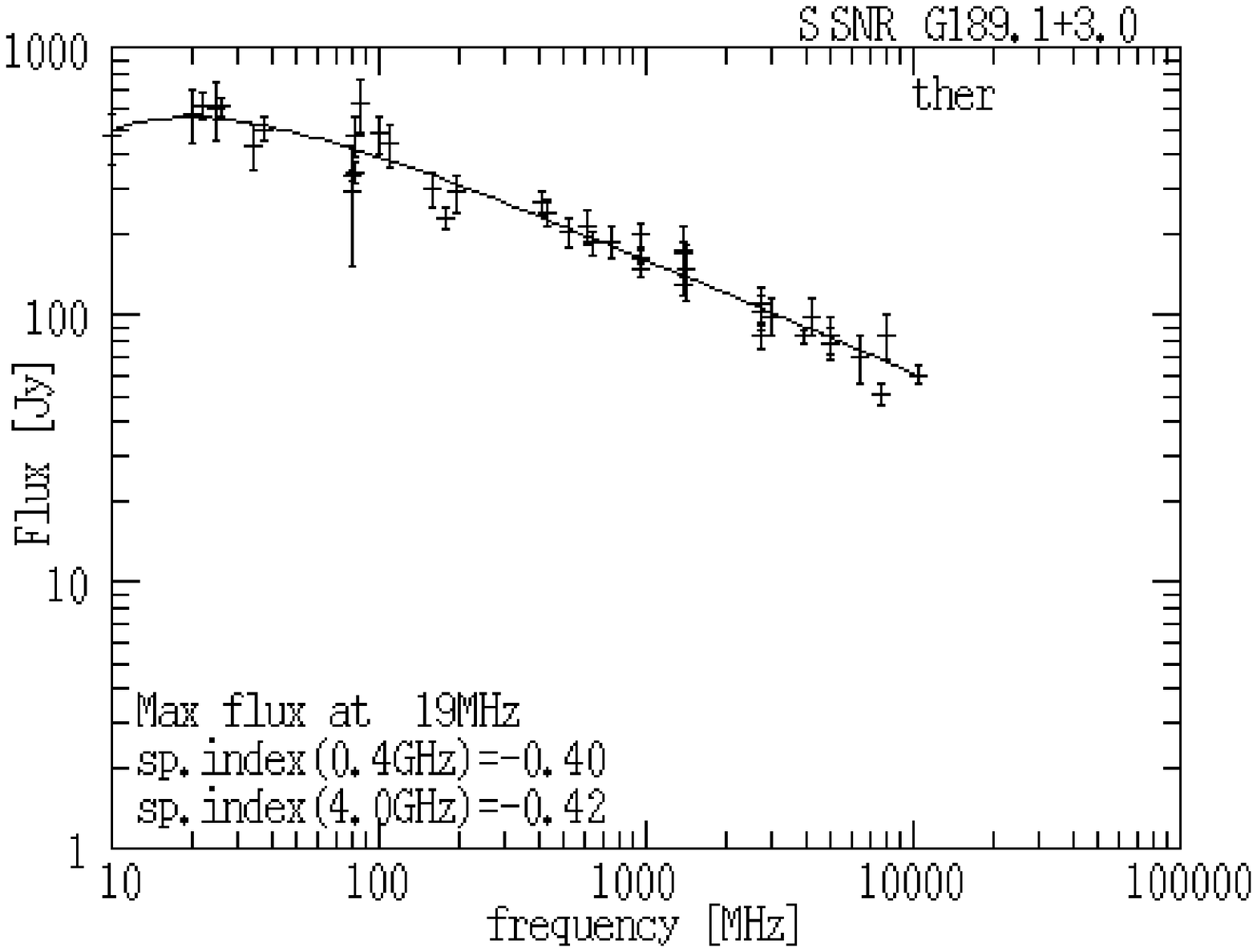,width=7.4cm,angle=0}}}\end{figure}
\begin{figure}\centerline{\vbox{\psfig{figure=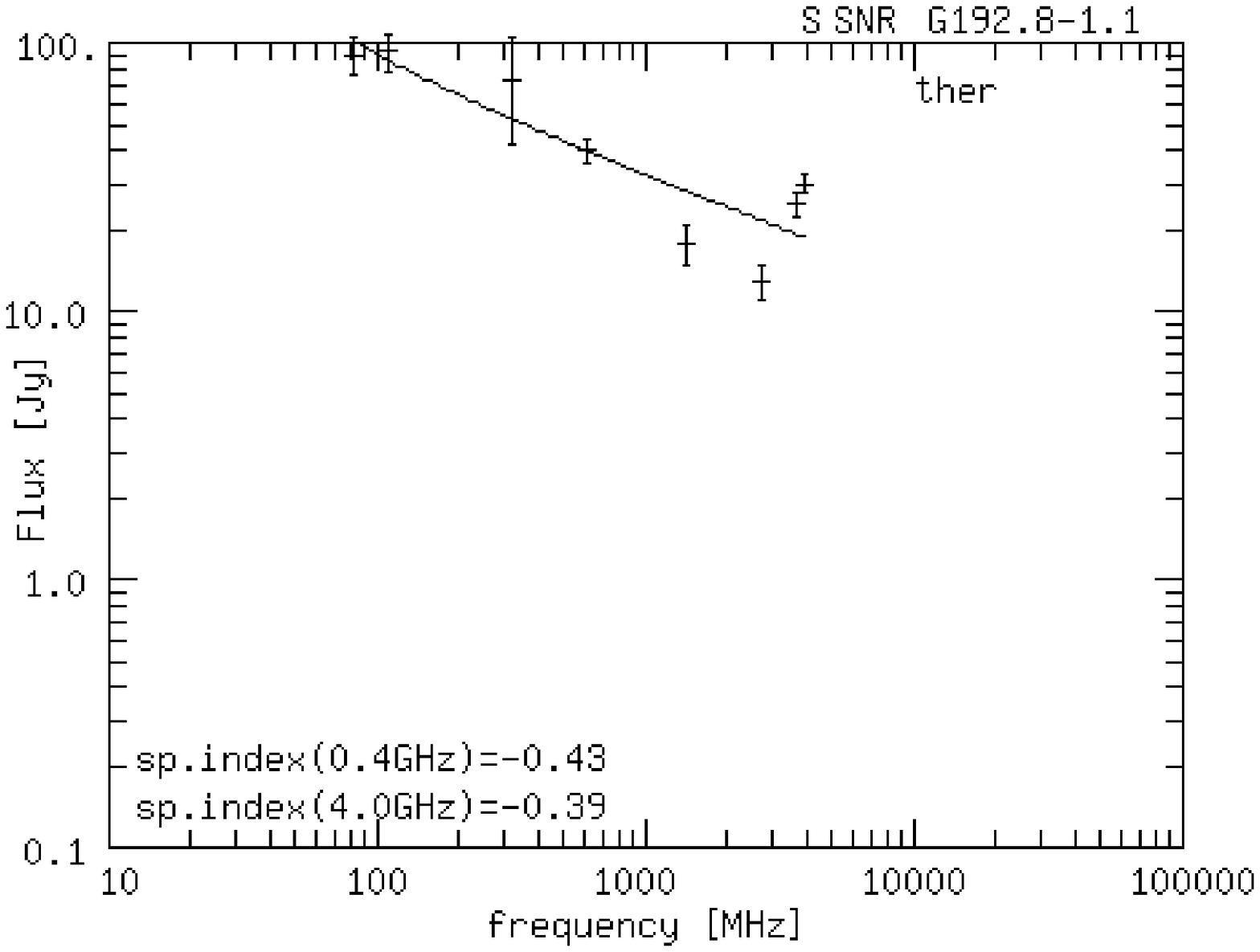,width=7.4cm,angle=0}}}\end{figure}
\begin{figure}\centerline{\vbox{\psfig{figure=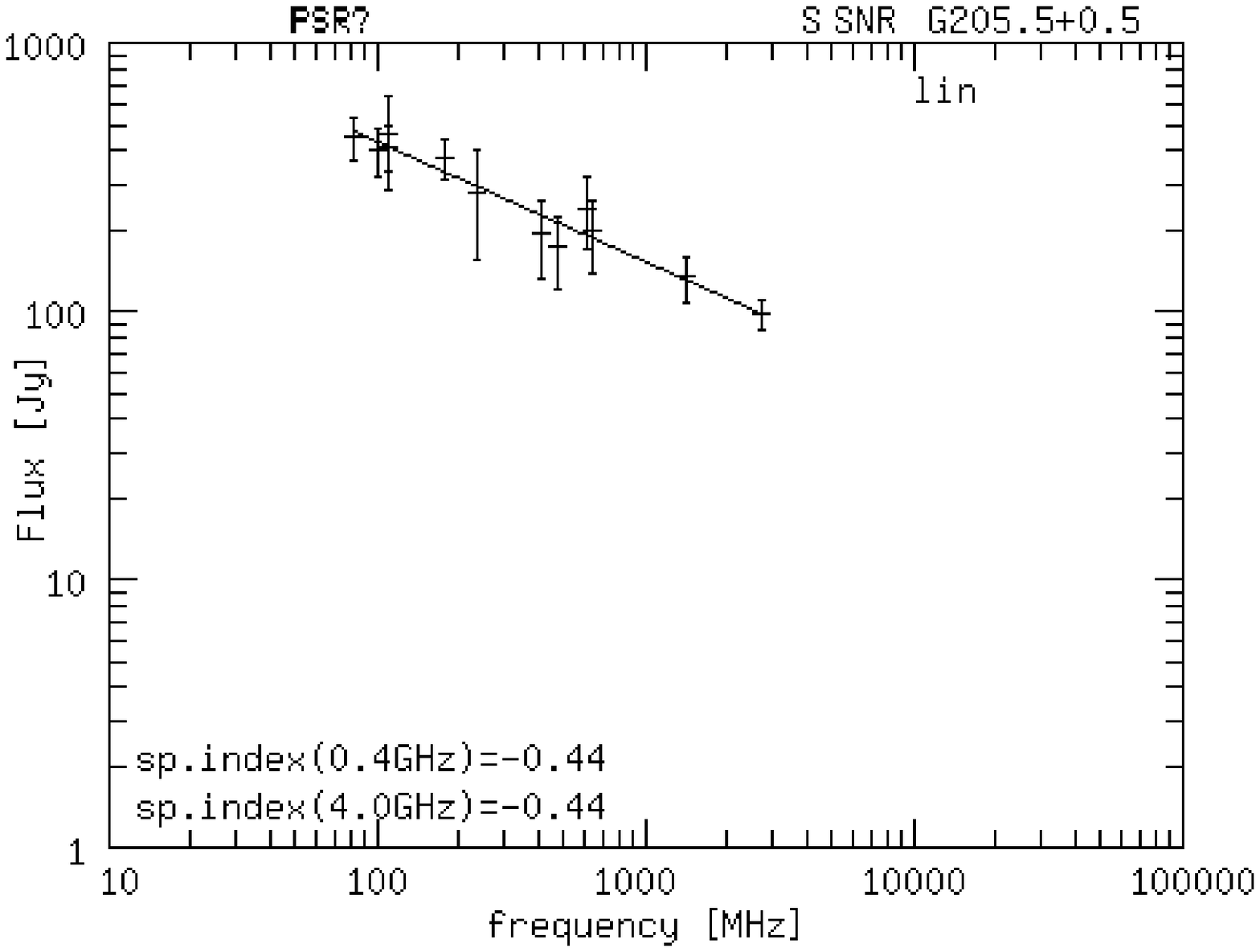,width=7.4cm,angle=0}}}\end{figure}
\begin{figure}\centerline{\vbox{\psfig{figure=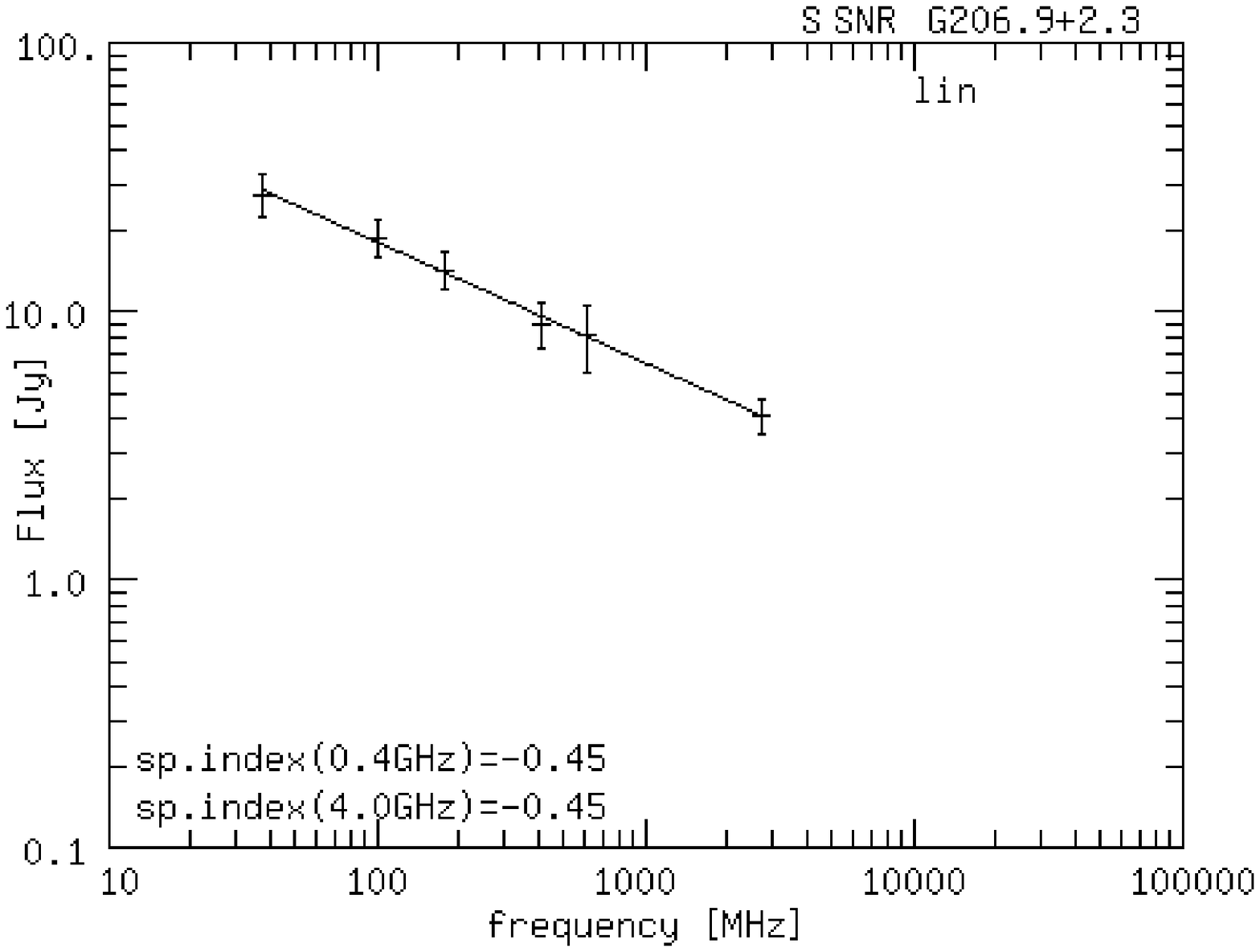,width=7.4cm,angle=0}}}\end{figure}
\begin{figure}\centerline{\vbox{\psfig{figure=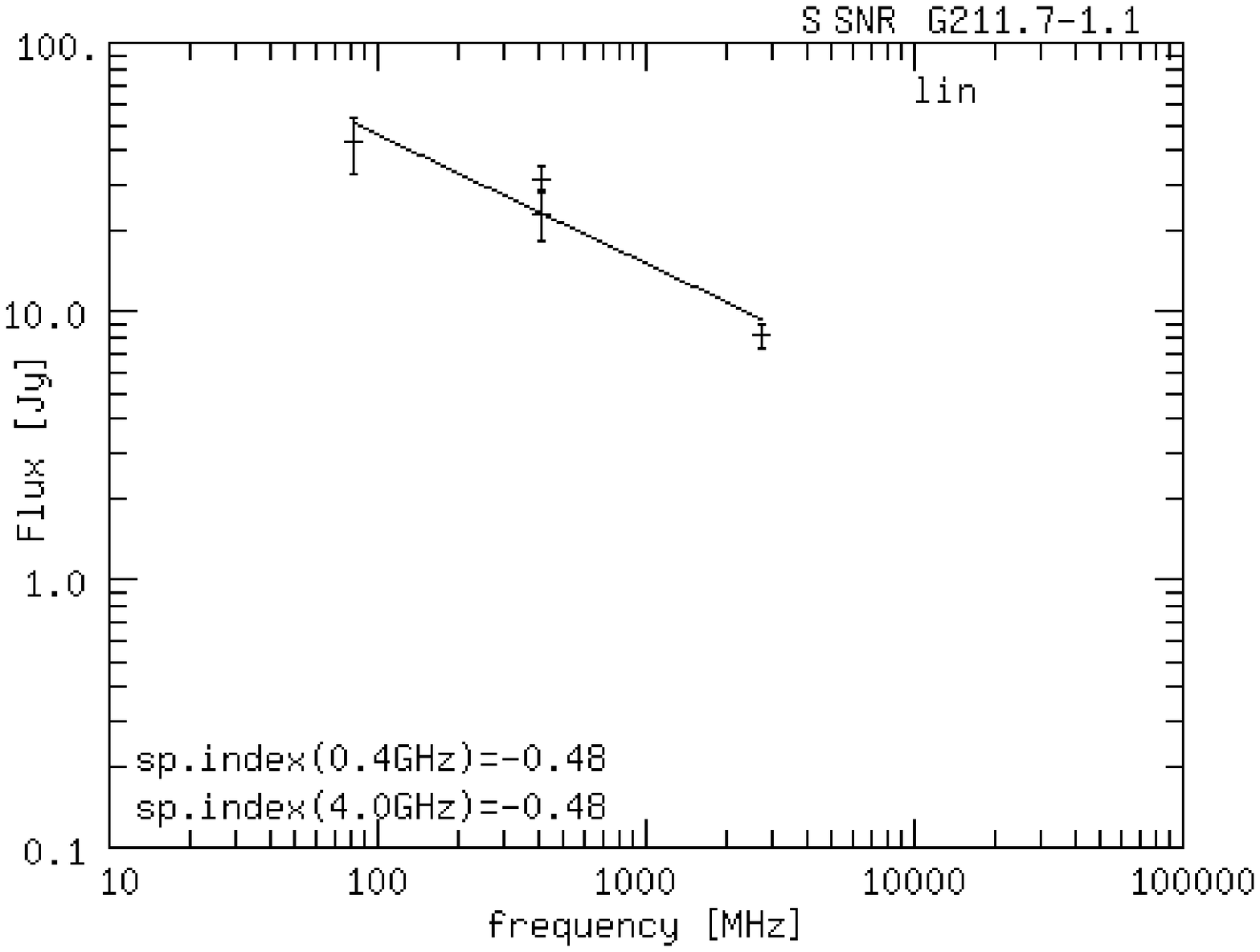,width=7.4cm,angle=0}}}\end{figure}
\begin{figure}\centerline{\vbox{\psfig{figure=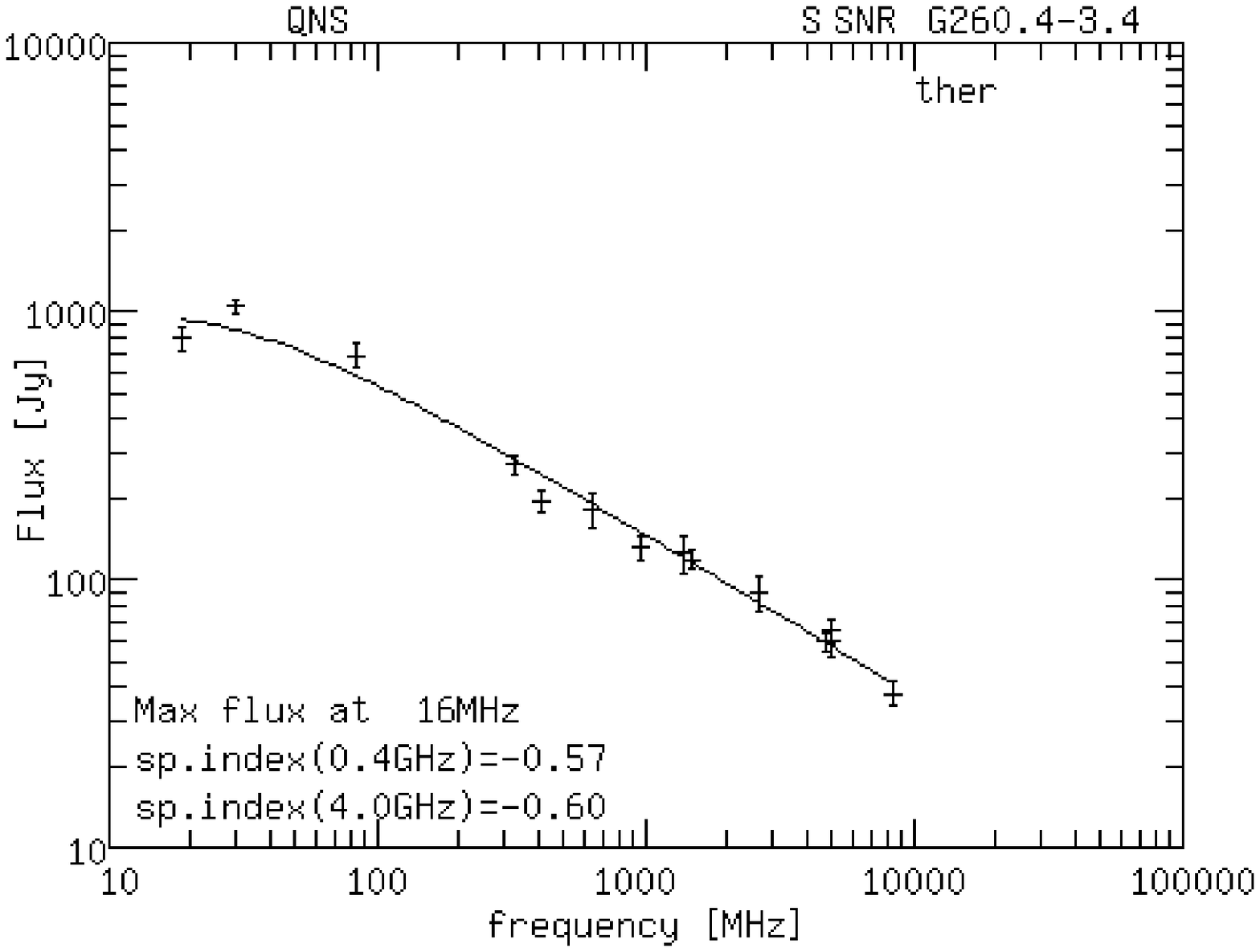,width=7.4cm,angle=0}}}\end{figure}\clearpage
\begin{figure}\centerline{\vbox{\psfig{figure=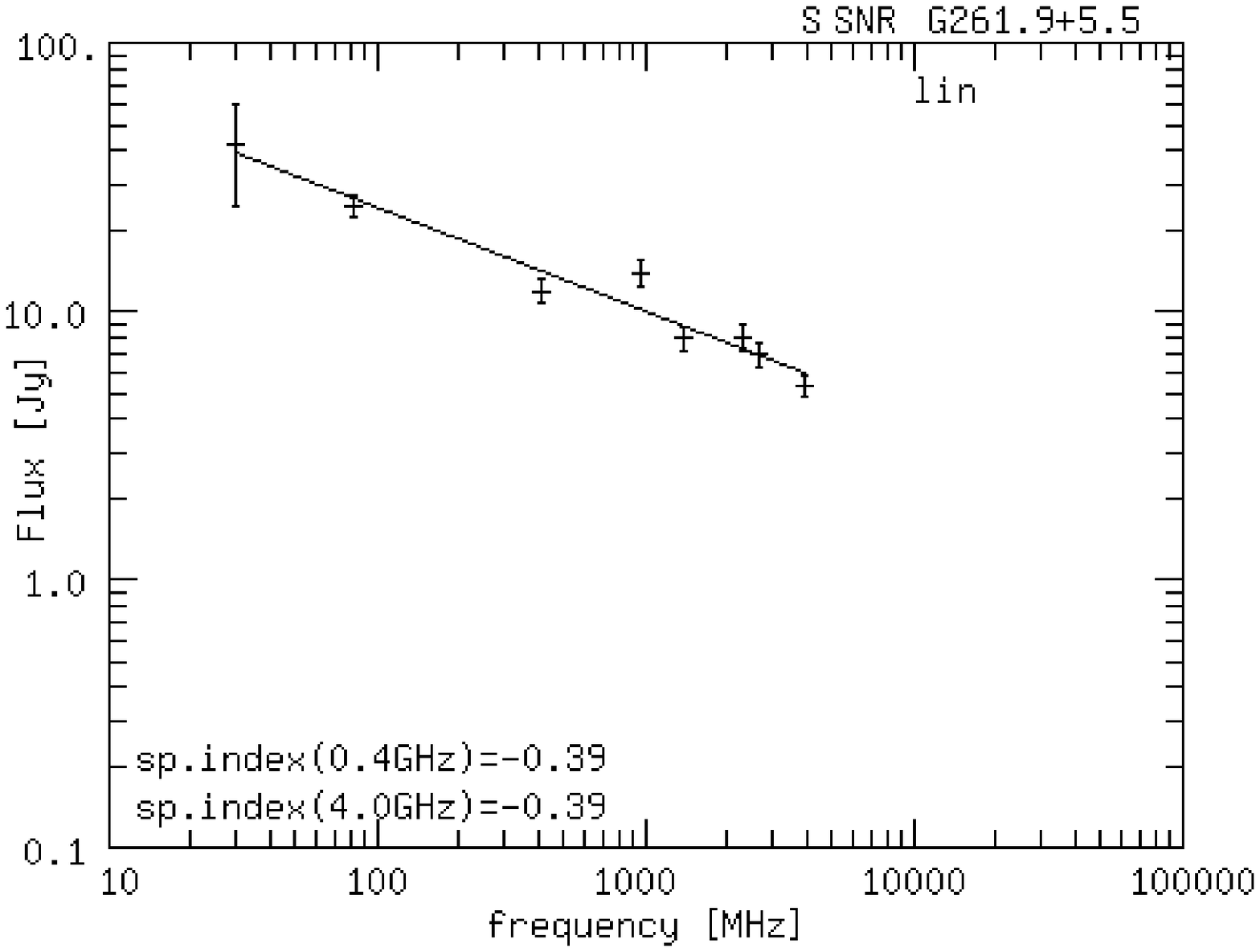,width=7.4cm,angle=0}}}\end{figure}
\begin{figure}\centerline{\vbox{\psfig{figure=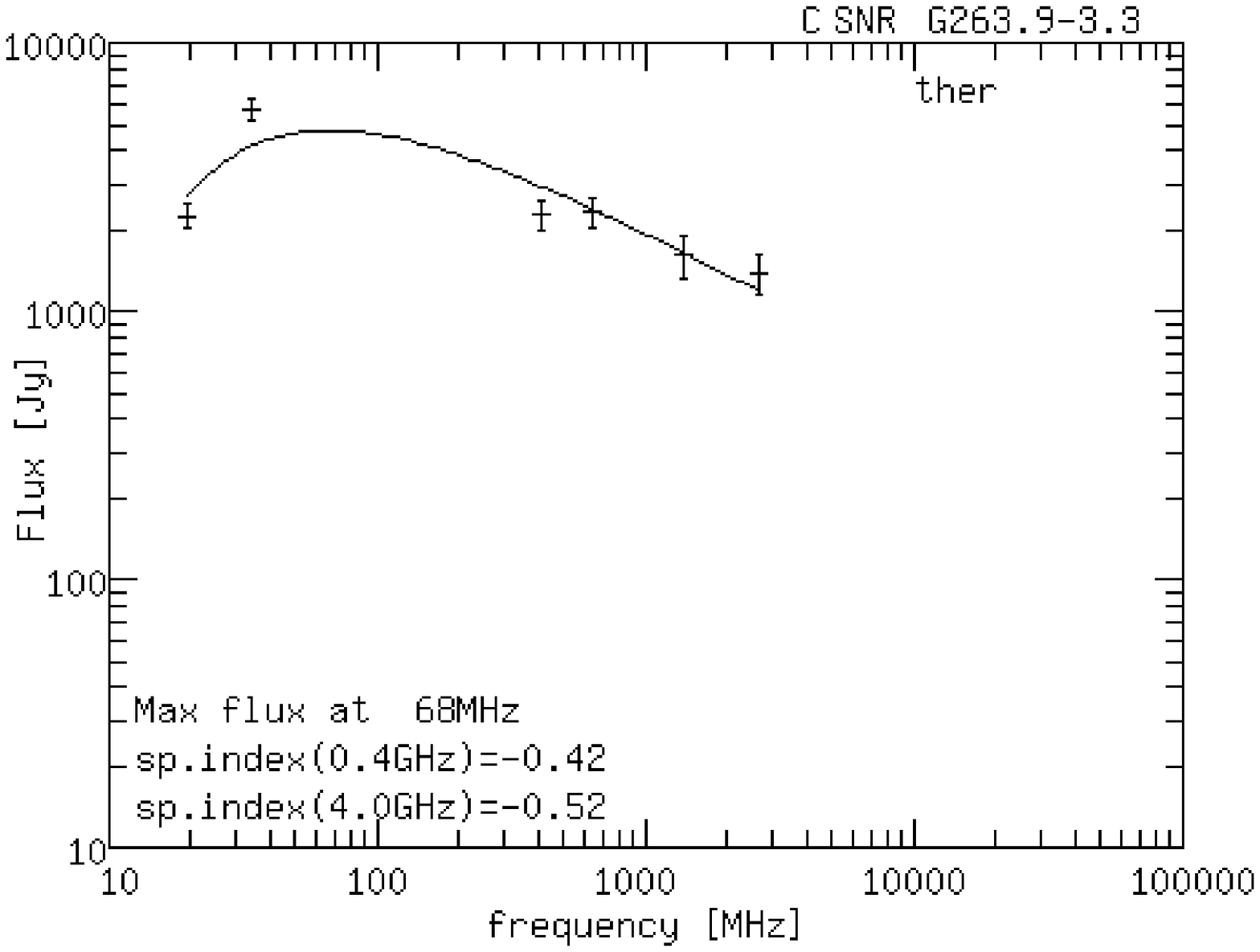,width=7.4cm,angle=0}}}\end{figure}
\begin{figure}\centerline{\vbox{\psfig{figure=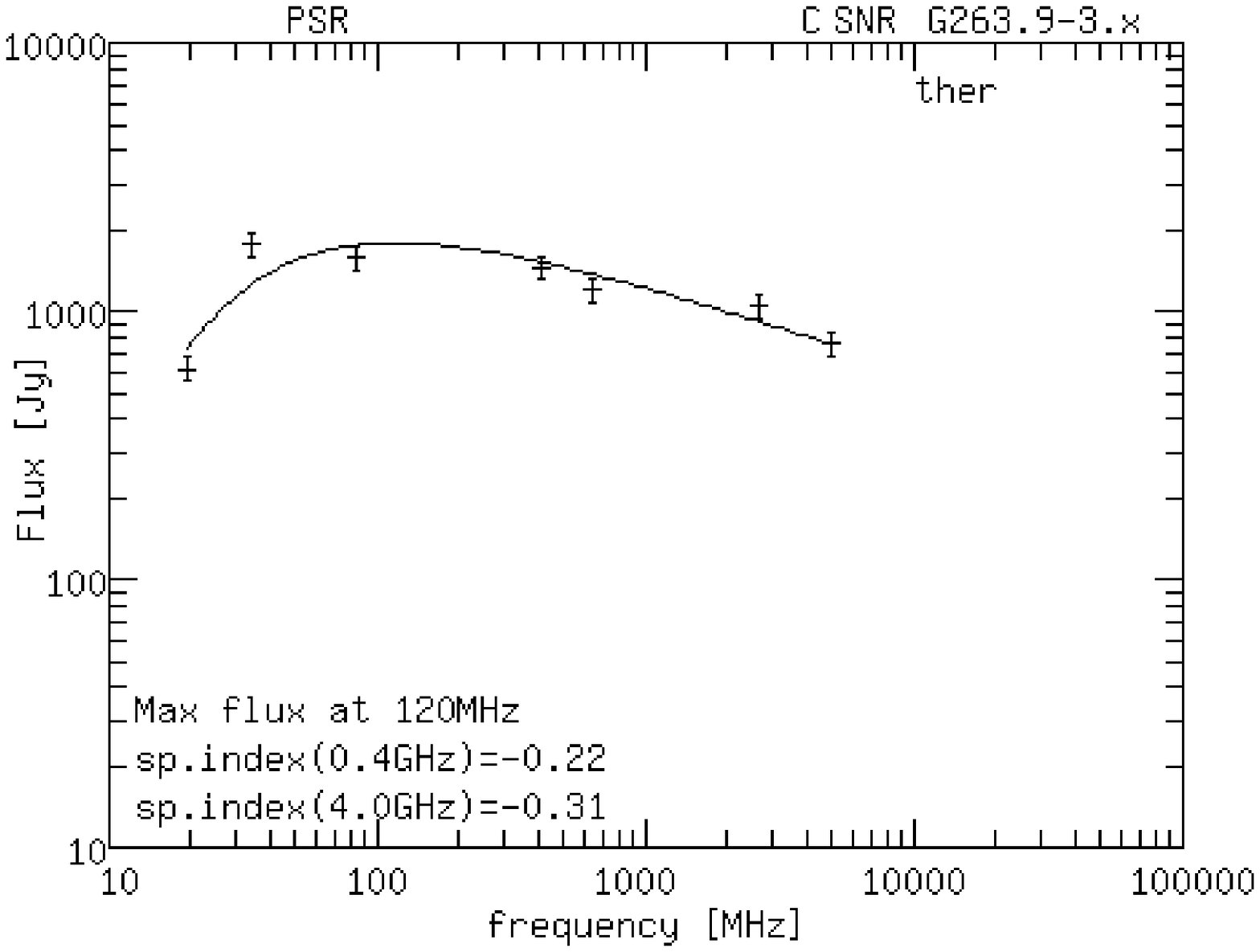,width=7.4cm,angle=0}}}\end{figure}
\begin{figure}\centerline{\vbox{\psfig{figure=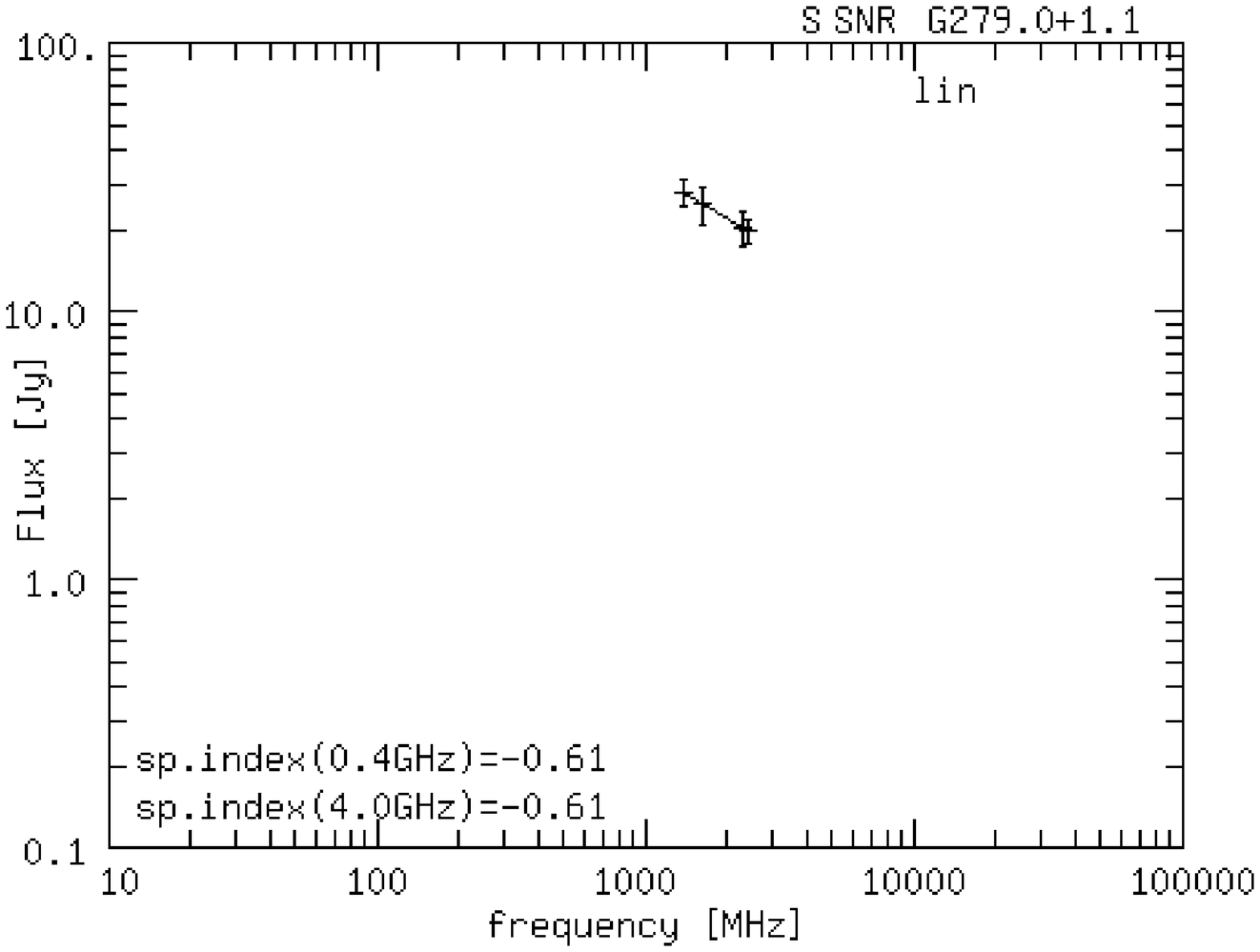,width=7.4cm,angle=0}}}\end{figure}
\begin{figure}\centerline{\vbox{\psfig{figure=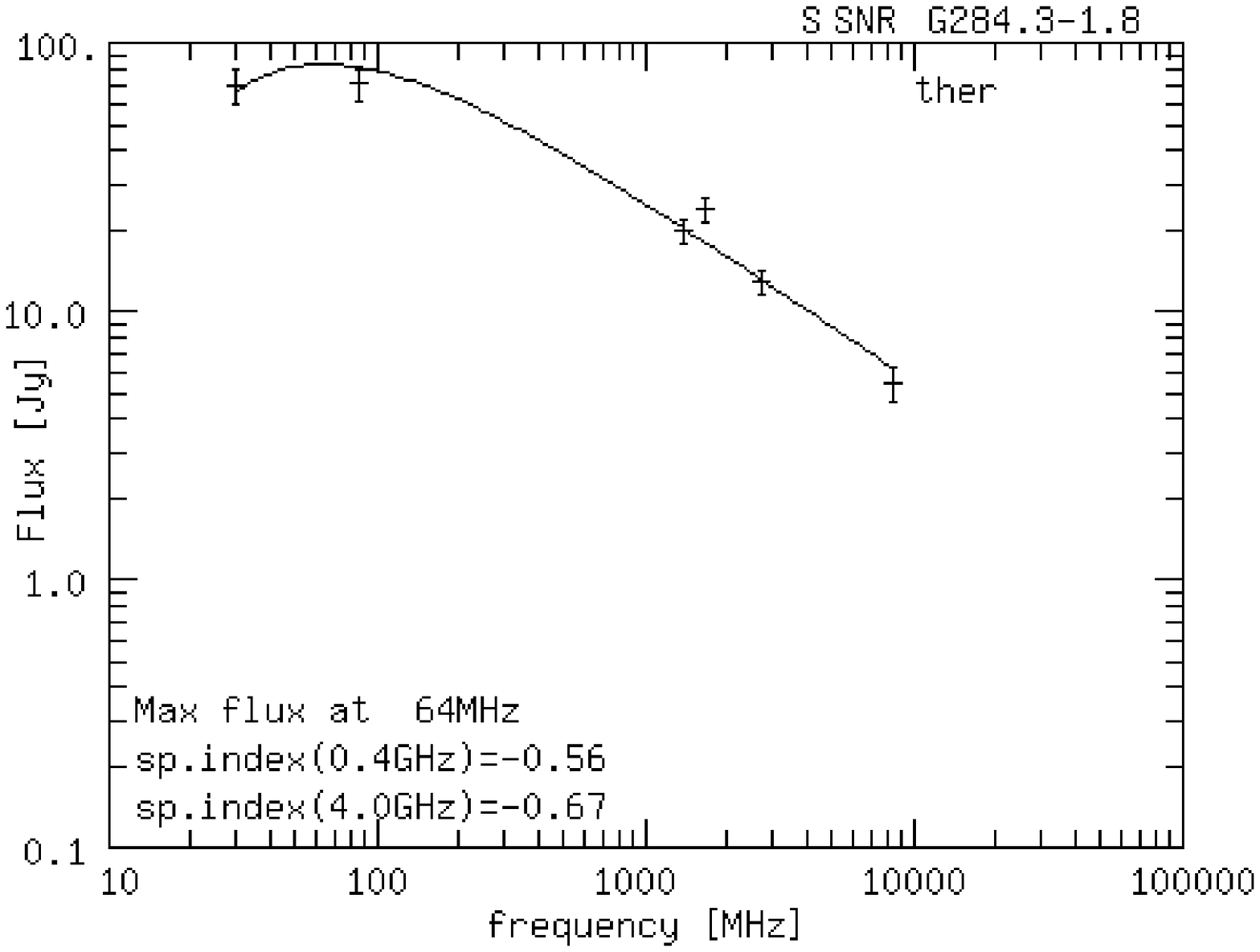,width=7.4cm,angle=0}}}\end{figure}
\begin{figure}\centerline{\vbox{\psfig{figure=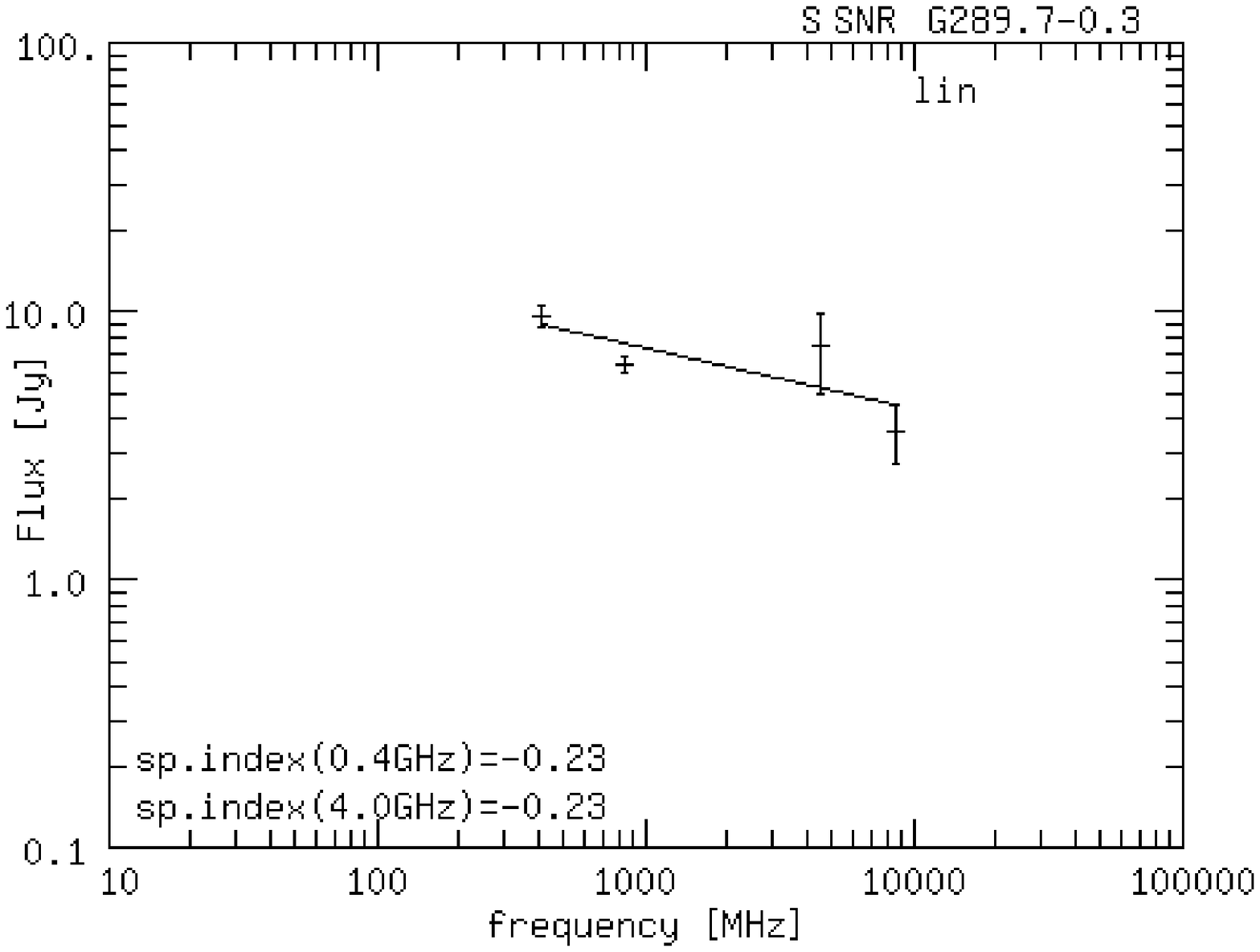,width=7.4cm,angle=0}}}\end{figure}
\begin{figure}\centerline{\vbox{\psfig{figure=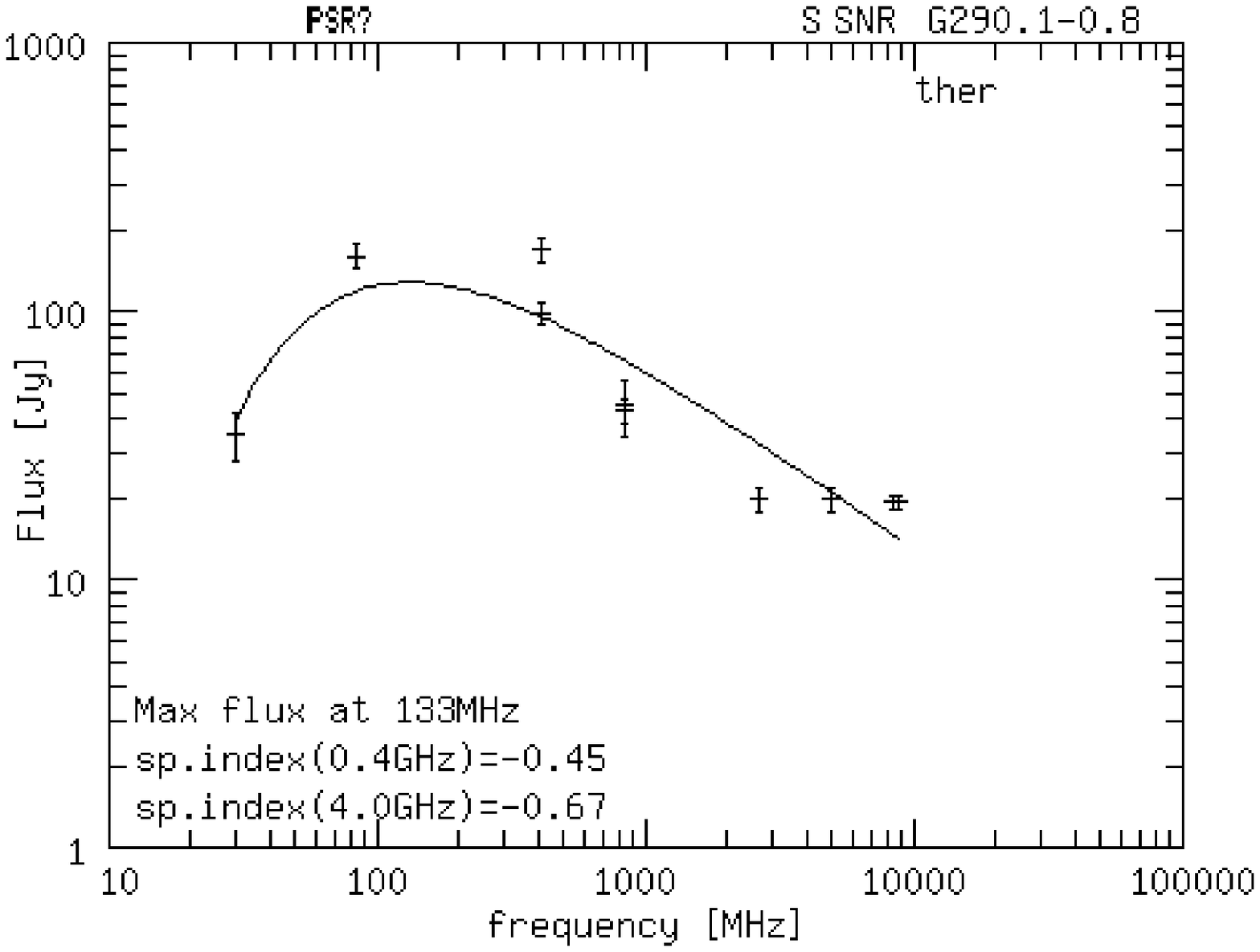,width=7.4cm,angle=0}}}\end{figure}
\begin{figure}\centerline{\vbox{\psfig{figure=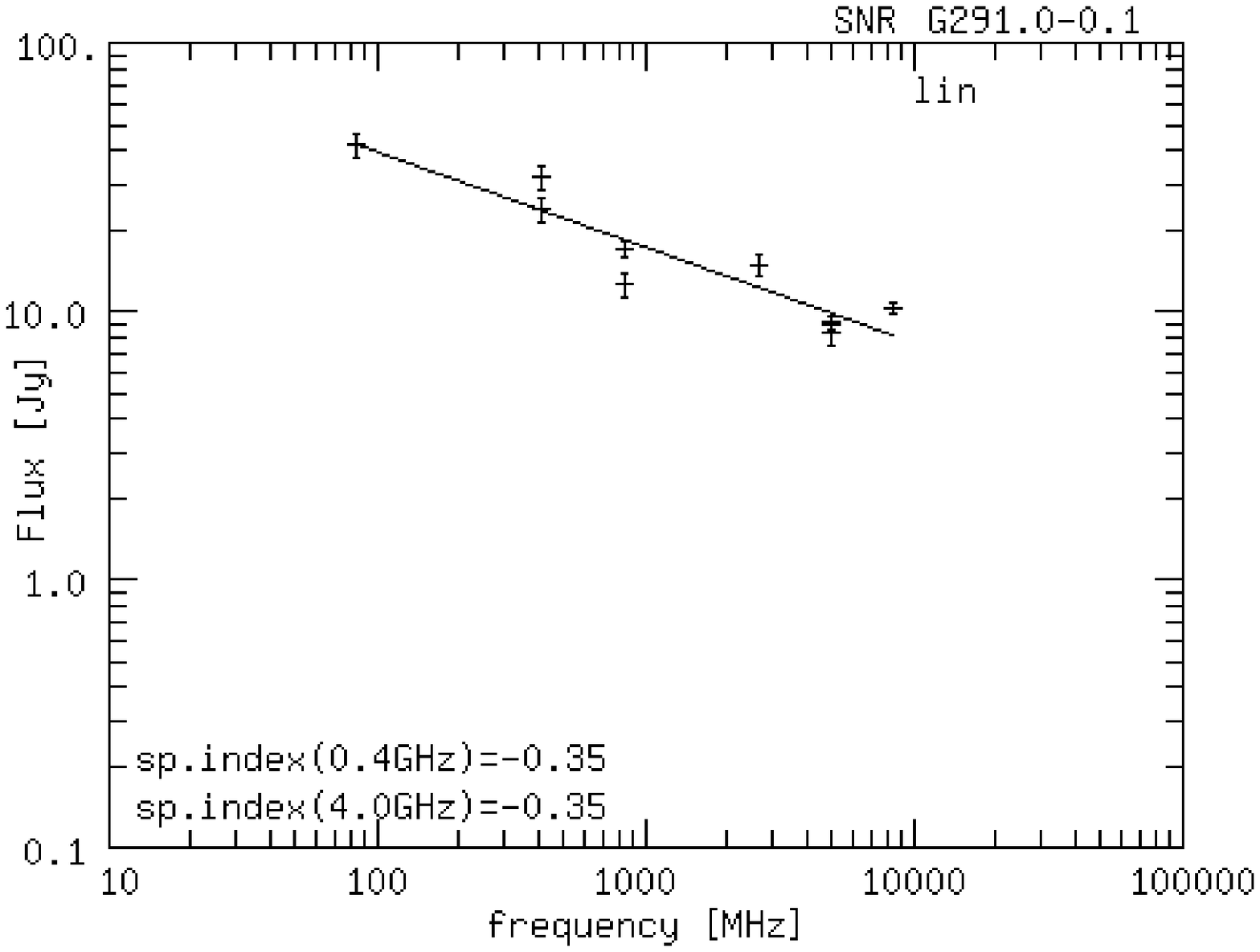,width=7.4cm,angle=0}}}\end{figure}\clearpage
\begin{figure}\centerline{\vbox{\psfig{figure=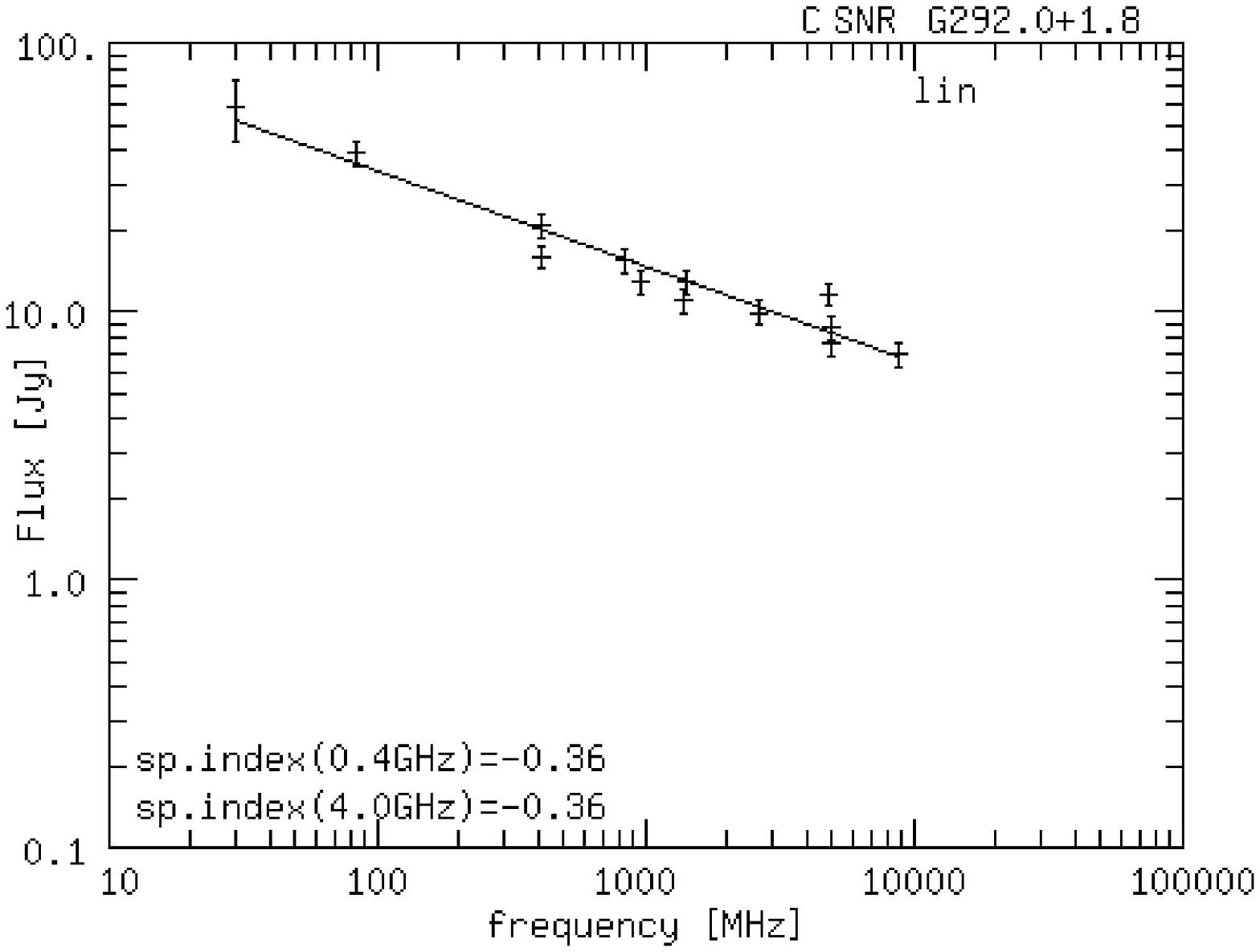,width=7.4cm,angle=0}}}\end{figure}
\begin{figure}\centerline{\vbox{\psfig{figure=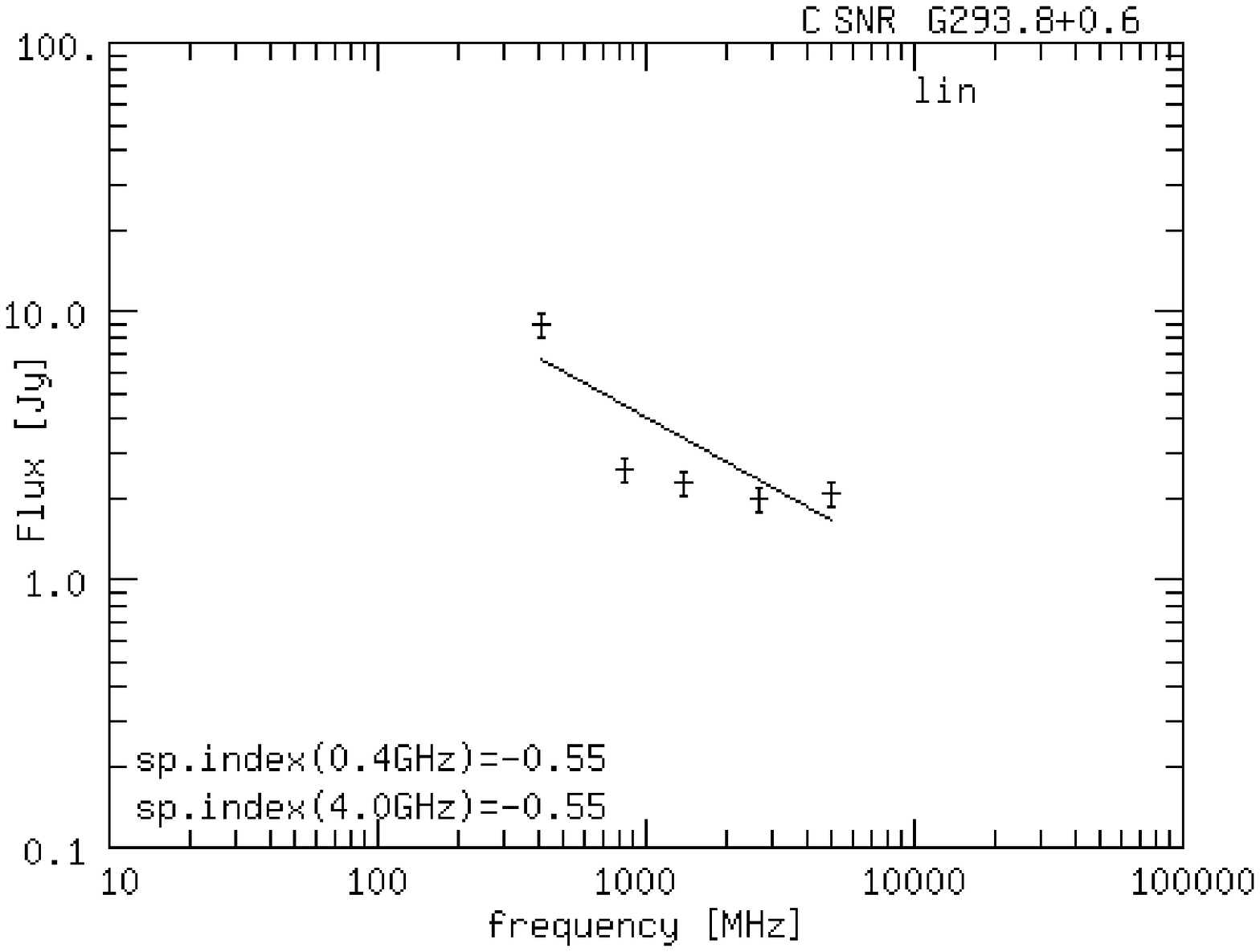,width=7.4cm,angle=0}}}\end{figure}
\begin{figure}\centerline{\vbox{\psfig{figure=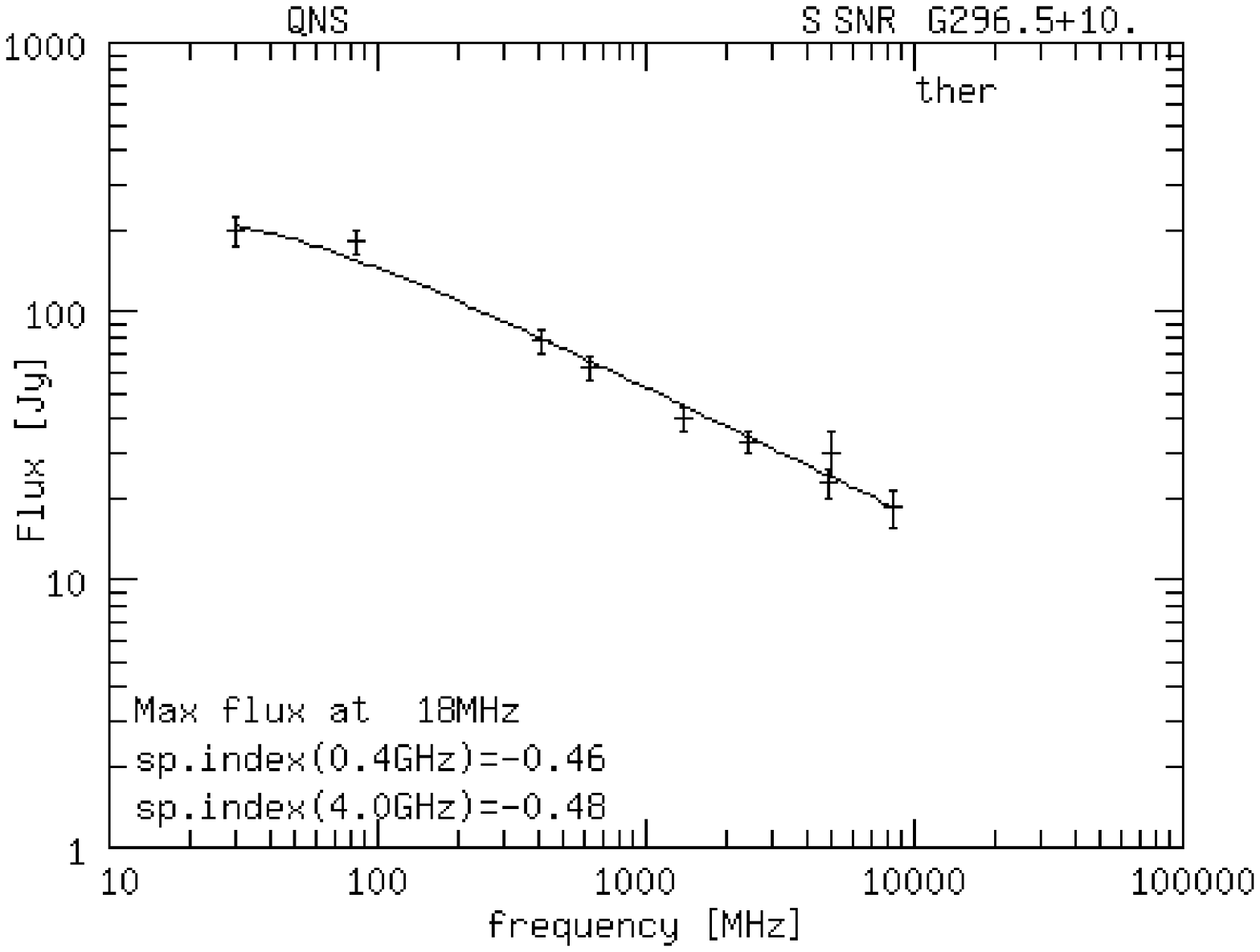,width=7.4cm,angle=0}}}\end{figure}
\begin{figure}\centerline{\vbox{\psfig{figure=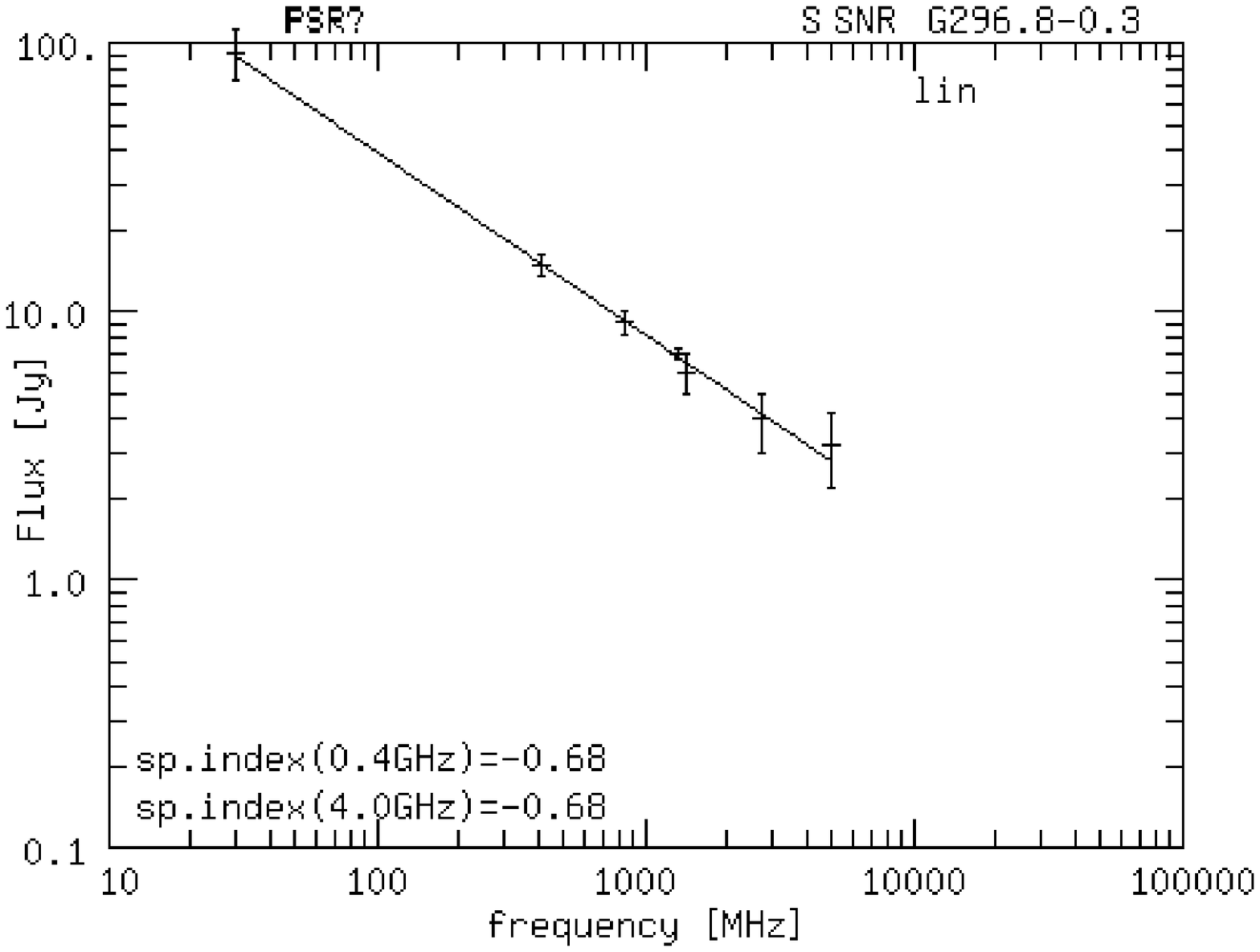,width=7.4cm,angle=0}}}\end{figure}
\begin{figure}\centerline{\vbox{\psfig{figure=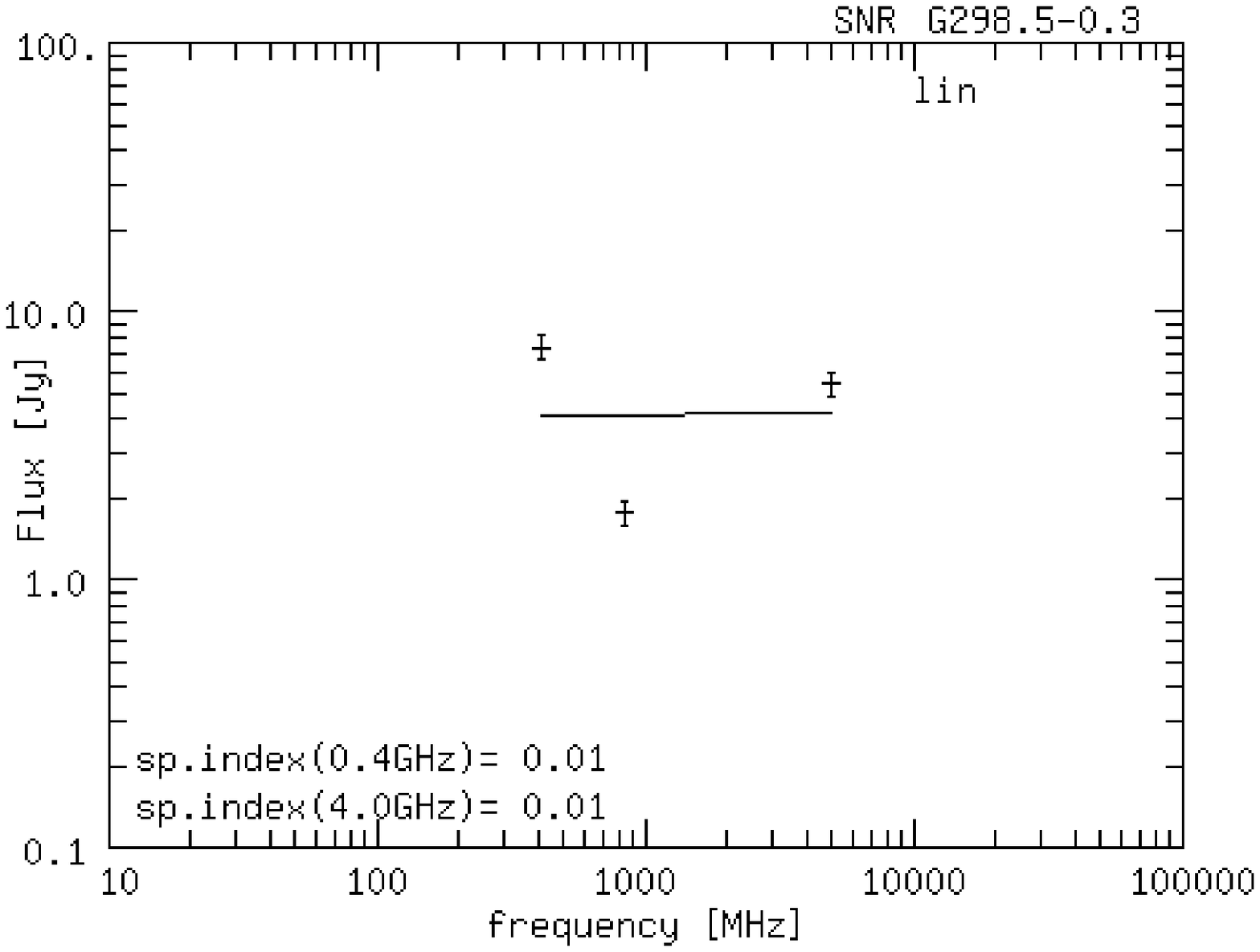,width=7.4cm,angle=0}}}\end{figure}
\begin{figure}\centerline{\vbox{\psfig{figure=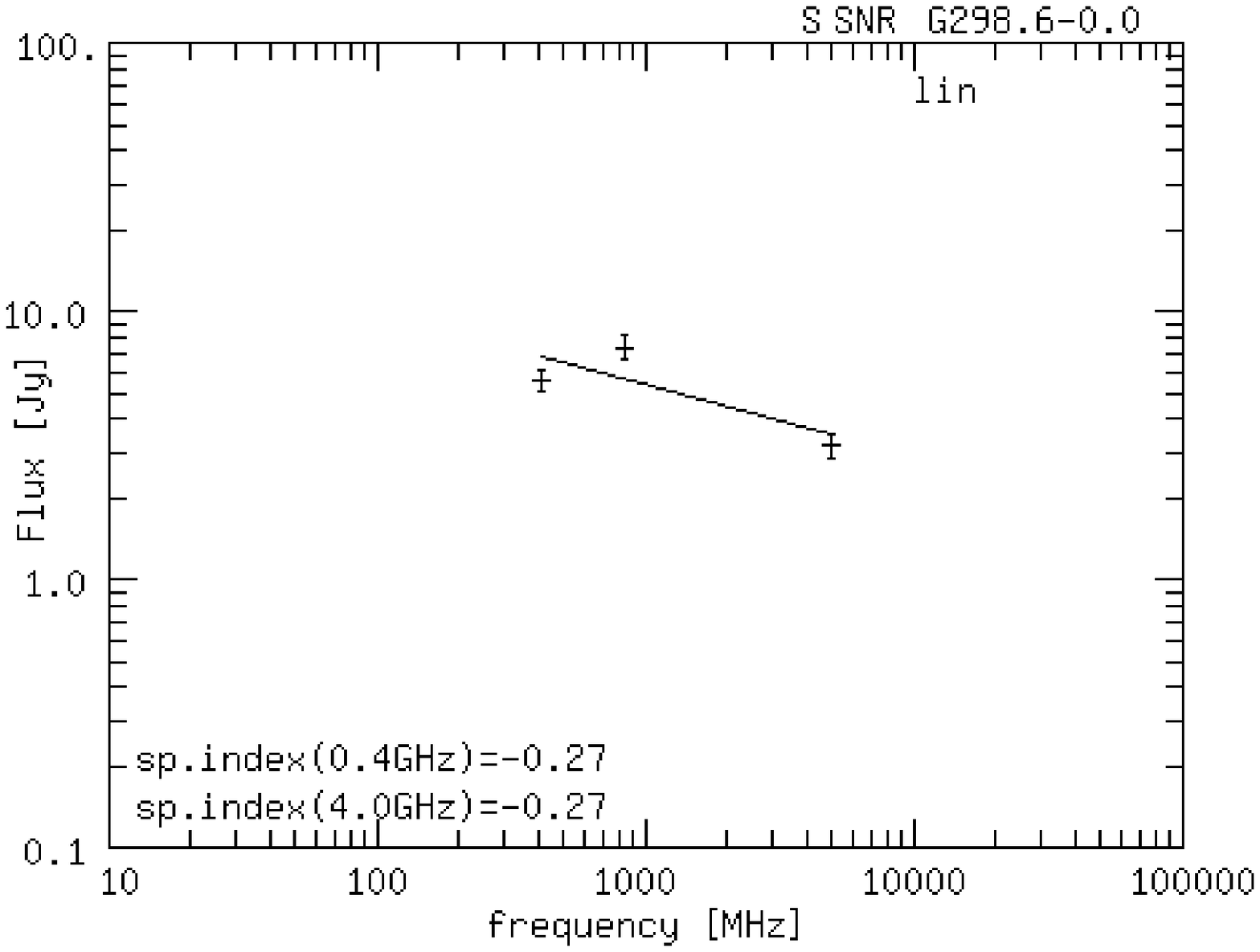,width=7.4cm,angle=0}}}\end{figure}
\begin{figure}\centerline{\vbox{\psfig{figure=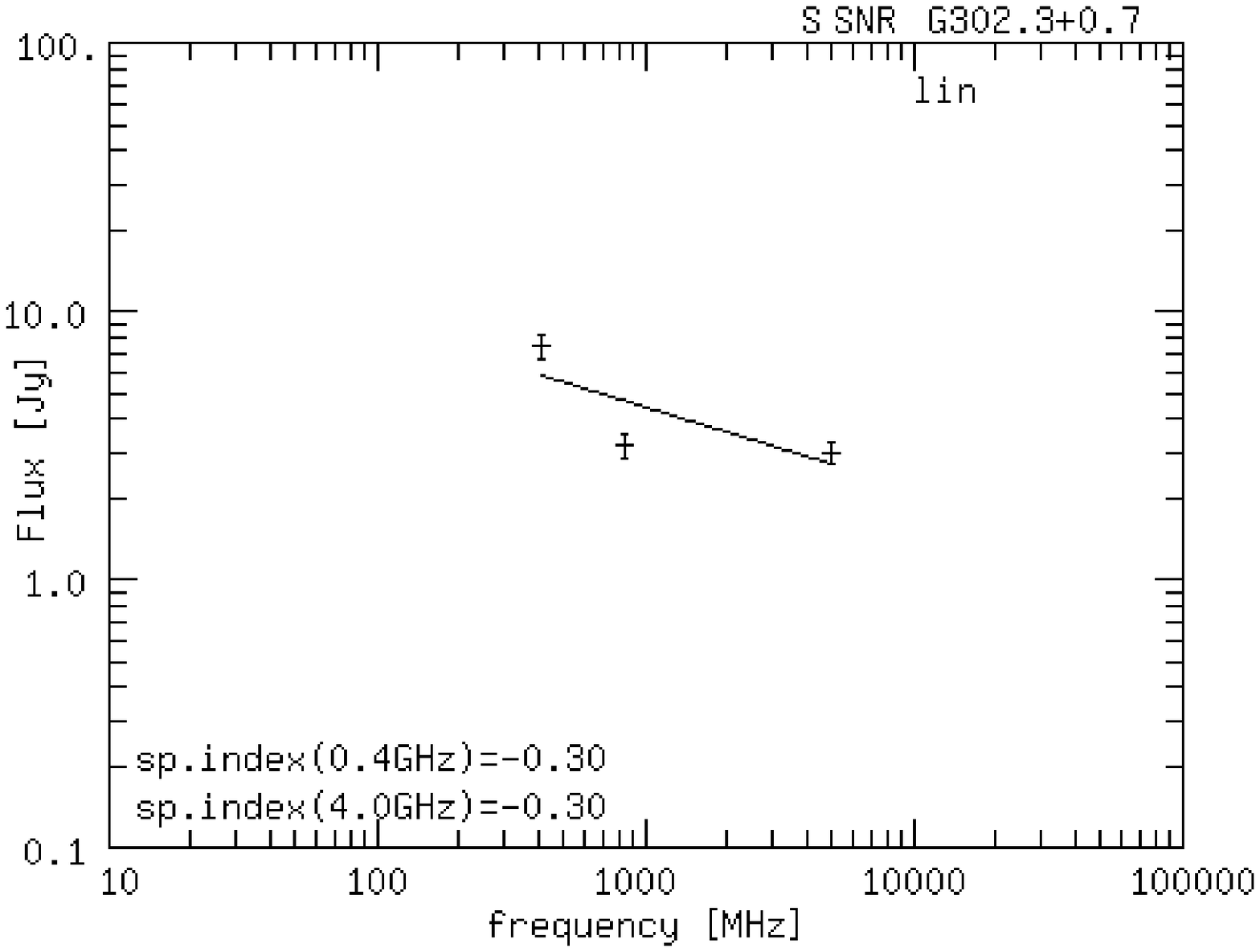,width=7.4cm,angle=0}}}\end{figure}
\begin{figure}\centerline{\vbox{\psfig{figure=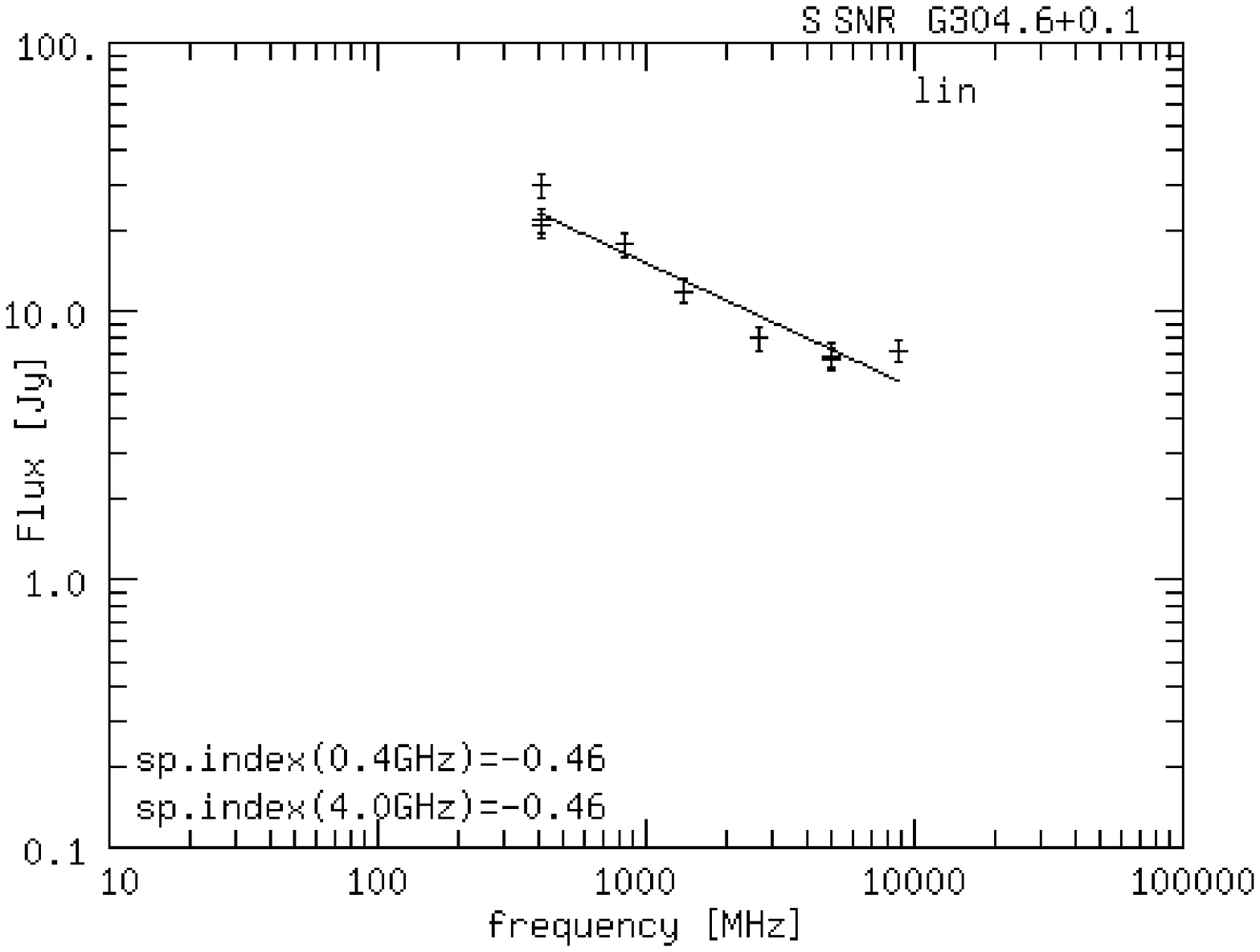,width=7.4cm,angle=0}}}\end{figure}\clearpage
\begin{figure}\centerline{\vbox{\psfig{figure=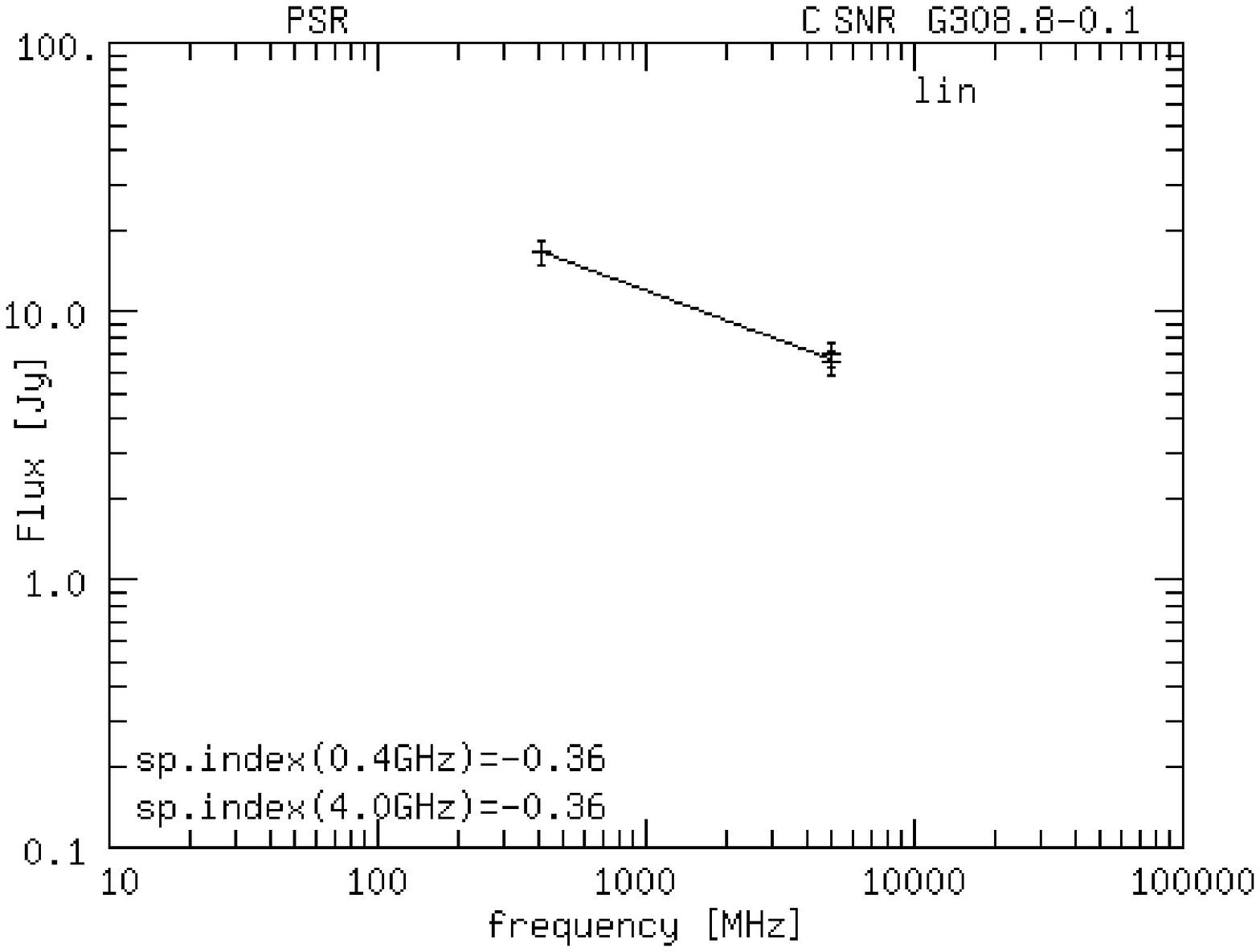,width=7.4cm,angle=0}}}\end{figure}
\begin{figure}\centerline{\vbox{\psfig{figure=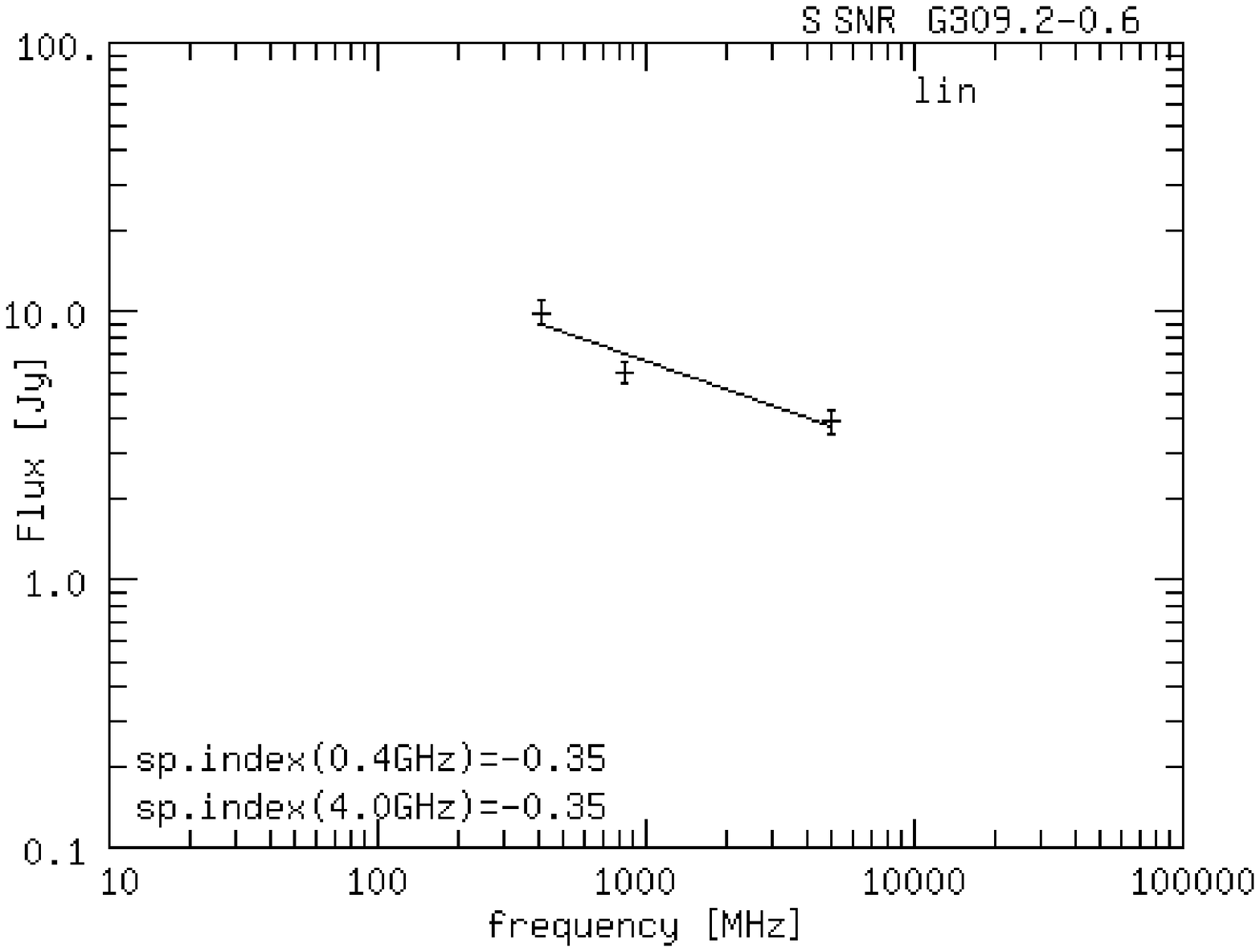,width=7.4cm,angle=0}}}\end{figure}
\begin{figure}\centerline{\vbox{\psfig{figure=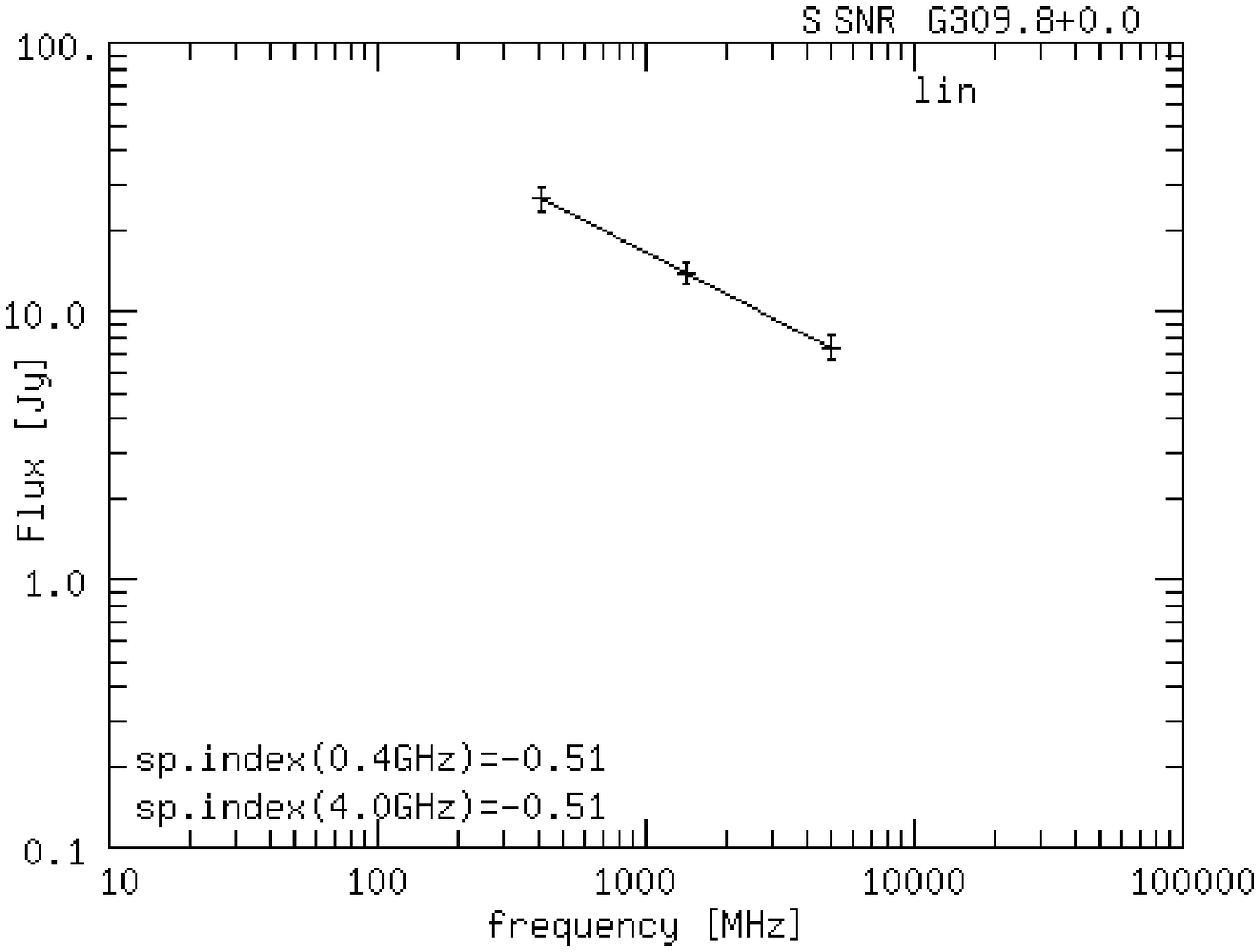,width=7.4cm,angle=0}}}\end{figure}
\begin{figure}\centerline{\vbox{\psfig{figure=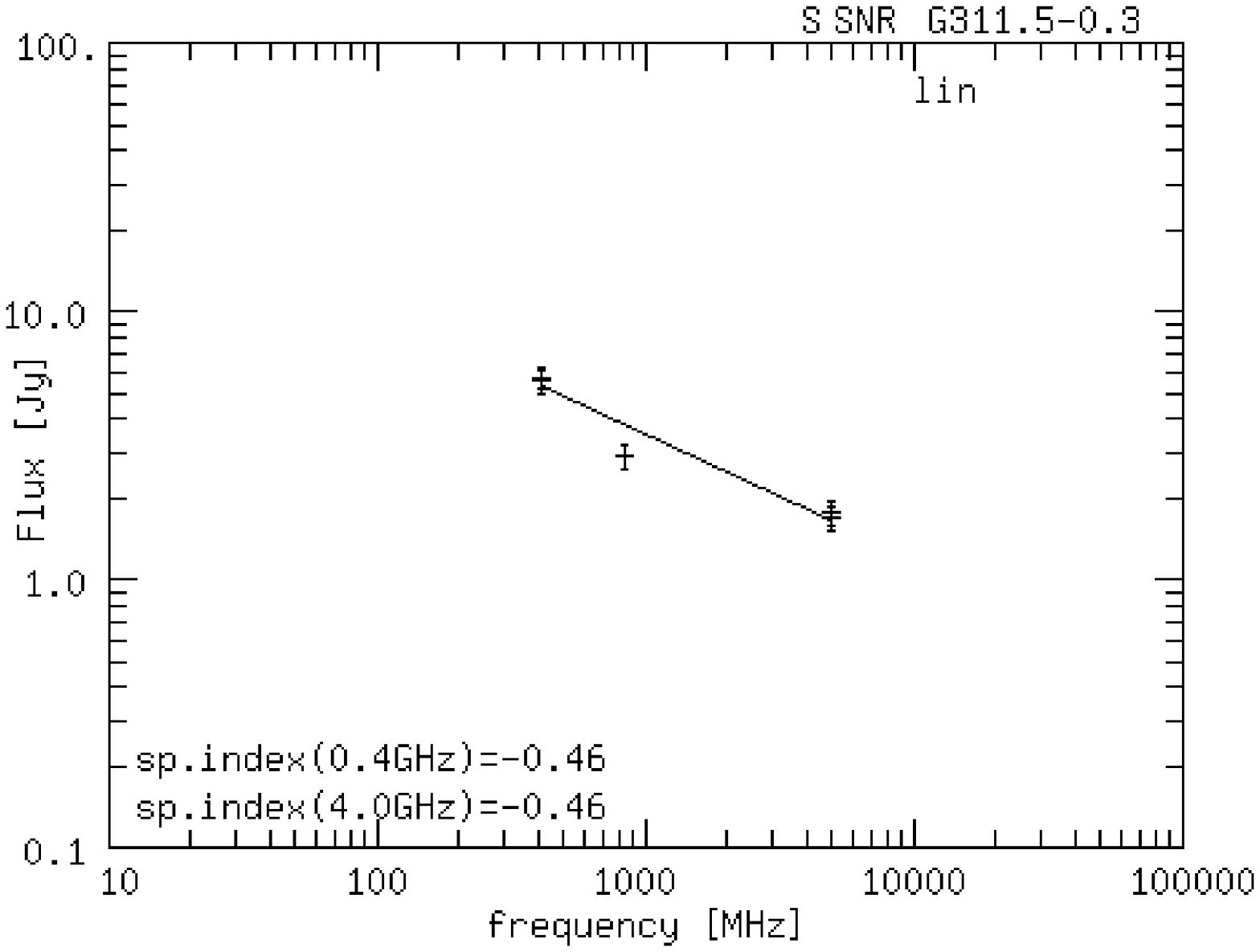,width=7.4cm,angle=0}}}\end{figure}
\begin{figure}\centerline{\vbox{\psfig{figure=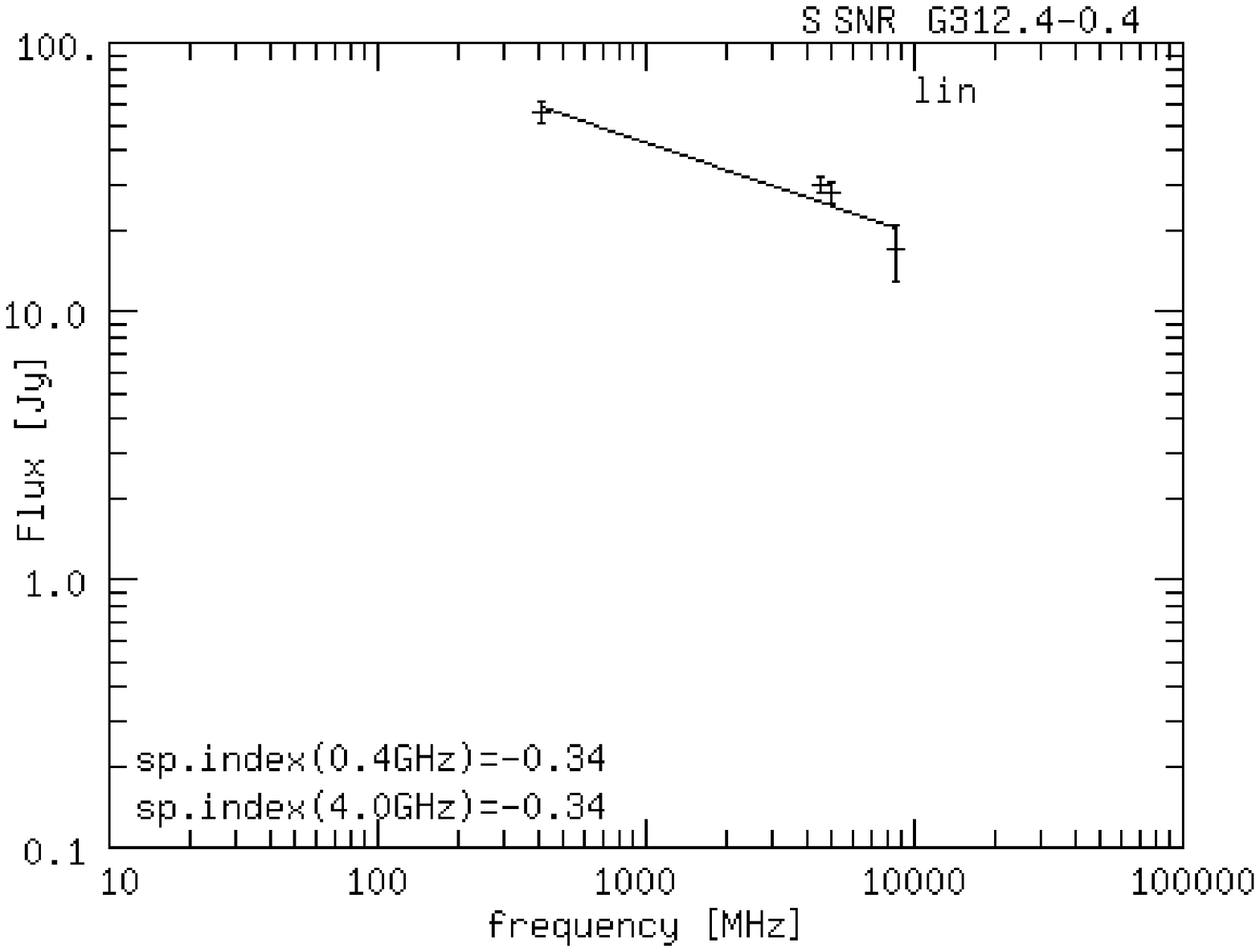,width=7.4cm,angle=0}}}\end{figure}
\begin{figure}\centerline{\vbox{\psfig{figure=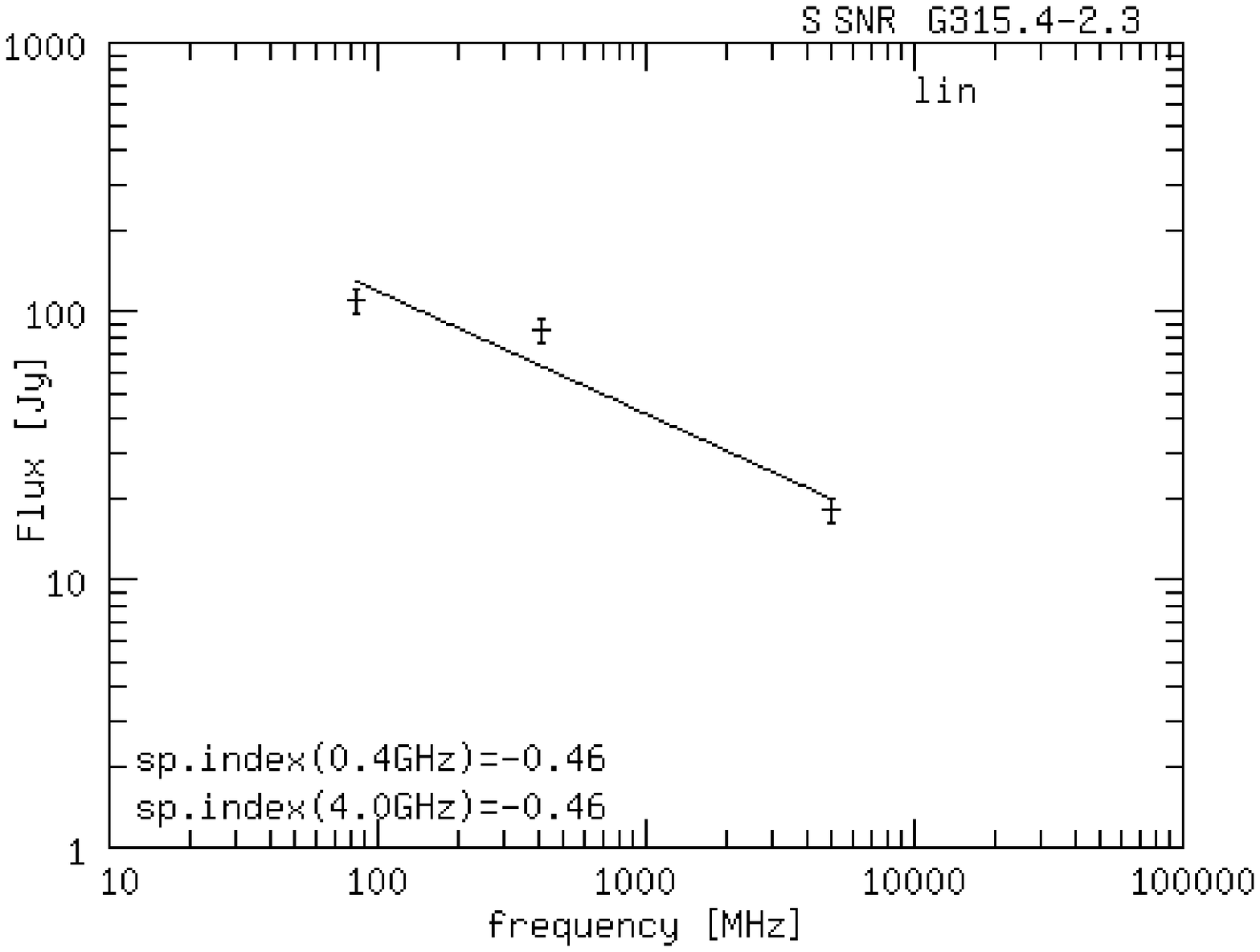,width=7.4cm,angle=0}}}\end{figure}
\begin{figure}\centerline{\vbox{\psfig{figure=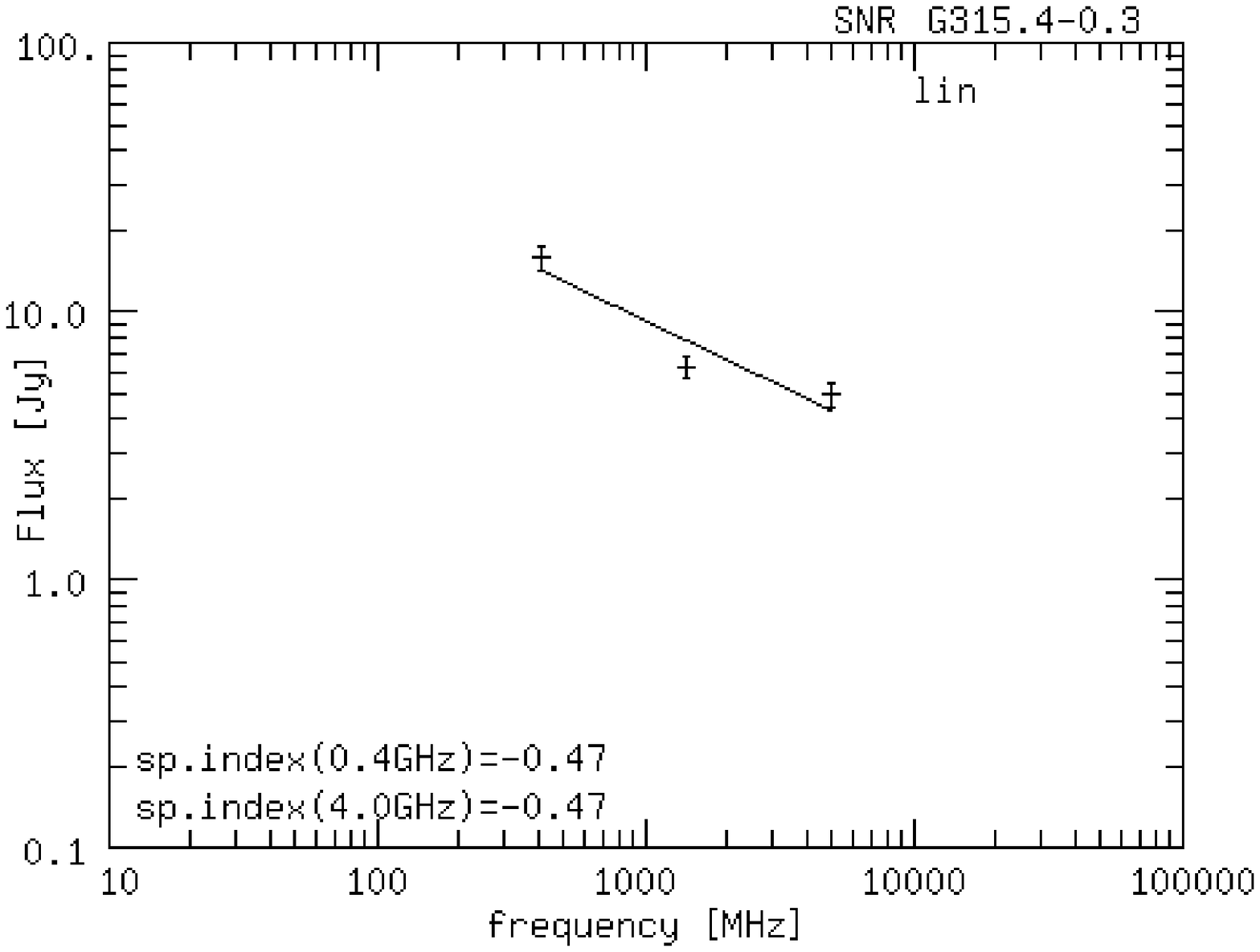,width=7.4cm,angle=0}}}\end{figure}
\begin{figure}\centerline{\vbox{\psfig{figure=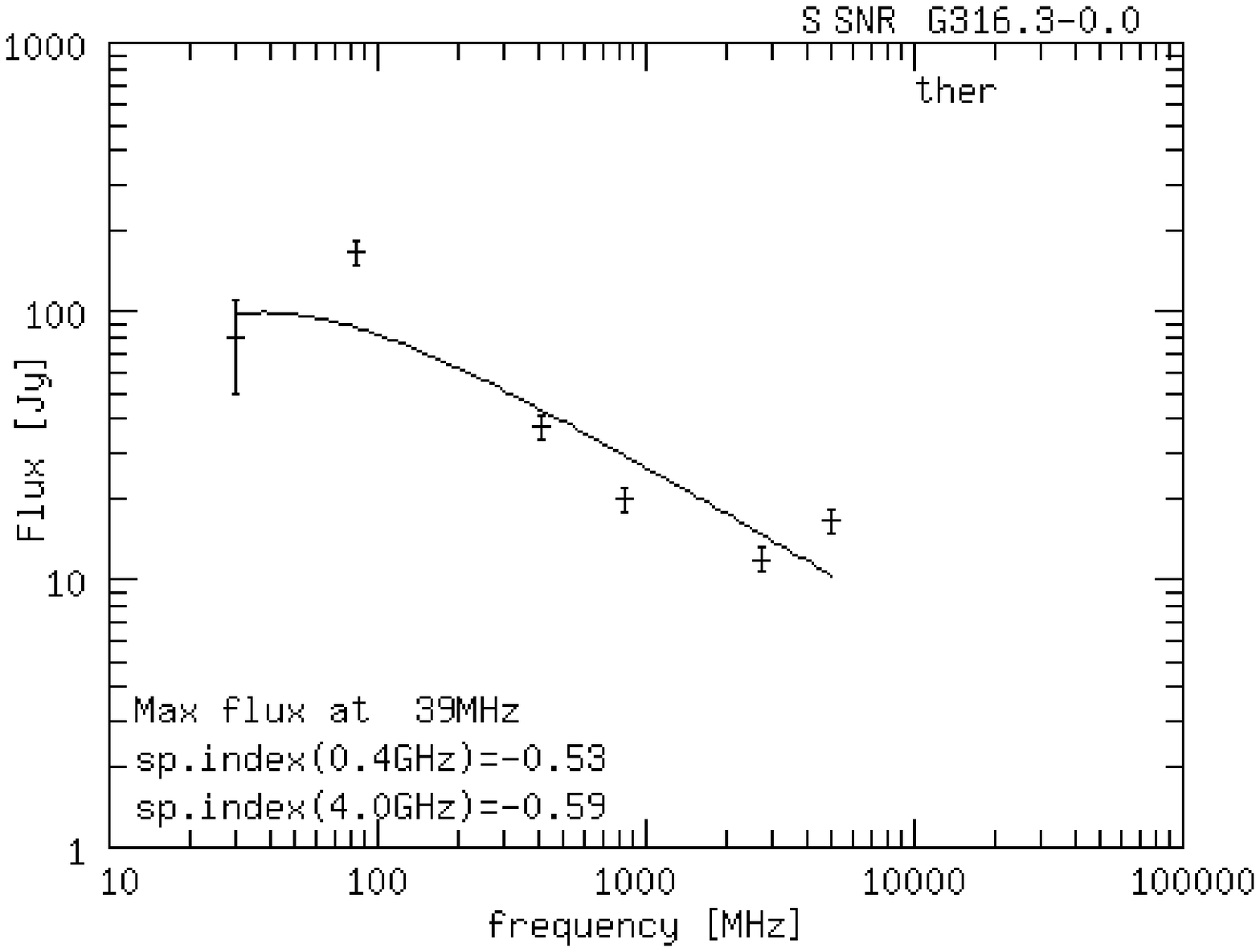,width=7.4cm,angle=0}}}\end{figure}\clearpage
\begin{figure}\centerline{\vbox{\psfig{figure=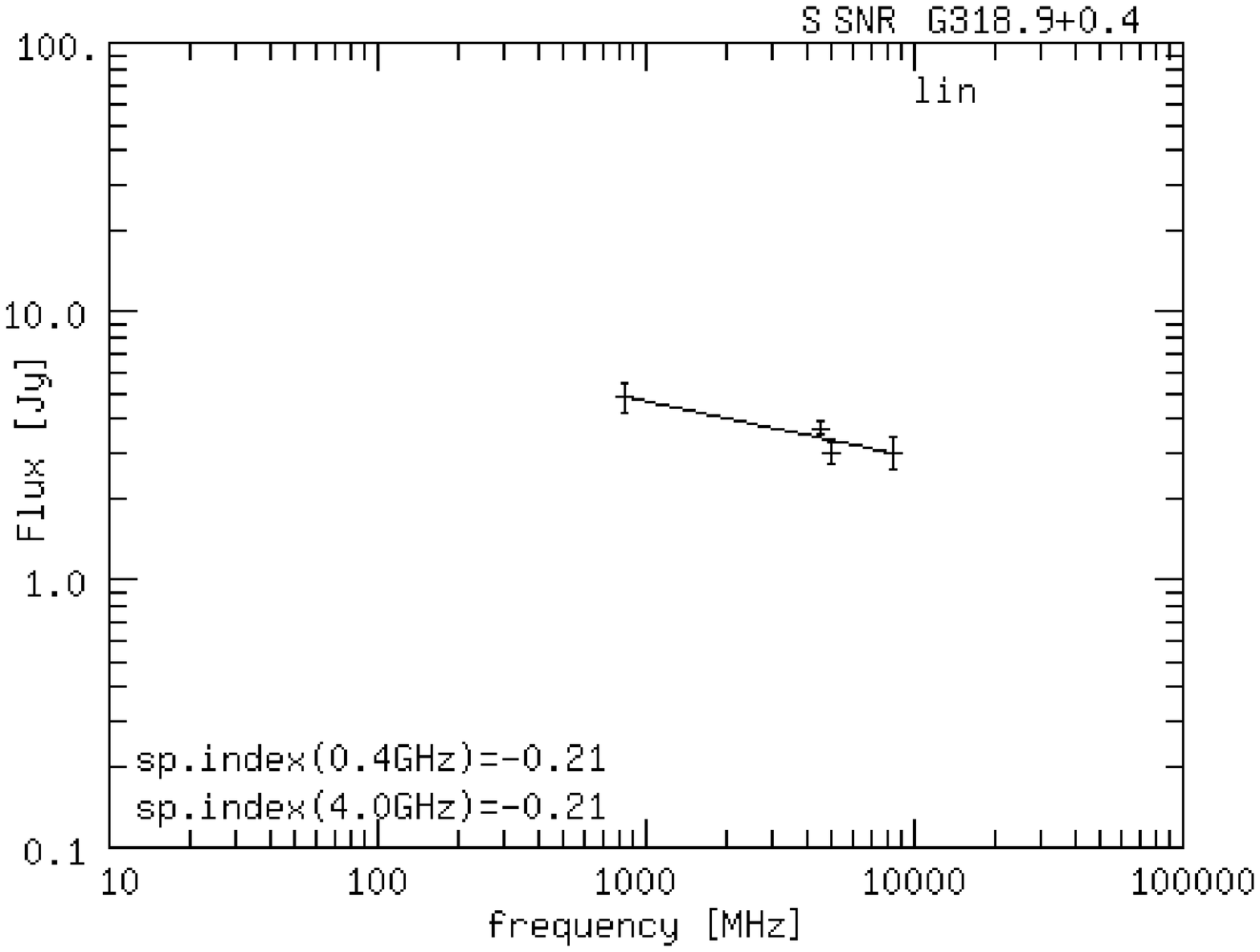,width=7.4cm,angle=0}}}\end{figure}
\begin{figure}\centerline{\vbox{\psfig{figure=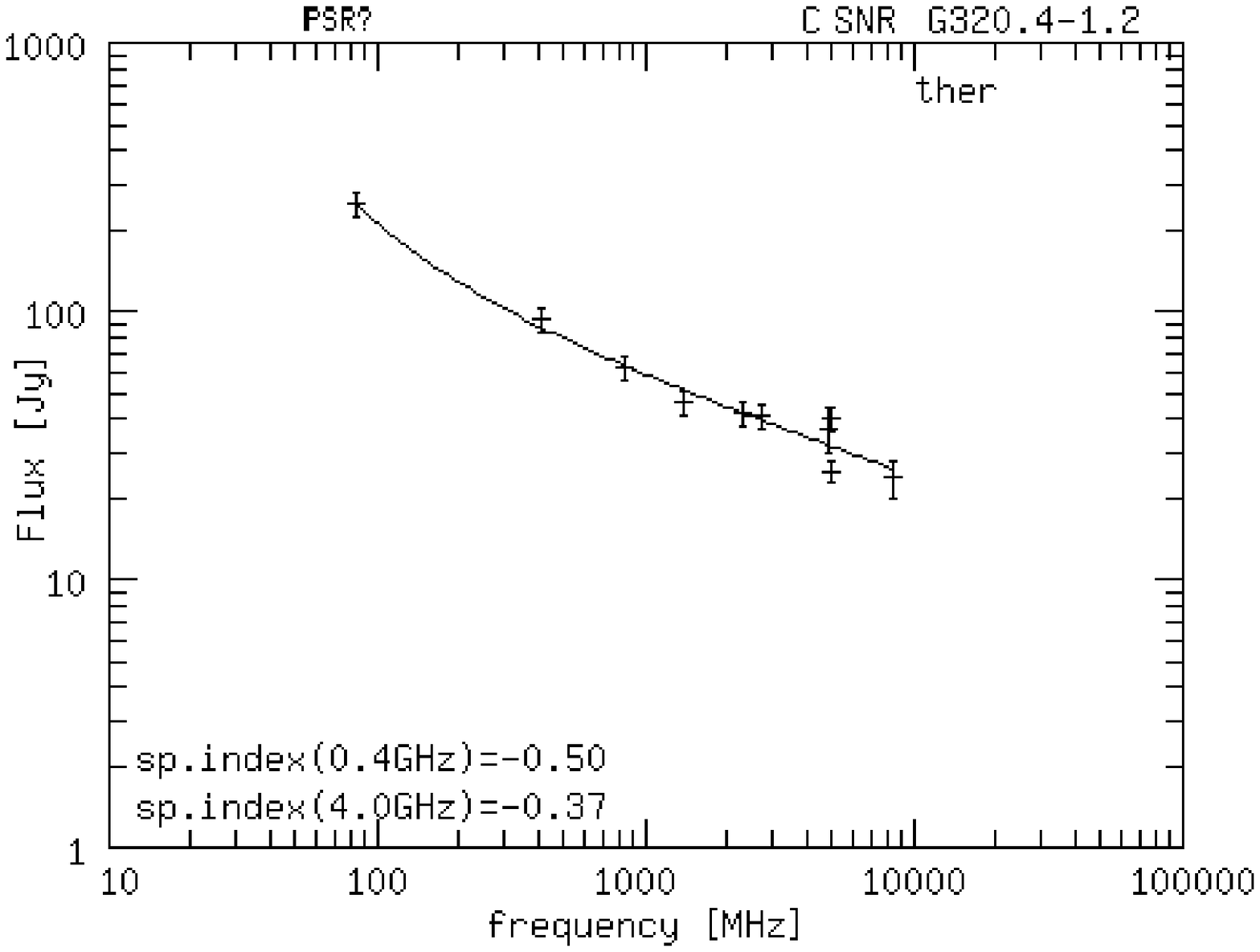,width=7.4cm,angle=0}}}\end{figure}
\begin{figure}\centerline{\vbox{\psfig{figure=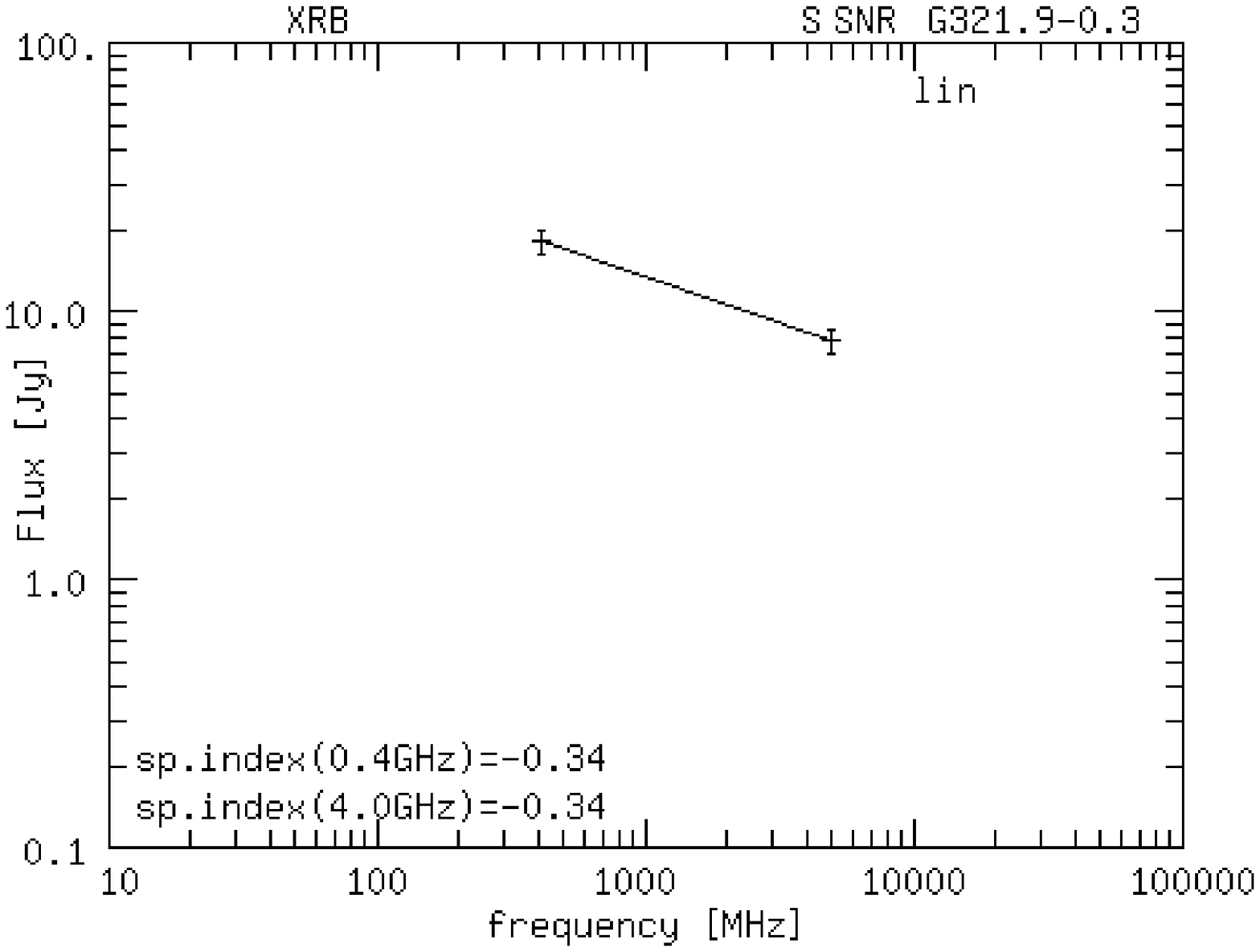,width=7.4cm,angle=0}}}\end{figure}
\begin{figure}\centerline{\vbox{\psfig{figure=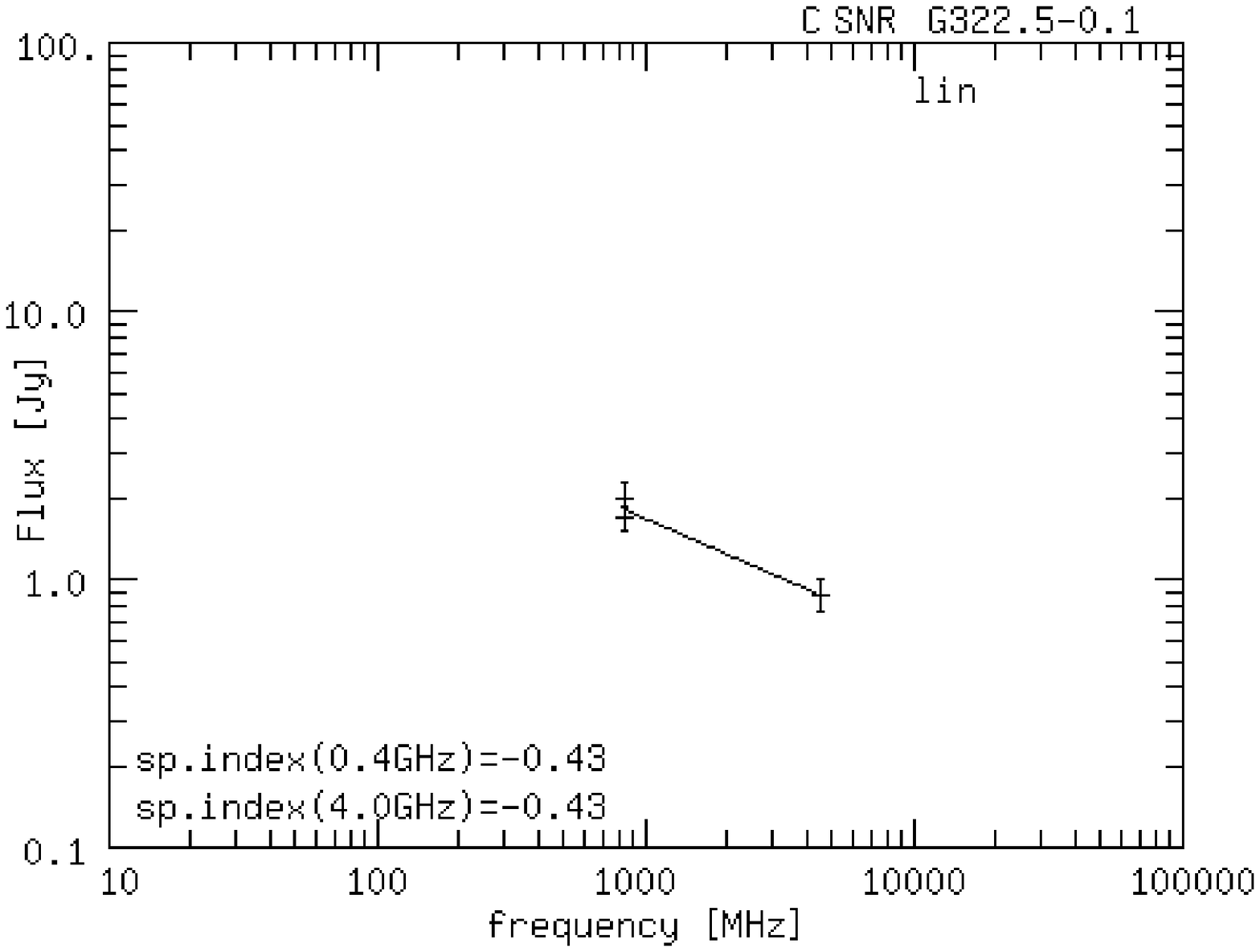,width=7.4cm,angle=0}}}\end{figure}
\begin{figure}\centerline{\vbox{\psfig{figure=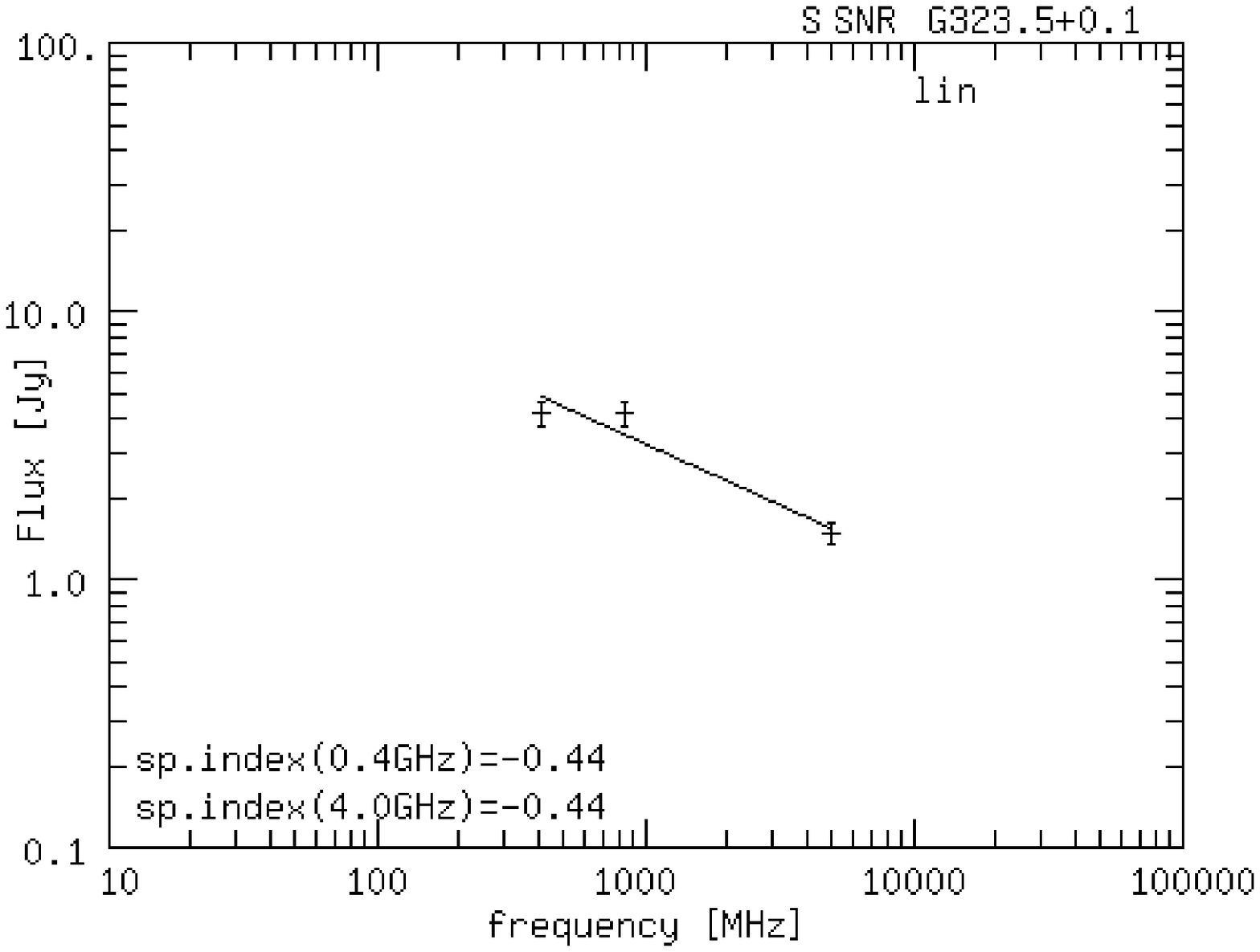,width=7.4cm,angle=0}}}\end{figure}
\begin{figure}\centerline{\vbox{\psfig{figure=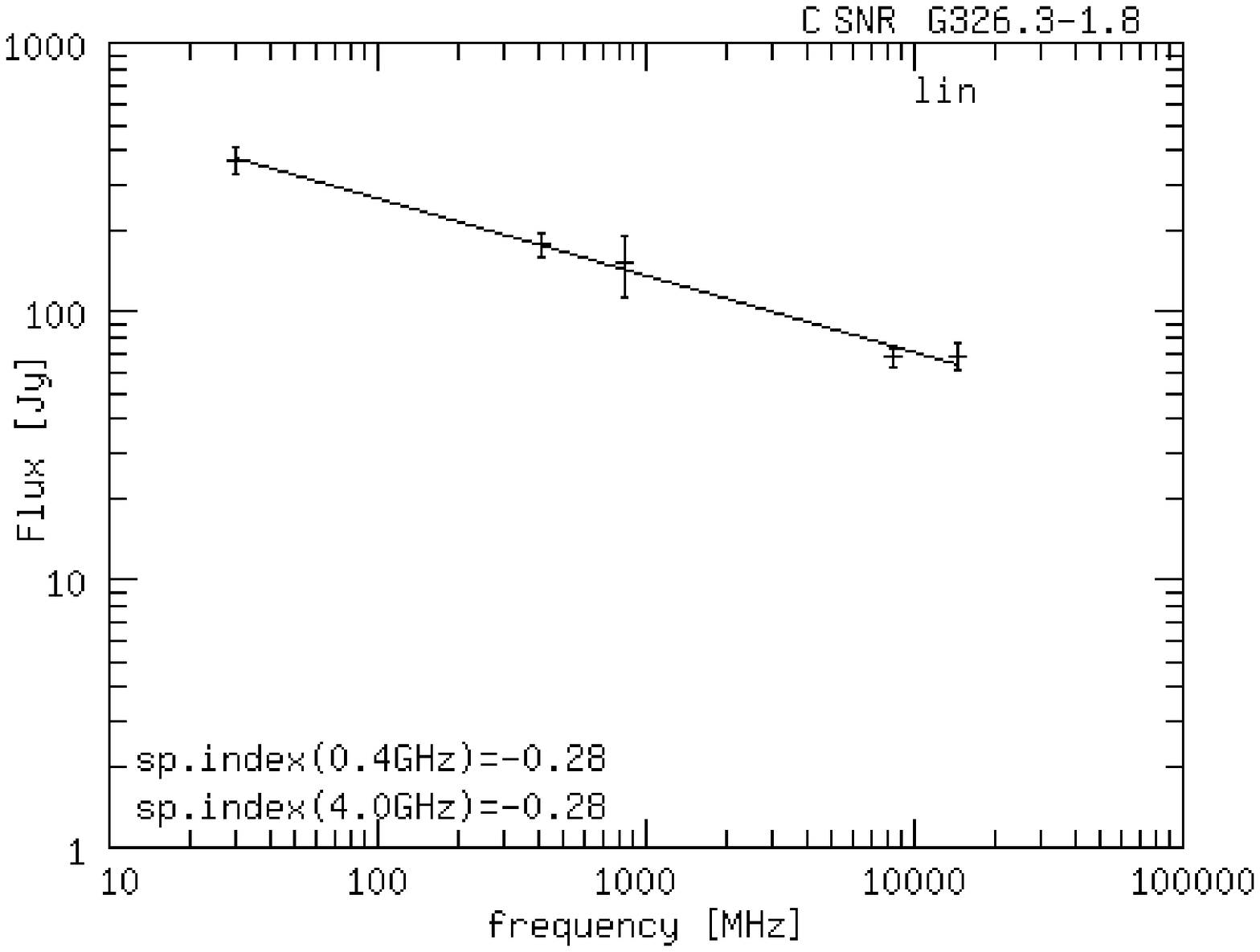,width=7.4cm,angle=0}}}\end{figure}
\begin{figure}\centerline{\vbox{\psfig{figure=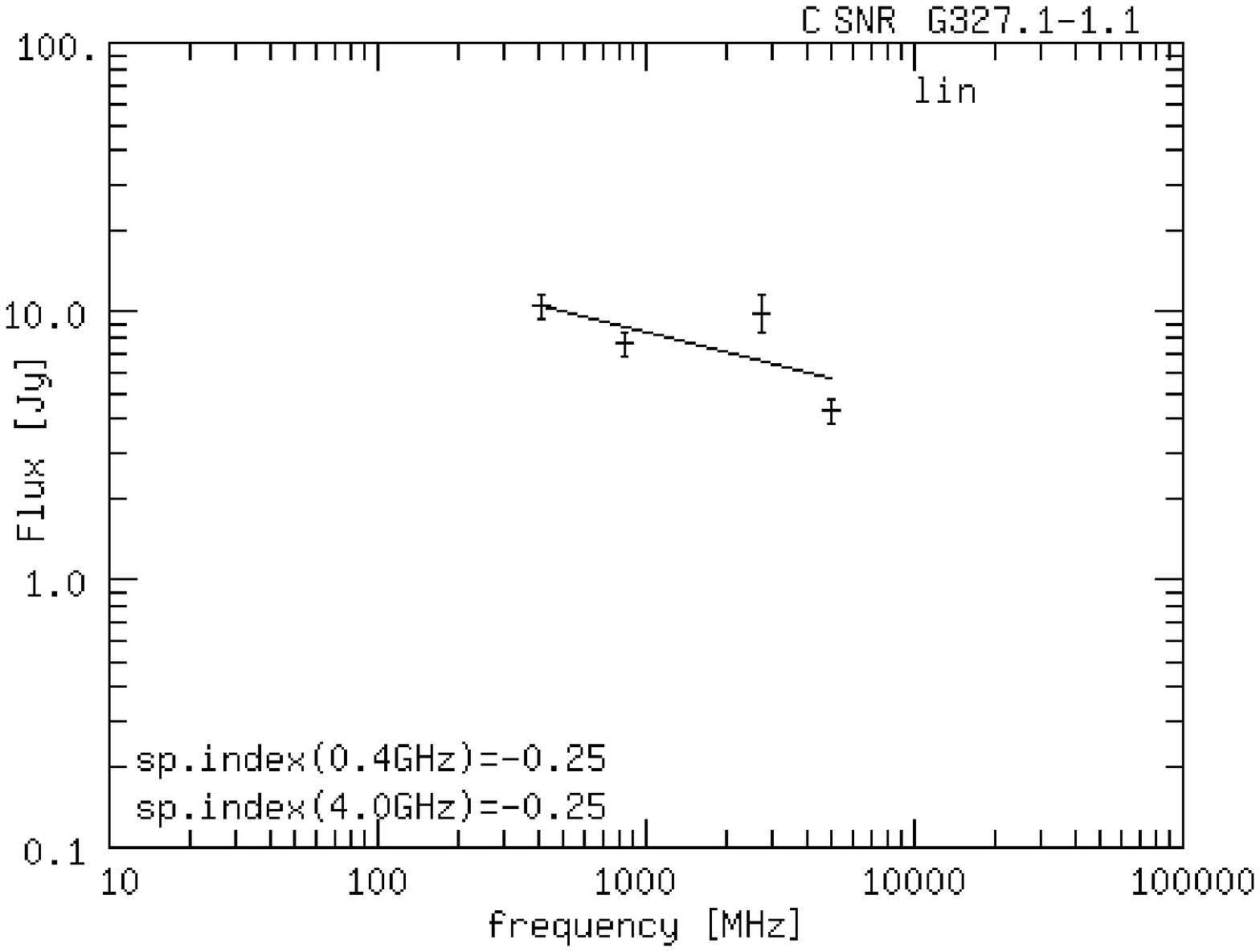,width=7.4cm,angle=0}}}\end{figure}
\begin{figure}\centerline{\vbox{\psfig{figure=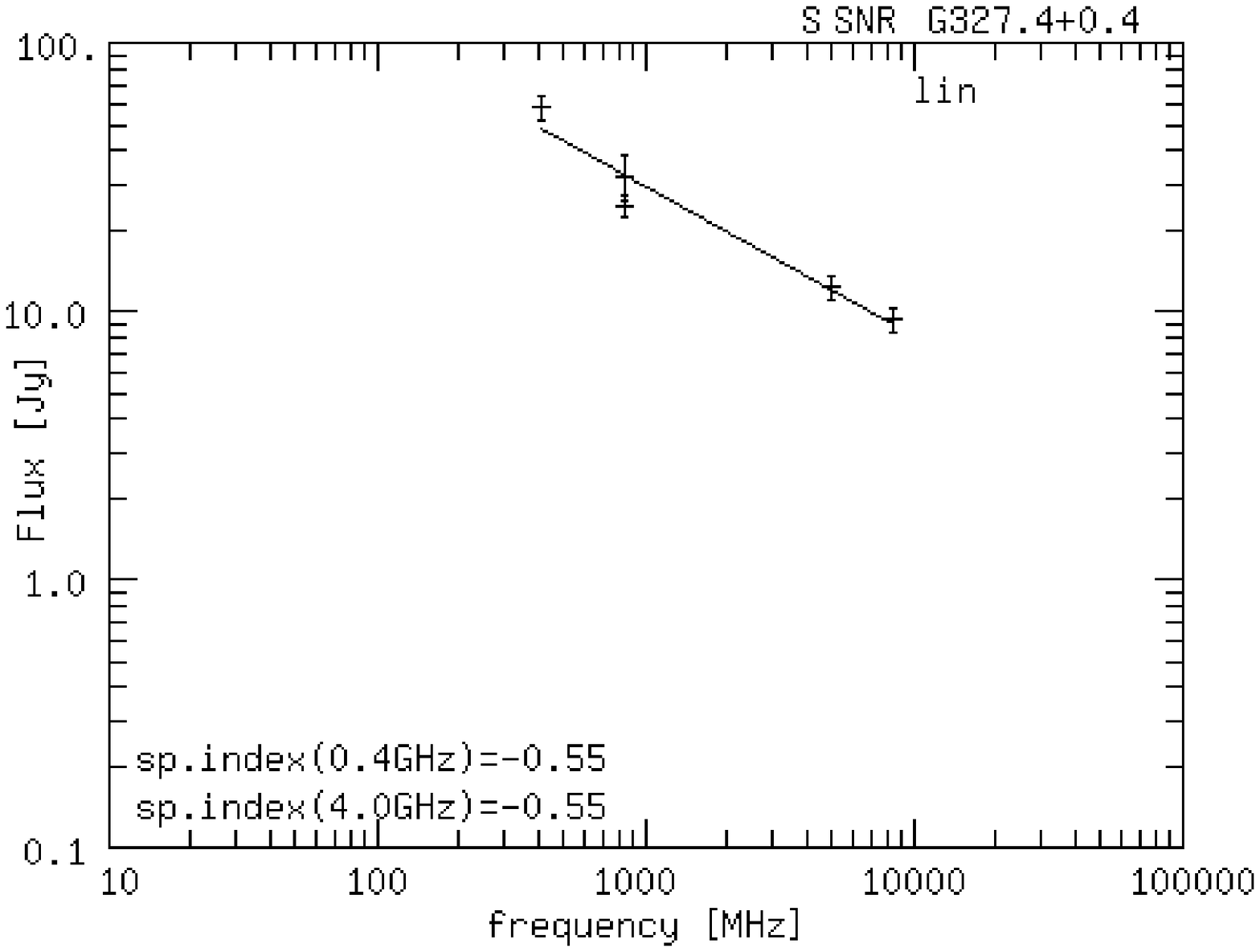,width=7.4cm,angle=0}}}\end{figure}\clearpage
\begin{figure}\centerline{\vbox{\psfig{figure=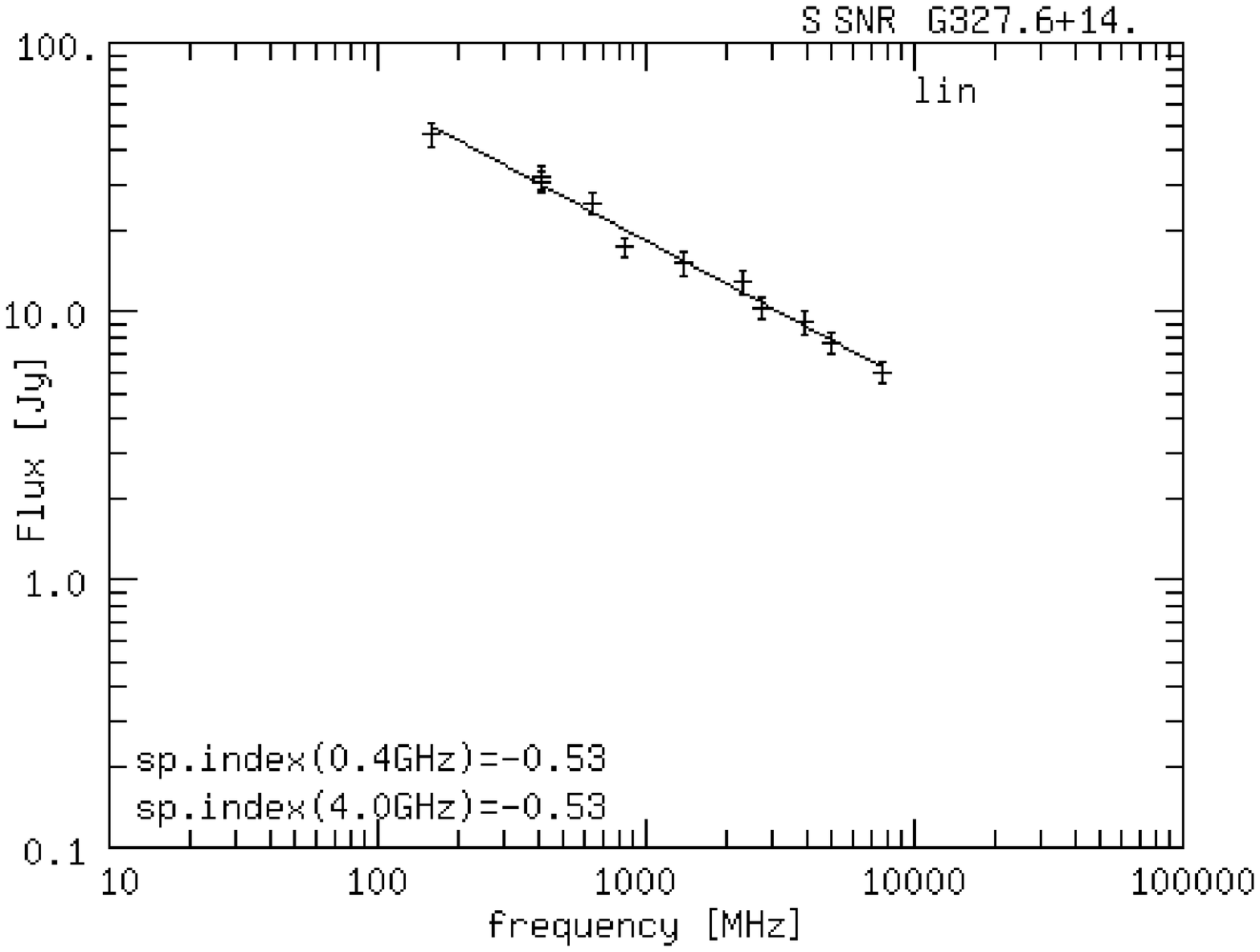,width=7.4cm,angle=0}}}\end{figure}
\begin{figure}\centerline{\vbox{\psfig{figure=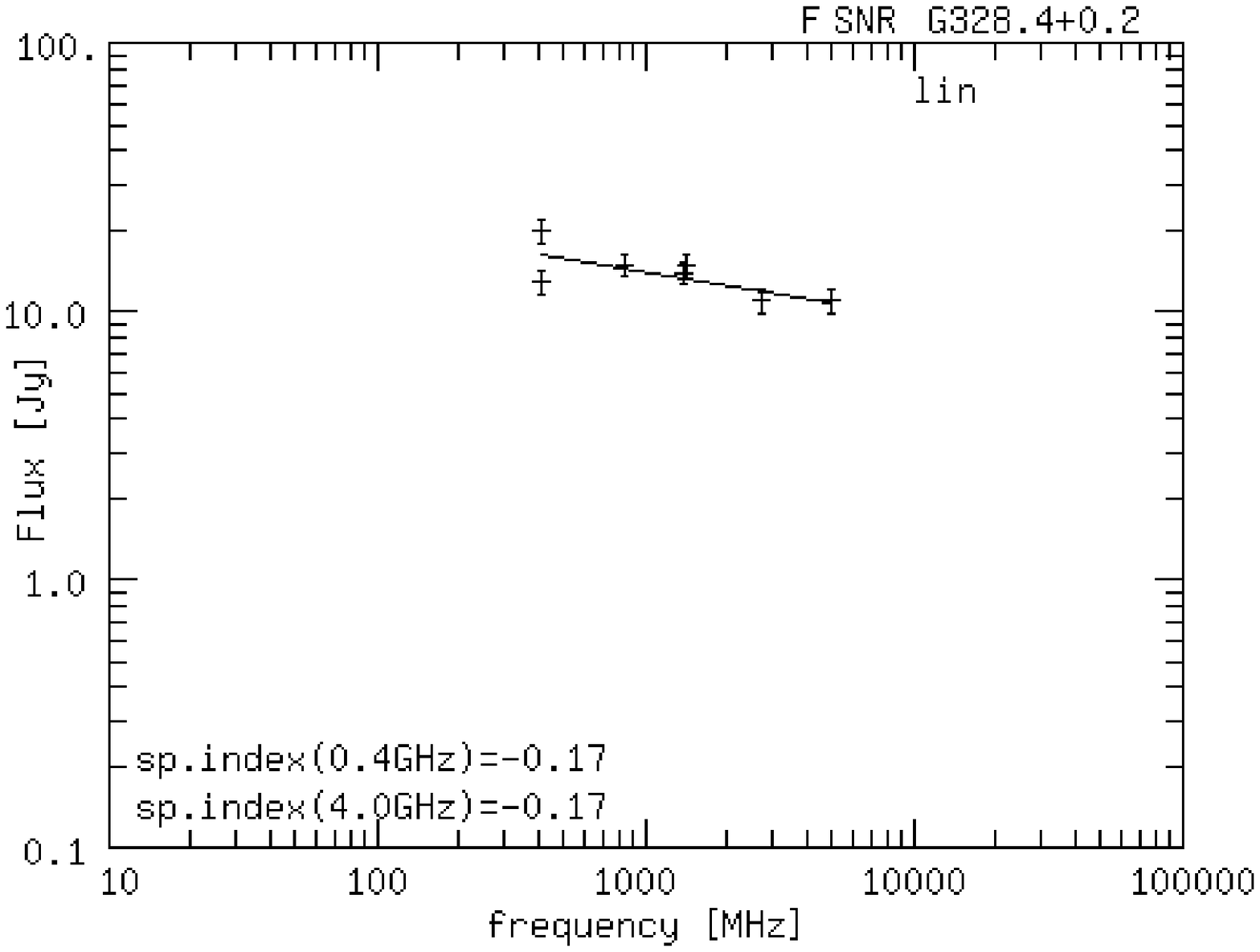,width=7.4cm,angle=0}}}\end{figure}
\begin{figure}\centerline{\vbox{\psfig{figure=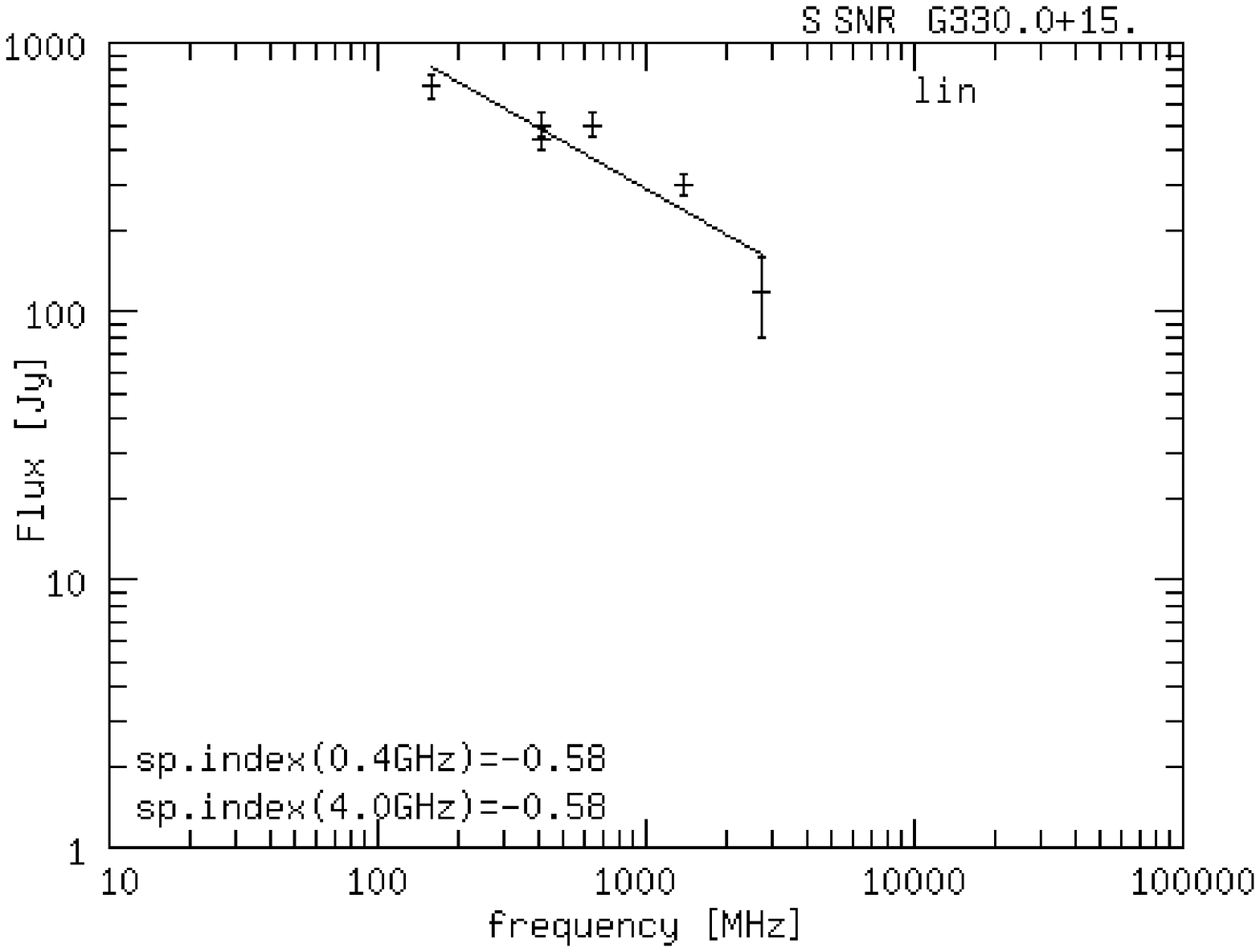,width=7.4cm,angle=0}}}\end{figure}
\begin{figure}\centerline{\vbox{\psfig{figure=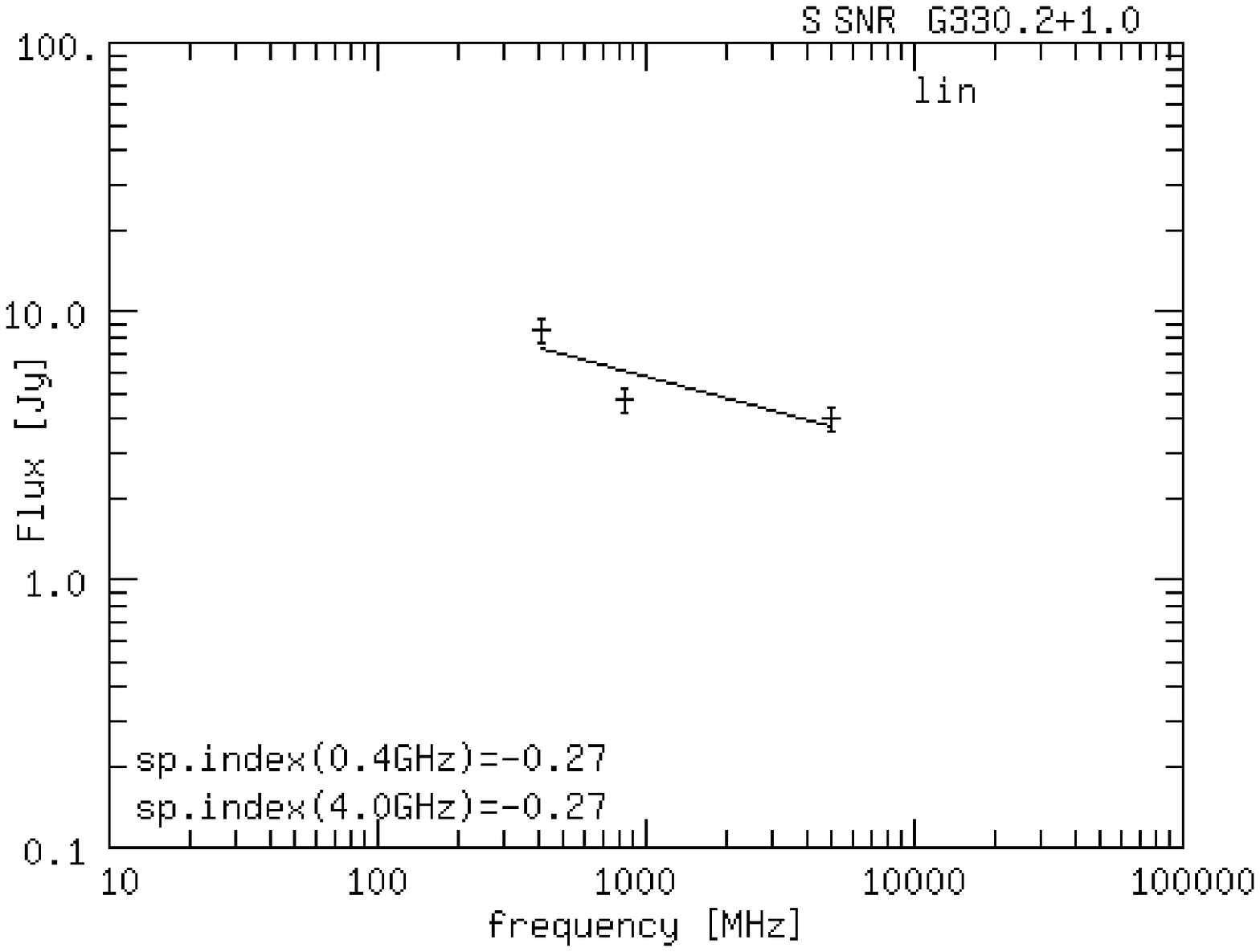,width=7.4cm,angle=0}}}\end{figure}
\begin{figure}\centerline{\vbox{\psfig{figure=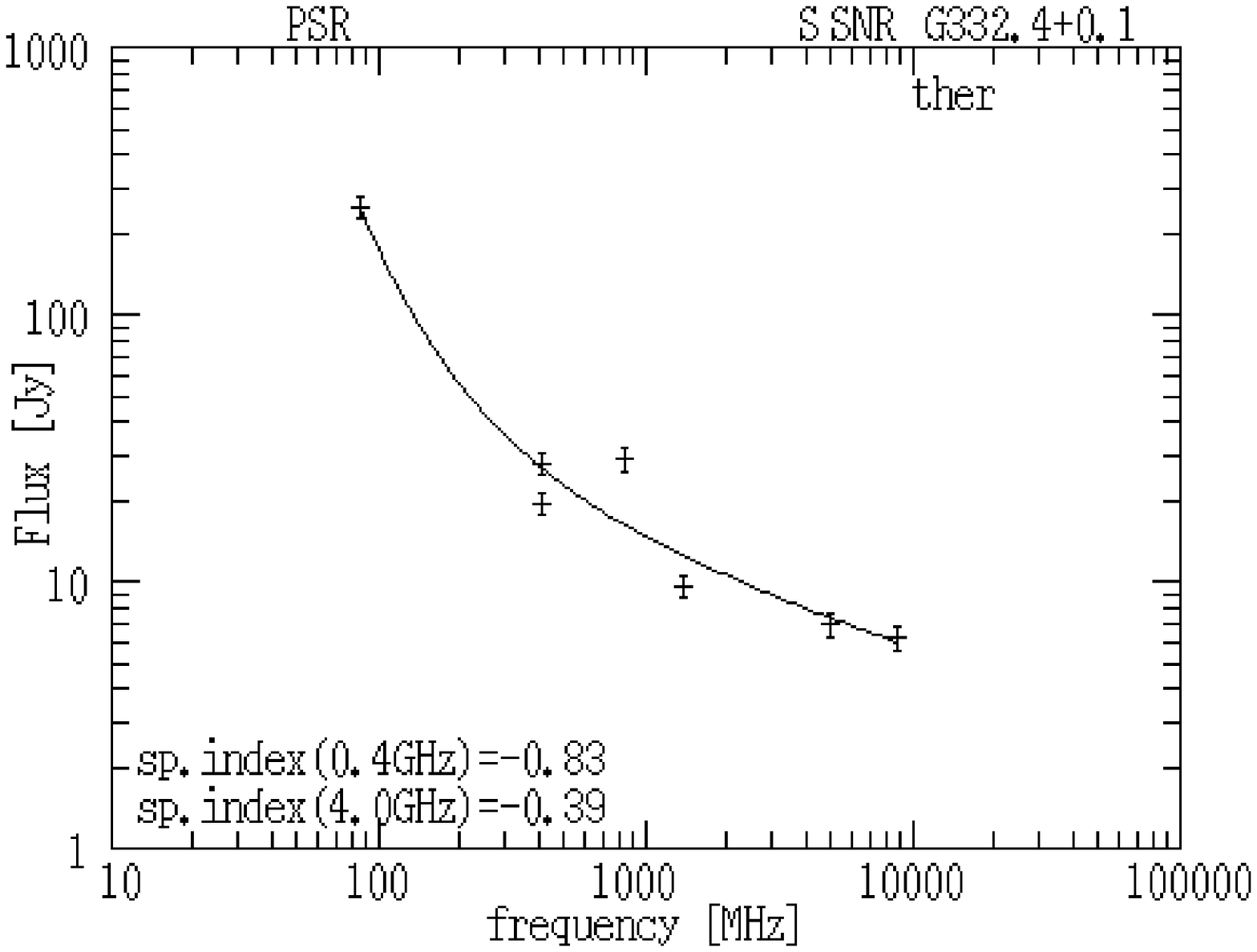,width=7.4cm,angle=0}}}\end{figure}
\begin{figure}\centerline{\vbox{\psfig{figure=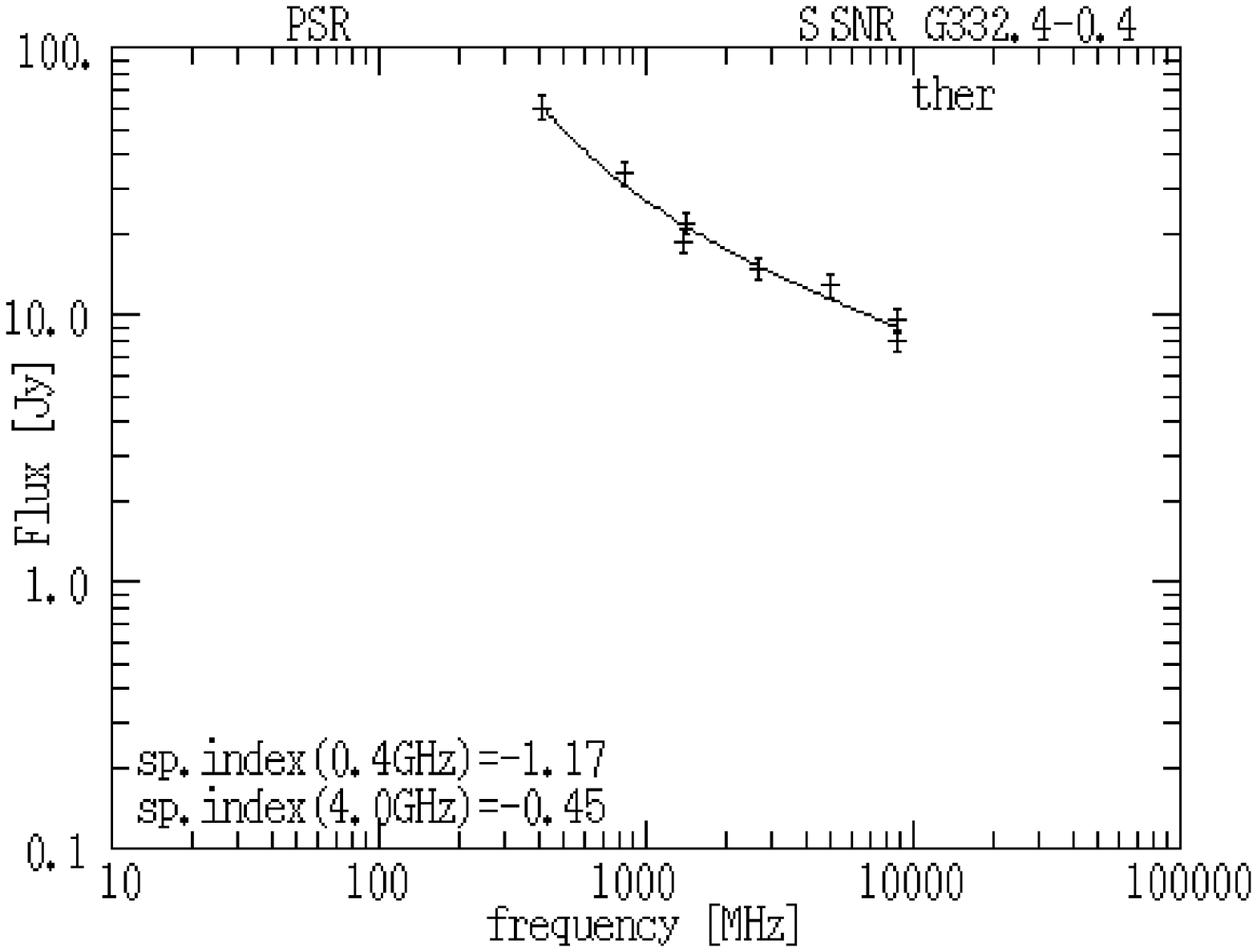,width=7.4cm,angle=0}}}\end{figure}
\begin{figure}\centerline{\vbox{\psfig{figure=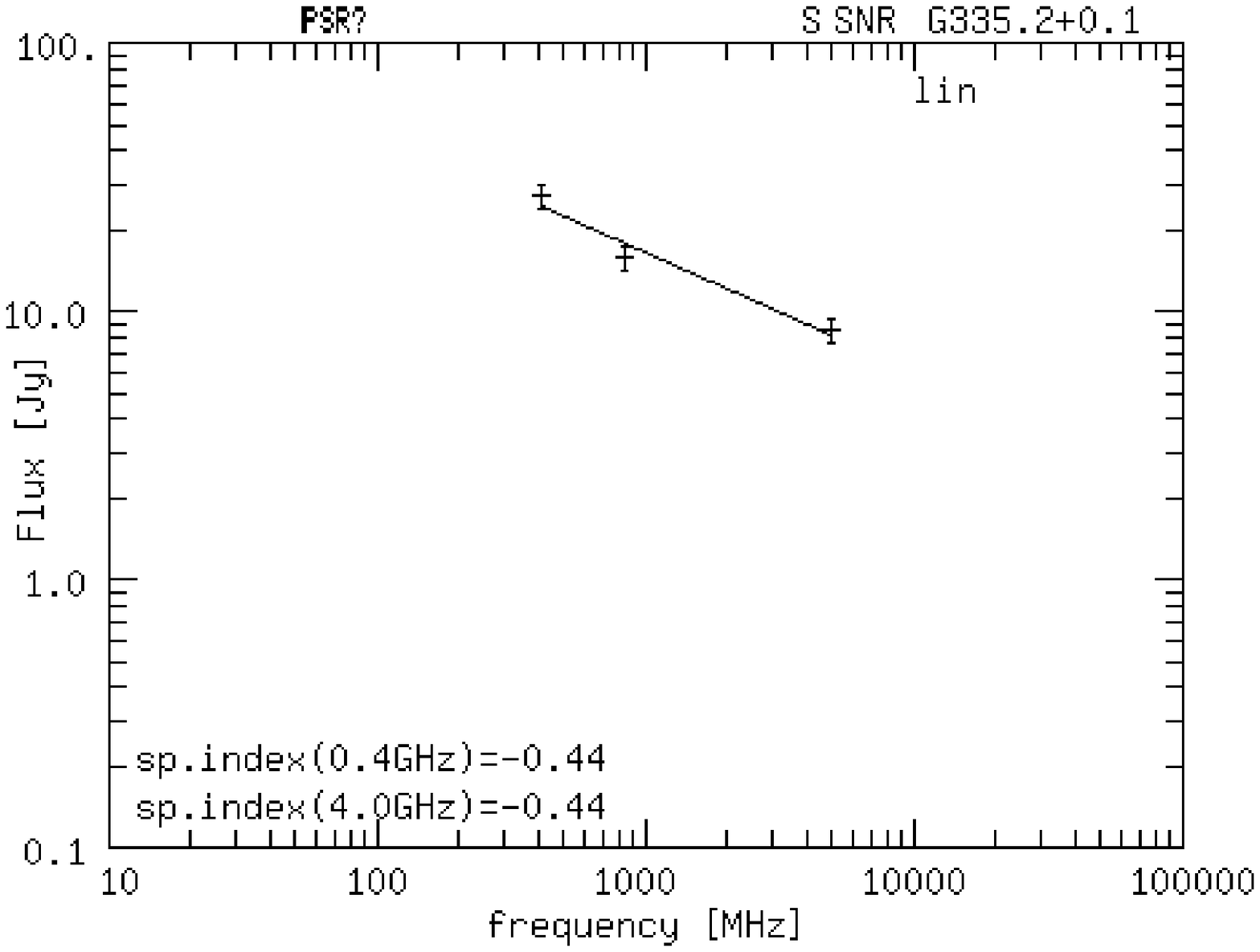,width=7.4cm,angle=0}}}\end{figure}
\begin{figure}\centerline{\vbox{\psfig{figure=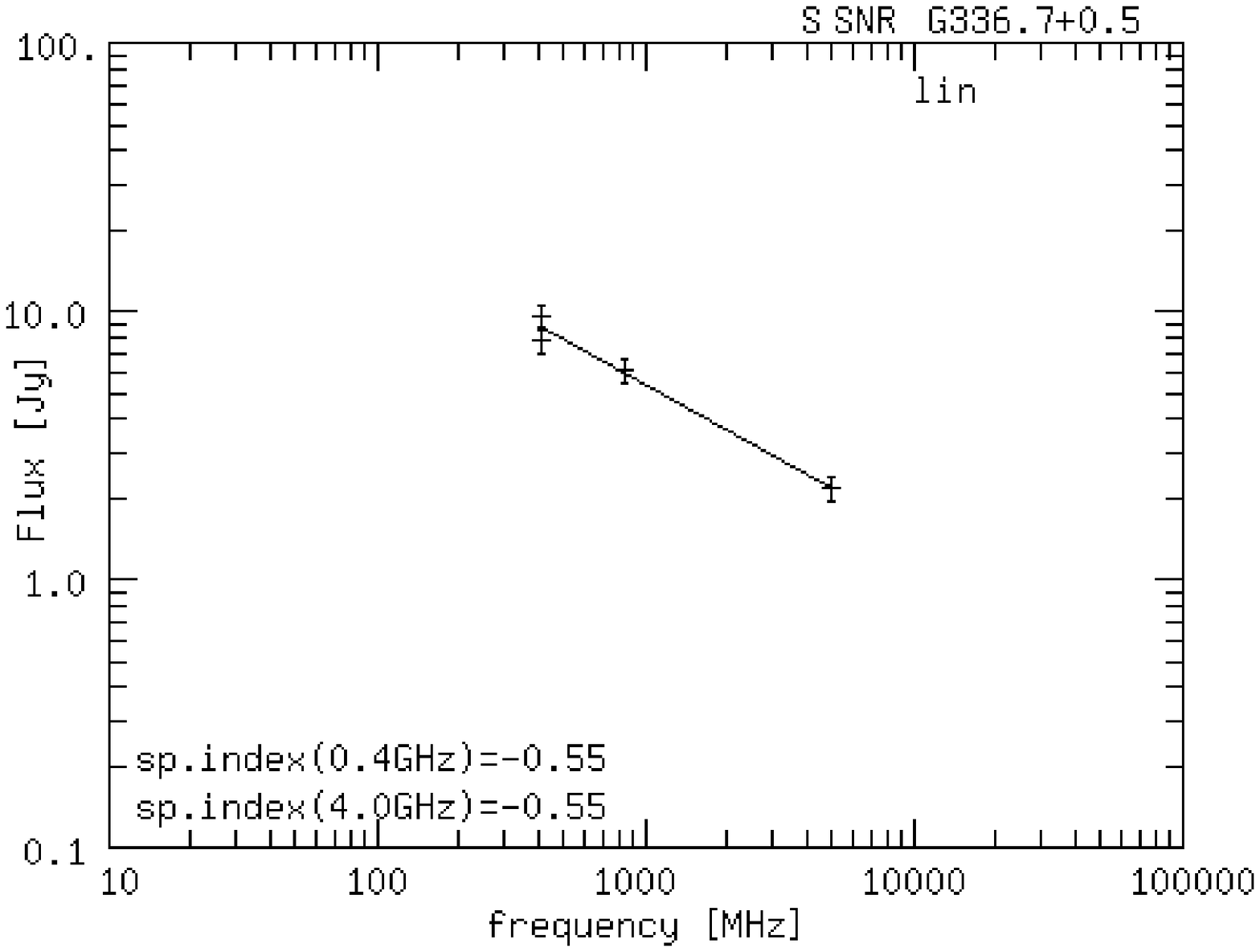,width=7.4cm,angle=0}}}\end{figure}\clearpage
\begin{figure}\centerline{\vbox{\psfig{figure=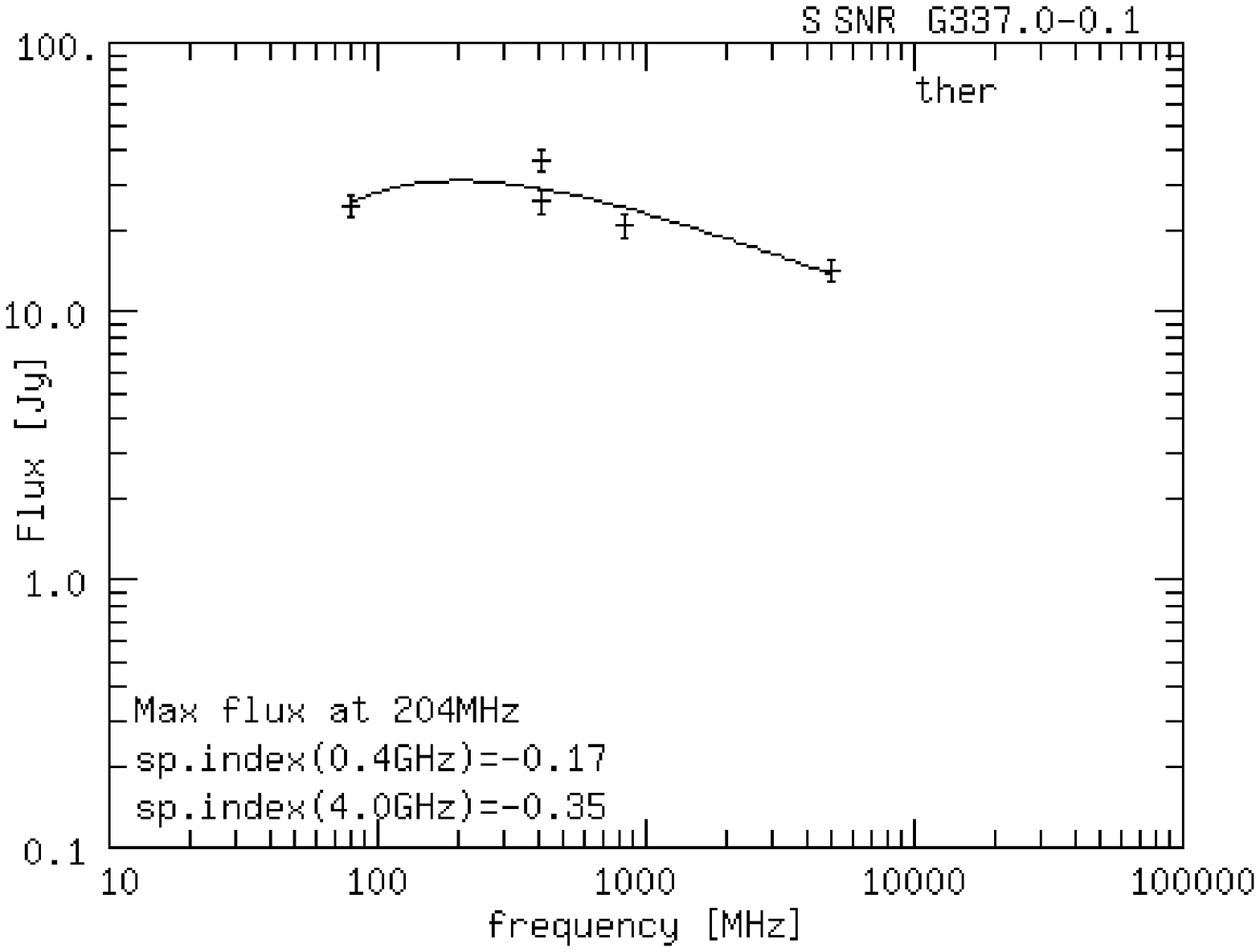,width=7.4cm,angle=0}}}\end{figure}
\begin{figure}\centerline{\vbox{\psfig{figure=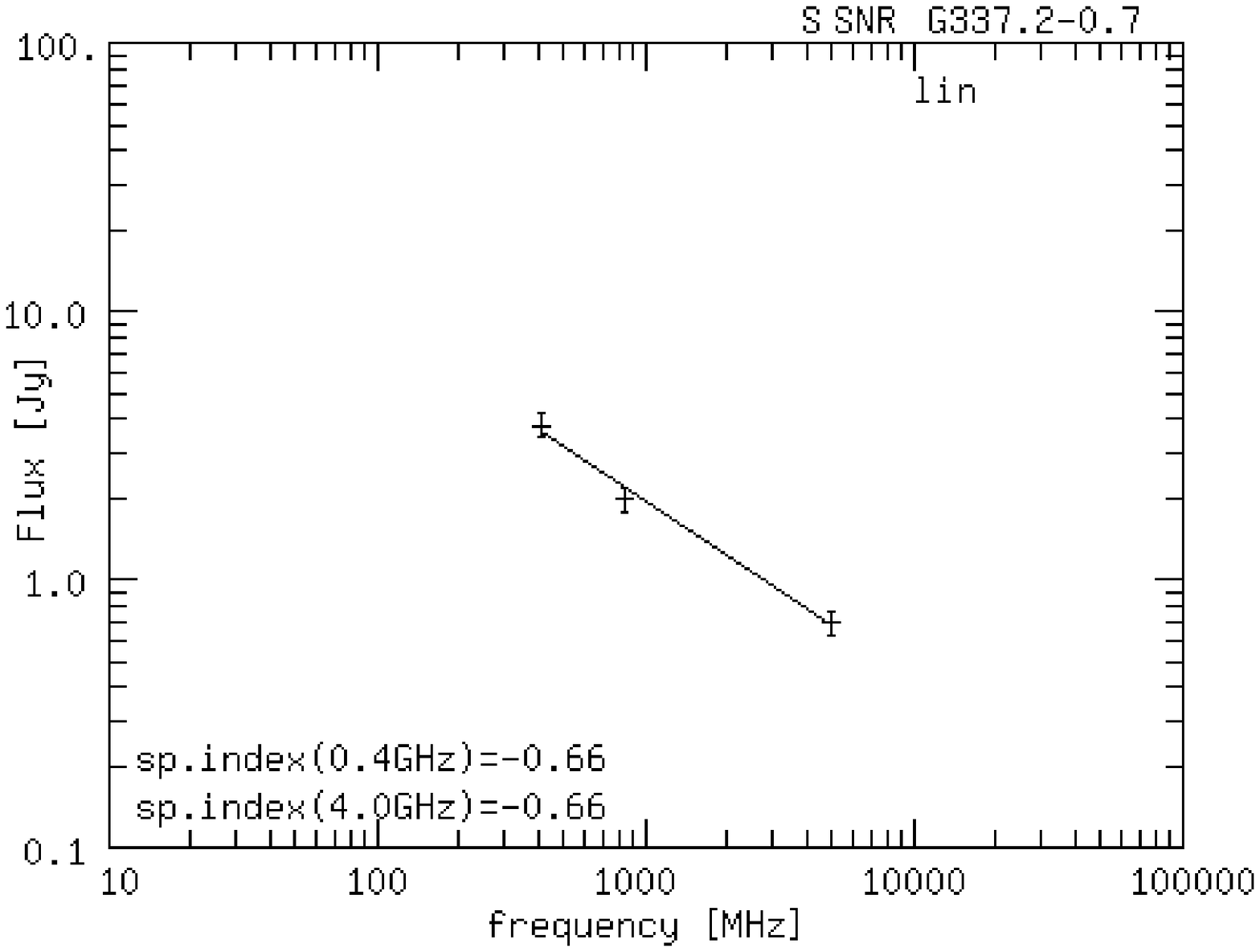,width=7.4cm,angle=0}}}\end{figure}
\begin{figure}\centerline{\vbox{\psfig{figure=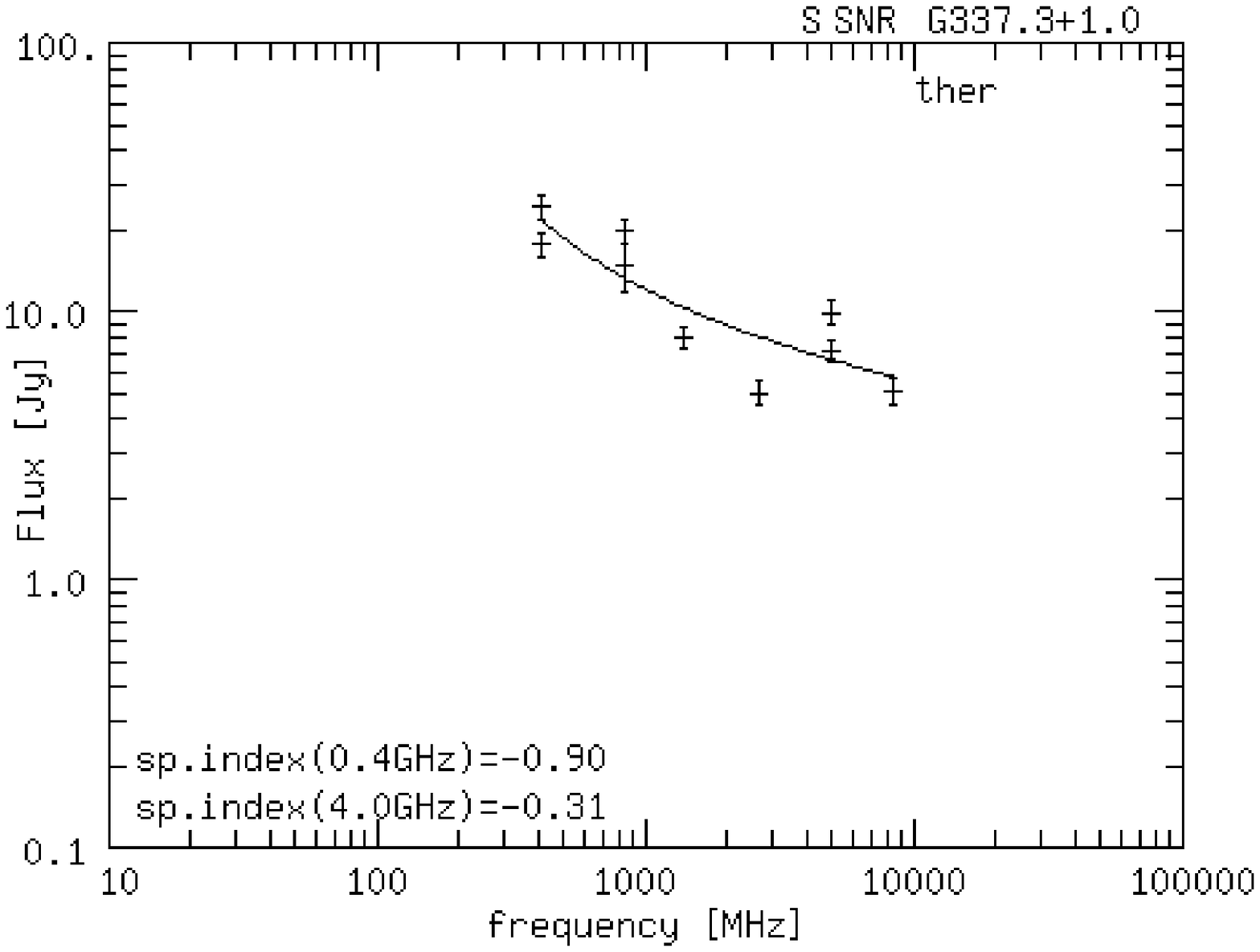,width=7.4cm,angle=0}}}\end{figure}
\begin{figure}\centerline{\vbox{\psfig{figure=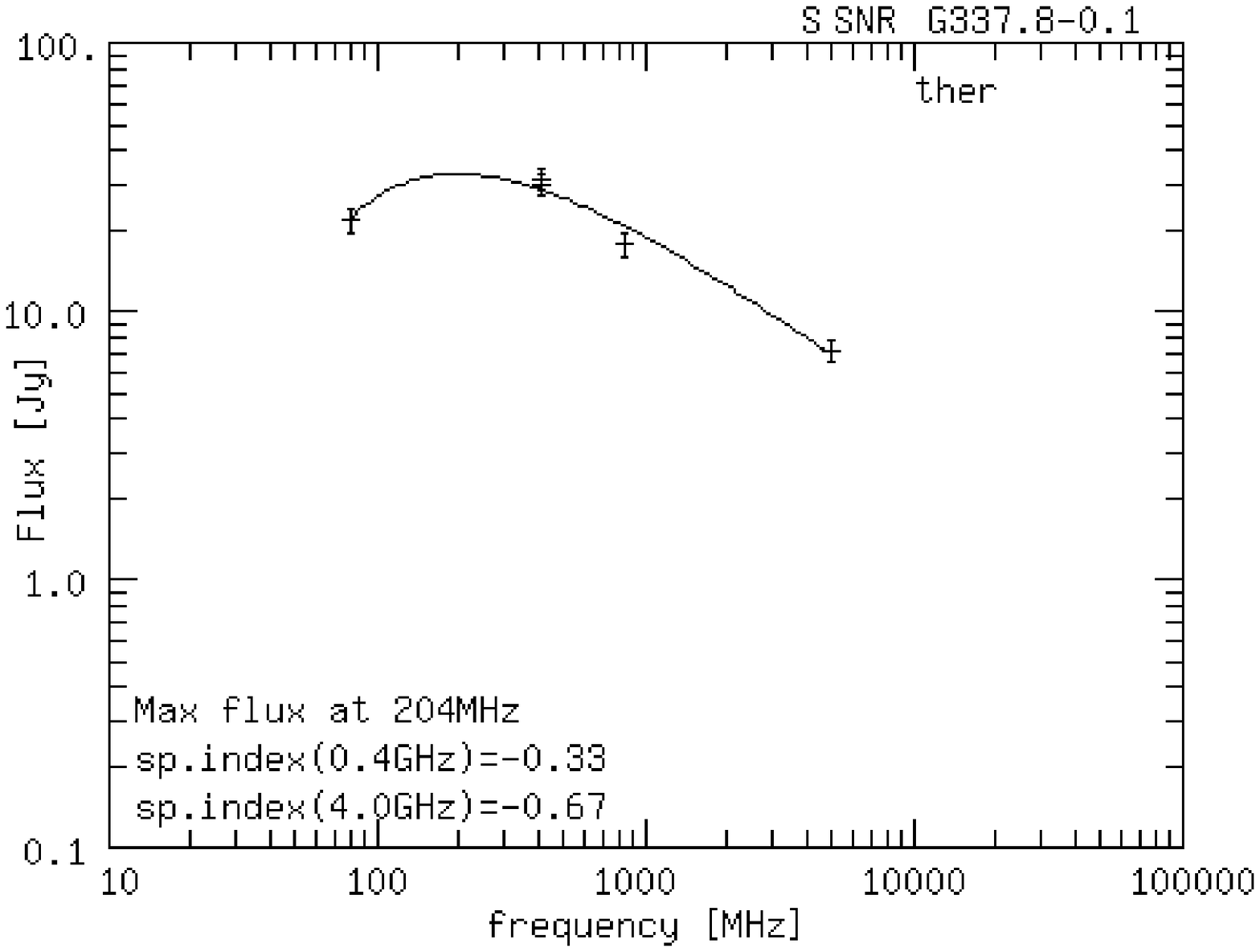,width=7.4cm,angle=0}}}\end{figure}
\begin{figure}\centerline{\vbox{\psfig{figure=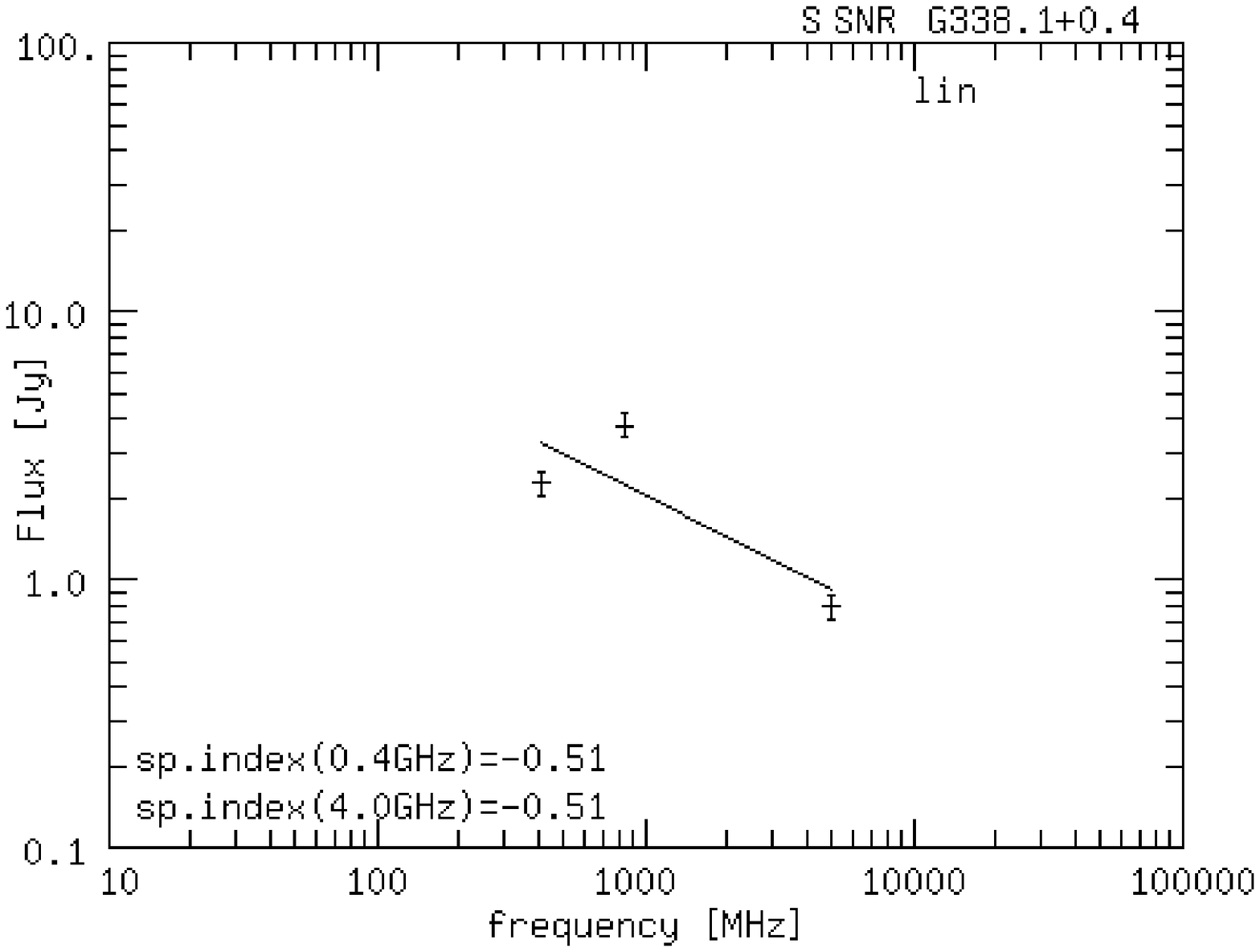,width=7.4cm,angle=0}}}\end{figure}
\begin{figure}\centerline{\vbox{\psfig{figure=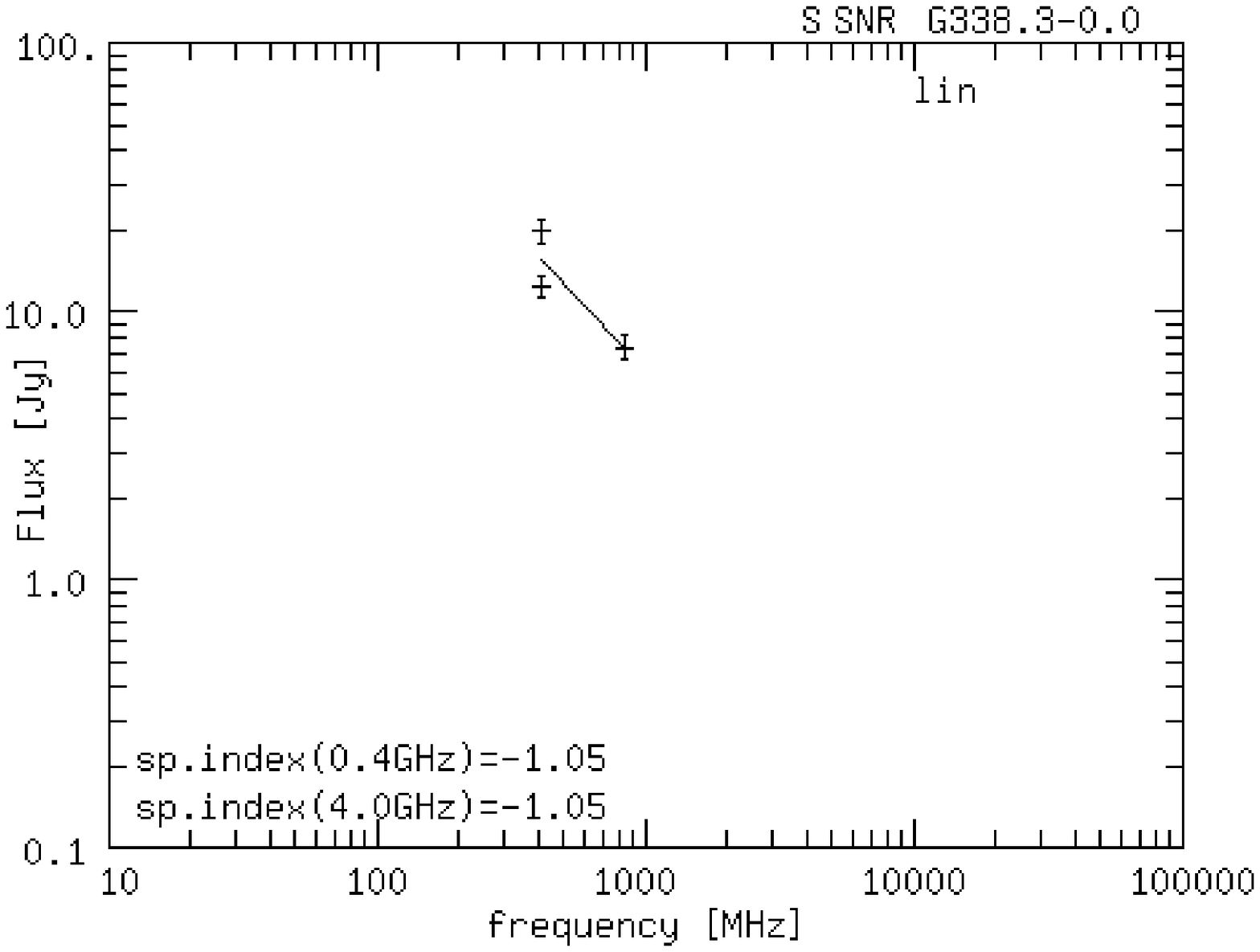,width=7.4cm,angle=0}}}\end{figure}
\begin{figure}\centerline{\vbox{\psfig{figure=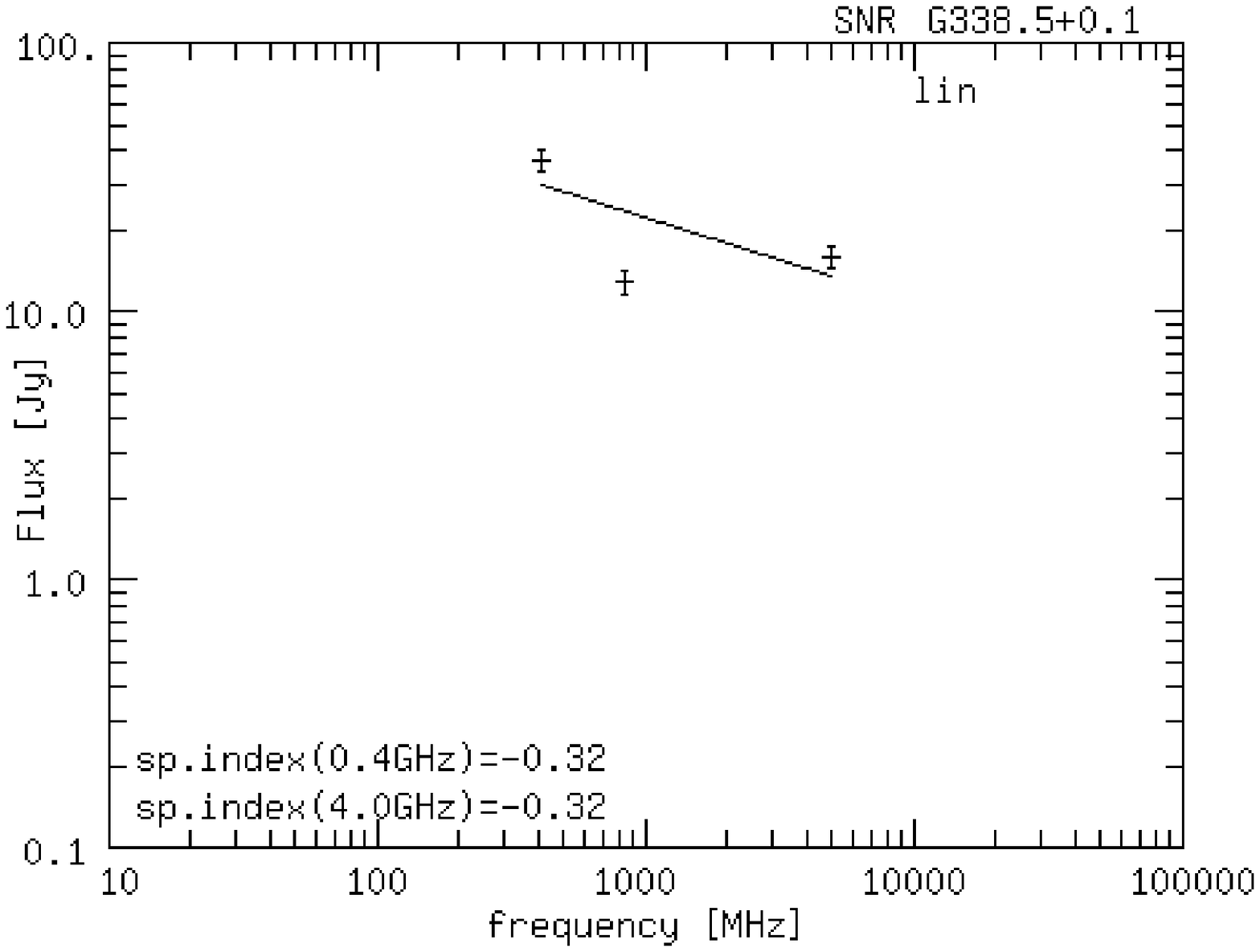,width=7.4cm,angle=0}}}\end{figure}
\begin{figure}\centerline{\vbox{\psfig{figure=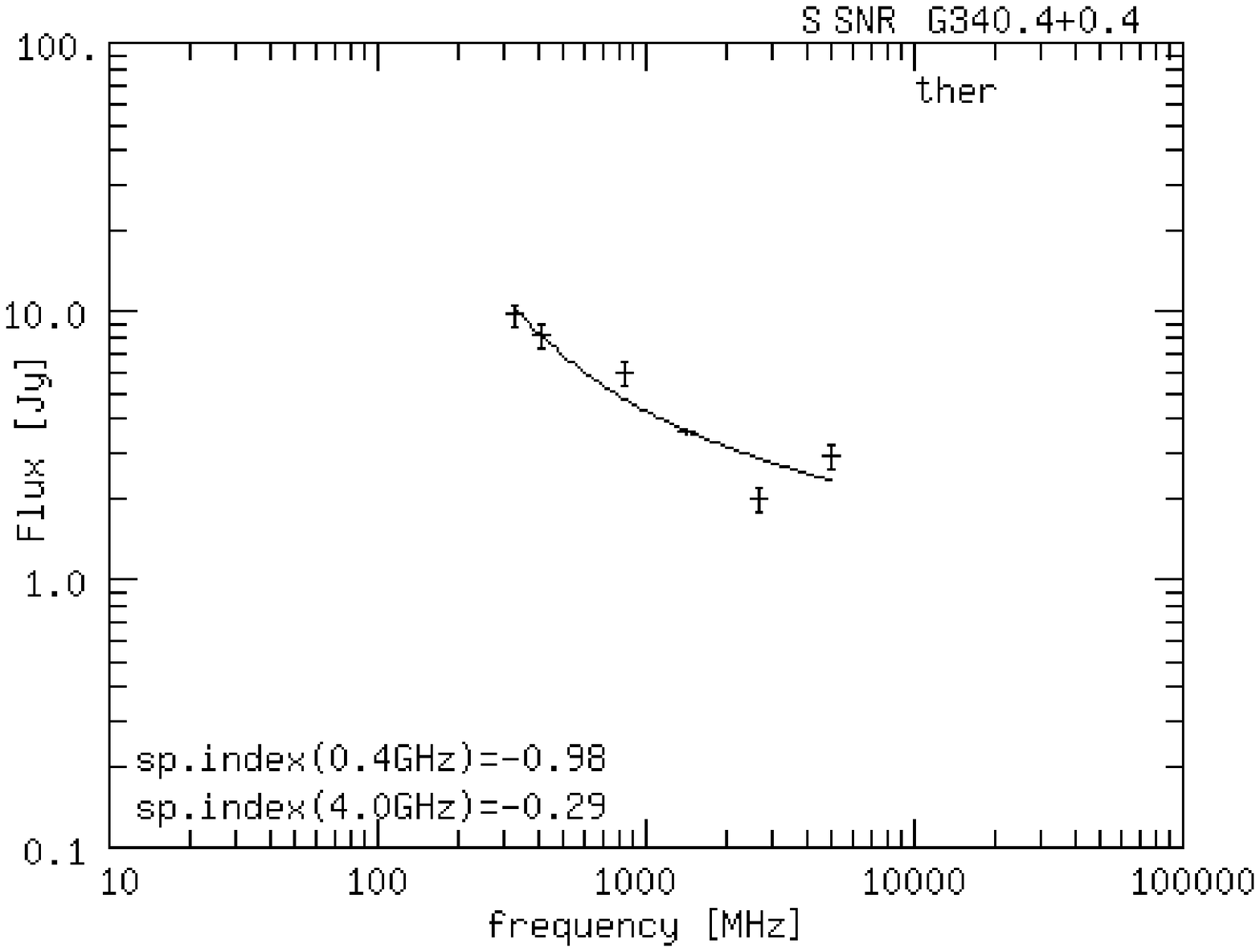,width=7.4cm,angle=0}}}\end{figure}\clearpage
\begin{figure}\centerline{\vbox{\psfig{figure=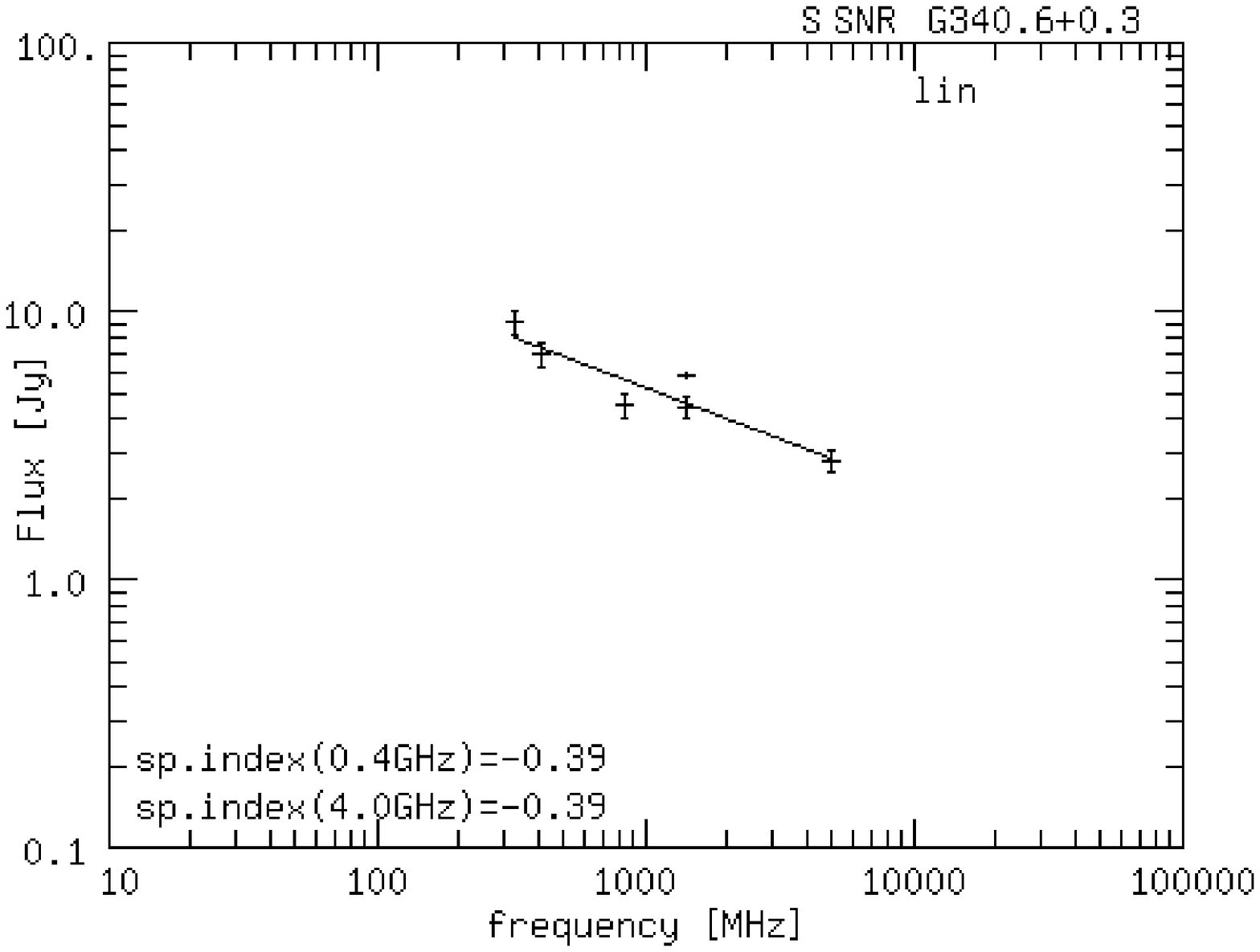,width=7.4cm,angle=0}}}\end{figure}
\begin{figure}\centerline{\vbox{\psfig{figure=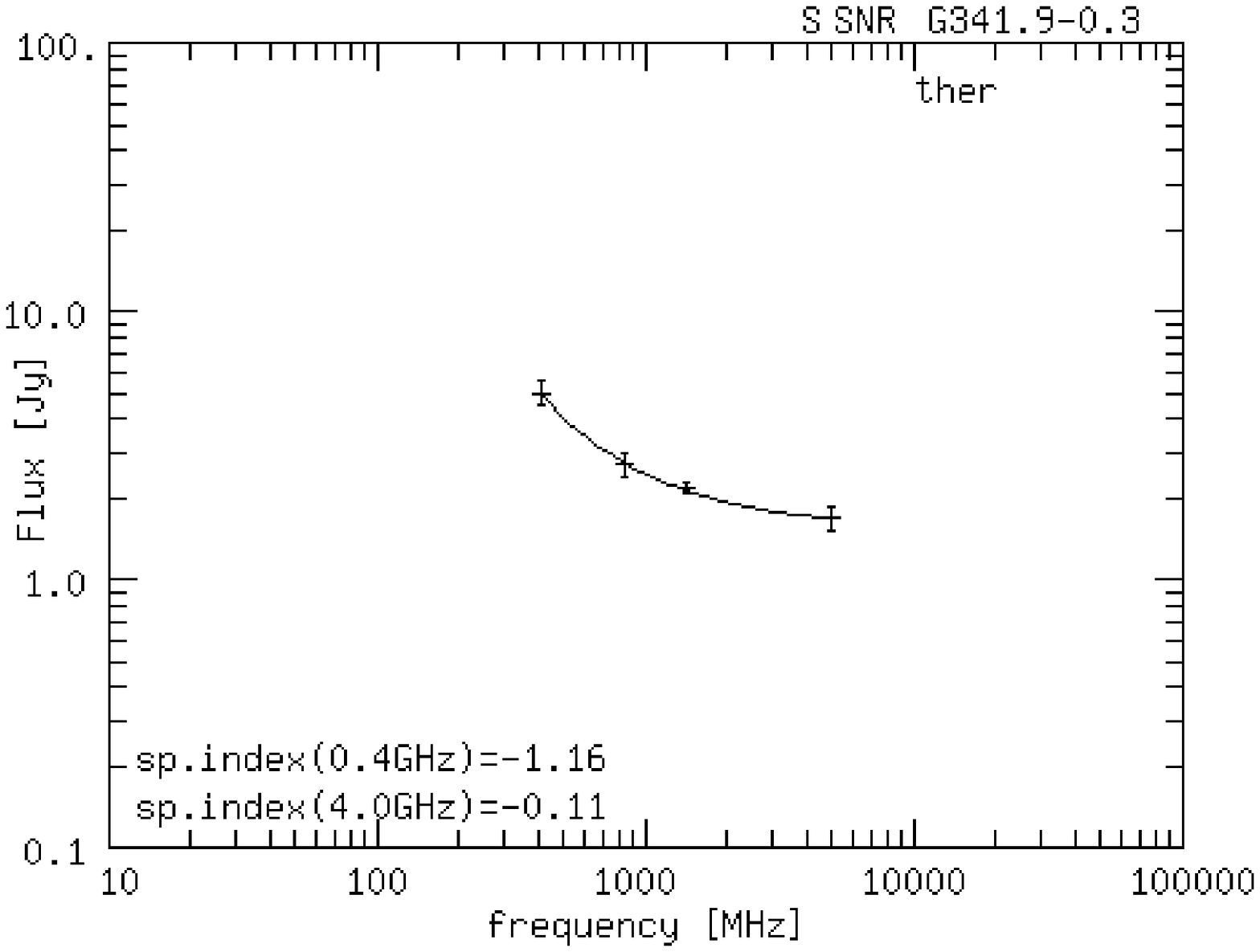,width=7.4cm,angle=0}}}\end{figure}
\begin{figure}\centerline{\vbox{\psfig{figure=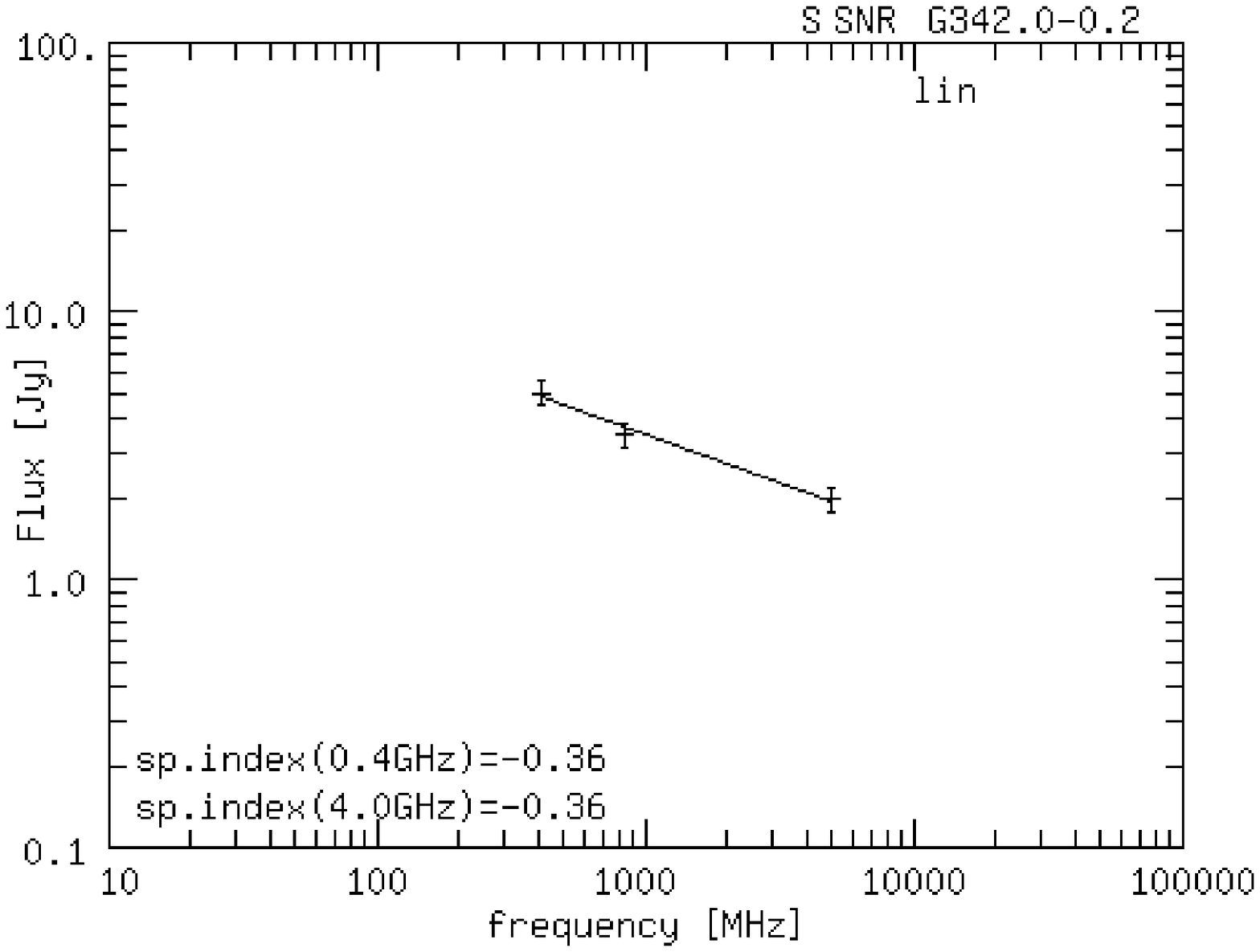,width=7.4cm,angle=0}}}\end{figure}
\begin{figure}\centerline{\vbox{\psfig{figure=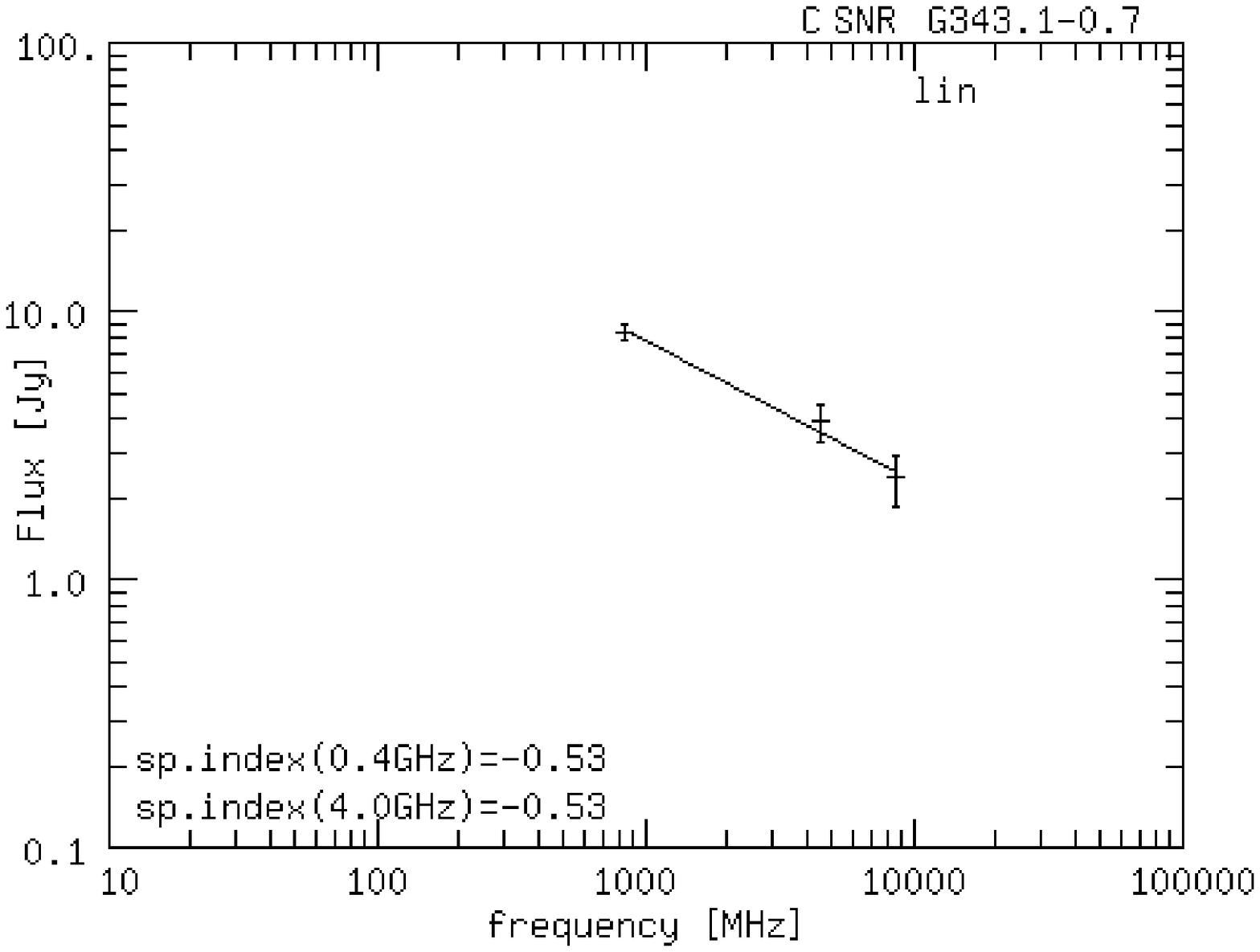,width=7.4cm,angle=0}}}\end{figure}
\begin{figure}\centerline{\vbox{\psfig{figure=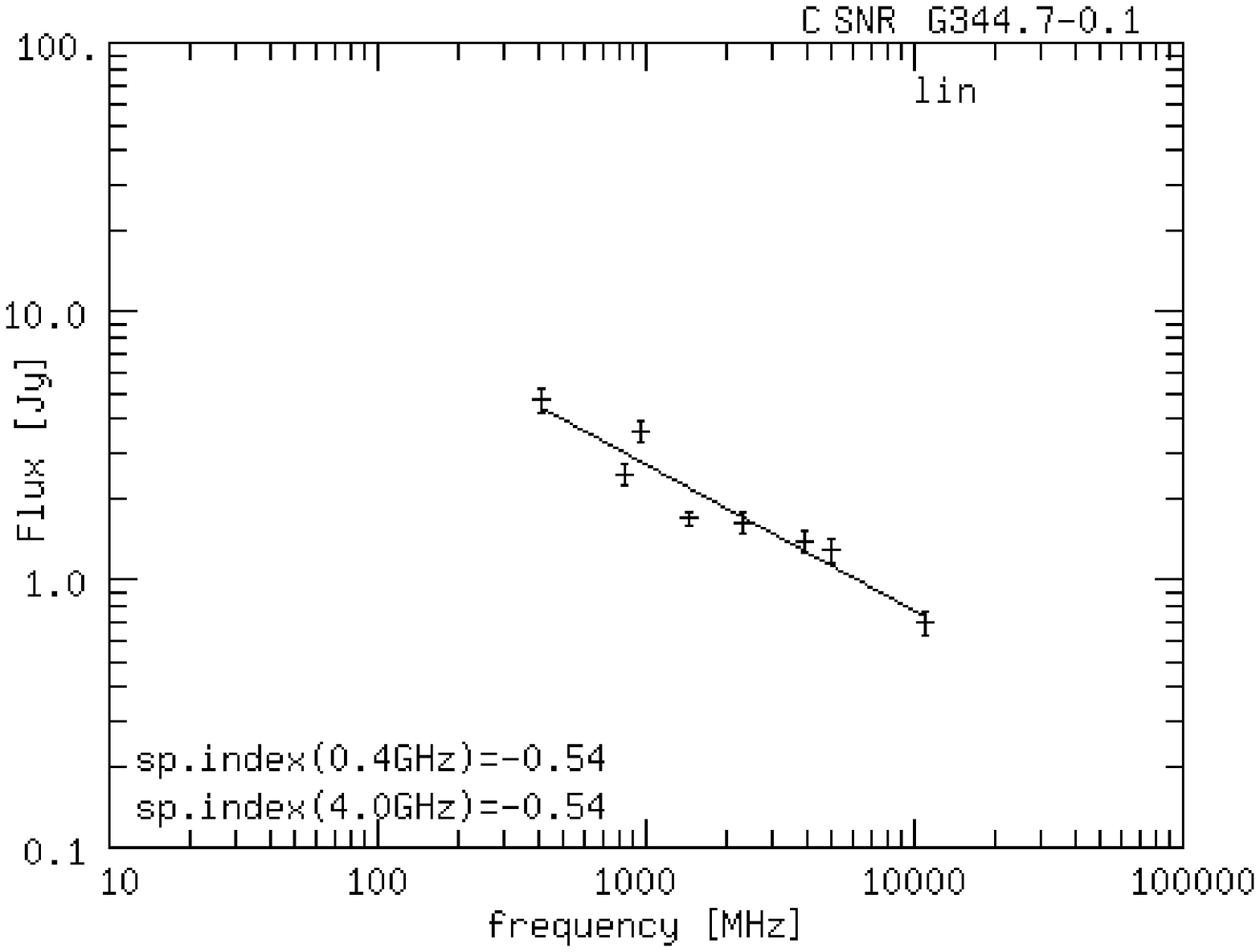,width=7.4cm,angle=0}}}\end{figure}
\begin{figure}\centerline{\vbox{\psfig{figure=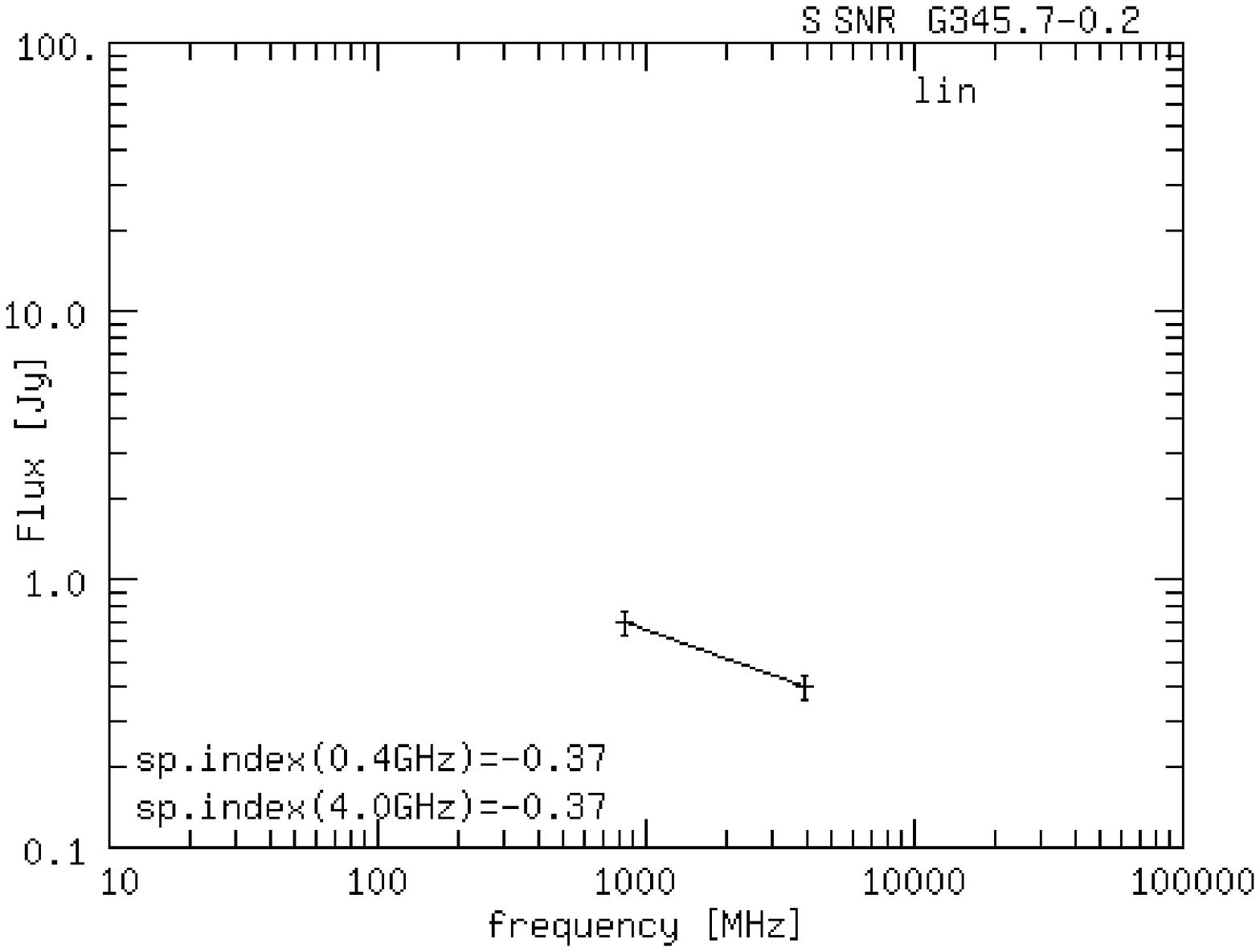,width=7.4cm,angle=0}}}\end{figure}
\begin{figure}\centerline{\vbox{\psfig{figure=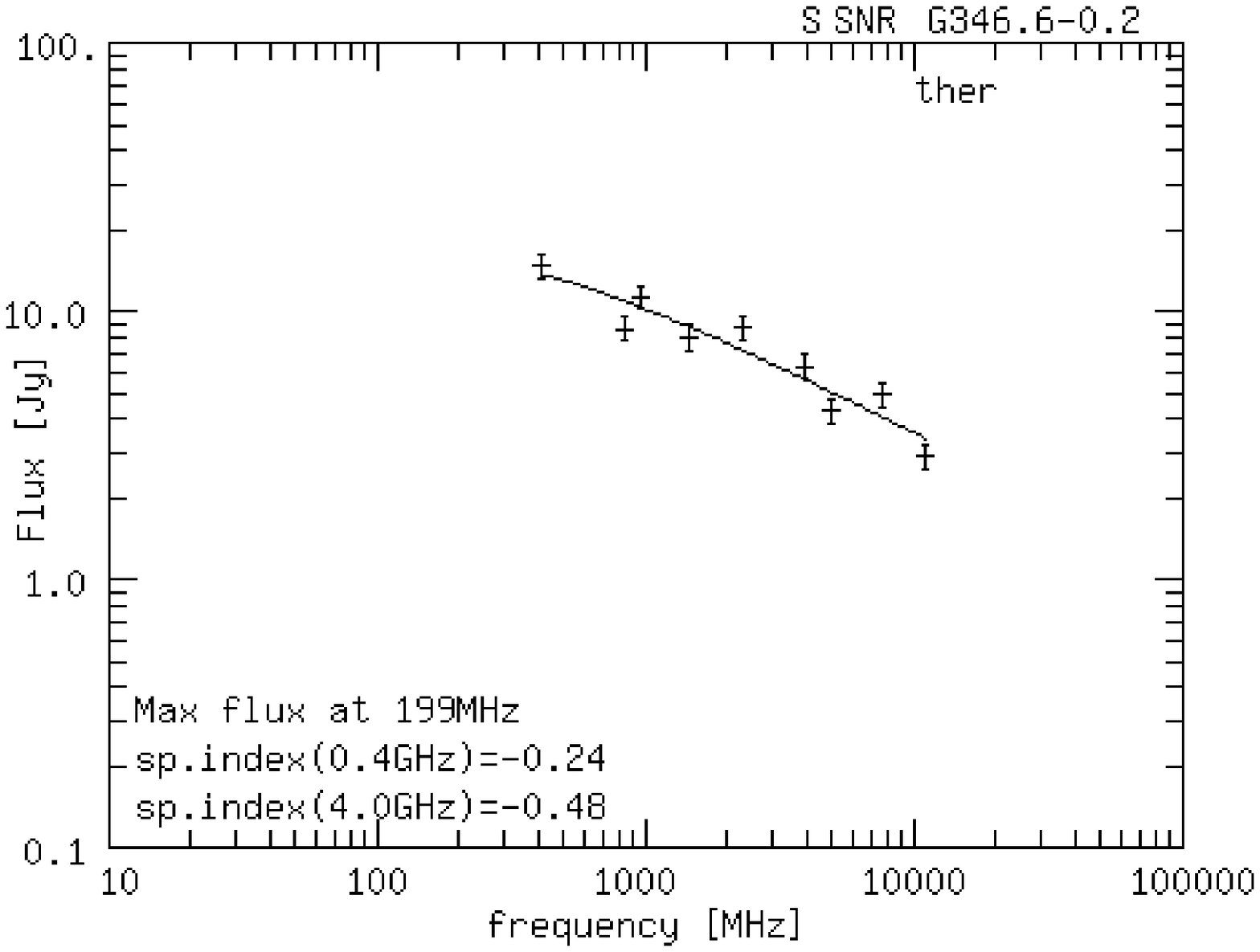,width=7.4cm,angle=0}}}\end{figure}
\begin{figure}\centerline{\vbox{\psfig{figure=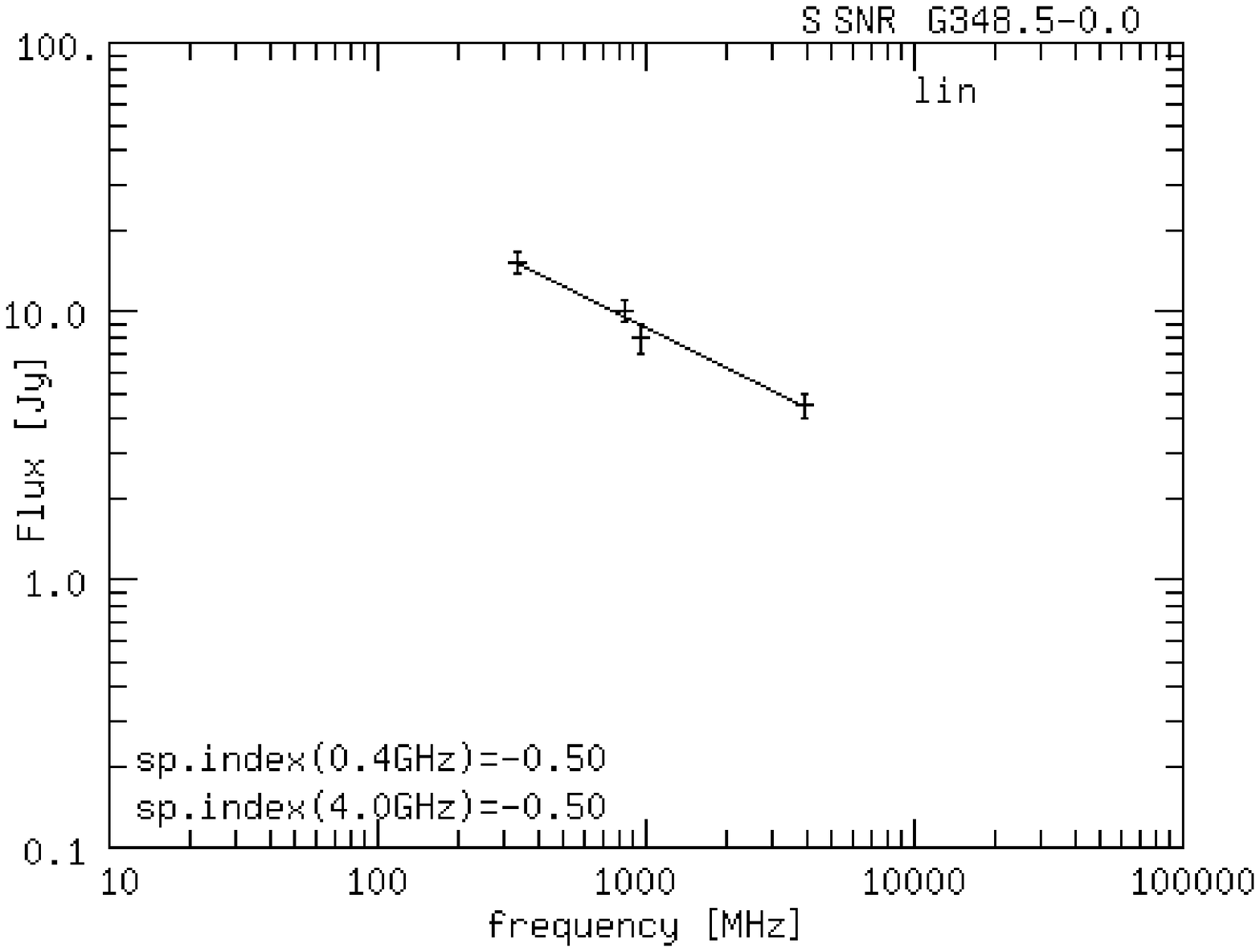,width=7.4cm,angle=0}}}\end{figure}\clearpage
\begin{figure}\centerline{\vbox{\psfig{figure=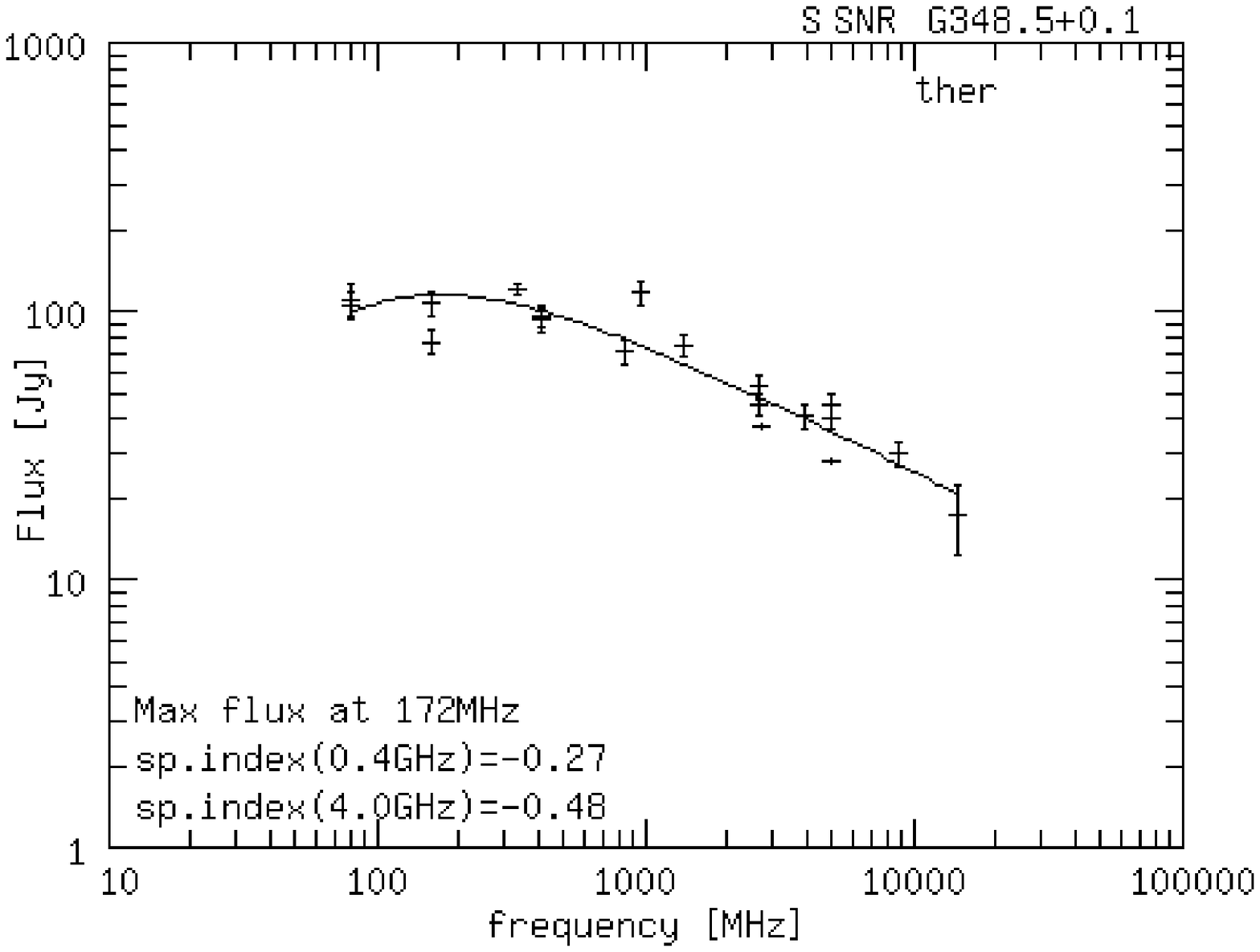,width=7.4cm,angle=0}}}\end{figure}
\begin{figure}\centerline{\vbox{\psfig{figure=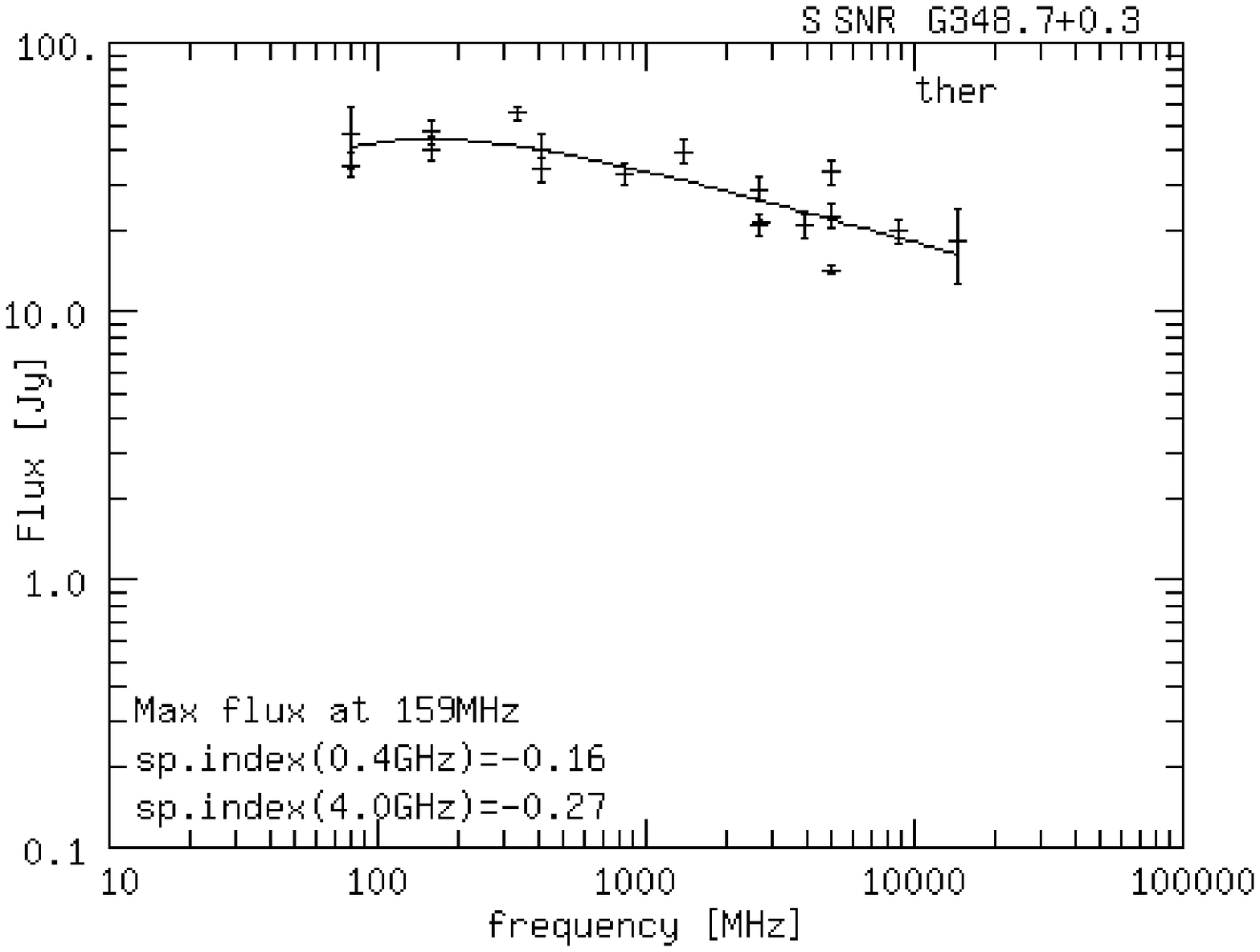,width=7.4cm,angle=0}}}\end{figure}
\begin{figure}\centerline{\vbox{\psfig{figure=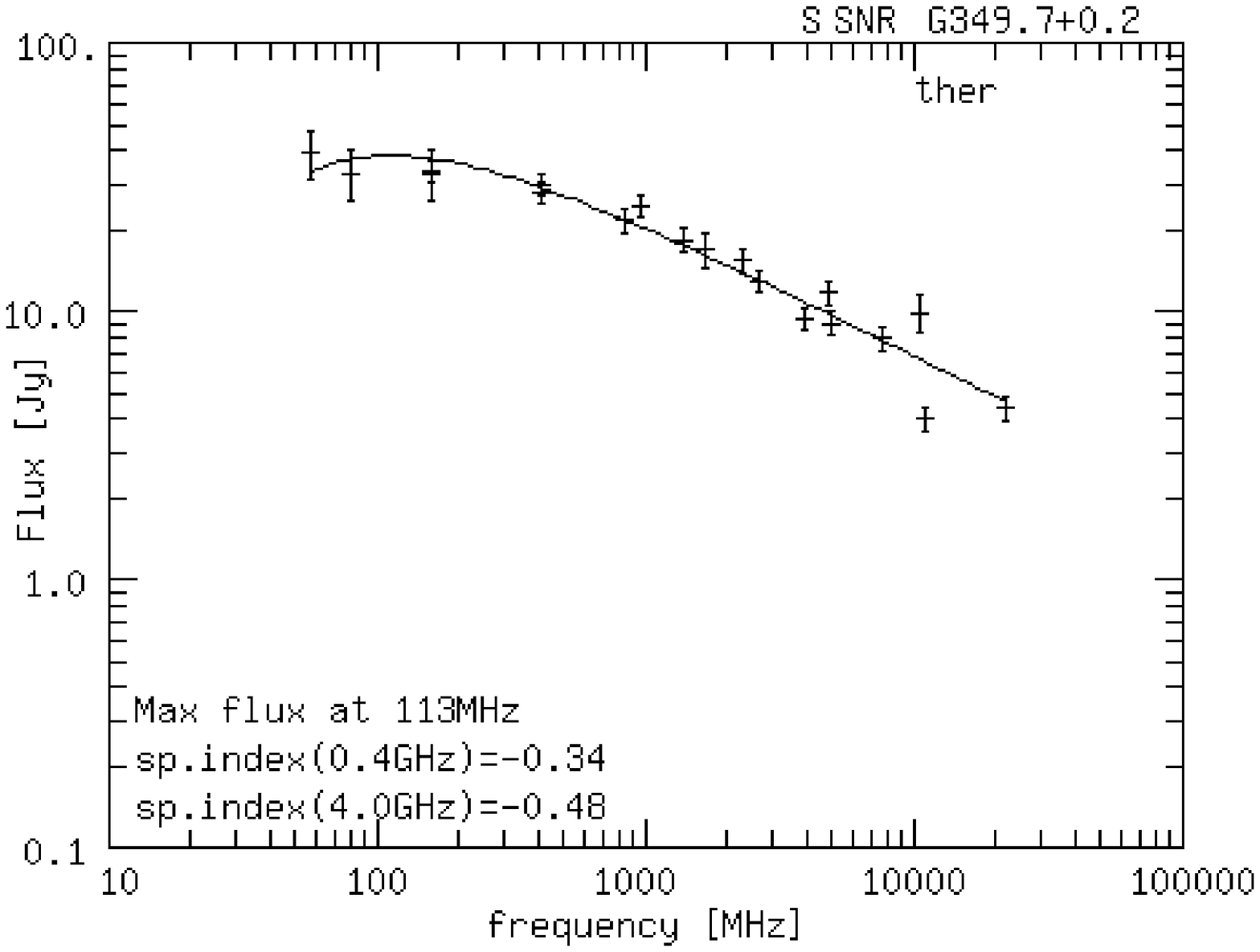,width=7.4cm,angle=0}}}\end{figure}
\begin{figure}\centerline{\vbox{\psfig{figure=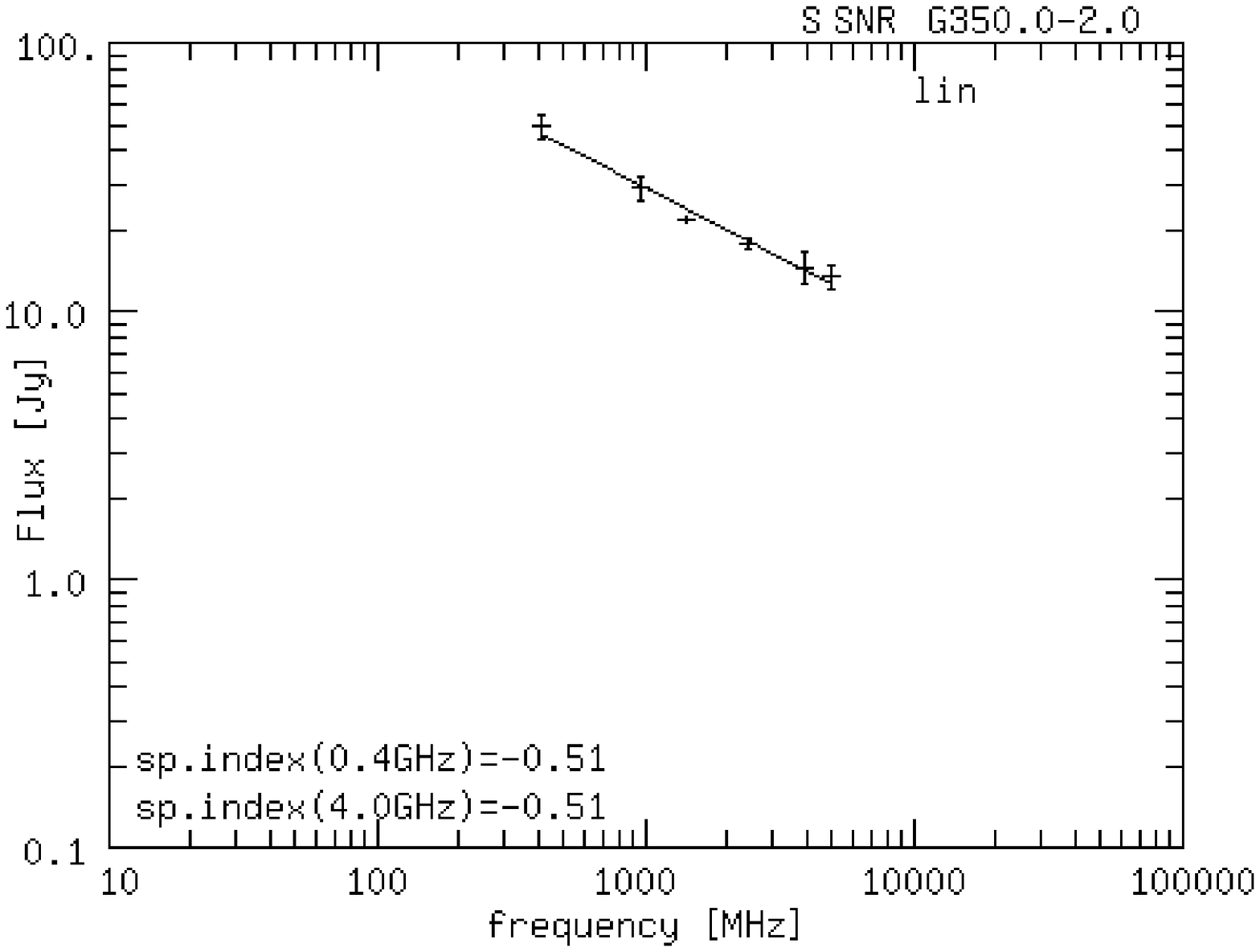,width=7.4cm,angle=0}}}\end{figure}
\begin{figure}\centerline{\vbox{\psfig{figure=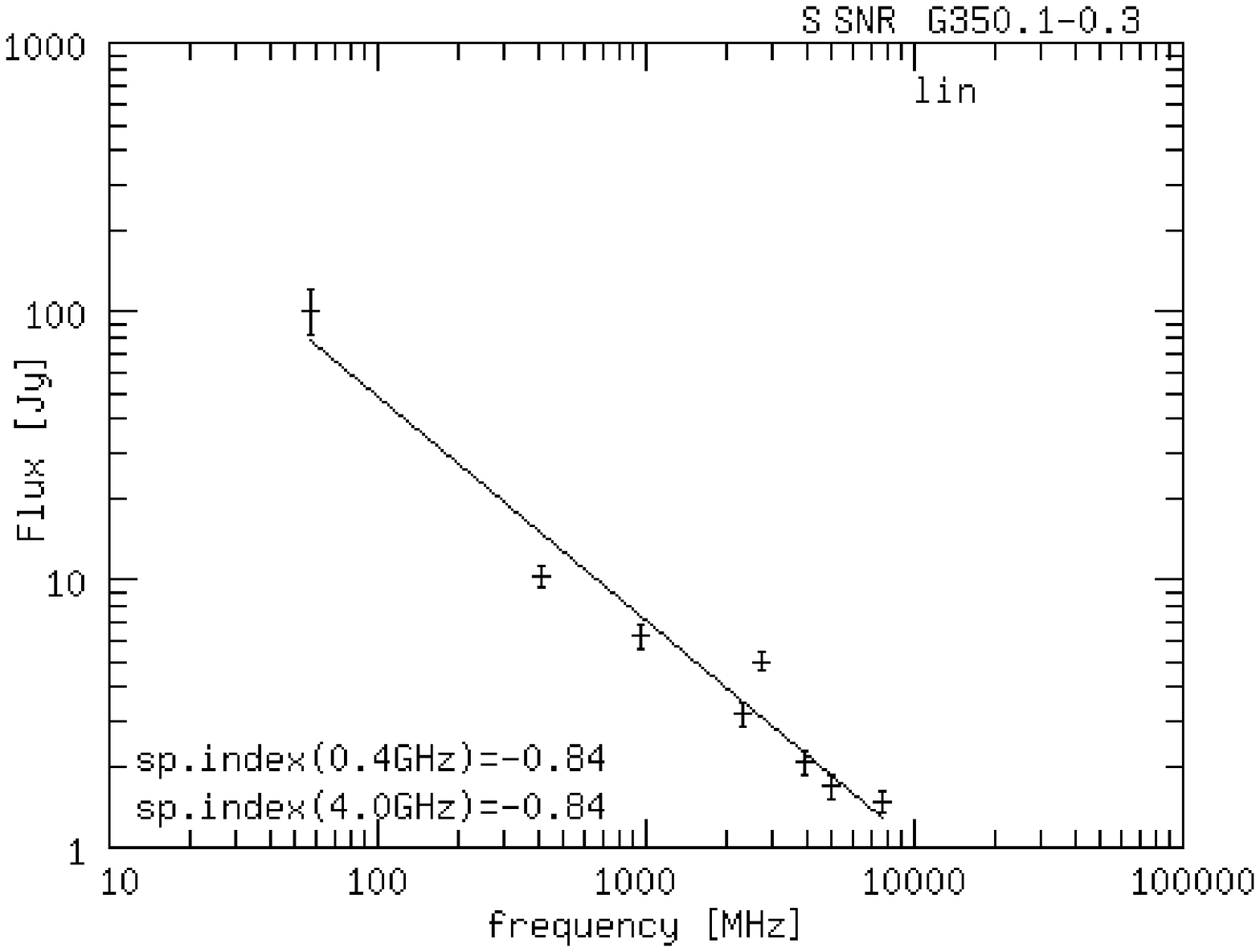,width=7.4cm,angle=0}}}\end{figure}
\begin{figure}\centerline{\vbox{\psfig{figure=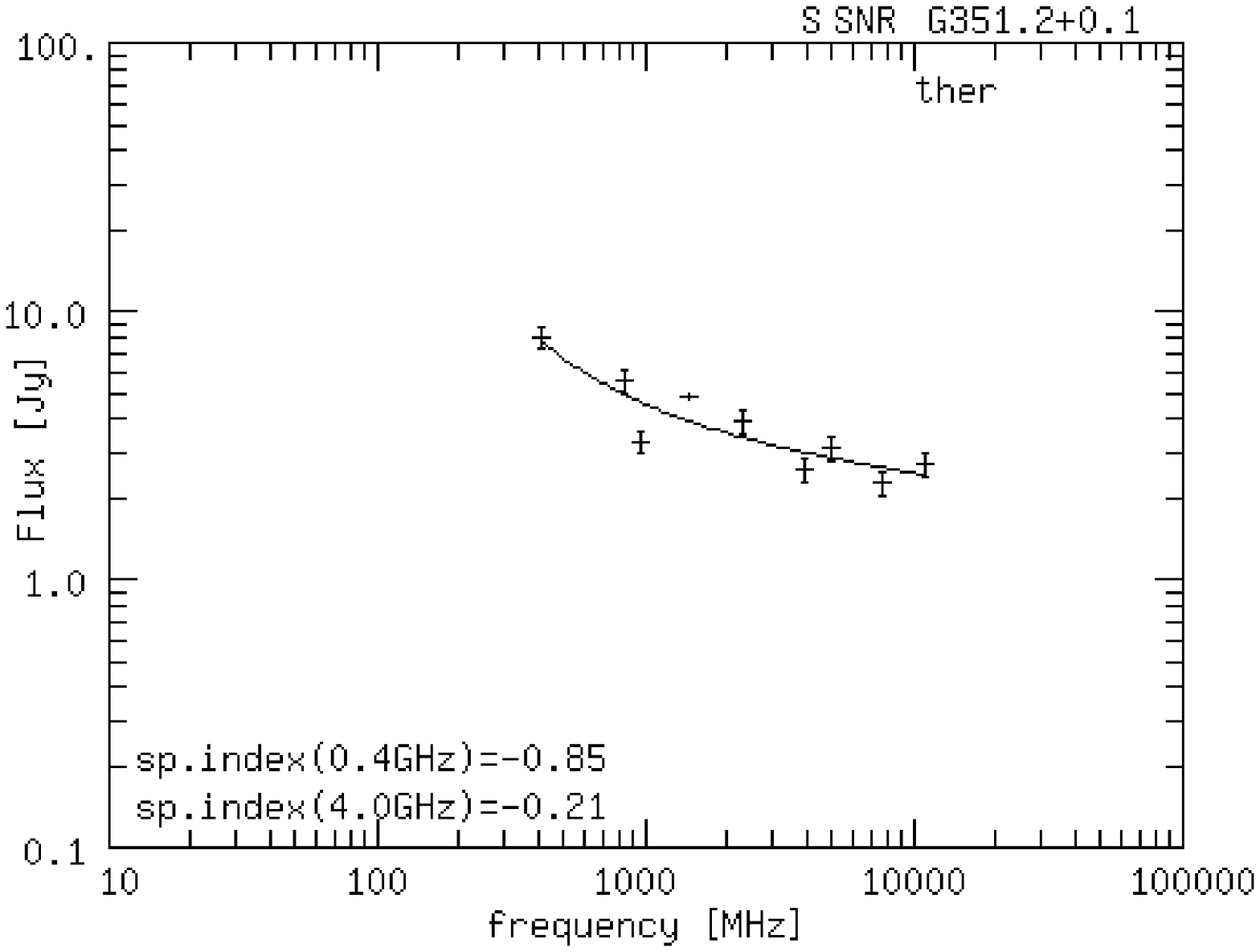,width=7.4cm,angle=0}}}\end{figure}
\begin{figure}\centerline{\vbox{\psfig{figure=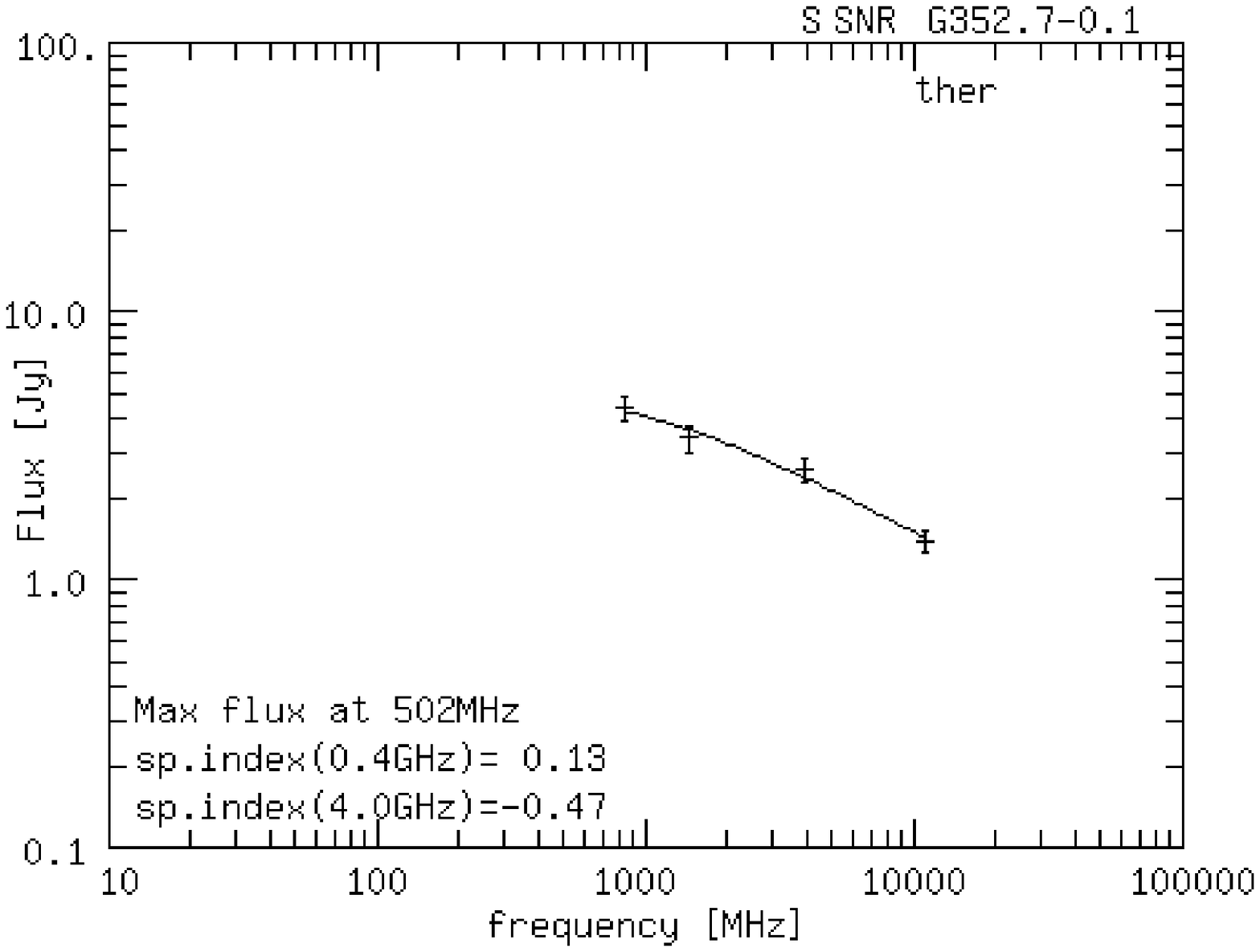,width=7.4cm,angle=0}}}\end{figure}
\begin{figure}\centerline{\vbox{\psfig{figure=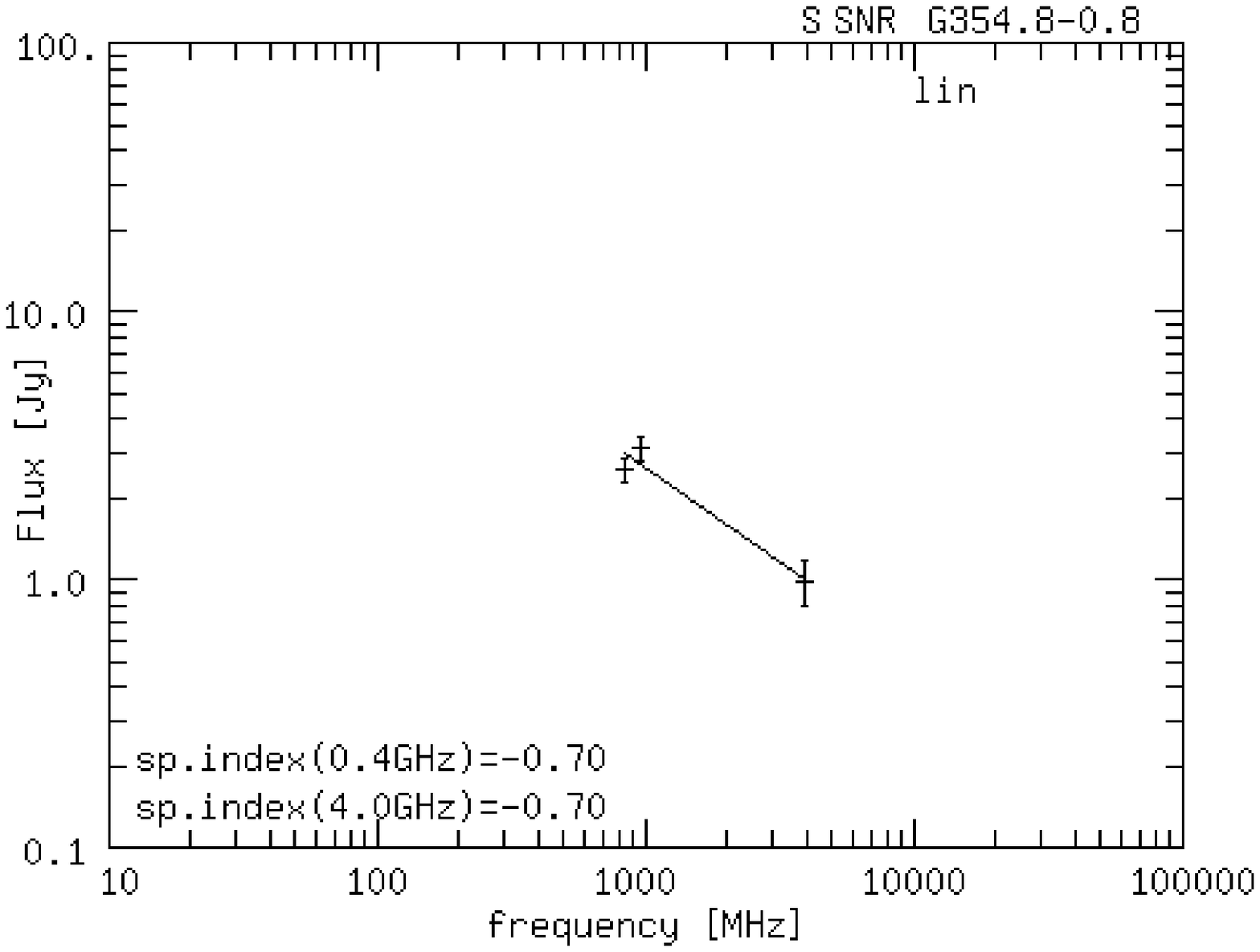,width=7.4cm,angle=0}}}\end{figure}\clearpage
\begin{figure}\centerline{\vbox{\psfig{figure=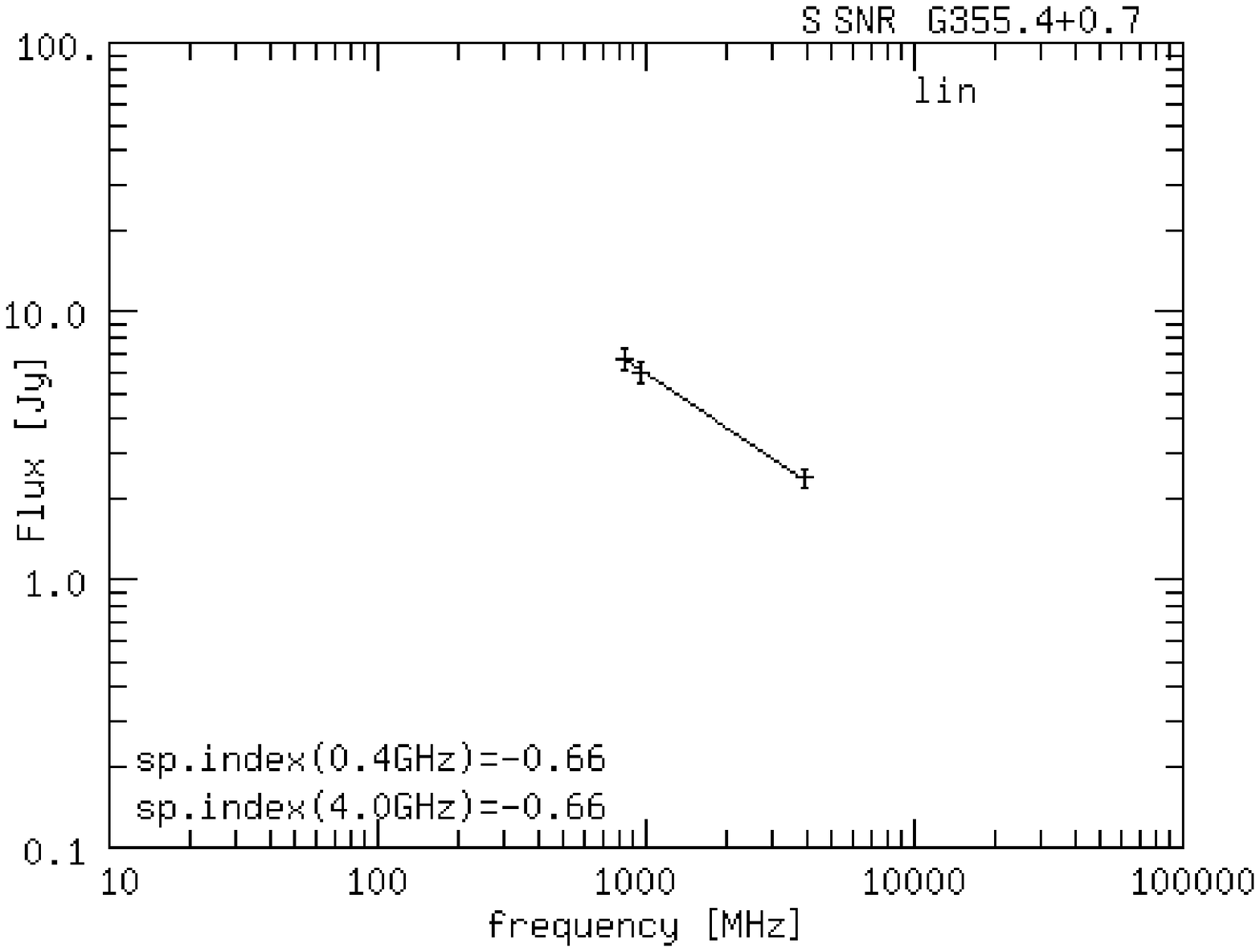,width=7.4cm,angle=0}}}\end{figure}
\begin{figure}\centerline{\vbox{\psfig{figure=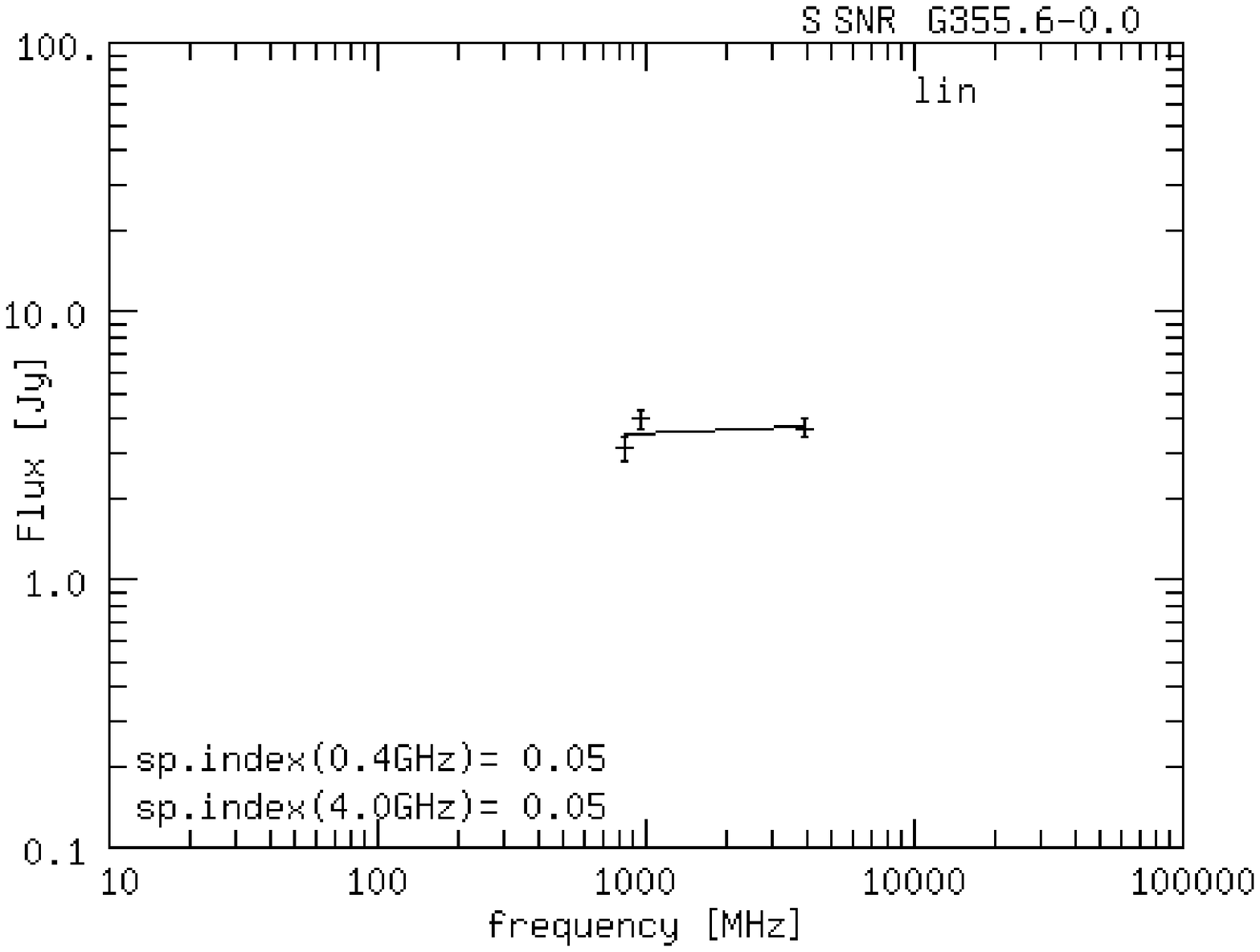,width=7.4cm,angle=0}}}\end{figure}
\begin{figure}\centerline{\vbox{\psfig{figure=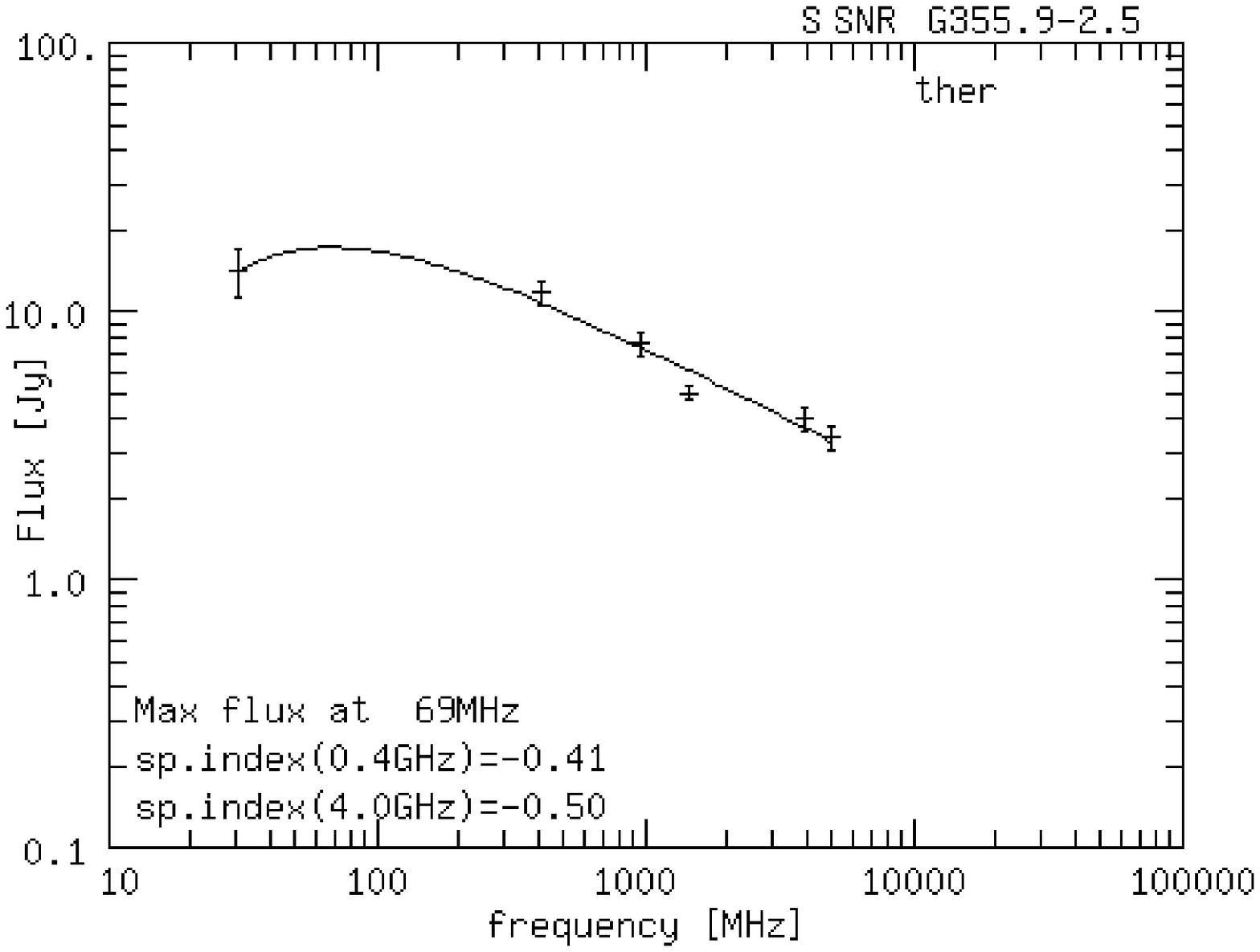,width=7.4cm,angle=0}}}\end{figure}
\begin{figure}\centerline{\vbox{\psfig{figure=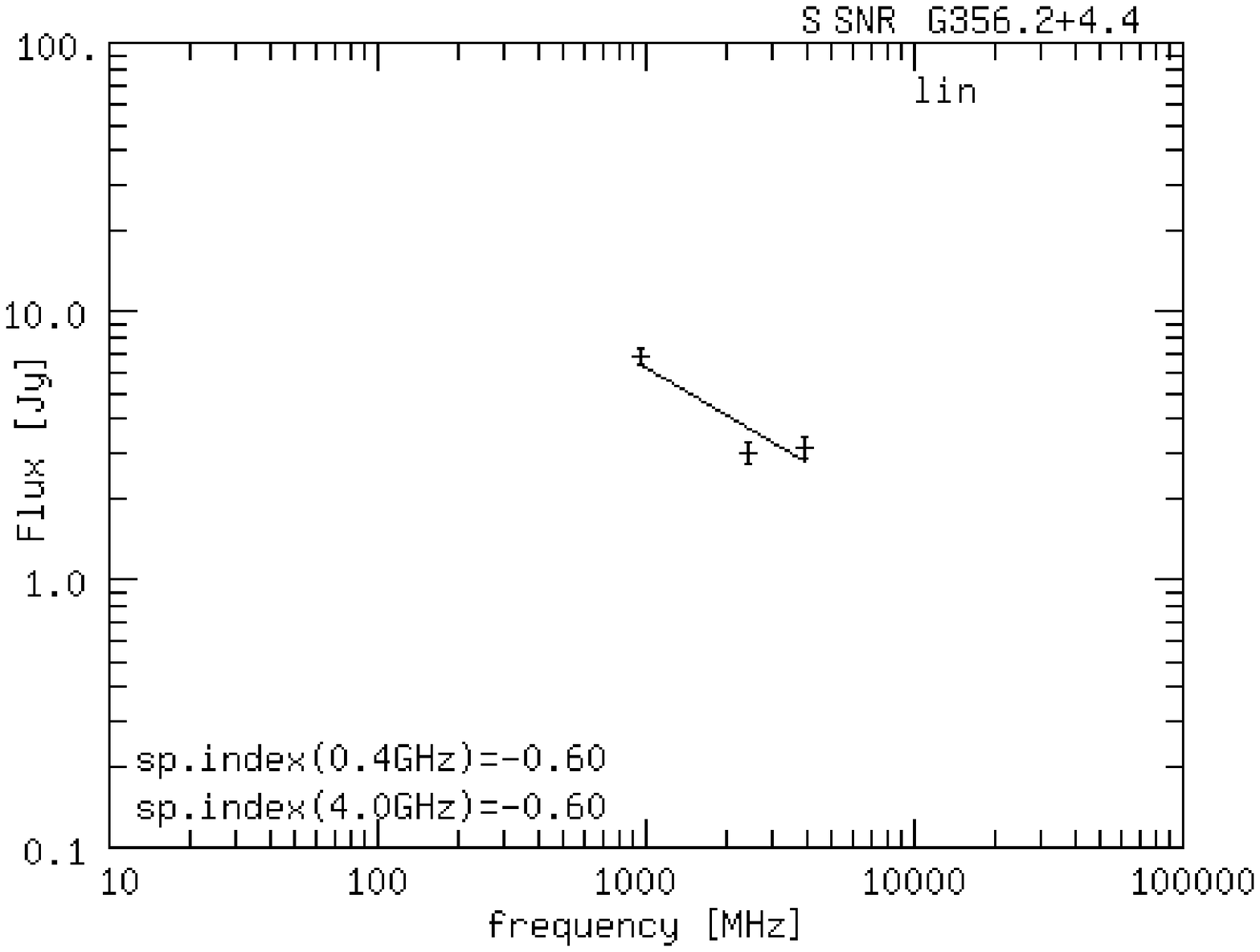,width=7.4cm,angle=0}}}\end{figure}
\begin{figure}\centerline{\vbox{\psfig{figure=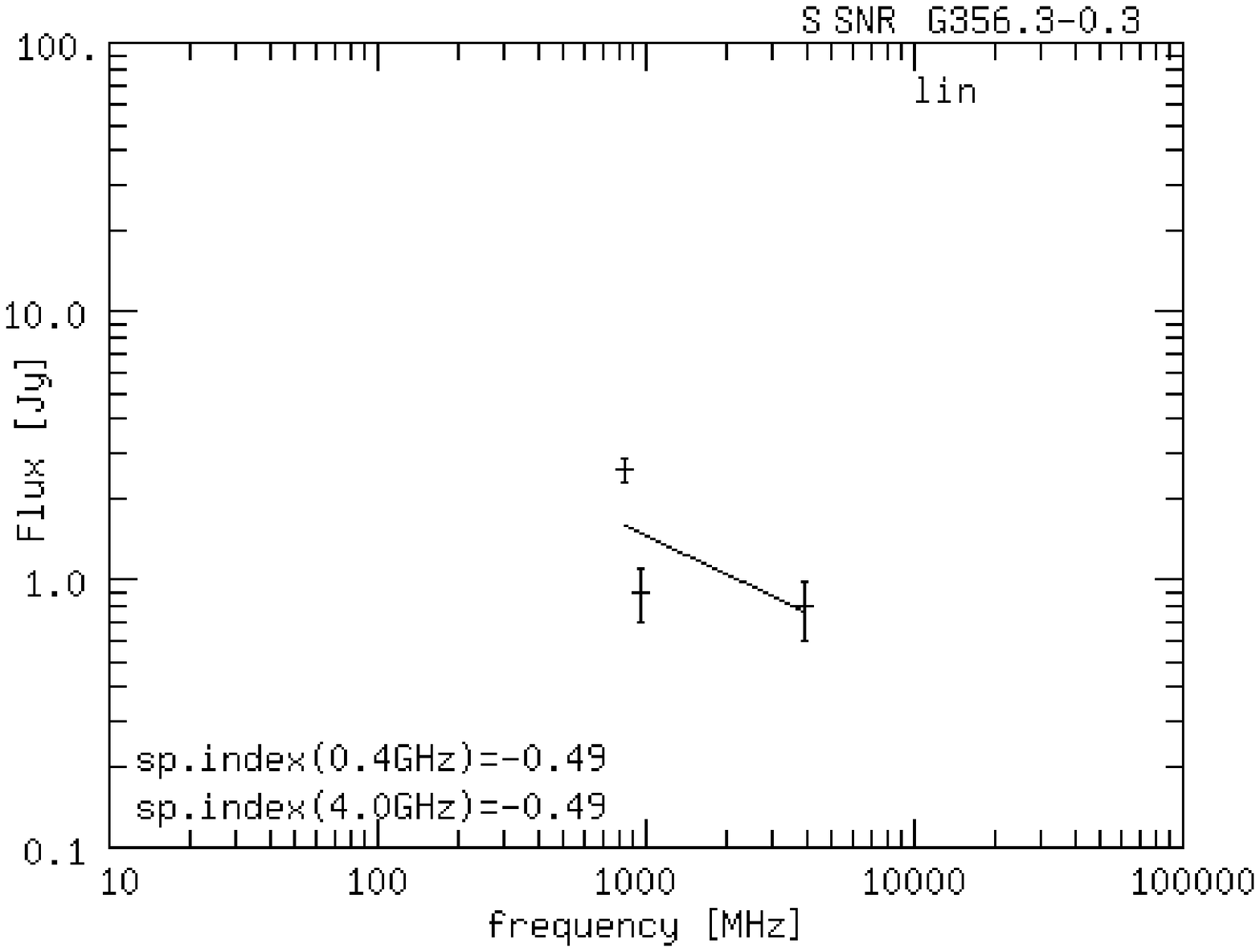,width=7.4cm,angle=0}}}\end{figure}
\begin{figure}\centerline{\vbox{\psfig{figure=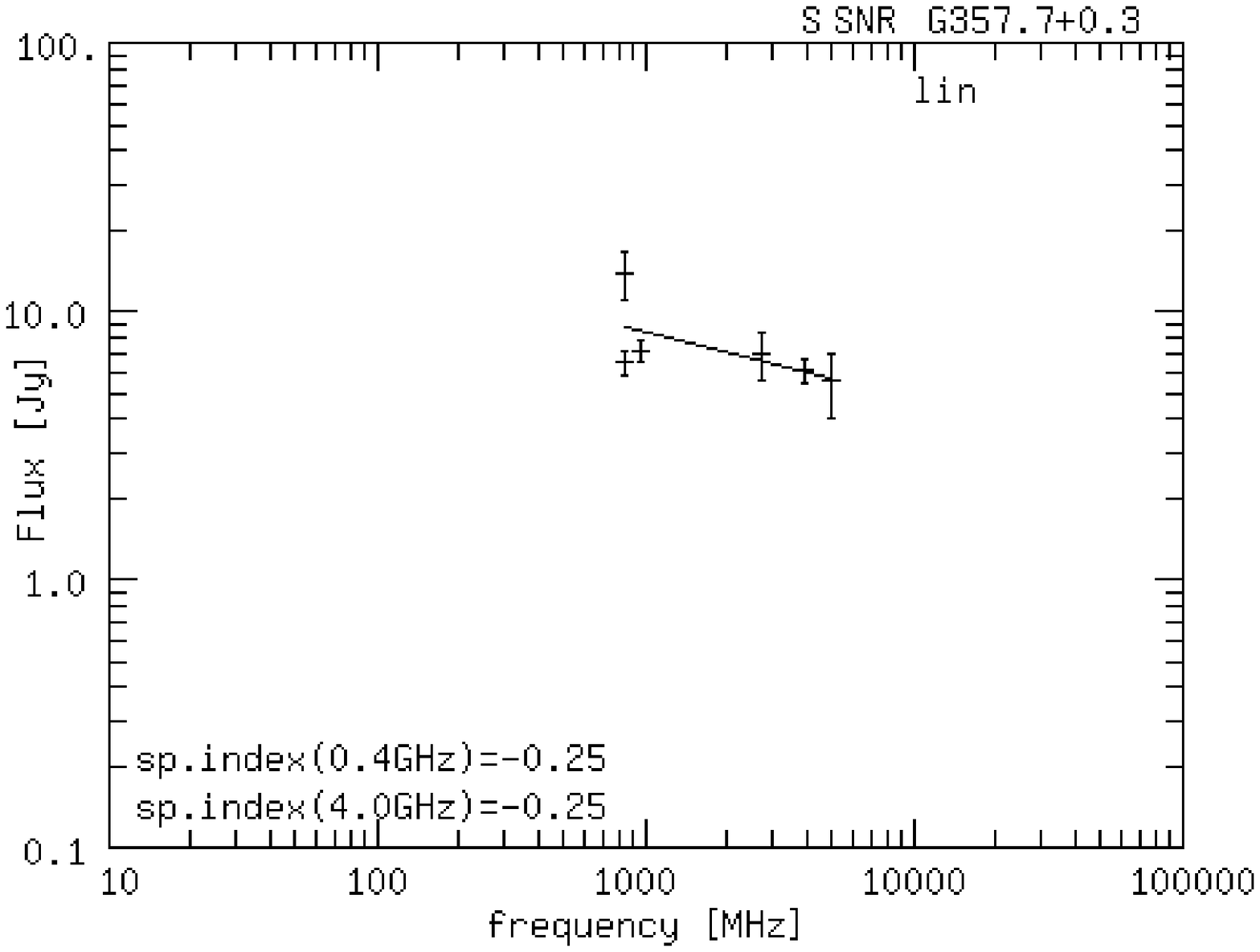,width=7.4cm,angle=0}}}\end{figure}
\begin{figure}\centerline{\vbox{\psfig{figure=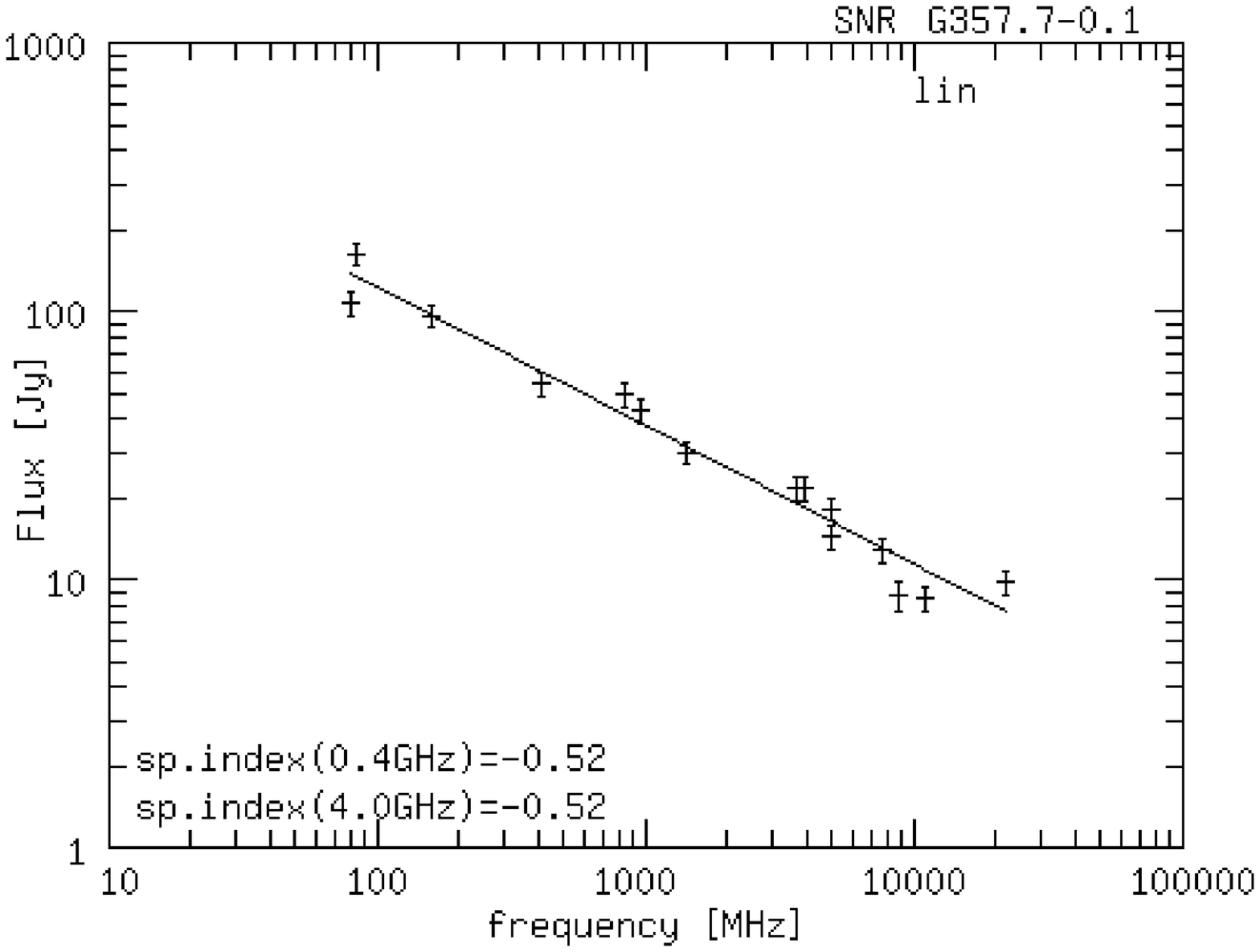,width=7.4cm,angle=0}}}\end{figure}
\begin{figure}\centerline{\vbox{\psfig{figure=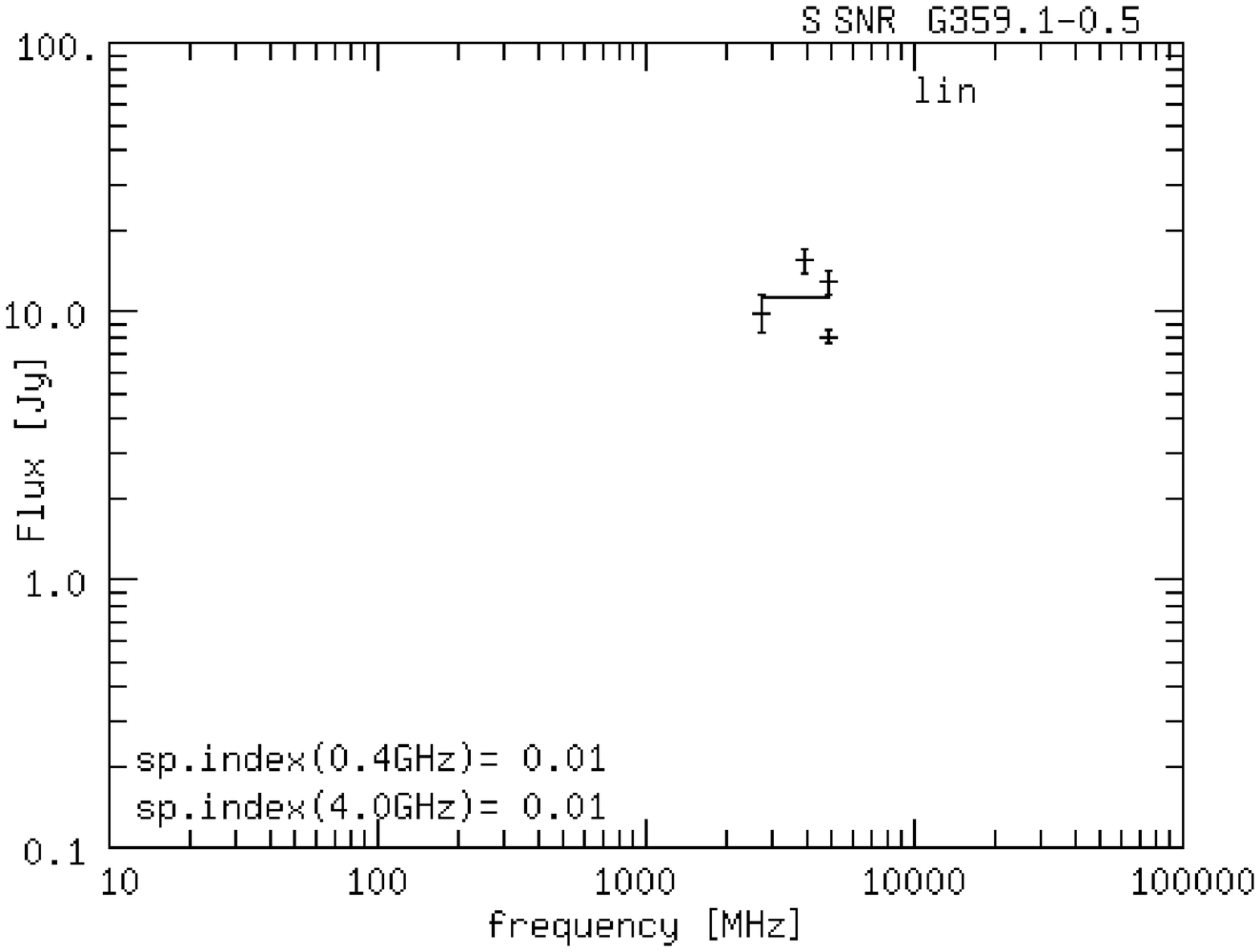,width=7.4cm,angle=0}}}\end{figure}
\end{document}